\documentclass[twoside]{report}
 \pdfoutput=1
\DeclareFontShape{OT1}{cmtt}{bx}{n}{<5><6><7><8><9><10><10.95><12><14.4><17.28><20.74><24.88>cmttb10}{}

\usepackage{atbegshi,picture}
\AtBeginShipout{\AtBeginShipoutUpperLeft{%
  \put(\dimexpr\paperwidth-1cm\relax,-1.5cm){\makebox[0pt][r]{\includegraphics[width=3cm]{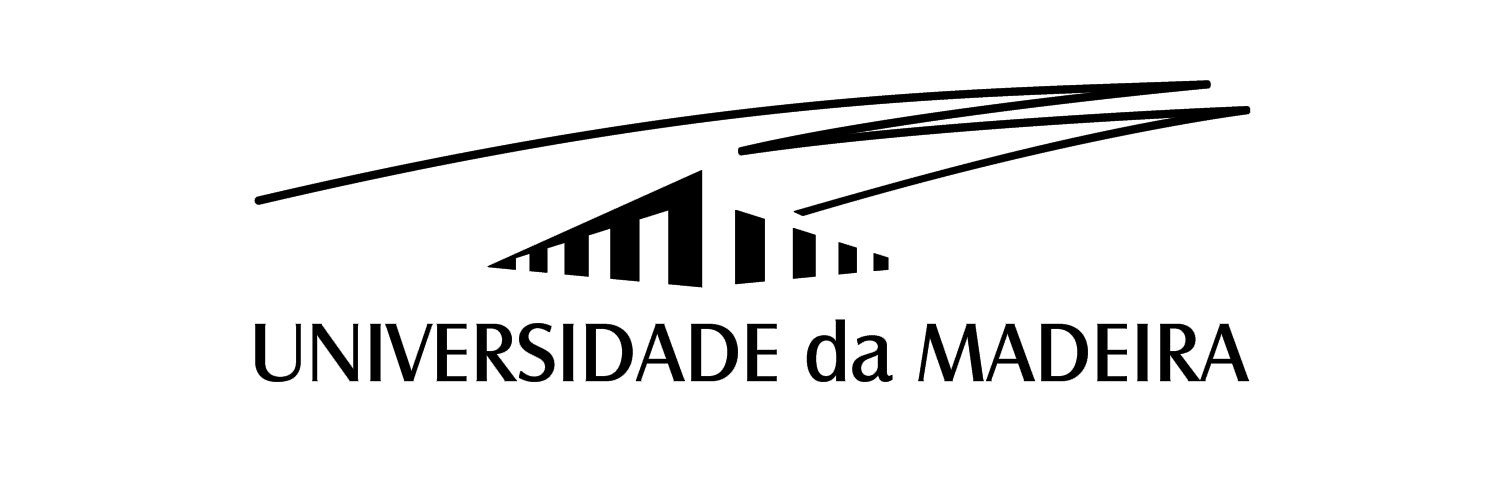}}}%
}}

\usepackage{float}
\usepackage{graphicx}
\usepackage{array}
\usepackage{multirow,array}
\usepackage{caption}
\usepackage{subcaption}
\usepackage {bsymb}
\usepackage{alltt}
\usepackage{paralist}
\usepackage{listings}
\usepackage{zed-csp}
\usepackage{fancyheadings} 
\newdimen\proofrulebreadth \proofrulebreadth=.05em
\newdimen\proofdotseparation \proofdotseparation=1.25ex
\newdimen\proofrulebaseline \proofrulebaseline=2ex
\newcount\proofdotnumber \proofdotnumber=3
\let\then\relax
\def\hfi{\hskip0pt plus.0001fil}
\mathchardef\squigto="3A3B
\newif\ifinsideprooftree\insideprooftreefalse
\newif\ifonleftofproofrule\onleftofproofrulefalse
\newif\ifproofdots\proofdotsfalse
\newif\ifdoubleproof\doubleprooffalse
\let\wereinproofbit\relax
\newdimen\shortenproofleft
\newdimen\shortenproofright
\newdimen\proofbelowshift
\newbox\proofabove
\newbox\proofbelow
\newbox\proofrulename
\def\shiftproofbelow{\let\next\relax\afterassignment\setshiftproofbelow\dimen0 }
\def\shiftproofbelowneg{\def\next{\multiply\dimen0 by-1 }%
\afterassignment\setshiftproofbelow\dimen0 }
\def\setshiftproofbelow{\next\proofbelowshift=\dimen0 }
\def\setproofrulebreadth{\proofrulebreadth}

\def\prooftree{
\ifnum  \lastpenalty=1
\then   \unpenalty
\else   \onleftofproofrulefalse
\fi
\ifonleftofproofrule
\else   \ifinsideprooftree
        \then   \hskip.5em plus1fil
        \fi
\fi
\bgroup
\setbox\proofbelow=\hbox{}\setbox\proofrulename=\hbox{}%
\let\justifies\proofover\let\leadsto\proofoverdots\let\Justifies\proofoverdbl
\let\using\proofusing\let\[\prooftree
\ifinsideprooftree\let\]\endprooftree\fi
\proofdotsfalse\doubleprooffalse
\let\thickness\setproofrulebreadth
\let\shiftright\shiftproofbelow \let\shift\shiftproofbelow
\let\shiftleft\shiftproofbelowneg
\let\ifwasinsideprooftree\ifinsideprooftree
\insideprooftreetrue
\setbox\proofabove=\hbox\bgroup$\displaystyle 
\let\wereinproofbit\prooftree
\shortenproofleft=0pt \shortenproofright=0pt \proofbelowshift=0pt
\onleftofproofruletrue\penalty1
}

\def\eproofbit{
\ifx    \wereinproofbit\prooftree
\then   \ifcase \lastpenalty
        \then   \shortenproofright=0pt  
        \or     \unpenalty\hfil         
        \or     \unpenalty\unskip       
        \else   \shortenproofright=0pt  
        \fi
\fi
\global\dimen0=\shortenproofleft
\global\dimen1=\shortenproofright
\global\dimen2=\proofrulebreadth
\global\dimen3=\proofbelowshift
\global\dimen4=\proofdotseparation
\global\count255=\proofdotnumber
$\egroup  
\shortenproofleft=\dimen0
\shortenproofright=\dimen1
\proofrulebreadth=\dimen2
\proofbelowshift=\dimen3
\proofdotseparation=\dimen4
\proofdotnumber=\count255
}

\def\proofover{
\eproofbit 
\setbox\proofbelow=\hbox\bgroup 
\let\wereinproofbit\proofover
$\displaystyle
}%
\def\proofoverdbl{
\eproofbit 
\doubleprooftrue
\setbox\proofbelow=\hbox\bgroup 
\let\wereinproofbit\proofoverdbl
$\displaystyle
}%
\def\proofoverdots{
\eproofbit 
\proofdotstrue
\setbox\proofbelow=\hbox\bgroup 
\let\wereinproofbit\proofoverdots
$\displaystyle
}%
\def\proofusing{
\eproofbit 
\setbox\proofrulename=\hbox\bgroup 
\let\wereinproofbit\proofusing
\kern0.3em$
}

\def\endprooftree{
\eproofbit 
  \dimen5 =0pt
\dimen0=\wd\proofabove \advance\dimen0-\shortenproofleft
\advance\dimen0-\shortenproofright
\dimen1=.5\dimen0 \advance\dimen1-.5\wd\proofbelow
\dimen4=\dimen1
\advance\dimen1\proofbelowshift \advance\dimen4-\proofbelowshift
\ifdim  \dimen1<0pt
\then   \advance\shortenproofleft\dimen1
        \advance\dimen0-\dimen1
        \dimen1=0pt
        \ifdim  \shortenproofleft<0pt
        \then   \setbox\proofabove=\hbox{%
                        \kern-\shortenproofleft\unhbox\proofabove}%
                \shortenproofleft=0pt
        \fi
\fi
\ifdim  \dimen4<0pt
\then   \advance\shortenproofright\dimen4
        \advance\dimen0-\dimen4
        \dimen4=0pt
\fi
\ifdim  \shortenproofright<\wd\proofrulename
\then   \shortenproofright=\wd\proofrulename
\fi
\dimen2=\shortenproofleft \advance\dimen2 by\dimen1
\dimen3=\shortenproofright\advance\dimen3 by\dimen4
\ifproofdots
\then
        \dimen6=\shortenproofleft \advance\dimen6 .5\dimen0
        \setbox1=\vbox to\proofdotseparation{\vss\hbox{$\cdot$}\vss}%
        \setbox0=\hbox{%
                \advance\dimen6-.5\wd1
                \kern\dimen6
                $\vcenter to\proofdotnumber\proofdotseparation
                        {\leaders\box1\vfill}$%
                \unhbox\proofrulename}%
\else   \dimen6=\fontdimen22\the\textfont2 
        \dimen7=\dimen6
        \advance\dimen6by.5\proofrulebreadth
        \advance\dimen7by-.5\proofrulebreadth
        \setbox0=\hbox{%
                \kern\shortenproofleft
                \ifdoubleproof
                \then   \hbox to\dimen0{%
                        $\mathsurround0pt\mathord=\mkern-6mu%
                        \cleaders\hbox{$\mkern-2mu=\mkern-2mu$}\hfill
                        \mkern-6mu\mathord=$}%
                \else   \vrule height\dimen6 depth-\dimen7 width\dimen0
                \fi
                \unhbox\proofrulename}%
        \ht0=\dimen6 \dp0=-\dimen7
\fi
\let\doll\relax
\ifwasinsideprooftree
\then   \let\VBOX\vbox
\else   \ifmmode\else$\let\doll=$\fi
        \let\VBOX\vcenter
\fi
\VBOX   {\baselineskip\proofrulebaseline \lineskip.2ex
        \expandafter\lineskiplimit\ifproofdots0ex\else-0.6ex\fi
        \hbox   spread\dimen5   {\hfi\unhbox\proofabove\hfi}%
        \hbox{\box0}%
        \hbox   {\kern\dimen2 \box\proofbelow}}\doll%
\global\dimen2=\dimen2
\global\dimen3=\dimen3
\egroup 
\ifonleftofproofrule
\then   \shortenproofleft=\dimen2
\fi
\shortenproofright=\dimen3
\onleftofproofrulefalse
\ifinsideprooftree
\then   \hskip.5em plus 1fil \penalty2
\fi
}

\usepackage{color}

\newcommand{\ebtojava}{{\it Event\-B\-2\-Java}}
\newcommand{\btojml}{{\it B\-2\-Jml}}
\newcommand{\ebtojml}{{\it Event\-B\-2\-Jml}}
\newcommand{\ebtodafny}{{\it Event\-B\-2\-Dafny}}
\newcommand{\jmle}{jmle}
\newcommand{\csharp}[1]{C{\scriptsize \raisebox{.6ex}{\#}}}
\newcommand{\specsharp}[1]{Spec{\scriptsize \raisebox{.6ex}{\#}}} 

\newcommand{\zkeyw}[1]{\texttt{#1}}

\newcommand{\bmethod}{B}
\newcommand{\bkeyw}[1]{\textsf{#1}}

\newcommand{\eb}{Event-B}
\newcommand{\ebkeyw}[1]{\textsf{#1}}
\newcommand{\ebtag}[1]{\textsf{\bf{#1}}}

\newcommand{\jml}{JML}
\newcommand{\jmlkeyw}[1]{\texttt{\bfseries #1}}
\newcommand{\jmlcode}[1]{\texttt{#1}}

\newcommand{\javakeyw}[1]{\texttt{\bfseries #1}}
\newcommand{\javacode}[1]{\texttt{#1}}
\newcommand{\javacomment}[1]{\textit{#1}}

\newcommand{\dafny}{Dafny}
\newcommand{\dafnykeyw}[1]{\texttt{\bfseries #1}}
\newcommand{\dafnycode}[1]{\texttt{#1}}

\def \BTOJMLN {\textsf{B2Jml}}
\newcommand{\BTOJML}[1]{\BTOJMLN\textsf{(}\ensuremath{{#1}}\textsf{)}}

\def \MODN {\textsf{MOD}}
\newcommand{\MOD}[1]{\MODN\textsf{(}\ensuremath{{#1}}\textsf{)}}

\def \TypeN {\textsf{TypeOf}}
\newcommand{\Type}[1]{\TypeN\textsf{(}\ensuremath{{#1}}\textsf{)}}

\def \EBTOJMLN {\textsf{EB2Jml}}
\newcommand{\EBTOJML}[1]{\EBTOJMLN\textsf{(}\ensuremath{{#1}}\textsf{)}}

\def \PREDN {\textsf{Pred}}
\newcommand{\PRED}[1]{\PREDN\textsf{(}\ensuremath{{#1}}\textsf{)}}

\def \EBTODAFNYN {\textsf{ToDafny}}

\def \DTypeN {\textsf{TypeOfToDafny}}
\newcommand{\DType}[1]{\DTypeN\textsf{(}\ensuremath{{#1}}\textsf{)}}

\def \DPREDN {\textsf{PredToDafny}}
\newcommand{\DPRED}[1]{\DPREDN\textsf{(}\ensuremath{{#1}}\textsf{)}}

\newcommand{\EBTODAFNY}[1]{\EBTODAFNYN\textsf{(}\ensuremath{{#1}}\textsf{)}}

\def \CtxN {\textsf{Ctx}}
\newcommand{\Ctx}[1]{\CtxN\textsf{(}\ensuremath{{#1}}\textsf{)}}
\def \InvConcreteN {\textsf{INV\_Concrete}}

\def \MRGN {\textsf{MRG}}

\def \GRDN {\textsf{GRD}}

\def \SIMN {\textsf{SIM}}

\def \InvAbstractN {\textsf{INV\_Abstract}}

\def \NATN {\textsf{NAT}}

\def \FISN {\textsf{FIS}}

\def \WFISN {\textsf{WFIS}}

\def \VARConvN {\textsf{VAR\_Conv}}

\def \VARAntN {\textsf{VAR\_Ant}}

\def \EBTOJAVAN {\textsf{EB2Java}}
\newcommand{\EBTOJAVA}[1]{\EBTOJAVAN\textsf{(}\ensuremath{{#1}}\textsf{)}}
\def \EBTOPROGN {\textsf{EB2Prog}}
\newcommand{\EBTOPROG}[1]{\EBTOPROGN\textsf{(}\ensuremath{{#1}}\textsf{)}}
\def \FreeN {\textsf{FreeVar}}
\newcommand{\Free}[1]{\FreeN\textsf{(}\ensuremath{{#1}}\textsf{)}}
\def \STATONEN {\textsf{Stat1}}
\newcommand{\STATONE}[1]{\STATONEN\textsf{(}\ensuremath{{#1}}\textsf{)}}
\def \STATTWON {\textsf{Stat2}}
\newcommand{\STATTWO}[1]{\STATTWON\textsf{(}\ensuremath{{#1}}\textsf{)}}
\def \ASGJAVAN {\textsf{AssgJava}}
\newcommand{\ASGJAVA}[2]{\ASGJAVAN\textsf{(}\ensuremath{{#1}},\ensuremath{{#2}}\textsf{)}}

\newcommand{\same}{same as JML}
\newcommand{\sameJ}{same as Java}
\newcommand{\noS}{not supported yet!}

\def \bsl {\symbol{92}}

\renewcommand{\chaptermark}[1]%
	{\markboth{\chaptername~\thechapter~--~#1}{}}

\renewcommand{\sectionmark}[1]%
	{\markright{\thesection\ #1}}

\newcommand{\dedication}[1]
   {\thispagestyle{empty}
     
   \begin{flushleft}\raggedleft #1\end{flushleft}
}

\begin{document}\sloppy

\begin{titlepage}

\begin{center}
{\bf\Large Universidade da Madeira} \\
\vspace{0.40cm}
{\bf\Large Centro de Ciencias Exactas e da Engenharia}

\vspace{0.80cm}

{\bf\Large PhD THESIS}

\vspace{0.60cm}

{\Large presented in fulfilment of the requirement for\\ the degree of Doctor of Philosophy}

\vspace{0.60cm}

{\Large Major: {\bf Software Engineering}}

\vspace{0.80cm}

{\bf\Large Code Generation for \eb}

\vspace{0.60cm}

{\Large presented by}

\vspace{0.40cm}

{\bf\Large V\'ICTOR ALFONSO RIVERA Z\'U\~NIGA}
\vspace{0.40cm}

{\Large supervised by}\\
\vspace{0.40cm}
{\bf\Large N{\'E}STOR CATA{\~N}O COLLAZOS}\\
\vspace{0.60cm}
{\Large June 2014}

\vspace{0.8cm}
{\bf \Large JURY}\\
\vspace{0.40cm}
{\Large
\begin{tabular}{lll}
Timothy Wahls & Dickinson College, USA &Reviewer   \\ 
Camilo Rueda & Javeriana University, Colombia & Reviewer  \\
Pedro Campos & University of Madeira, Portugal & Reviewer\\
Jos{\'e} Carmo & University of Madeira, Portugal & Rector \\
N{\'e}stor Cata{\~n}o & Madeira-ITI, Portugal & Supervisor\\

\end{tabular}
}
\end{center}
\end{titlepage}

\newpage
\thispagestyle{empty}

\mbox{}

\vspace{7cm}

\begin{flushright}

{\bf\LARGE Code Generation for \eb}

\vspace{0.1cm}

\rule{\linewidth}{0.15cm}%
\end{flushright}

\newpage
\dedication{To my parents and Laura Nogueira Mazaira.}
\newpage
\newpage
\thispagestyle{empty}

\begin{titlepage}

\begin{flushleft}
{\Large \bf
Acknowledgments 
}

\vspace{0.01cm}

\rule{\linewidth}{0.05cm}%

\end{flushleft}

{
The fulfilment of this thesis would not have been possible without the
contribution of N\'estor Cata\~no, my thesis supervisor. I want to
thank him for his guidance during all this process, his encouragement,
and dedication. Specially thanks for his support and
patience. I also want to thank Camilo Rueda and Tim Wahls for their
ideas, and discussions. Thanks to Madeira Interactive Technologies
Institute (M-ITI - Portugal) and {\it Departamento Administrativo de
Ciencia, Tecnolog\'ia e Innovaci\'on} (Colciencias - Colombia) for
funding my research over
these years. Many thanks, from the bottom of my heart, to my mother
Esperanza Z\'u\~niga (R.I.P.) and my father Luis Rivera for their love
and their enthusiasm that have carried me out this far.

I apologize to all those not mentioned that have helped me in one way
or another. Even though I have unconsciously omitted their names, I am
very grateful for the help they provided.

This work was supported by the Portuguese Foundation for
Science and Technology (FCT) grant PTDC/EIA-CCO/105034/2008 (FAVAS: A
FormAl Verification for real-time Systems), and Colciencias, The
Colombian Agency for Science and Technology Development.

}

\end{titlepage}

\newpage
\setcounter{page}{5}

\newpage
\thispagestyle{empty}
\mbox{}
\newpage

\tableofcontents
\listoffigures
\newpage
\chapter*{Abstract}
Stepwise refinement and Design-by-Contract are two formal approaches
for modelling systems. These approaches are widely used in the
development of systems. 
Both approaches have (dis-)advantages: in stepwise refinement a model
starts with an abstraction of the system and more details are added
through refinements. Each refinement must be provably consistent with the
previous one. Hence, reasoning about abstract models is
possible. A high level of expertise is necessary in
mathematics to have a good command of the underlying languages,
techniques and tools, making  this approach less
popular. Design-by-Contract, on the other hand, works on the program
rather
than the program model, so developers in the software industry are more
likely to have expertise in it. However, the benefit of reasoning over more
abstract models is lost.

A question arises: is it possible to combine both approaches in the
development of systems, providing the user with the benefits of both?
This thesis answers this question by translating the stepwise
refinement method with
\eb\ to Design-by-Contract with Java and \jml, so users can take full
advantage of both formal approaches without losing their benefits. This
thesis presents a set of syntactic rules that translates \eb\ 
to \jml-annotated Java code. It also
presents the implementation of the syntactic rules as the \ebtojava\
tool. We used \ebtojava\ to translate several \eb\ models. The tool
generated \jml-annotated Java code for all the considered \eb\ models
that serve as
final implementation. We also used \ebtojava\ for the development of
two software applications. Additionally, we compared \ebtojava\ against
two other tools that also generate Java code from \eb\ models. \ebtojava\
enables users
to start the software development process in \eb, where users can model the
system and prove its consistency, to then transition to \jml-annotated
Java code, where users can continue the development process.

\textbf{\textit{Key Words---}} Modelling system by stepwise refinement, \eb,
Design-by-Contract, Java, \jml, \ebtojava.
\chapter*{Resumo Portugu\^es}
{\it Stepwise Refinement} e {\it Design-by-Contract} s\~ao dois
m\'etodos formais. Estes m\'etodos muito
utilizados no desenvolvimento de sistemas. Ambos t\^em (des-)vantagens:
em {\it Stepwise Refinement} um modelo come\c{c}a com uma abstra\c{c}\~ao do sistema e
mais detalhes s\~ao adicionados atrav\'es de refinamentos. Cada
refinamento \'e provavelmente consistente com o anterior. Assim, o
racioc\'inio sobre modelos abstratos \'e poss\'ivel. Um alto
n\'ivel de conhecimento \'e necess\'ario em matem\'atica para ter um bom
dom\'inio da sintaxe da linguagem, t\'ecnicas e ferramentas  tornando esta
metodologia menos popular; {\it Design-by-Contract} trabalha no programa, e
n\~ao no modelo do programa de modo que os desenvolvedores na ind\'ustria
de software s\~ao mais predispostos a ter conhecimento sobre ele. No
entanto, os benef\'icios do racioc\'inio em rela\c{c}\~ao aos modelos mais
abstratos s\~ao perdidos.

Surge uma quest\~ao, \'e poss\'ivel combinar ambos m\'etodos no
desenvolvimento de sistemas, fornecendo ao utilizadores com os
beneficios dos dois? Esta tese responde essa quest\~ao,
traduzindo {\it Stepwise Refinement} com Event-B para o {\it Design-by-Contract}
com Java e JML assim \'e poss\'ivel tirar o m\'aximo proveito de ambos
m\'etodos formais. Apresenta-se aqui um conjunto de regras sint\'aticas
que traduz Event-B para o c\'odigo Java com anota\c{c}\~oes JML. Apresenta
tamb\'em a implementa\c{c}\~ao das regras sint\'aticas como a ferramenta
EventB2Java. Usamos EventB2Java para traduzir v\'arios modelos em
Event-B. A ferramenta gerou c\'odigo Java com anota\c{c}\~oes JML para todos
os modelos considerados que servem de implementa\c{c}\~ao final. Tamb\'em
usamos EventB2Java para o desenvolvimento de duas aplica\c{c}\~oes de
software. Al\'em disso, comparamos EventB2Java com duas outras
ferramentas que tamb\'em geram c\'odigo Java do Event-B. EventB2Java
permite que os utilizadores iniciem o desenvolvimento de software em
Event-B onde podem modelar o sistema, prov\'a-lo de forma consistente e
fazer a transi\c{c}\~ao para c\'odigo Java com anota\c{c}\~oes JML utilizando a
nossa ferramenta, onde os utilizadores podem continuar o
desenvolvimento.

\textbf{\textit{Palavras-chave---}} Refinement Calculus, \eb,
Design-by-Contract, Java, \jml, \ebtojava.
\newpage
\thispagestyle{empty}
\mbox{}
\newpage
\chapter{Introduction}
\label{chapter:introduction}
Information systems have become essential to people. As an example,
people use web systems to search for things to buy, and use bank
transaction systems to make payments, and even trust their lives
to critical software systems, such as control software used by
airplanes. Often, people are unaware of the consequences that
malfunctioning software can  have on their lives. Hence, software must be
built in a correct fashion.

Concepts such as robustness and reliability are important in software
today: people expect software systems
to work as expected. Several approaches for software reliability and robustness
exist~\cite{Duvall07,Peled01}, each can be supplemented with
Software Testing~\cite{Beizer90,Beizer95}. However,
testing techniques alone are not adequate to ensure the correctness of
 critical software (or any other software). As Edsger Dijkstra said:
``Testing shows the presence, not the absence of bugs''. Tests can
only show the situations where a system will fail, but cannot say
anything about the behaviour of the system outside the testing
scenarios. While it is true that validating the code against certain
properties, as in testing, makes software testing popular and 
important, it is also true that testing does not validate the system
 as a whole: system behaviours beyond the ones considered by the tests can
 produce casualties. 

A way to ensure the correctness of critical software is using formal
methods~\cite{formalmethods}, that are mathematically based on rigorous
techniques for the specification of systems (well-formed statements in a
mathematical logic) based on requirements, the verification (rigorous
deductions in that logic), and the implementation of software (and
hardware) systems.  Formal methods enable users to express properties over
the system that must be proven true for all possible inputs. Formal
methods are concerned with the
system as a whole, proving that each component of the system 
interacts with each other in a correct way. It seems right then to
think that formal methods are the key for
the construction of correct software. Simulating can
be seen as a complement of formal methods: one can model a system using
formal methods ensuring the correctness of the system w.r.t. some
requirements, and then simulating the system by running the model to be
sure that what one proved mathematically was indeed what one
wanted. 

Refinement Calculus techniques
\cite{Dijkstra75,Morris87,Morgan88,BackW89} are techniques to
implement software systems based on formal methods. In
Refinement Calculus, users write an
abstract model of a program and define properties over it, and then
transform the model into an
implementation via a series of refinement steps. Each refinement
adds more detail and properties to the system. The behaviour of each
refinement is
provably consistent with the behaviour of the previous step. The final
refinement is the actual implementation of the system modelled. This
technique is known as correctness-by-construction, as it allows to
reasoning about the model ensuring it is correct by the preservation of
the given properties.

\bmethod\ method \cite{TheBBook} is an example of a formal technique based on
Refinement Calculus. It is a method for specifying, designing and coding
software systems introduced by J.-R. Abrial. Atelier \bmethod\
\cite{atelierb} is an IDE that enables users to work with \bmethod\
method. The correctness of models in \bmethod\
is achieved by discharging proof obligations. Proof
obligations are correctness conditions on the model of the system
that need to be proven
satisfied. Several provers exist helping the process of discharging proof
obligations. For instance, Atelier B comes with its own automatic
prover.

Another example of a formal technique based on Refinement Calculus is
\eb\ \cite{EB:Book}. \eb\ models are complete developments of discrete
transition systems. \eb\ was also introduced by J.-R. Abrial and it is
derived from the \bmethod\ method. Rodin \cite{rodin} is an Eclipse IDE
that provides support for \eb. The correctness of models in \eb\ is
also achieved by discharging proof obligations as in \bmethod. Rodin
comes with its own automatic prover. Other provers could be used in the
process of discharging proof obligations. For example, \dafny\
\cite{dafny2,dafny} is an imperative object-based language with built-in
specification constructs that comes with automatic provers, e.g. Z3
\cite{Z3:overview}. One
could express proof obligations in the input language of \dafny\ and
then use its automatic provers to discharge the proof obligations.

Limitations of Refinement Calculus techniques stem from the level of
expertise required in mathematics, using the underlying languages and
tools. As J. Bowen
and V. Stavridou describe in
\cite{Bowen93safety-criticalsystems}, one of the principal issues of
the wide adoption of formal methods is that they require mathematical
expertise. Software developers lack the expertise required (as
states by J.-R. Abrial in \cite{EB:Book}), since they have little or
no mathematical background. This issue could be avoided with 
the use of techniques for the verification of formal systems such as
Model Checking
\cite{modelchecking}. Model Checking is a formal verification method that
automatically checks whether a model meets a given
specification. However, this technique has another issue. Model
Checking can only handle finite systems, and  suffers of
the state-explosion problem. Another limitation of refinement calculus is
to find a good system
structuring for the development of the system. For
instance, one might include a lot of details in an abstract model of
the system which will probably be difficult to prove correct. Similarly if a
refinement is very concrete. 

Another technique to implement software systems based on formal methods
is Design-by-Contract (DbC) \cite{Meyer92}. The general idea about
Design-by-Contract is that a software contract exists between a method
and a client code. The client code must ensure that the pre-condition
of the method is satisfied when it is called, and the method
implementation must ensure that the post-condition holds when the
method terminates (assuming that the pre-condition was
satisfied). Design-by-Contract techniques work on the program rather
than on the
program model, so developers in the software industry
are more likely to have expertise in this technique. Java
Modeling Language (JML)
\cite{JML:Preliminary:Design:2006,Burdy-etal05,JMLRefManual} is an
example of this technique. \jml\ is based
on Design-by-Contract in which code is verified against a formal
specification. Like Refinement Calculus, DbC techniques also have some limitations.
Design-by-Contract does not have
the mathematical based rigorousness as Refinement Calculus does, so the ability to reason
over more abstract models is lost. 

The main goal of this thesis is to bridge the gap between Refinement
Calculus with \eb\ and Design-by-Contract with Java and \jml. Thus,
allowing software developers to benefit from both formal methods in the
software development of applications. This thesis presents a code
generator for \eb\ that generates \jml-annotated Java code.  This
allows the development process of a software application to start
with a formal model in Event-B. The user decides the level of model
abstraction and defines the properties over the model. These
properties can be proven correct by discharging the proof obligations
within the Event-B tool (Rodin). Next, the user translates the model
to a JML-annotated Java code using our code generator. Once the code
is generated, the user can continue the system development of the
application in Java.


The work presented in this thesis allows users of different levels of expertise
to work together in
the development of systems. For instance, a user who is an expert in the
underlying notation of \eb\ and an expert in mathematics can work at early stages
of the system development, then transition to \jml-annotated Java
code of the \eb\ model, using our code generator, where an expert
software developer can continue with the final implementation of the
system. 

The code generator for \eb\ 
  is formally defined by means of translation rules. This thesis
  presents the translation rules and the implementation of those rules as the
  \ebtojava\ tool. \ebtojava\ 
  automates the process of code generation from \eb\ models. The code
  generator, in addition to generating Java
  code of  \eb\ models, also generates \jml\ specifications. Thus,
  users can customise the Java code and verify it against the \jml\
  specifications to make sure the customisation does not invalidate 
  the initial \eb\ model.

We have validated \ebtojava\ by using the tool to generate \jml-annotated
Java code for an ample set of \eb\ models. This thesis presents these
\eb\ models and the code generated by \ebtojava,
and also presents a benchmark in which \ebtojava\ is compared against
other tools for
generating Java code from \eb\ models. We used \ebtojava\ in the
development of two case studies: the
  first case study is on the development of an Android \cite{Android}
  application that follows the MVC
  (Model-View-Controller) design pattern  \cite{Patterns:Gamma:95};
  the second case study is on testing an \eb\ model by translating it
  to Java and performing Java Unit (JUnit) testing of the generated
  Java code.

This thesis also introduces a tool that generates \dafny\ code from
\eb\ proof obligations (PO), to assist users in the process of proving the
system correct.  The translation of \eb\ POs to
\dafny\ is defined by means of translation rules. Rules were implemented as
the \ebtodafny\ tool. \ebtodafny\ helps users to discharge \eb\ POs
 by translating them into the input language of \dafny\ where
the user can use \dafny's automatic provers (e.g. Simplify
\cite{simplify}, Zap \cite{zap}) to prove the PO. In
this sense, \ebtodafny\ equips Rodin with other theorem provers
for discharging proof obligations, e.g. Simplify or Zap theorem
provers. It also gives the opportunity to developers to reasoning
about the \eb\ model using the feedback that \dafny\ gives rather than
discharging the proof obligation in Rodin, that requires more
mathematical expertise that the developer might not have.

\paragraph{Thesis Summary: } The development of the work presented in this thesis
started by proposing a translation from \bmethod\ machines to \jml\ specifications
(described in Chapter \ref{chapter:b2jml}). We saw the \bmethod\
method as a
good starting candidate for the development of systems, since systems
are first modelled in an abstract way. Next, the model is proven to
satisfy certain safety and security properties, and then transformed
to code via a series of property preserving refinement steps. We
proposed the \btojml\ tool
that generates \jml\ specifications from \bmethod, where users can
manually write Java code. Then we realised that the \eb\ method (an evolution of
the \bmethod\ method) is a better starting candidate for the
development of systems, so we proposed a
translation from \eb\ machines to \jml\ specifications. Chapter
\ref{chapter:eb2jml} discusses how the \eb\ method is better than
\bmethod, the chapter also discusses the translation and the
implementation of the \ebtojml\ tool. Once we
developed this tool (\ebtojml),  we used it to generate \jml\
specifications from the
\eb\ model MIO, a model for a transportation system. Then we manually
generated Java code for those
specifications. We then realised that it is more useful to have a tool
that automatically generates Java code and we saw the importance of
embedding \jml\ specifications into the code since users might want to customise the
generated code without invalidating the initial model. We proposed and
implemented (discussed it in Chapter \ref{chapter:eb2java}) a
translation from \eb\ models to \jml-annotated Java code. We
implemented it as the \ebtojava\ tool. Having the \jml\ specifications
embedded into the Java code also gives an insight into the documentation of
the code that can be read easily. In modelling a system in \eb, one
needs to prove the correctness of the model. A series of proof
obligations are generated and needed to be discharged to gain
confidence of the model. Discharging proof obligation can be a
difficult task, so we proposed a translation from \eb\ proof obligations
to the input language of \dafny, thus users can use \dafny's
provers. Our intention was to provide tools to help users in the
process of proving an \eb\ model correct. Chapter
\ref{chapter:eb2dafny} describes this translation and presents the
\ebtodafny\ tool that automates the process of translation.

\section{Thesis overview} 

\paragraph{Chapter~\S\ref{chapter:introduction}. Introduction } This
chapter describes the
problem addressed by this thesis and describes the importance of using
more-formal methods to build systems

\paragraph{Chapter~\S\ref{chapter:background}. Background } This
chapter provides
the background knowledge required to understand the work done in this
thesis.

\paragraph{Chapter~\S\ref{chapter:b2jml}. Translating \bmethod\
  Machines to \jml\ Specifications. } This thesis work
started with the idea of generating \jml\ specifications from
\bmethod. This chapter presents the work based on this initial idea. The translation is defined
using
syntactic rules and it is implemented as the \btojml\ tool, and 
 integrated into the ABTools \cite{Boulanger2003} (an open source
environment for developing \bmethod). \btojml\
enables users to use \bmethod's strong
support for model verification during early stages of software
development to generate a fully verified model of an application, and
then transition to \jml\ specifications to simplify the task of
generating a Java
implementation and to take advantage of \jml\ (semi-) automatic tools
such as runtime or static assertion checkers. 

\paragraph{Contributions. } The main contributions of this chapter are
\begin{inparaenum}[\itshape i\upshape)]
\item the definition of a set of rewriting rules to translate \bmethod\
  to \jml, and
\item the implementation of the rules as the \btojml\ tool.
\end{inparaenum}
Work done in this chapter has been published in
\cite{conference:B2Jml:12,e-eb:to:jml-java}. I participated in the
design and testing of the syntactic rules and in the design of
\btojml\ and its integration to the ABTools suite.

\paragraph{Chapter~\S\ref{chapter:eb2jml}. Translating \eb\ Machines
  to \jml\ Specifications. } This chapter presents a translation from
\eb\ to \jml. This work
goes in the same direction as the work presented in Chapter
\ref{chapter:b2jml} as they both generate \jml\ specifications from
a formal model. I decided to change the initial formal method from
\bmethod\ to \eb\ mainly because the \bmethod\
method is devoted to the development of correctness-by-construction
software, while the purpose of \eb\ is to model full systems
(including hardware, software and environment of operation). Chapter
\ref{chapter:eb2jml} justifies the decision for
translating from
\eb\ instead of \bmethod\ into \jml, it also presents  a set of syntactic rules to generate
\jml\
specifications from \eb. It presents the
implementation of the rules as the \ebtojml\ tool. \ebtojml\ is implemented as a
 Rodin \cite{rodin} plug-in. This chapter also shows the application
 of the tool to a model in \eb.


\paragraph{Contributions. } The main contributions of this chapter are the definition of
 translation rules from \eb\ models to \jml\ specifications, and the
implementation of the translation rules as the \ebtojml\ tool. This
translation tool allows experts in \eb\ to work together with software
developers, usually experts in main stream programming languages like
Java. For instance, an expert in \eb\ notation starts
the development of a system, then uses \ebtojml\ to transition to
\jml\ where a software developer writes Java code from the \jml\ specifications. I
participated in the definition of the syntactic rules of the translation and I
fully implemented the \ebtojml\ tool. 

\paragraph{Chapter~\S\ref{chapter:eb2java}. Translating \eb\ Machines
  to \jml-annotated Java Code. } We decided to extend the
work described in Chapter \ref{chapter:eb2jml} to not just generate
\jml\ specifications, but also to generate Java code from \eb\
models. This chapter
presents the core work of this Ph.D. thesis. It describes the
translation of \eb\ models to \jml-annotated Java code. The
translation is achieved through syntactic rules and it is implemented as the
\ebtojava\ tool. 

\paragraph{Contributions. }  The main contributions of this chapter are
\begin{inparaenum}[\itshape i\upshape)]
\item the definition of a set of translation rules to translate \eb\
  models to \jml-annotated Java code, and
\item the implementation of the rules as the \ebtojava\ tool.
\end{inparaenum}
Users can benefit from the work accomplished in this chapter for the
following reasons 
\begin{itemize}
\item the \ebtojava\ tool generates both sequential and multithreaded
  Java code,
\item \ebtojava\ can be applied to both abstract and refinement \eb\
  models, and
\item the generation of \jml\ specifications enable users to
  write customised code that replaces the code
  generated by \ebtojava, and then to use existing \jml\ tools 
  \cite{jmle-jcard:09,jmle} to verify that the customised code
  is correct.
\end{itemize}

Work done in this chapter has been published in
\cite{conference:EB2Java:14} and in a book chapter
\cite{e-eb:to:jml-java}, and submitted to a journal paper
\cite{EventB2Java2014}. My participation was to define the translation rules
for the translation of \eb\ models to \jml-annotated Java code, and to
fully implement the \ebtojava\ tool.

\paragraph{Chapter~\S\ref{chapter:eb2dafny}. Translating \eb\ Machine
  POs to \dafny.}  This chapter presents a translation
of \eb\ proof obligations into the input language of \dafny\ by means of
syntactic rules. \dafny\
\cite{dafny2,dafny} is an imperative object-based language with built-in
specification constructs. The rules were implemented as the \ebtodafny\ Rodin
plug-in. To prove an \eb\ model is consistent, it is necessary to discharge a
series of proof obligations. Typically, proof obligations are
automatically discharged by Rodin provers. However, there are some
proofs that need a user's assistance to be discharged. \ebtodafny\ assists
users in the process of discharging proof obligations by translating
them into \dafny. \ebtodafny\ generates \dafny\ code that is
correct if and only if the \eb\ refinement-based proof obligations
hold.

\paragraph{Contributions. } The main contributions of this chapter are
the definition of translation rules from \eb\ proof obligations to \dafny\ and
the implementation of the translation rules as the \ebtodafny\ tool. Work
done in this chapter has been published in  \cite{conference:TOPI:12}. I
participated in the definition of the translation rules from \eb\ proof
obligations to \dafny\ programming language and in the
implementation of \ebtodafny.

\paragraph{Chapter~\S\ref{chapter:case-studies}. Case Studies. } 
We have validated the implementation of \ebtojava\ by applying it
to several \eb\ models. This chapter presents those \eb\ models
and the \jml-annotated Java code generated by \ebtojava. This chapter
also presents two case studies using \ebtojava: 
\begin{inparaenum}[\itshape 1\upshape)]
\item the first case study is on the
development of an Android \cite{Android} application. This development
demonstrates how \ebtojava\ can be used as part of a software
development methodology to generate the functionality (the Model)
of an Android application that is organised following the MVC (Model-
View-Controller) design pattern \cite{Patterns:Gamma:95};
\item the second
case study
is on testing the behaviour of the Tokeneer
\cite{tok} \eb\ model, a security-critical access control system. This
development demonstrates how
\ebtojava\ and Java Unit (JUnit) testing \cite{Link2003} can be used
to refine (improve) an \eb\ model to conform to an existing System
Test Specification (STS) document.
\end{inparaenum}
This chapter also presents a
benchmark that compares the \ebtojava\ tool with
existing tools for generating Java code from \eb\ models. We compared
 \ebtojava\ against EB2J \cite{Mery:2011} and Code
Generation \cite{CodeGen10}  tools for nine \eb\ models and six
comparison criteria.

\paragraph{Contributions. }  The main contributions of this chapter
are
\begin{inparaenum}[\itshape i\upshape)]
\item the presentation of two case studies using the \ebtojava\ tool, and
\item the presentation of a benchmark comparing \ebtojava\ against two
  existing tools that generate Java code from \eb\ models.
\end{inparaenum}
Work done in this chapter has been published in
\cite{EventB2Java2014,e-eb:to:jml-java}. My participation on this work
was:
regarding the first case study, I modelled the system in \eb\ (and
discharged all proof obligations). The system is an extension of an
existing \eb\ model of a Social Network. I also implemented the
Controller of the system in Java, and implemented the View using
Android API. Regarding the second case study, I participated 
in modelling the \eb\ model for Tokeneer and discharging the proof
obligations of the model. I implemented the Java Unit test cases. I
participated in the definition of the criteria for the benchmark, and
I undertook the comparison of the existing tools for generating Java
code from \eb\ models against \ebtojava.
\chapter{Background}
\label{chapter:background}

\section{The \bmethod\ method}
\label{background:bmethod}
The \bmethod\ method~\cite{TheBBook} is a strategy for software
development in which an abstract model of a system is transformed into
an implementation via a series of steps that progressively concretise
the abstract model. These steps or stages are referred to as
refinement steps, and a model $M_{i+1}$ of a system at stage $i + 1$
is said to \textit{refine}  the model $M_i$ at stage $i$. Each
refinement step adds more details to the system. The behaviour of each
refinement has to be provably consistent with the behaviour of the
model in the previous step, keeping a palpable behavioural relation
with its abstraction. This relation is
modelled through a ``gluing invariant'' property, that relates the
states between the concrete and abstract models. Refinement steps
generate proof obligations to guarantee that the system works
correctly. Roughly speaking, a refinement model should be such that it
can replace the refined model without the user noticing any
change.


\bmethod\ models are called machines, and are composed of (1) a static part:
variables, constants, parameters and invariants; and, (2) a dynamic part:
operations, that describe how the system evolves. \bmethod\ machines
use predicate calculus (essentially predicate logic and set theory) to
model properties. Machine operations are defined using various forms
of substitutions. Figure \ref{fig:background:bmethod:substitutions}
shows the syntax and the semantics of substitutions in B (taken from
\cite{TheBBook}). The following explains these substitutions (more
detailed information can be found in \cite{conference:B2Jml:12}):

In Figure~\ref{fig:background:bmethod:substitutions}, $P$,
$Q$, and $R$ are predicates for which the semantics are to be valid,
and $S$ and $T$ are substitutions of variables by expressions. $[S]R$,
with $S$ equals $x := E$, denotes the predicate resulting from the
substitution of any free occurrence of variable $x$ in $R$ by
expression $E$.

A {\it preconditioned} substitution $P | S$ denotes the substitution
$S$ under the operation pre-condition $P$. Hence, the correct
behaviour of the substitution $S$ is only ensured when it is activated
in a state in which $P$ holds. When $P$ does not hold, $P | S$ is
not guaranteed to verify any predicate $R$ and a crash of the system
occurs. A {\it guarded} substitution $P\Longrightarrow{S}$ executes a
substitution $S$ under the assumption $P$, hence if $P$ does not hold,
the substitution is able to establish any predicate $R$. 

A {\it bounded choice} substitution $S [] T$ non-deterministically
implements a substitution between either of $S$ or $T$. The semantics
of a bounded choice substitution ensures that whichever of the two
substitutions is implemented, it must satisfy $R$.  

An {\it unbounded choice} substitution $\forall x \qdot S$ generalises a bounded choice
substitution for any substitution
$S$. Figure~\ref{fig:background:bmethod:substitutions}
presents a particular unbounded substitution that further requires $x$
to make predicate $P$ true. $P$ and $S$ both depend on $x$ and the
machine variables. {\it Guarded bounded} substitutions combine {\it bounded
choice} and {\it guarded } substitutions. {\it Var~ Choice} is a
syntactic extension to the {\it bounded choice} substitution.

\begin{figure}[htp]
  {\small
    \[
    \begin{array}{|l|l|l|l|}
      \hline 
      {\bf Substitution} & {\bf Syntax} & {\bf Definition}  & {\bf Semantics} \\  \hline 
      
      {\it Preconditioned} & 
      \begin{array}{l}
        \bkeyw{PRE } P\\
        \bkeyw{THEN } S\\
        \bkeyw{END} 
      \end{array} & 
      P | S & 
      \begin{array}{l}
        [P | S]R\\
        \Leftrightarrow \\
        P \wedge [S]R
      \end{array}
      \\ \hline
      
      {\it Guarded} &
      \begin{array}{l}
        \bkeyw{SELECT } P \\
        \bkeyw{THEN } S\\ 
        \bkeyw{END} 
      \end{array} &
      P\:\Longrightarrow\:S &
      \begin{array}{l}
        [P  \Longrightarrow S](R) \\
        \Leftrightarrow  \\
        (P \Rightarrow [S]R)
      \end{array}
      \\ \hline
      
      {\it Bounded~ Choice} & 
      \begin{array}{l}
        \bkeyw{CHOICE } S \\
        \bkeyw{OR } T \\
        \bkeyw{END}
      \end{array} & 
      S [] T & 
      \begin{array}{l}
        (S [] T)(R) \\
        \Leftrightarrow  \\
        ([S]R \wedge [T]R)
      \end{array}
      \\ \hline
      
      {\it Unbounded~ Choice} &
      \begin{array}{l}
        \bkeyw{ANY } x \\
        \bkeyw{WHERE } P\\ 
        \bkeyw{THEN } S \\
        \bkeyw{END}
      \end{array} &
      \begin{array}{l}
      \forall x \qdot \\
      ~(P\:\Longrightarrow\:S) 
      \end{array}&
      \forall x \qdot P \Longrightarrow [S]R
      \\ \hline
      
      {\it Guarded~ Bounded}  &
      \begin{array}{l}
        \bkeyw{SELECT } P \\
        \bkeyw{THEN } S \\
        \bkeyw{WHEN } Q \\
        \bkeyw{THEN } T  \\
        \bkeyw{END}
      \end{array} &
      \begin{array}{l}
        \textsf{CHOICE}\\
        ~~P \Longrightarrow S \\
        \textsf{OR } \\
        ~~Q \Longrightarrow T\\
        \textsf{END}
      \end{array}  &
      \begin{array}{l}
        (P\:\Longrightarrow\:S\;[]\;Q\:\\
        ~~~~\Longrightarrow\:T) (R) \\
        \Leftrightarrow \\
        (P \Rightarrow [S]R)\:\wedge\:\\
        ~~~~(Q \Rightarrow [T]R)
      \end{array}
      \\ \hline

       {\it Var~ Choice} &
      \begin{array}{l}
        \bkeyw{VAR } x \\
        \bkeyw{IN } S \\
        \bkeyw{END}
      \end{array} &
      \begin{array}{l}
      \forall x \qdot \\
      ~(true\:\Longrightarrow\:S) 
      \end{array}&
      \forall x \qdot true \Longrightarrow [S]R
      \\ \hline
    \end{array}
    \]
  }
  \caption{Substitutions in B (taken from \cite{TheBBook}).}
  \label{fig:background:bmethod:substitutions}
\end{figure}

\subsection{An Example in \bmethod}
\label{b:example}
We presented a B \bmethod\ model of a social
networking site taken from~\cite{matelas:2010} that models social
network content, social network friendship relations, and privacy on
contents. Figure \ref{fig:b:machine} presents an excerpt of
the \bmethod\ model. Machine
$SOCIAL\_\-NETWORK$ declares two sets, \bkeyw{PERSON} and
\bkeyw{RAWCONTENT}, representing the set of all possible persons and
the set of all possible content (text, video, photographs, etc.) in a
social network respectively.  Variables $person$
and $rawcontent$ are
the sets of all persons and content that are actually in the network,
and $content$ is a relation mapping people to their own content.

      

      
      

\begin{figure}
  \centering
  {\small
    \[
    \begin{array}{l}
      \bkeyw{MACHINE}~SOCIAL\_NETWORK\\
      ~\bkeyw{SETS}~~~\bkeyw{PERSON};~\bkeyw{RAWCONTENT}\\
      ~\bkeyw{VARIABLES} ~~person,rawcontent,content\\
      ~\bkeyw{INVARIANT}\\
      ~~~~~persons \subseteq \bkeyw{PERSON}~ \wedge~~~rawcontent \subseteq \ebkeyw{RAWCONTENT}~ \wedge\\ 
      ~~~~~content \in person \rel rawcontent ~\wedge\\
      ~~~~~dom(content) = person ~\wedge~~~ran(content) = rawcontent\\ 
      ~\bkeyw{INITIALISATION} \\
      ~~~~~person := \emptyset \parallel~~~rawcontent := \emptyset \parallel
      ~~~content := \emptyset \\
      ~\bkeyw{OPERATIONS} \\
      ~~~~~transmit\_rc ( ow , rc, pe ) \hat{\;=} \\
      ~~~~~~~~~\bkeyw{PRE} \\
      ~~~~~~~~~~~~~rc \in rawcontent ~\wedge~~~ow \in person ~ \wedge \\
      ~~~~~~~~~~~~~pe \in person ~ \wedge ~~~ow \neq pe ~\wedge \\
      ~~~~~~~~~~~~~pe \mapsto rc \not\in content ~~~\bkeyw{THEN} ~~~\bkeyw{ANY} prs\\
      ~~~~~~~~~~~~~\bkeyw{WHERE}~~~~prs \subseteq person \\
      ~~~~~~~~~~~~~\bkeyw{THEN} \\
      ~~~~~~~~~~~~~~~~~content := content \bunion \{pe \mapsto rc\} \bunion prs \cprod \{rc\} \\
      ~~~~~~~~~~~~~\bkeyw{END} \\
      ~~~~~\bkeyw{END} \\
      ~\ebkeyw{END} \\
    \end{array}
    \]
  }
  \caption{A \bmethod-machine for a social networking site.}
  \label{fig:b:machine}
\end{figure}

A common operation in social networking sites is sharing content with
people in the social network. The \bmethod\ example models
this by transmitting raw content to a set of persons in the social
network (see operation $transmit\_rc$). The operation 
publishes a raw content $rc$
(e.g. a photo) from the page of $ow$ (i.e. the owner of
$rc$) on the page of $pe$. If
$transmit\_rc$ is invoked when its pre-condition (following
\bkeyw{PRE}) is true, the meaning of the operation is the meaning of
its substitution (the code following \bkeyw{THEN}).  The operation is
not guaranteed to achieve any result if invoked when its pre-condition
is false.

In the definition of $transmit\_rc$,
$pe \mapsto rc$ represents the pair of elements 
$(pe,rc)$, so that the content $rc$ is explicitly
transmitted to person $pe$.
The construct \bkeyw{ANY} models unbounded choice substitution; it gives
the implementer the opportunity to choose any value
for the bound variable $prs$ that satisfies the 
\bkeyw{WHERE} condition $prs \subseteq person$.
This gives a refining or implementation machine the flexibility to 
additionally transmit the content $rc$ to all of an as yet 
unspecified set of people.

\subsection{Tool support for B}
There are two tools that enable users to work with the B method. The
B-Toolkit \cite{btoolkit} developed by B-Core provides a suite of
tools to
support formal development of software systems using B method. Among
the tools in the B-Toolkit there is a tool for the specification, design, and code
configuration of B models. It also contains a syntax and type checkers,
and a specification animator. The B-Toolkit is available but it is not
actively maintained. Another tool that enables users to work with the
B method is Atelier B \cite{atelierb} developed by the ClearSy company. It comes
with an interactive mode that uses a graphical interface based on
windows and buttons, with an editor for writing abstract and
refinement machines, with a proof obligation generator, and with its
own automatic prover.

\section{The \eb\ Method}
\label{background:eb}

\eb~\cite{EB:Book} is another formal method for modelling complete
developments of
discrete transition systems. \eb\ was introduced by J-R. Abrial, and
is derived from the \bmethod\ method. Unlike in \bmethod\ models, the
static part of \eb\ models is separated from the dynamic part, and is
referred to as ``contexts''. Thus, \eb\ models are composed of machines
(the dynamic part. e.g. variables, invariants, events), and contexts
(the static part. e.g. carrier sets, constants). Three basic
relationships between machines and contexts are used to structure a
model:
\begin{itemize}
\item A machine \ebtag{sees} a context.
\item A machine can \ebtag{refine} another machine. 
\item A context can \ebtag{extend} another context. 
\end{itemize}

Figure~\ref{fig:background:eb:ebmachine} shows a general
structure of an \eb\ machine. It contains a list of machines that this
machine \ebtag{refine}s, and a list of context that it \ebtag{see}s. It also
contains a list of \ebtag{variables} used in the machine, a list of
\ebtag{invariants} restricting the possible values the variables can
take, and a \ebtag{variant} that is a numeric expression. The purpose
of the \ebtag{variant} is to ensure that certain events, called
convergent events, do not monopolize the system. This is done by
decreasing the \ebtag{variant} when a convergent event is triggered,
and when the value of the \ebtag{variant} is negative, these events
are not allowed to be triggered. The \ebtag{events} of a machine
determine the way the system evolves. It does so via a series of
substitutions of variables whenever an event is
triggered. \ebtag{events} contain a clause \ebtag{status} that defines
the event as \ebtag{ordinary}, \ebtag{convergent} (the event has to
decrease the \ebtag{variant}), or \ebtag{anticipated} (the event must
not increase the \ebtag{variant} and can only be refined by
\ebtag{convergent} events). It also contains a list of local
variables (that can be seen as parameters of the event) under the
clause \ebtag{any}. Events contain a guard (under clause \ebtag{where}) that needs
to be true in order for the event to be triggered. If the guard is
true, the event might perform its actions. Actions are under the
clause \ebtag{then} and they define how the system evolves by means of
substitutions. In \eb\ there are two kind of substitutions:
deterministic assignment, that takes the form of
${<}variable\_\-identifier{>} ~:=~{<}expression{>}$; and
non-deterministic assignment, that takes the form of
${<}variable\_\-identifier\_\-list{>}~ :|~
{<}before\_\-after\_\-predicate{>}$. Here, the
$before\_\-after\_\-predicate$ is the relationship that exists between
the value of a variable {\it just  before} and {\it just after} the
assignment. It may contain machine variables. Non-deterministic
assignments generalise deterministic
assignments. For example, $v := v + w$ can be expressed as 
$v :| v' = v + w$, where $v'$ is the value of $v$ after the
assignment.

\begin{figure}[htp]
  \begin{center}
    {\normalsize
      \begin{tabular}{|l|}
        \hline
        \textsf{$<$machine\_identifier $>$} \\
        ~~\ebtag{refines} \\
        ~~~~$<$ machine\_identifier $>$ \\
        ~~\ebtag{sees} \\
        ~~~~$<$ context\_identifier\_list $>$\\
        ~~\ebtag{variables} \\
        ~~~~$<$ variable\_identifier\_list $>$\\
        ~~\ebtag{invariants} \\
        ~~~~$<$ label $>$:~$<$ predicate $>$\\    
        ~~\ebtag{variants} \\
        ~~~~$<$ variant $>$\\    
        ~~\ebtag{events} \\
        ~~~~$<$ event\_list $>$\\    
        \hline
      \end{tabular}
    }
  \end{center}
  \caption{General structure of \eb\ machine (taken from \cite{EB:Book}).}
  \label{fig:background:eb:ebmachine}
\end{figure}

Figure~\ref{fig:background:eb:ebcontext} shows a general structure of
an \eb\ context. It contains a list of \ebtag{extends} that defines
which contexts this context is extending. It also contains a list of
carrier sets (\ebtag{sets}) and a list of \ebtag{constant}s. Finally,
it defines \ebtag{axiom}s that assert properties of sets and
constants.

\begin{figure}[htp]
  \begin{center}
    {\normalsize
      \begin{tabular}{|l|}
        \hline
        \textsf{$<$context\_identifier $>$} \\
        ~~\ebtag{extends} \\
        ~~~~$<$ context\_identifier\_list $>$ \\
        ~~\ebtag{sets} \\
        ~~~~$<$ set\_identifier\_list $>$\\
        ~~\ebtag{constants} \\
        ~~~~$<$ constant\_identifier\_list $>$\\
        ~~\ebtag{axioms} \\
        ~~~~$<$ label $>$:~$<$ predicate $>$\\    
        \hline
      \end{tabular}
    }
  \end{center}
  \caption{General structure of \eb\ context (taken from \cite{EB:Book}).}
  \label{fig:background:eb:ebcontext}
\end{figure}

\subsection{An example in \eb}
An excerpt of a social network abstract \eb\
model, adapted from the B model in \cite{matelas:2010}, is depicted in
Figure \ref{fig:ebnet}. The context $c$ (not shown here), that the
machine \ebkeyw{sees}, defines carrier sets \ebkeyw{PERSON} (the set of all
possible people in the network) and \ebkeyw{CONTENTS} (the set of all
possible images, text, ... in the network). The abstract machine
declares variables $persons$ (the set of people actually in the
network), $contents$ (the set of content actually in the network),
$owner$ (a total surjection function mapping each content item to its owner),
and $pages$ (a total {surjective} relation indicating
which content items are visible to which people). Invariant
\ebtag{inv5} ensures that each content item is visible to its
owner. \eb\ provides notations $\tsur$ for a
total surjective function, and $\strel$ for a total surjective
relation.

\begin{figure}[t]
{\small
\[
\begin{array}{c@{\hspace*{15pt}}c}
\begin{array}{l}
\ebkeyw{machine}~abstract~~\ebkeyw{sees}~c\\
~\ebkeyw{variables}~persons~contents~owner~pages \\
~\ebkeyw{invariants}\\
~~\ebtag{inv1 } persons \subseteq \ebkeyw{PERSON}\\
~~\ebtag{inv2 } contents \subseteq \ebkeyw{CONTENTS}\\
~~\ebtag{inv3 } owner \in contents \tsur persons\\
~~\ebtag{inv4 } pages \in contents \strel persons\\
~~\ebtag{inv5 } owner \subseteq pages\\
~\ebkeyw{events}\\
~~initialisation \\
~~~\ebkeyw{begin} \\ 
~~~~\ebtag{in1 }persons,contents,owner,pages := \emptyset,\emptyset,\emptyset,\emptyset \\
~~~\ebkeyw{end} \\
\\
~~transmit\_rc \\
~~~\ebkeyw{any}~prs~rc~ow~\ebkeyw{where}\\
~~~~\ebtag{grd1 }prs \subseteq persons ~~~\ebtag{grd2 }rc \in contents\\
~~~~\ebtag{grd3 }owner(rc) = ow ~~~\ebtag{grd4 }owner(rc) \not = prs \\
~~~\ebkeyw{then}\\
~~~~\ebtag{act1 }pages := pages \bunion (\{rc\} \cprod prs)\\
~~\ebkeyw{end} \\
\ebkeyw{end}
\end{array}
\end{array}
\]
} 
\caption{\eb\ machine for a social networking site.}
\label{fig:ebnet}
\end{figure}

The $initialisation$ event ensures that all of these sets, functions
and relations are initially empty. The symbol $\emptyset$ represents
the empty set. The abstract machine further defines the event
$transmit\_\-rc$ that allows a user of the network to share his own
content to people in the social network. Events can be
executed/triggered when their guards (the
part after the \ebkeyw{where}) hold. Hence,  $transmit\_\-rc$ 
can execute whenever there is a person $ow$ in the network that owns a
content item $rc$, and the people in $prs$ do not already own the item
$rc$. The
meaning of an event is the meaning of the actions in its body (the
part after the \ebkeyw{then}). $transmit\_\-rc$ event's action adds
the content item $rc$ to the page of each person in the set $prs$. The
symbol $\cprod$ represents cross-product.  

\section{Tool support for \eb}
\label{background:rodin}
The Rigorous Open Development Environment for Complex Systems (RODIN)
\cite{rodin} is an open-source Eclipse IDE that provides support for
\eb\ and that provides a set of tools for working with \eb\ models,
e.g. an editor, a proof generator, and provers. Existing Rodin
plug-ins provide extended functionality such as model checking and
animation \cite{prob}.

Rodin comes with an API that offers a series of Java interfaces for
manipulating  \eb\ components called the data model. It also comes
with a persistence layer (called the Rodin database) that uses
\verb+XML+ files to store these components. It is intended to abstract
the concrete persistence implementation from the data model. The
database API is located in the \javacode{org.rodinp.core} package. Full
source code for Rodin is available in~\cite{rodinSources}.

\subsection{Rodin Proof Obligations}
\label{background:pos}
In modelling in \eb, users transform an abstract machine to code via a
series of refinements, where the behaviour of each refinement is
provably consistent with the behaviour of the previous step. Each
refinement adds more details to the system. A
refinement generates proof obligations that must be formally verified
in order to assert that a model $M_{i+1}$ is indeed a refinement of a
previous model $M_i$. Hence a set
of Proof Obligations (PO) is generated. POs are sequents that need to
be proven true in order for the underlying system to be correct. Rodin
automatically generates them. Rodin provides the tool proof
generator, and several provers. The provers provided by Rodin help
users to discharge POs. However, when POs are not discharged
automatically, the assistance of the user is necessary to discharge
them. Generated POs are described in this section.

The Rodin proof-obligation generator automatically generates proof
obligations based on both the machine and the context. As explained
above, there are three kinds of
relations between machines and contexts: 
\begin{inparaenum}[\itshape i\upshape)]
\item a machine \ebtag{sees} a context,
\item a concrete machine can \ebtag{refine} an abstract machine, and
\item a context can \ebtag{extend} another context.
\end{inparaenum}  Given the abstract
event $evt_0$ and the concrete event $evt$ in Figure
\ref{fig:background:rodin:evt_abs_con}, and given an
abstract and a concrete machine declaring the events respectively,
Rodin generates several proof obligations to ensure that the machines
are models of the same system yet at a different level of
abstraction. In Figure
\ref{fig:background:rodin:evt_abs_con}, $s$ and $c$ are the sets and constants seen
by the abstract and concrete machines, $v$ is the set of abstract
variables, $w$, that may include $v$, is the set of concrete variables,
predicates $G$ and $H$ are the abstract and concrete guards, $BA_0$
and $BA$ are \emph{before-after} predicates that relate the state of
variables before and after actions occur\footnote{Primed variables
  refer to after-states.}.

An abstract machine can declare an abstract invariant
$I$ and a concrete machine can additionally  declare an invariant $J$
(also called ``gluing'' invariant since it relates the states between
the concrete and abstract machines) that depends on the context and
the local machine
variables respectively. Contexts can further declare a set of theorems
and axioms.

\begin{figure}[t]
  \centering
  {\normalsize
    \begin{tabular}{|l|}
      \hline
      $evt_0$ \\
      ~\ebkeyw{any} $x$ \ebkeyw{where} $G(s,c,v,x)$ \\
      ~\ebkeyw{then} \\
      ~~~\ebkeyw{act} $v :| BA_0(s,c,v,x,v')$ \\
      \ebkeyw{end} \\
      \\
      $evt$ \ebkeyw{refines} $evt_0$ \\
      ~\ebkeyw{any} $y$ \ebkeyw{where} $H(y,c,w)$ \\
      ~\ebkeyw{then} \\
      ~~~\ebkeyw{act} $w :| BA(s,c,w,y,w')$ \\
      \ebkeyw{end} \\
      \hline
    \end{tabular}
  }
  \caption{Events of an \eb\ abstract and refinement machines.}
  \label{fig:background:rodin:evt_abs_con}
\end{figure}

Rodin generates invariant preservation proof obligations
(\textsf{INV}) for every abstract (concrete) event of the abstract
(concrete) machine  expressing that
given the axioms and theorems, the abstract (gluing) invariant, the guard of the
event, and the before-after predicate, the abstract (the
concrete) invariant holds in the after state. Rodin generates a guard
strengthening proof obligation (\textsf{GRD}) for every event
expressing that the guard of the concrete event must be as least as
strong as the guard of the abstract event. It generates a
feasibility proof-obligation (\textsf{FIS}) for the action of every
event stating that a solution to the before-after predicates
exists. For every event merging two abstract events, Rodin generates a
merging proof obligation (\textsf{MRG}) that ensures that the guard of the merging
event is stronger than the disjunction of the guards of the abstract
events. For every refining event, Rodin generates a simulation proof obligation
(\textsf{SIM}) that ensures that abstract actions are correctly
simulated by the concrete actions. That is, the result produced by the
concrete action does not contradict the result produced by the
abstract action. Rodin also generates numeric proof
obligations to ensure that variable declarations are well-defined
(\textsf{WD}).

In Rodin, one can further declare a machine \textsf{variant}. A
\textsf{variant}  can be defined as a set in which case Rodin
generates a finite proof obligation that ensures that the variant is a
finite set (\textsf{FIN}). A \textsf{variant}  can also be defined as
a numeric expression in which case Rodin generates a numeric variant
proof obligation that ensures that the expression is a
positive integer expression (\textsf{NAT}). Machine events can be
declared \ebkeyw{convergent} or \ebkeyw{anticipated}: for
\ebkeyw{convergent} events, Rodin generates proof obligation
\textsf{NAT} expressing that the modified \textsf{variant} (evaluated after
executing the event action) must remove elements to the
\textsf{variant} (if this is defined as a set), or must decrease the
\textsf{variant} (if this is defined as a numeric expression); for
\ebkeyw{anticipated} events, Rodin generates proof obligation
\textsf{NAT} expressing that the modified \textsf{variant} must not
add elements to
the \textsf{variant} (if this is defined as a set), or must not
increase the \textsf{variant} (if this is defined as a numeric
expression).

One can declare a \emph{witness} of a refinement event with the aid of
the \ebkeyw{with} clause of Rodin. A witness expression relates
bounded variables of an abstract event with bounded variables of the
concrete event, e.g. one could have added the expression \ebkeyw{with} $x:
x' = y$ to the declaration of the concrete event $evt$ in Figure
\ref{fig:background:rodin:evt_abs_con}, meaning that bounded variable
$x$ in the
abstract event is renamed as bounded variable $y$ in the concrete
event. Rodin generates a witness proof obligation (\textsf{WFIS}) for
every event witness expressing that a solution for the witness
expression exists.

Theorems must be provable from contexts or machines (\textsf{THM}). In
\eb, Theorems are used to simplify complex proof-obligations.

\section{The Java Modeling Language  (\jml)}
\label{background:jml}

\jml\ \cite{JML:Preliminary:Design:2006,Burdy-etal05,JMLRefManual}  is an
interface specification language
for Java -- it is designed for specifying the behaviour of Java
classes, and is included directly in Java source files using special
comment markers \texttt{//@} and \texttt{/*@ */}.  \jml's type system
includes all built-in Java types and additional types representing
mathematical sets, sequences, functions and relations, that are
represented as \jml\ specified Java classes in the
\texttt{org.jmlspecs.models} package.  Similarly, \jml\ expressions
are a superset of Java expressions, with the addition of notations
such as \texttt{==>} for logical implication, \jmlkeyw{\bsl exists} for
existential quantification, and \jmlkeyw{\bsl forall} for universal
quantification.

\jml\ class specifications can include \jmlkeyw{invariant} clauses
(assertions that must be satisfied in every visible state of the
class), \jmlkeyw{initially} clauses (specifying conditions that the
post-state of every class constructor must satisfy), and history
constraints (specified with the keyword \jmlkeyw{constraint}, that
are similar to invariants, with the additional ability to relate pre-
and post-states of a method).  Concrete \jml\ specifications can be
written directly over fields of the Java class, while more abstract
ones can use specification-only \jmlkeyw{model} and \jmlkeyw{ghost}
fields. \jmlkeyw{ghost} fields are not related to the concrete state of the
class and can be declared \javakeyw{final}, while \jmlkeyw{model}
fields are related to Java implementation fields via a
\jmlkeyw{represents} clause, that acts much like a gluing invariant
in \eb\ refinement.

\jml\ provides pre-post style specifications for Java methods describing
software contracts~\cite{Meyer92}: if the caller of a method
meets the pre-condition, the method must ensure the post-condition.
\jml\ uses keywords \jmlkeyw{requires} for method
pre-\-conditions, \jmlkeyw{ensures} for normal method
post-\-conditions, \jmlkeyw{signals} and \jmlkeyw{exsures} for method
exceptional post-conditions, and \jmlkeyw{assignable} and
\jmlkeyw{modifies} for frame conditions (lists of locations whose
values may change from the pre-state to the post-state of a
method). An \jmlkeyw{assignable} clause of $\bsl\jmlkeyw{nothing}$
prevents any location from being modified from the pre- to the
post-state, and an \jmlkeyw{assignable} clause of
$\bsl\jmlkeyw{everything}$ allows any side-effect. Declaring a method as
\jmlkeyw{pure} has the same effect as \jmlkeyw{assignable} 
$\bsl\jmlkeyw{nothing}$. The pre-state is the state on
method entry and the post-state is the state on method exit.
A \jmlkeyw{normal\_behavior} method specification states that if the
method pre-condition holds in the pre-state of the method, then it
will always terminate in a normal state, and the normal post-condition will
hold in this state. 
A \jml\ 
\jmlkeyw{exceptional\_behavior} method specification states that if
the method pre-condition holds in the pre-state of the method, then it
will always terminate in an exceptional state, throwing a
\javacode{java.lang.Exception}, and the corresponding exceptional
post-condition will hold in this state.
In an \jmlkeyw{ensures} or \jmlkeyw{signals} clause, the keyword
\jmlkeyw{\bsl old} is used to indicate expressions that must be
evaluated in the pre-state of the method -- all other expressions are
evaluated in the post-state.  The \jmlkeyw{\bsl old} keyword can also
be used in history constraints, providing a convenient way to specify
(for example) that the post-state value of a field is always equal to
the pre-state value, thus making the field a constant.

\begin{center}
  \begin{figure}
    \begin{lstlisting}[frame=none]
      //@ model import org.jmlspecs.models.JMLEqualsSet;
      //@ model import org.jmlspecs.models.JMLEqualsToEqualsRelation;

      public abstract class SOCIAL_NETWORK {
        //@ public final ghost JMLEqualsSet<Integer> PERSON;
        //@ public final ghost JMLEqualsSet<Integer> RAWCONTENT;
        //@ public model JMLEqualsSet<Integer> person;
        //@ public model JMLEqualsSet<Integer> rawcontent;
        //@ public model JMLEqualsToEqualsRelation<Integer,Integer>  content;

        /*@ public invariant person.isSubset(PERSON) 
            && rawcontent.isSubset(RAWCONTENT)
            && (new Relation<Integer, Integer> (person, rawcontent)).has(content)
            && content.domain().equals(person) 
            && content.range().equals(rawcontent);*/

        /*@ public initially person.isEmpty() &&
              rawcontent.isEmpty() && content.isEmpty();*/

        /*@ public normal_behavior
            requires rawcontent.has(rc) && person.has(ow)
              && person.has(pe) && !ow.equals(pe)
              && !content.has(ModelUtils.maplet(pe,rc));
            assignable content;
            ensures (\exists JMLEqualsSet<Integer> prs;\old(prs.isSubset(person)); 
                content.equals(\old(ModelUtils.toRel(content.union(ModelUtils.toRel(ModelUtils.maplet(pe,rc)))).union(ModelUtils.cartesian(prs, ModelUtils.toSet(rc))))));
          also 
            public exceptional_behavior
            requires !(rawcontent.has(rc) && person.has(ow)
              && person.has(pe) && !ow.equals(pe)
              && !content.has(ModelUtils.maplet(pe,rc)));
            assignable \nothing; signals (Exception) true;*/ 
        public abstract void transmit_rc(Integer rc, Integer ow,
        Integer pe);
      } 
    \end{lstlisting}
    \caption{A \jml\ specification of a social networking class.}
    \label{fig:background:jml:social-JML}
  \end{figure}
\end{center}



Figure \ref{fig:background:jml:social-JML} presents a simple example
of a \jml\ specified Java \javakeyw{abstract} class. This class
defines an excerpt of a Social Network. Classes
\texttt{JML\-Equals\-Set} and
\texttt{JML\-Equals\-To\-Equals\-Relation} (in the
\texttt{org.\-jmlspecs.\-models} package) are built-in to \jml\ and
represent mathematical sets and relations, respectively.  

The specification of the \texttt{transmit\_rc} method uses two
specification cases (the keyword \jmlkeyw{also}) - the first
specifying that content \texttt{rc} is transmitted to person
\texttt{pe} if \texttt{pe} and \texttt{ow} are actually persons of the
network, \texttt{rc} belongs to the content of the network,
belongs to person \texttt{ow}, and \texttt{rc} has not been
transmitted yet to person \texttt{pe}. The second specifying that
attempting to transmit \texttt{rc} to a person \texttt{pe} when
either \texttt{pe} or \texttt{ow} are not actually people of the
network, or \texttt{rc} does not belong either to the content of the
network or to person \texttt{ow}, or \texttt{rc} has been already
transmitted to person \texttt{pe}, has no effect. The specification of
the \texttt{transmit\_rc} method demonstrates the syntax of
existentially quantified assertions in \jml, and the use of
\jmlkeyw{exceptional\_\-behavior} specification cases to specify when
exceptions are to be thrown.

\subsection{Tool support for \jml}
\label{background:jml:tool}
There are different techniques that work with \jml\ specifications
along with the proper tooling. The
most basic technique is parsing and type-checking the \jml\
specifications as done by the \jml\ checker \textsf{jml}.  
Several tools have been developed to help users with the correct
specification of \jml\ clauses. For instance, the CHASE tool
\cite{conference:CatanoHuisman:03}
checks the assignable clause of a \jml-annotated Java program. It
checks the assignable clause for every method by checking for every
assignment and for every method call in the body. It determines whether it agrees
with the assignable clause of the method that is checked. The tool
gives feedback to the user on the forgetting variables that may be
modified by a specific method. Another tool that helps users on
specifying \jml\ clauses is Daikon \cite{Ernst:Daikon}. Daikon
automatically infers invariants from \jml-annotated Java programs. 

Another more specialised technique that works with \jml\ consists of
testing specifications by executing them. This is done by \jmle\
\cite{jmle} that translates \jml\ specifications to Java programs that
are executed using the Java Constraint Kit. An alternative technique is to
check the
correctness of the method with respect to its \jml\ specification
 by run-time checking. Such run-time
checking is done by the \jml\ compiler \textsf{jmlc}
\cite{Cheon02aruntime}. \textsf{jmlc} is an extension of the Java compiler that
compiles \jml-annotated Java programs into Java bytecode. The compiled
bytecode contains safety properties as pre-conditions,
normal and exceptional post-conditions, invariants, and history
constraints. If an assertion is violated, an error message 
arises. Another tool for assertion checking during testing phase is
\textsf{jmlunit} \cite{jmlunit10}. \textsf{jmlunit} uses the \jml\
specification as a testing oracle, automating the process of
generating JUnit tests. The tests created by this tool catches any
assertion given by the \jml\ run-time assertion, thus it checks if,
for instance, a pre-condition or an invariant is violated, meaning
the Java code does not meet the \jml\ specification. 

The major problem with run-time checking is that it is limited by the
execution paths done by the test suite (since it is executed in
run-time). Another technique that works with \jml\ is the static
verification of the Java
code. This can give more assurance in the correctness of the Java code
as it establishes the correctness for all possible execution
paths. Typically, this technique generates proof obligations from the
\jml\ specification and uses a theorem prover to discharge them. There
are several tools for working with technique: 

\begin{itemize}
\item The LOOP tool
\cite{Berg01theloop} works over sequential
Java implementations. It translates proof obligations and uses PVS
\cite{cade92-pvs} or Isabelle \cite{Isabelle} to discharge the proof
obligations. 
\item Krakota \cite{Marche03thekrakatoa} is well suited for
Java Cards Applets. Krakota receives a \jml-annotated
Java program as an input and translates it to the input language of the WHY tool
\cite{WhySpecAndTool03}. WHY is used to automatically generate proof
obligations and uses the COQ \cite{coq} prover to discharge the proofs.
\item ESC/JAVA2 \cite{escjava2} is intended to detect simpler errors,
like null pointers or, out of bound array access. ESC/JAVA2 uses provers
like Z3 \cite{Z3:overview} to discharge proofs obligations.
\item  OpenJML
\cite{Cok2011} translates \jml\ specifications into SMTLIB
\cite{BarST-SMTLIB} (Satisfiability Modulo Theories Library) format
and passes the proof problems implied to backend SMT solvers.
\item  The JACK
tool \cite{Antoine02jack:java} works on Java cards applets. The input
for JACK is a \jml-annotated Java program where user needs to express
the proof obligation (property) that he wants to prove. The tool
translates the program to the input language of \bmethod\ method, and
translates the proof obligations to lemmas in \bmethod. Lemmas are
proven using a prover developed within Atelier B \cite{atelierb}.
\end{itemize}

There are several more tools for checking the \jml\ specification
against the Java code. \cite{Burdy-etal05} contains an overview of
these tools.

One limitation of these tools is that none of them has been developed to work
with Java 7, that introduces generic types. A generic type is a
generic class that is parameterised over types, so it is not possible
to know the specific type in a static manner.

\section{\dafny}
\label{background:dafny}
\dafny\ \cite{dafny2,dafny} is an imperative object-based language with built-in
specification constructs. The \dafny\ static program verifier can be
used to verify the functional correctness of programs. \dafny\ runs
under Microsoft Visual Studio, and from the command line,
that requires a .NET virtual machine. \dafny\ provides
support for the
annotation of programs as {contracts}: pre- and
post-conditions. It also provides support for abstract specifications through the
definition and use of \dafnykeyw{ghost} variables and methods, and for
the definition and specification of
mathematical functions. Functions are
specification-only constructs; they exist for verification-only
purposes and are \dafnykeyw{ghost} by default. The
\dafnykeyw{requires} specification of the function (that goes in the
same
direction of \jmlkeyw{requires} \jml\ clause) may be used to define its
domain (partial functions). Post-conditions are written as
\dafnykeyw{ensures} specifications (that also goes in the same direction of
\jmlkeyw{ensures} \jml\
clause). It represents the post-state of the function
and must hold when the caller of the function meets the
pre-state. Pre- and post-conditions must be written at the beginning
of
the function. The \dafnykeyw{assert} clause can be written somewhere in the
middle of the function. It tells that a particular expression always
holds when control reaches that part of the code. The
\dafnykeyw{reads} part declares the
function's frame condition, that is all the memory locations that the
function is allowed to read \cite{dafny2,dafny}. Finally, the
\dafnykeyw{decreases} part states 
the termination metrics of the function.

Program verification with \dafny\ works by translating the program
written in \dafny\ to the Boogie 2 proving engine \cite{boogie-leino-06} in
such a way that the correctness of the Boogie program implies
the correctness of the \dafny\ program. 

\section{Software Design Patterns}
Code patterns are common to many software solutions. According to
\cite{appletonPatter},
a software pattern ``creates a common structure to help software
developers to resolve recurring problems encountered throughout
software development''. 

Gamma et. al. popularised the term Software Design
Patterns \cite{Patterns:Gamma:95}, that is a reusable solution to a
common problem within a given context. A Design Pattern is not an
implementation of a solution but a template for addressing the
problem. One of these Software Design Patters is Model-View-Controller
(MVC). MVC separates the internal representation of the information
from the user's perspective. It is composed of three components
namely the Model (M), the Controller (C), and Views (V).  All
components interact with each other: 

\begin{itemize}
\item the Controller is a bridge between the Model and
the Views. It sends commands to the Model for it to update its
state. The Controller also sends commands to the Views to change the
presentation of the Model,

\item the Model contains the logic of the system. It sends
information to the Controller and Views every time the Model
changes its state,

\item the Views, that are the graphical representation of
the information, request information from the Model necessary to update
the information presented to the user. 
\end{itemize}

MVC design pattern is commonly used for the development of
Graphical User Interfaces (GUI) since the separation of the internal
presentation of information allows developers to change just the Views.

\chapter{Translating \bmethod\ Machines to \jml\ Specifications}
\label{chapter:b2jml}

\paragraph{This chapter.} This chapter presents a translation from
\bmethod\ machines to \jml\ specifications defined through syntactic rules. It
also presents the implementation of this translation as the \btojml\
tool.  The tool enables \bmethod\ experts to use Refinement
Calculus techniques to develop critical components in \bmethod, and
then translate the result to \jml\ for developers with less
mathematical expertise to be able to implement code that respects the
\jml\ specification. \btojml\  enables developers to use lightweight
\jml\ tools such as the \jmle\ tool for executing \jml\
specifications~\cite{jmle,jmle-jcard:09}, runtime assertion
checkers and static analysers \cite{Burdy-etal05}. \btojml\ fully
supports the \bmethod\ syntax except for the
\bmethod\ constructs for multiple incremental specifications of
machines, e.g. for including, importing, seeing, or extending other
machines. We
integrated
\btojml\ to the ABTools suite~\cite{Boulanger2003}. We have validated
the tool by applying it to a moderately complex \bmethod\ model of a
social networking site. We further executed the resulting \jml\
specifications against a suite of test cases developed for a
hand-translation of a \bmethod\ model. The full code of the
implementation of the \btojml\ tool is available
at~\cite{Boulanger2011}.

The work presented in this chapter inspired the work presented in the
rest of this thesis. It has been published in
\cite{conference:B2Jml:12,e-eb:to:jml-java}. I participated in this
work at the end of the implementation of the \btojml\ tool. 

The rest
of this chapter is organised as follows. Section
\ref{btojml:translation}  describes
the set of transition rules that translate \bmethod\ models to \jml\
abstract class. These rules are implemented as the \btojml\ tool and
integrated into the ABTools suite. Section \ref{btojml:implementation}
presents the implementation. We have validated \btojml\ by
applying it to the \bmethod\ model presented in Section
\ref{btojml:apply}. The \bmethod\ model is a moderately
complex model of a social networking site. That section also presents
the resulting \jml\ abstract Java class. Finally, Section
\ref{btojml:conclusion} gives conclusions.

\paragraph{Contributions.} 
\begin{inparaenum}[\itshape i\upshape)]
\item We present the definition of the
  translation of \bmethod\ machines into \jml\ specifications via
syntactic rules, and
\item we present the implementation of this translation as
\btojml\ tool. 
\end{inparaenum}
Users might prefer to use \jml\ since 
\begin{inparaenum}[\itshape a\upshape)]
\item implementing a \jml\ specification in Java is
  much more straightforward than implementing an equivalent \bmethod\
  machine in Java. And
\item the user may be more familiar with \jml\ syntax
  than \bmethod\ notation.
\end{inparaenum}

\paragraph{Related work.} Jin and Yang \cite{jin2008} outline an
approach for translating VDM-SL to \jml. Their motivations are similar
to the \btojml\ tool in viewing VDM-SL as a better language for
modelling at an abstract level (as in \bmethod), and \jml\ as a better
language  for working closer to an implementation level.  Their
approach has not been automated -- they only describe a strategy for
translating specifications by hand.

Boulanger~\cite{Boulanger2006} describes partially automated
translations between \bmethod\ and VHDL (in both directions).  This
approach permits co-design -- verified \bmethod\ implementation
machines can be translated to VHDL for realisation in hardware, and
translations of VHDL libraries can be used by \bmethod\ machines.
Hence, these translations allow designers to verify models of
hardware components.

Bouquet et.~al.~have defined a translation from \jml\ to
\bmethod \cite{groslambert-jml2b-05} and implemented their approach as
the JML2B tool~\cite{groslambert-jml2b-07}.  Although their
translation goes in the opposite direction of the work presented
here, their motivation is quite similar -- they view translation as a
way to gain access to more appropriate tools for the task at hand,
which in this case is verifying the correctness of an abstract model
without regard to code. \jml\ verification tools are primarily
concerned with verifying the correctness of code with respect to
specifications, while \bmethod\ has much stronger tool support for
verifying models.

In some ways, translating from \jml\ to \bmethod\ is a more difficult
problem  than the reverse, as \jml\ includes many concepts (objects,
inheritance, exceptions, etc.) that do not appear in \bmethod.
Hence, Bouquet et al. were required to build representations of these
concepts in \bmethod\ for use by their translated machines.
Distinguishing pre- and post-state values required considerable
effort, while in \btojml\  translation it was relatively clear which
parts of a \bmethod\ machine should be evaluated in the pre-state,
and which parts needed to be evaluated in the post-state.  The
translations are also similar because the correspondence between
\bkeyw{PRE} substitutions and \jmlkeyw{requires} clauses; invariants
in both languages; \bmethod\ operations and \jml\ methods and so on
is straightforward.  One significant difference is that Bouquet
et. al. translate a \jml\ class specification to a \bmethod\ machine
that has a set variable containing all instances of the class.
Additional variables of the machine represent each \jml\ field as a
function from this set of instances to the value of the field for
that instance.  This provides a mechanism for distinguishing pre- and
post-state values (by making copies of these functions), but also
forces the \bmethod\ operation representing a \jml\ method to take the
calling object as an explicit parameter, rather than referring
directly to the machine variables in the usual way. This makes the
correspondence between the \jml\ specification and the \bmethod\
machine more difficult to see.

\section{The Translation from \bmethod\ to \jml}
\label{btojml:translation}
The translation from \bmethod\ to \jml\ is implemented with the aid of a \BTOJMLN\
operator. It is defined (via syntactic rules), it takes \bmethod\
syntax as input and returns the corresponding \jml\ specification. 
To assist in this translation, we defined a \MODN\ operator. It
calculates the set of variables modified by an operation. 
The definition of \MODN\
is inspired by the syntactic rules backing the analysis performed by
the Chase tool in \cite{conference:CatanoHuisman:03}.
Further, a \TypeN\  operator is employed (without definition) to denote
the inference of the type of a \bmethod\ variable and its translation into a 
corresponding \jml\ type.
Correspondence
between \bmethod\ and \jml\ types is briefly described at the end of
section \ref{btojml:translation:bey_subs}.

\BTOJMLN\ translates a \bmethod\ abstract or refinement machine
to a \jml\ annotated abstract Java class.  The machine variables
become \jmlkeyw{model}
fields, and operations are translated to abstract methods with \jml\
specifications.  A preconditioned substitution in \bmethod\ generates
\jml\ (normal and exceptional) method specification cases. Additionally,
although substitutions in \bmethod\ have no explicit notion of post-condition,
\BTOJMLN\ translates other substitutions to \jml\ post-conditions
that relate the pre- and post-state values of the variables modified
by the substitution.  The rules
defining \BTOJMLN\ are deterministic so one
cannot apply two different rules at the same time. In this sense,
these rules define a calculus that computes the translation of
\bmethod\ into \jml. \BTOJMLN\ does not fully consider the language B0
(B0 contains constructs closer to an implementation e.g. $WHILE$)

Section \ref{btojml:translations:substitutions} presents the translation of general substitutions,
Section \ref{btojml:translation:bey_subs} presents the translation of other
\bmethod\ syntaxes.  

\subsection{Translating Substitutions}
\label{btojml:translations:substitutions}

Rule \textsf{Sel} translates a guarded substitution to a \jml\
implication (\jmlcode{==>}) in which the antecedent is obtained from the
translation of the guard and the consequent is obtained from the
translation of the nested substitution. This matches the definition of
a guarded substitution in \bmethod\ presented in Figure
\ref{fig:background:bmethod:substitutions} where in a guarded
substitution $\bkeyw{WHEN } P \bkeyw{ THEN } S \bkeyw{ END}$,
substitution $S$ is executed under
the assumption of $P$. Rules \textsf{When} generalise rule
\textsf{Sel}. The first rule \textsf{When} is a synonym of rule
\textsf{Sel}, in \bmethod, the construct for \bkeyw{WHEN} in the form 
$\bkeyw{WHEN } P \bkeyw{ THEN } S \bkeyw{ END}$ can be seen as 
$\bkeyw{SELECT } P \bkeyw{ THEN } S\bkeyw{ END}$ so it is translated in
the same way as \bkeyw{SELECT}. The second rule \textsf{When}
considers two guards\footnote{The rule for the simultaneous substitution
  $S\:||\:\mathit{SS}$ is presented later in this section.}.

{\small
  \[
  \begin{prooftree}
    \PRED{P} = \texttt{P} ~~ \BTOJML{S} = \texttt{S}
    \using \textsf{(Sel)}
    \justifies
    \begin{array}{c}
      \BTOJML{\bkeyw{SELECT } P \bkeyw{ THEN } S \bkeyw{ END}}\\ = \\
      \jmlcode{\bsl} \jmlkeyw{old}\texttt{(P)}\jmlcode{ ==> }\texttt{S}
    \end{array}
  \end{prooftree}
  \]
}

{\small
  \[
  \begin{prooftree}
    \PRED{P} = \texttt{P} ~~ \BTOJML{S} = \texttt{S}
    \using \textsf{(When)}
    \justifies
    \begin{array}{c}
      \BTOJML{\bkeyw{WHEN } P \bkeyw{ THEN } S \bkeyw{ END}}\\ = \\
      \jmlcode{\bsl} \jmlkeyw{old}\texttt{(P)}\jmlcode{ ==> }\texttt{S}
    \end{array}
  \end{prooftree}
  \]
}

{\small
  \[
  \begin{prooftree}
      \begin{array}{ll}
        \PRED{P} = \texttt{P} & \BTOJML{S} = \texttt{S}\\
        \PRED{Q} = \texttt{Q} & \BTOJML{T} = \texttt{T}\\
      \end{array}
    \using \textsf{(When)}
    \justifies
    \begin{array}{c}
      \BTOJMLN{(\bkeyw{SELECT } P \bkeyw{ THEN } S}
      \\~~~~~~~~~~~~\bkeyw{WHEN } Q \bkeyw{ THEN } T \bkeyw{ END)} 
      \\ = \\
      \jmlcode{(\bsl} \jmlkeyw{old} \texttt{(P)} \jmlcode{ ==> }
      \texttt{S} \jmlcode{) \&\&}\\
      \jmlcode{(\bsl} \jmlkeyw{old} \texttt{(Q)} \jmlcode{ ==> }
      \texttt{T} \jmlcode{)}\\

    \end{array}
  \end{prooftree}
  \]
}

Rules \textsf{If} and \textsf{IfElse} translate the \bkeyw{IF} and
\bkeyw{IF ELSE} substitutions to \jml.  

{\small
  \[
  \begin{prooftree}
    \PRED{P} = \texttt{P} ~~ \BTOJML{S} = \texttt{S}
    \using \textsf{(If)}
    \justifies
    \begin{array}{c}
      \BTOJML{\bkeyw{IF } P \bkeyw{ THEN } S \bkeyw{ END}}
      \\ = \\
      \jmlcode{\bsl} \jmlkeyw{old}\texttt{(P)}\jmlcode{ ==> }\texttt{S}
    \end{array}
  \end{prooftree}
  \]
}

{\small
  \[
  \begin{prooftree}
    \PRED{P} = \texttt{P} ~~ \BTOJML{S} = \texttt{S}~~ \BTOJML{T} = \texttt{T}
    \using \textsf{(IfElse)}
    \justifies
    \begin{array}{c}
      \BTOJML{\bkeyw{IF } P \bkeyw{ THEN } S \bkeyw{ ELSE } T \bkeyw{ END}}
      \\ = \\

      \jmlcode{\bsl} \jmlkeyw{old}\texttt{(P)}\jmlcode{?}\texttt{ S}
      \jmlcode{ : } \texttt{T}
    \end{array}
  \end{prooftree}
  \]
}

Rule \textsf{Pre} presents the translation of preconditioned
substitutions.  A preconditioned substitution is conceptually
different from a guarded substitution. While a guarded substitution
imposes a condition on the internal behaviour of the machine, a
preconditioned substitution imposes a condition (the pre-condition) on
the caller. Hence, a preconditioned substitution aborts if the
pre-condition does not hold. This matches the definition of
preconditioned substitutions presented in Figure
\ref{fig:background:bmethod:substitutions} where in order to execute
the substitution $S$ in a preconditioned substitution $\bkeyw{PRE } P
\bkeyw{ THEN } S \bkeyw{ END}$ one must prove $P$. In \jml, the
behaviour of preconditioned substitution is modelled by throwing an
exception. 

{\small
  \[
  \begin{prooftree}
    \PRED{P} = \texttt{P} ~~ \MOD{S} = \texttt{A}~~ \BTOJML{S} = \texttt{S}
    \using \textsf{(Pre)}
    \justifies
    \begin{array}{c}
      \begin{array}{c}
        \BTOJML{\bkeyw{PRE } P \bkeyw{ THEN } S \bkeyw{ END}} =\\
      \end{array}
      \\
      \begin{array}{l}
        \jmlcode{/*@}~\jmlkeyw{public normal\_behavior}\\
        \quad\quad\jmlkeyw{requires } \texttt{P} \jmlcode{; } \jmlkeyw{assignable } \texttt{A} \jmlcode{;}  \\
        \quad\quad\jmlkeyw{ensures } \texttt{S}\jmlcode{;}\\
        \quad~ \jmlkeyw{also public exceptional\_behavior}\\
        \quad\quad\jmlkeyw{requires } !\texttt{P; }
        \jmlkeyw{assignable}~\jmlcode{\bsl} \jmlkeyw{nothing}\jmlcode{;}\\
        \quad\quad\jmlkeyw{signals(Exception) } \jmlkeyw{true}\jmlcode{; */}
      \end{array}
    \end{array}
  \end{prooftree}
  \]
}

Rule \textsf{Choice} below introduces the translation for bounded choice
substitutions, whose meaning is the meaning of any of the nested
substitutions.

{\small
  \[
  \begin{prooftree}
    \BTOJML{S} = \texttt{S} ~~ \BTOJML{T} = \texttt{T}
    \using \textsf{(Choice)}
    \justifies
    \begin{array}{c}
      \BTOJML{\bkeyw{CHOICE } S \bkeyw{ OR } T \bkeyw{ END}}~ = \texttt{
        S } || \texttt{
        T }
    \end{array}
  \end{prooftree}
  \]
}

Rule \textsf{Any} generalises rule \textsf{When} for unbounded choice
substitutions.
The type of the variable $x$ is inferred from
its usage in the predicate $P$ and substitution $S$.  
If at least one value of $x$ satisfies $P$,  any value 
can be chosen for use in $S$.
If no $x$ satisfies $P$, the substitution is equivalent to
\textsf{skip}.

{\small
  \[
  \begin{prooftree}
    \PRED{P} = \texttt{P} ~~ \BTOJML{S} = \texttt{S}~~ \Type{x} = \texttt{Type}
    \using \textsf{(Any)}
    \justifies
    \begin{array}{c}
      \begin{array}{c}
        \BTOJML{\bkeyw{ANY } x \bkeyw{ WHERE } P \bkeyw{ THEN } S
          \bkeyw{ END}} =\\
      \end{array}
      \\
      \begin{array}{l}
        (\jmlcode{\bsl} \jmlkeyw{exists } \texttt{Type } x \jmlcode{;
        }\jmlcode{\bsl} \jmlkeyw{old} \texttt{(P) }\jmlcode{\&\& }
        \texttt{S)}~||\\
        (\jmlcode{\bsl} \jmlkeyw{forall } \texttt{Type } x \jmlcode{;
        }!\jmlcode{\bsl} \jmlkeyw{old} \texttt{(P))}
      \end{array}
    \end{array}
  \end{prooftree}
  \]
}

The \textsf{VAR} construct in B introduces a local variable $x$ in the
scope of a substitution $S$, and so is equivalent to an \textsf{ANY}
substitution that does not constrain its bound 
variable~\cite{TheBBook}.

{\small
  \[
  \begin{prooftree}
    \using \textsf{(Loc)}
    \justifies
    \begin{array}{c}
      \BTOJML{\bkeyw{VAR } x \bkeyw{ IN } S \bkeyw{ END}}
      \\ = \\
      \BTOJML{\bkeyw{ANY } x \bkeyw{ WHERE TRUE} \bkeyw{ THEN } S
        \bkeyw{ END}}
    \end{array}
  \end{prooftree}
  \]
}

Rule \textsf{Asg} presents the translation of an assignment
from an expression $E$ to a variable $v$, the simplest nontrivial
substitution in B. This substitution is mapped to a \jml\ predicate in
which the value of the variable in the post-state equals the value of
the expression evaluated in the pre-state. If the variable $v$ is of a
primitive type, the translation will use \jmlcode{==} rather than the
\jmlcode{equals} method.

{\small
  \[
  \begin{prooftree}
    \PRED{E} = \texttt{E}
    \using \textsf{(Asg)}
    \justifies
    \begin{array}{c}
      \BTOJML{v~:=~E}~=~v\jmlcode{.equals(}\jmlcode{\bsl} \jmlkeyw{old}\texttt{(E))}
    \end{array}
  \end{prooftree}
  \]
}

Rule \textsf{Sim} presents the rule for the simultaneous substitution
$S || SS$, in which $SS$ could be another 
simultaneous substitution.
As the name indicates, the nested substitutions occur simultaneously.
Note that our rules translate $x := y~ || ~y := x$ to
$x\jmlcode{.equals(}\jmlcode{\bsl} \jmlkeyw{old} (y)) 
~\&\&~y\jmlcode{.equals(}\jmlcode{\bsl} \jmlkeyw{old} (x))$,
that matches the B semantics.  B does not allow simultaneous assignments
to the same variable.

{\small
  \[
  \begin{prooftree}
    \BTOJML{S} = \texttt{S} ~~~
    \BTOJML{SS} = \texttt{SS}
    \using \textsf{(Sim)}
    \justifies
    \begin{array}{c}
      \BTOJML{S~||~SS}
      \\=\\
      \texttt{S} \jmlcode{ \&\& } \texttt{SS}
    \end{array}
  \end{prooftree}
  \]
}

Further, frame conditions are checked; the only variables modified by a general
substitution are those modified by assignments within the
substitution. Hence, a set of \textsf{Mod} rules are defined to
calculate the set of these variables.  In rule \textsf{ModAsg}
below, the assigned variable is added to the frame-condition
set. Other rules for \bmethod\ substitutions are \textsf{ModSel}, \textsf{ModGua},
\textsf{ModAny}, \textsf{ModCho} and \textsf{ModSim}. These rules are
similar to the ones underpinning the checking performed by the Chase
tool \cite{conference:CatanoHuisman:03}.

{\small
  \[
  \begin{prooftree}
    \using \textsf{(ModAsg)}
    \justifies
    \begin{array}{c}
      \MOD{v := E}~=~\{v\}
    \end{array}
  \end{prooftree}
  \]
}

{\small
  \[
  \begin{prooftree}
    \MOD{S} = \texttt{S}
    \using \textsf{(ModSel)}
    \justifies
    \begin{array}{c}
      \MOD{\bkeyw{SELECT } P \bkeyw{ THEN } S \bkeyw{ END}}
      \\ = \\
      \texttt{S}
    \end{array}
  \end{prooftree}
  \]
}

{\small
  \[
  \begin{prooftree}
    \MOD{S} = \texttt{S} ~~~ \MOD{T} = \texttt{T}
    \using \textsf{(ModGua)}
    \justifies
    \begin{array}{c}
      \MOD{\bkeyw{SELECT } P \bkeyw{ THEN } S \bkeyw{ WHEN } Q \bkeyw{ THEN } T  \bkeyw{ END}}
      \\ = \\
      \texttt{S } \bunion \texttt{ T}
    \end{array}
  \end{prooftree}
  \]
}

{\small
  \[
  \begin{prooftree}
    \MOD{S} = \texttt{S}
    \using \textsf{(ModAny)}
    \justifies
    \begin{array}{c}
      \MOD{\bkeyw{ANY } x \bkeyw{ WHERE } P \bkeyw{ THEN } S \bkeyw{ END}}
      \\ = \\
      \texttt{S}
    \end{array}
  \end{prooftree}
  \]
}

{\small
  \[
  \begin{prooftree}
    \MOD{S} = \texttt{S} ~~~ \MOD{T} = \texttt{T}
    \using \textsf{(ModCho)}
    \justifies
    \begin{array}{c}
      \MOD{\bkeyw{CHOICE } S \bkeyw{ OR } T \bkeyw{ END}}
      \\ = \\
      \texttt{S } \bunion \texttt{ T}
    \end{array}
  \end{prooftree}
  \]
}

{\small
  \[
  \begin{prooftree}
    \MOD{S} = \texttt{S}~~~\MOD{T} = \texttt{T}
    \using \textsf{(ModSim)}
    \justifies
    \begin{array}{c}
      \MOD{S~||~T}
      \\ = \\
      \texttt{S } \bunion \texttt{ T}
    \end{array}
  \end{prooftree}
  \]
}

As the variable introduced by a \textsf{VAR} substitution is local
to that substitution, it should not appear in an
\jmlkeyw{assignable} clause and so is removed via rule \textsf{ModVar}.

{\small
  \[
  \begin{prooftree}
    \MOD{S} = \texttt{S}
    \using \textsf{(ModVar)}
    \justifies
    \begin{array}{c}
      \MOD{\bkeyw{VAR } x \bkeyw{ IN } S \bkeyw{ END}}
      \\ = \\
      \texttt{S } -~\{x\}
    \end{array}
  \end{prooftree}
  \]
}

\subsection{Beyond Substitutions}
\label{btojml:translation:bey_subs}
First the translation of an entire B machine into a \jml-annotated
Java class is presented, followed by the translation of the components of that
machine.  As presented here, the translation considers only a single
carrier set, only a single variable and so on, but can easily be extended
to multiple carrier sets, variables, etc.

{\small
  \[
  \begin{prooftree}
    \begin{array}{ll}
      \BTOJML{\bkeyw{SETS } s} = \texttt{S} & \BTOJML{\bkeyw{CONSTANTS }
        c} = \texttt{C}\\
      \BTOJML{\bkeyw{VARIABLES } v} = \texttt{V} & \BTOJML{\bkeyw{PROPERTIES }
        P} = \texttt{P}\\
      \BTOJML{\bkeyw{INVARIANT } I} = \texttt{I} & \BTOJML{\bkeyw{ASSERTIONS }
        A} = \texttt{A}\\
      \BTOJML{\bkeyw{INITIALISATION } B} = \texttt{B} & \BTOJML{Q} =
      \texttt{Q}
    \end{array}
    \using \textsf{(M)}
    \justifies
    \begin{array}{l}
      \BTOJMLN{(}
      \bkeyw{MACHINE } M\\
      \quad \bkeyw{SETS } s\\
      \quad \bkeyw{CONSTANTS } c\\
      \quad \bkeyw{PROPERTIES } P\\
      \quad \bkeyw{VARIABLES } v\\
      \quad \bkeyw{INVARIANT } I\\
      \quad \bkeyw{ASSERTIONS } A\\
      \quad \bkeyw{INITIALISATION } B\\
      \quad \bkeyw{OPERATIONS } Q\\
      \quad \bkeyw{ END}) ~ =\\
      \jmlkeyw{public abstract class}~M\jmlcode{ \{}\\
      \quad
      \texttt{S}~~\texttt{C}~~\texttt{V}~~\texttt{P}~~\texttt{I}~~\texttt{A}~~\texttt{B}
      \\ \\
      \quad \texttt{Q}\\
      \}
    \end{array}
  \end{prooftree}
  \]
}

\bmethod\ operations can contain input and output parameters as shown by
Rules \textsf{Oper}.

{\small
  \[
  \begin{prooftree}
    \begin{array}{l}
      \BTOJML{Q} = \texttt{Q}
    \end{array}
    \using \textsf{(Oper)}
    \justifies
    \begin{array}{c}
    \begin{array}{l}
      \BTOJML{op = Q} ~ =\\
      \quad \texttt{Q}\\
      \quad\jmlkeyw{public abstract void}~op\jmlcode{();}
    \end{array}
    \end{array}
  \end{prooftree}
  \]
}
{\small
  \[
  \begin{prooftree}
    \begin{array}{ll}
      \Type{r} = \texttt{Tr} & \Type{par} = \texttt{Tp}\\
      \BTOJML{Q} = \texttt{Q}
    \end{array}
    \using \textsf{(Oper)}
    \justifies
    \begin{array}{c}
\begin{array}{l}
      \BTOJML{r \leftarrow op(par) = Q} ~ =\\
      \quad\texttt{Q}\\
      \quad\jmlkeyw{public abstract }
      \texttt{Tr}~op\jmlcode{(}\texttt{Tp } par\jmlcode{);}
    \end{array}
    \end{array}
  \end{prooftree}
  \]
}

The return type of the corresponding method is either the
translated type of the single output parameter, or \jmlcode{Object []}
in order to contain the values of multiple output parameters.

As there is not information about carrier sets, they are simply modelled as sets of 
integers as shown in Rule \textsf{Set}.
All constants are being translated as \jmlkeyw{final ghost} variables, using
\jmlkeyw{ghost} variables because \jml\ \jmlkeyw{model} variables
can not be declared \jmlkeyw{final}.
Like \jmlkeyw{model} variables,
\jmlkeyw{ghost} variables are specification only and so do not appear
directly in implementations.  Note that different instances of the class
could use different carrier sets, so the field should not be static.

{\small
  \[
  \begin{prooftree}
    \using \textsf{(Set)}
    \justifies
    \begin{array}{c}
      \BTOJML{\bkeyw{SETS } s}
      \\=\\
      \jmlcode{/*@ } \jmlkeyw{public final ghost } \jmlcode{JMLEqualsSet<Integer> } s \jmlcode{ = }\\
 \quad \quad \quad  \quad       \jmlcode{ new Range(INT.Min, INT.Max); */}
    \end{array}
  \end{prooftree}
  \]
}

Rule \textsf{Enum} translates enumerated sets to 
sets of strings, where each string is the name of an enumeration
constant. Method \jmlcode{convertFrom} returns a set cointaing all
elements of the given array.

{\small
  \[
  \begin{prooftree}
    \using \textsf{(Enum)}
    \justifies
    \begin{array}{l}
      \BTOJML{\bkeyw{VARIABLES } v~=~\{s1, \ldots, sn\}}
      ~ =\\
      \jmlcode{/*@ } \jmlkeyw{public static final ghost}\\
      \quad \quad \quad\jmlcode{JMLEqualsSet<String> v}\\
      \quad \quad \quad =~ \jmlcode{JMLEqualsSet.convertFrom(}\\
      \quad \quad\jmlkeyw{new }\jmlcode{String [] } \{``s1'', \ldots, ``sn''\})\jmlcode{;}\jmlcode{*/} \\
    \end{array}
  \end{prooftree}
  \]
}

Rule~\textsf{Cons} uses \TypeN\ to infer the type of a constant
from the \bkeyw{PROPERTIES} section of the machine.

{\small
  \[
  \begin{prooftree}
    \Type{c}~=~\textsf{Type}
    \using \textsf{(Cons)}
    \justifies
    \begin{array}{c}
      \BTOJML{\bkeyw{CONSTANTS } c}
      \\=\\
      \jmlcode{//@ } \jmlkeyw{public static final ghost } \textsf{Type }
      c\jmlcode{;}
    \end{array}
  \end{prooftree}
  \]
}

As \bmethod\ \bkeyw{PROPERTIES} clauses specify properties of constants,
Rule \textsf{Prop} translates them as 
\jmlkeyw{static invariants}.

{\small
  \[
  \begin{prooftree}
    \PRED{P}~=~\textsf{P}
    \using \textsf{(Prop)}
    \justifies
    \begin{array}{c}
      \BTOJML{\bkeyw{PROPERTIES } P}
      \\=\\
      \jmlcode{//@ } \jmlkeyw{public static invariant } \textsf{P}\jmlcode{;}
    \end{array}
  \end{prooftree}
  \]
}

Ordinary machine variable declarations are translated to \jml\
\jmlkeyw{model} variables.
The type of the variable is inferred from the machine invariant.

{\small
  \[
  \begin{prooftree}
    \Type{v}~=~\textsf{Type}
    \using \textsf{(Var)}
    \justifies
    \begin{array}{c}
      \BTOJML{\bkeyw{VARIABLES } v}
      \\=\\
      \jmlcode{//@ } \jmlkeyw{public model } \textsf{Type } v\jmlcode{;}
    \end{array}
  \end{prooftree}
  \]
}


\bmethod\ \bkeyw{invariant} are translated to \jml\
  \jmlkeyw{invariants} and \bmethod\
assertions are translated as redundant invariants, as both assertions
and redundant invariants are implied by ordinary invariants.

{\small
  \[
  \begin{prooftree}
    \PRED{I}~=~\textsf{I}
    \using \textsf{(Inv)}
    \justifies
    \BTOJML{\bkeyw{INVARIANT } I} ~=~\jmlcode{//@ } \jmlkeyw{public invariant } \textsf{I}\jmlcode{;}
  \end{prooftree}
  \]
}

{\small
  \[
  \begin{prooftree}
    \PRED{I}~=~\textsf{I}
    \using \textsf{(Ass)}
    \justifies
    \BTOJML{\bkeyw{ASSERTIONS } I} ~=~\jmlcode{//@ } \jmlkeyw{public invariant\_redundantly } \textsf{I}\jmlcode{;}
  \end{prooftree}
  \]
}

A \bmethod\ \bkeyw{INITIALISATION} clause is translated to a \jml\ \jmlkeyw{initially}
clause, as both provide initial values for variables.  The assertion
within the \jmlkeyw{initially} clause uses \texttt{==} rather than calling
the \jmlcode{equals} method if the type of \texttt{v} is primitive.

{\small
  \[
  \begin{prooftree}
    \PRED{E}~=~\textsf{E}
    \using \textsf{(Init)}
    \justifies
    \BTOJML{\bkeyw{INITIALISATION } v~:=~E} ~=~\jmlcode{//@ } \jmlkeyw{initially } v\jmlcode{.equals(}\textsf{E}\jmlcode{);}
  \end{prooftree}
  \]
}


The language used in \bmethod\ expressions is essentially predicate logic and set
theory.  In the translation, sets, binary relations and 
binary functions are being presented by the \jml\ library model classes \jmlcode{JML\-Equals\-Set}, 
\jmlcode{JML\-Equals\-To\-Equals\-Relation}, and \jmlcode{JML\-Equals\-To\-Equals\-Map}
respectively.  These classes test membership using the \jmlcode{equals}
method of the class that the elements belong to, rather than the 
Java \texttt{==} operator.
Several examples of rules for translating \bmethod\ operators on these
types are presented below, where $s_i$'s are sets 
and $r$ is a relation.

{\small
  \[
  \begin{prooftree}
    \PRED{s_1}~=~\textsf{s1} ~~\PRED{s_2}~=~\textsf{s2}
    \using \textsf{(Subset)}
    \justifies
    \BTOJML{s_1 \subseteq s_2} ~=~\textsf{s1}.\jmlcode{isSubset(} \textsf{s2}\jmlcode{)}
  \end{prooftree}
  \]
}

{\small
  \[
  \begin{prooftree}
    \PRED{x}~=~\textsf{x} ~~\PRED{s}~=~\textsf{s}
    \using \textsf{(Has)}
    \justifies
    \BTOJML{x \in s} ~=~\textsf{s}.\jmlcode{has(} \textsf{x}\jmlcode{)}
  \end{prooftree}
  \]
}

{\small
  \[
  \begin{prooftree}
    \PRED{r}~=~\textsf{r} ~~\PRED{s}~=~\textsf{s}
    \using \textsf{(Image)}
    \justifies
    \BTOJML{r[s]} ~=~\textsf{r}.\jmlcode{image(} \textsf{s}\jmlcode{)}
  \end{prooftree}
  \]
}

\TypeN\ maps a \bmethod\ set type to the \jml\ model class 
\jmlcode{JML\-Equals\-Set},
a relation type to \jmlcode{JML\-Equals\-To\-Equals\-Relation}, and a function
type to \jmlcode{JML\-Equals\-To\-Equals\-Map}.  
As the types of \bmethod\ variables are specified implicitly (by stating membership
in some possibly deferred set),
the type must be inferred from its usage within the machine.
This type inference was already implemented in ABTools (see 
Section~\ref{btojml:implementation}), so in the implementation is translated
from the
representation of \bmethod\ types used by ABTools to the corresponding \jml\ types.
A library code to capture additional properties of \bmethod\ types is being used.  For instance,
given the \bmethod\ expression: 
\[d \in \pow(\nat)~ \& ~r \in \pow(\nat) ~\& ~f \in d \tfun r\]
that states that $f$ is a total function from $d$ to $r$, the type of $f$
is translated as \jmlcode{JML\-Equals\-To\-Equals\-Map<Integer, Integer>}
and the following is generated as part of the class invariant:

\[ \jmlcode{(}\jmlkeyw{new }\jmlcode{Total<Integer, Integer>(d, r)).has(f)} \]

Library class \jmlcode{org.jmlspecs.b2jml.util.Total} represents the set of
all total functions from
the specified domain to the specified range, so the
\jmlcode{has} method  returns true
if and only if $f$ is a total function from $d$ to
$r$. 

\section{The \btojml\  Tool}
\label{btojml:implementation}

The \btojml\ tool is integrated into ABTools~\cite{Boulanger2003}, that
is an open source environment for developing \bmethod\ language processing
tools.  Full source code for ABTools and \btojml\ is available
in~\cite{Boulanger2011}.  ABTools uses ANTLR \cite{Parr2007} to
generate a parser for \bmethod.  The parser constructs abstract syntax trees,
that are then traversed (using an ANTLR tree walker) to infer and
attach type information to each node.  ABTools can currently generate
refinement proof obligations and translate \bmethod\ machines to ASCII,
\LaTeX, HTML and XML formats, and has some initial support for
generating C and Java implementations.  This functionality is also
implemented via ANTLR tree walkers.

The bulk of the \btojml\ implementation is realised as an additional
ANTLR tree
walker, that implements the \BTOJMLN, \MODN, and \TypeN\  operators
presented previously.  The tree walker traverses
the syntax tree constructed by ABTools to generate the \jml\ specification
as indicated by the rules for \BTOJMLN, collecting the variables that
are modified by each operation as a side effect.
Additional utility classes implement the \bmethod\ operators on
functions, relations and sequences that do not directly correspond to
methods of the \jml\ model classes, as well as providing support for \bmethod\
typing via classes such as \texttt{org.\-jmlspecs.\-b2jml.\-util.\-Total}
as previously described.  

\paragraph{Installing the \btojml\ tool: } \btojml\ is part of the ABTools
distribution, so to use it one needs to install ABTools from
eclipse. It can be installed in Eclipse downloading the sources from
the SVN repository
\url{https://abtools.svn.sourceforge.net/svnroot/abtools}. To run
\btojml\ one needs to add the argument {\tt -toJML}. More detailed
instructions on how to install and use the tool can be found at
\cite{B2JML_webpage}.

\section{Using the \btojml\ Tool}
\label{btojml:apply}
We have validated \btojml\ tool by applying it to a moderate complex
\bmethod\ model of a Social Networking Site, an excerpt of this model is
presented in Chapter \ref{chapter:background} (Section \ref{b:example}
page \pageref{b:example})

\subsection{Generating \jml-annotated Abstract Java Classes}
\label{b2jml:example:jml_generated}

We used \btojml\ tool to translate the most abstract \bmethod\ machine
for the social networking site described in
\cite{matelas:2010}. Then, the resulting \jml-annotated
Java abstract class was typed and syntax-checked with OpenJML
\cite{Cok2011}.
Figure \ref{fig:b2jml:example:social-JML} presents the output of
applying the \btojml\
tool to the \bmethod\ model in Figure \ref{fig:b:machine}.  Figure
\ref{fig:b2jml:example:social-JML} shows how \btojml\ tool
translates \bmethod\ carrier sets as \jmlkeyw{final ghost} variables
with type \jmlcode{JMLEqualsSet<Integer>}, and \bmethod\ variables as
\jmlkeyw{model} variables and the type is calculated from the
\bmethod\ \jmlkeyw{invariant} using the \TypeN\ operator previously
introduced. The \bmethod\ \bkeyw{invariant} is translated as a \jml\
\jmlkeyw{invariant}, and the \bkeyw{initialisation} in \bmethod\ is
translated as \jmlkeyw{initially} clause in \jml. Class
\jmlcode{ModelUtils} defines the method \jmlcode{maplet} that receives
two parameters and returns an instance of class
\jmlcode{JMLEqualsEqualsPair} with both values, the method
\jmlcode{to\-Rel}  converts a set of pair to an instance of
\jmlcode{JML\-Equals\-To\-Equals\-Relation}, the method
\jmlcode{to\-Set} receives several parameters and returns a
\jmlcode{JML\-Equals\-Set} containing all parameters, and the method
\jmlcode{cartesian} that returns the cross-product of two sets.

\begin{figure}
\begin{lstlisting}[frame=none]
import org.jmlspecs.models.*;

public abstract class SOCIAL_NETWORK{
 //@ public final ghost JMLEqualsSet<Integer> PERSON = new Range(INT.Min,INT.Max); 
 //@ public final ghost  JMLEqualsSet<Integer> RAWCONTENT = new Range(INT.Min,INT.Max); 
 //@ public model  JMLEqualsSet<Integer> person;
 //@ public model JMLEqualsSet<Integer> rawcontent;
 //@ public model JMLEqualsToEqualsRelation<Integer,Integer> content;

 /*@ public invariant person.isSubset(PERSON)
      && rawcontent.isSubset(RAWCONTENT)
      && new Relation<Integer, Integer>(person, rawcontent)).has(content)
      && content.domain().equals(person)
      && content.range().equals(rawcontent);*/

 /*@ public initially person.isEmpty() &&
   rawcontent.isEmpty() && content.isEmpty();*/

  /*@ public normal_behavior
    requires rawcontent.has(rc) && person.has(ow)
      && person.has(pe) && !ow.equals(pe)
      && !content.has(ModelUtils.maplet(pe,rc));
    assignable content;
    ensures (\exists JMLEqualsSet<Integer> prs;
    \old(prs.isSubset(person));
      content.equals(\old(ModelUtils.toRel(
            content.union(ModelUtils.toRel(
            ModelUtils.maplet(pe,rc)))).union(ModelUtils.cartesian(prs,
            ModelUtils.toSet(rc))))));
   also public exceptional_behavior
    requires !(rawcontent.has(rc) && person.has(ow)
      && person.has(pe) && !ow.equals(pe)
      && !content.has(ModelUtils.maplet(pe,rc)));
    assignable \nothing; signals(Exception) true;*/
    public abstract void transmit_rc(Integer rc, Integer ow, Integer pe);
} 
\end{lstlisting}
\caption{A \jml\ specification of a social networking class.}
\label{fig:b2jml:example:social-JML}
\end{figure}

In the specification of the  $transmit\_rc$ method, the
\jmlkeyw{normal\_behavior} case guarantees that if the
\jmlkeyw{requires} clause (pre-condition) is satisfied, no exception
will be thrown, only the locations listed in the \jmlkeyw{assignable}
clause can be modified by the method, and the post-state will satisfy
the \jmlkeyw{ensures} clause (post-condition).  In an
\jmlkeyw{ensures} clause, expressions in \jmlcode{\bsl}\jmlkeyw{old}
are evaluated in the pre-state, while all other expressions are
evaluated in the post-state. The \jmlkeyw{exceptional\_behavior} case
specifies that the method will throw  an exception and no locations
will be modified if its pre-condition is satisfied.

As a further validation step, the translated specification was
executed  using the \jmle\ tool~\cite{jmle,jmle-jcard:09}.
This tool translates \jml\ specifications to constraint programs,
that can then be run using the Java Constraint Kit (JCK)
\cite{abdennadher02constraint}. Methods in the generated constraint
programs can be called from
ordinary Java code, so the programs can be used directly as
(large and slow) Java implementations of the \jml\ specifications they
were generated from.  As the translation rules were being developed
for \btojml, they were used to produce a hand-translation of the
social networking machine.  \jmle\ was used to execute this
hand-translation against a suite of JUnit test cases designed to check
that the behaviour of this translation was as expected.  This
also provided a convenient way to check that \btojml\ behaved as
expected - when the implementation was mature enough,  \btojml\ was
used to translate the \bmethod\ machine to \jml, and then used \jmle\ to
translate the \jml\ specification to a constraint program.  The suite
of JUnit test cases were ran against this program, confirming that the
behaviour of the specification generated by \btojml\ matched the
behaviour of the hand translation for this set of test cases.

Finally, as all operations of the \bmethod\ model have been verified
to preserve the machine invariant using Atelier B~\cite{atelierb}, the
tool gives the confidence that all methods in the \jml\ specification 
preserve the invariant as well.  Note that the meanings of \bmethod\
machine  invariants and \jml\ class invariants are closely related -
both are assertions that the machine variables/class fields must
satisfy both before and after the execution of any operation/public
method.

\section{Conclusion}
\label{btojml:conclusion}
In this chapter we presented some translation rules to produce \jml\
specifications from \bmethod\ machines. We also introduced the 
implementation of the rules as the \btojml\ tool that is integrated
to the ABTools suite. We validated \btojml\ by applying it to a social
networking model written in \bmethod. The \bmethod\ model is composed
of an abstract machine that defines the core of a social network
and five refinements that add functionality to it. \btojml\ was able to
generate \jml\ specifications from the abstract machine and all five
refinements of the Social
Networking \bmethod\ model. As a further validation of \btojml\ we used
OpenJML \cite{Cok2011} to type-check the \jml\ specifications
generated for the Social Networking \bmethod\ model. OpenJML
uncovered some problems with our tool regarding type inference of
variables, we used OpenJML's feed-back to correct these problems.

\btojml\ bridges Refinement Calculus with \bmethod\ and
Design-by-Contract with \jml\ allowing people from different
backgrounds to work together in the development of software. \btojml\
enables experts in \bmethod\ methodology to model systems in \bmethod,
where the model can be
proven correct w.r.t. some properties. Then, the user decides the
level of abstraction in \bmethod\ so as to generate \jml\
specifications of the model. From the \jml\ specification, an expert
in \jml\ can use \jml\ machinery, such as \jmle\ \cite{jmle}, to
validate the \jml\ specification. Finally, to manually write Java code and use \jml\
machinery, such as OpenJML \cite{Cok2011}, to verify if the manually
written Java code meets the \jml\ specification. \btojml\ makes
the generation of Java code from \bmethod\ models easier than directly
implement Java code from the \bmethod\ machine, or refine the
\bmethod\ machine close to an implementation machine. 

The work presented in this chapter has some limitations: \btojml\ does
not fully support the syntax underlying \bmethod. This imposes
restrictions to the user to translate \bmethod\ models to \jml; the
translation has not been proven correct in the logic of a prover. We
have validated \btojml\ by
applying it to a \bmethod\ model, however, to gain full
confidence of the tool we need to prove the soundness of the
translation rules; and generated \jml\ specifications contains generic
types that cannot (yet) be handled by the current \jml\ tools.

We decided not to maintain this tool since we realised that \eb\
offers more benefits (discussed later on) than \bmethod. We decided to
put our effort in defining and implementing a tool that works over
\eb\ (as explained in Chapter \ref{chapter:eb2jml}). 

\chapter{Translating \eb\ Machines to \jml\ Specifications}
\label{chapter:eb2jml}

\paragraph{This chapter. } The work presented in this chapter goes in
the same direction of the previous chapter. The
previous chapter describes how \bmethod\ machines can be translated
into \jml\ specifications, whilst this chapter shows a translation of \eb\ machines to
\jml\ specifications. \eb\ method is a derivative formalism of the \bmethod\ method,
and it is also introduced by Abrial J.-R. \cite{EB:Book}. 


We considered \eb\ machines instead of \bmethod\ machines
mainly because \eb\ is considered a stronger language than
\bmethod. For instance, \eb\ enables users to define new events that
refine the \textbf{skip} event, whereas this is not allowed in
classical \bmethod.  Secondly, an event in \eb\ can be refined as
several events whereas this it not possible in \bmethod. Thirdly, the
\bmethod\ method is devoted to the
development of Correctness-by-Construction software, whereas the purpose of
\eb\ is used to model full systems (including hardware, software and
environment). Fourthly, \eb\ provides more flexibility to model
systems as it is composed of contexts and machines, that allows 
users to separate the static and dynamic parts of a system. A context
contains definitions and properties of types and constants. A machine
contains state variables, invariants and events that update the variables. 

This chapter presents the definition rules for a translation from \eb\
to \jml\ specifications and the implementation of this translation as
the \ebtojml\ tool \cite{EventB2JML_webpage}, that is a plug-in for
the Rodin platform. Many of the \bmethod\ constructs for multiple
incremental specification of machines (such as the \bkeyw{INCLUDES},
\bkeyw{IMPORTS}, and \bkeyw{SEES} keywords) that we did not  implement
in \btojml\ tool are not included in \eb, so their translation rules
do not need to be defined for machine composition. The
translation has been validated  by applying the \ebtojml\ tool to a
moderately complex \eb\ model MIO, a model for a transportation
system. The MIO model is presented in
\cite{TeachFM-09}. The tool generated a
\jml-annotated Java abstract class. We further manually added
Java code for this abstract class. The rest of this chapter is
organised as follows. Section \ref{ebtojml:translation} presents our
approach to the translation of \eb\ models into \jml\
specifications. 
Section \ref{ebtojml:implementation} presents the
\ebtojml\ tool that implements the translation, and
Section  \ref{eb2jml:apply} shows an example to demonstrate our
approach and tool. Section \ref{ebtojml:conclusion} is
devoted for conclusions.

\paragraph{Contributions.}  The main contributions of this chapter are 
\begin{inparaenum}[\itshape i\upshape)]
\item the definition of a translation of
  \eb\ models to \jml\ through a collection of rules. 
\item The implementation of this translation as
  the \ebtojml\ tool.
\end{inparaenum}


\paragraph{Related work.}   In \cite{Mery:2011}, M\'ery and Singh
present the EB2ALL tool-set that includes the EB2C, EB2C\verb|++|,
EB2J, and EB2C$^{\sharp}$ plug-ins, each translating \eb\ machines to
the indicated language. Unlike \ebtojml, EB2ALL supports only
a small subset of \eb's syntax, and users are required to write a
final \eb\ implementation refinement in the syntax supported by the
tool. In \cite{ehdl}, Ostroumov and Tsiopoulos present the EHDL
prototype tool that generates VHDL code from \eb\ models. The tool
supports a reduced subset of \eb's syntax and users are required to
extend the \eb\ model before it can be translated. In \cite{Wright09},
Wright defines a B2C extension to the Rodin platform that translates
\eb\ models to C code.  The Code Generation tool \cite{CodeGen10}
generates concurrent Java and Ada programs for a \emph{tasking}
extension \cite{Tasking11} of \eb. As part of the process of generating
code with the Code Generation tool, the model need to be in a concrete
refinement, and users are asked to model the flow of the
execution of events in the tasking extension.  \ebtojml\ differs
from all of these tools in that  \ebtojml\ does not require user
intervention before translation, and can translate a much larger
subset of \eb\ syntax.

Jin and Yang \cite{jin2008}
outline an approach for translating VDM-SL \cite{Jones90} to
\jml. Their motivations are similar to ours in that they view VDM-SL
as a better language for modelling at an abstract level (much the way
that we view \eb), and \jml\ as a better language for working closer
to the implementation level.  In fact, they translate VDM variables to Java
fields, thus dictating the fields of an implementation. 

\section{The Translation from \eb\ to \jml}
\label{ebtojml:translation}
The translation from \eb\ to \jml\ is implemented with the aid of an 
\EBTOJMLN\ operator, that translates an \eb\ machine and any context
that it ``\ebkeyw{sees}'' to a \jml\ annotated Java abstract class.  Operator
\EBTOJMLN\ uses three helper operators that work in the same direction
as the previous Chapter, namely, \PREDN, \MODN, and \TypeN. 

\begin{figure}
  {\small 
    \[
    \begin{prooftree}
      \begin{array}{ll}
        \begin{array}{l}
          \EBTOJML{\ebkeyw{sets}\;s} = \texttt{S}\\
          \EBTOJML{\ebkeyw{constants}\;c} = \texttt{C}\\
          \EBTOJML{\ebkeyw{axioms}\;X(s, c)} = \texttt{X}\\
          \EBTOJML{\ebkeyw{theorems}\;T(s, c)} = \texttt{T}
        \end{array}
        &
        \begin{array}{l}
          \EBTOJML{\ebkeyw{variables}\;v} = \texttt{V}\\
          \EBTOJML{\ebkeyw{invariants}\;I(s, c, v)} = \texttt{I}\\
          \EBTOJML{\ebkeyw{events}\: e} = \texttt{E}
        \end{array}
      \end{array}
      \using \textsf{(M)}
      \justifies
      \begin{array}{ll}
        \EBTOJMLN(\\
        \begin{array}{l}
          \quad\ebkeyw{context}~ctx \\
          \quad~~\ebkeyw{sets}~s \\
          \quad~~\ebkeyw{constants}~c\\
          \quad~~\ebkeyw{axioms}~X(s, c)\\
          \quad~~\ebkeyw{theorems}~T(s, c)\\
          \quad\ebkeyw{end}
        \end{array} &
        \begin{array}{l}
          \quad\ebkeyw{machine}\;M\;\ebkeyw{sees}\;ctx \\
          \quad~~\ebkeyw{variables}\;v \\
          \quad~~\ebkeyw{invariants}\;I(s,c,v) \\
          \quad~~\ebkeyw{events}\;e \\
          \quad\ebkeyw{end}
        \end{array} 
        \\\quad \quad) =\\
        \javakeyw{public abstract class}~M \javacode{\{}\\
        \quad \texttt{S} ~~ \texttt{C} ~~ \texttt{X} ~~\texttt{T} ~~\texttt{V} ~~ \texttt{I} ~~ \texttt{E}\\
        \javacode{\}}
      \end{array}
    \end{prooftree}
    \]
  }
  \caption{The translation of the \eb\ machine $M$, and the context
    $ctx$ that $M$ sees to \jml-annotated Java abstract class.}
  \label{fig:eb2jml:translation:eb_machine}
\end{figure}

Figure~\ref{fig:eb2jml:translation:eb_machine} presents Rule
\textsf{M}, that translates
a machine $M$ that \ebkeyw{sees} context $ctx$. All \eb\ proof
obligations are assumed to be discharged before a machine is translated, so that
proof obligations and closely associated \eb\ constructs (namely
\ebkeyw{witness}es and \ebkeyw{variant}s) need not be considered in the
translation. A \ebkeyw{witness} contains the value of a disappearing abstract
event variable, and a \ebkeyw{variant} is an expression that should be
decreased by all \ebkeyw{convergent} \eb\ events and should not be
incremented by any \ebkeyw{anticipated} \eb\ events. An \eb\ machine is
translated to a single \jml-annotated Java abstract class,
that can then be extended by a subclass that implements the abstract
methods\footnote{Rule defined in
    Chapter \ref{chapter:eb2java} for \eb\ machines does not translate
    an \eb\ machine to an abstract Java class but to a Java concrete class since
    rules in Chapter \ref{chapter:eb2java} generate an actual
    implementation of an \eb\ machine.}.  This might allow the
  translation to be re-run when the \eb\ model is updated without the risk
  of losing hand-written or generated Java code\footnote{Notice that
    big changes in the \eb\ model may imply big changes in the
    Java abstract class generated that might invalidate the existing
    subclass.}.  The
translation of the context $ctx$ is incorporated into the translation of
machines that ``see'' the context.

\jml\ \jmlkeyw{model} fields are appropriate for representing machine
variables. Carrier sets and constants are also translated as
\jml\ \jmlkeyw{model} fields with the addition of a history
\jmlkeyw{constraint} that prevents any change in the value of the
field. \jmlkeyw{model} fields can be attached to a
\jmlkeyw{represents} clause to the declaration of the associated
implementation field\footnote{Carrier sets, constants and variables
  in Chapter \ref{chapter:eb2java} are translated as concrete Java (\javakeyw{static
    final}) fields, so no \jmlkeyw{model} field specification is then
  generated.}. As
there is no type information about carrier
sets in \eb, they are simply translated as sets of integers as
depicted by rule \textsf{Set}.  

{\small
  \[
  \begin{prooftree}
    \using \textsf{(Set)}
    \justifies
    \begin{array}{c}
      \EBTOJML{\ebkeyw{sets}\;s}
      = \\
      \jmlcode{/*@ } \jmlkeyw{public model }
      \jmlcode{BSet<Integer> } s\jmlcode{;}\\
      \quad\quad\quad\quad\quad\quad \jmlkeyw{public constraint }
      s\jmlcode{.equals(\bsl} \jmlkeyw{old}\jmlcode{(}s\jmlcode{)); */}
    \end{array}
  \end{prooftree}
  \]
}

Translation of constants and machine variables is similar, except that
constants are constrained to be immutable as previously described.
If the constant $c$ is of a primitive type, the translation will use
\texttt{==} rather than the \jmlcode{equals} method. 
The helper operator \TypeN\ translates the type of an \eb\ variable or
constant to the \jml\ representation of that type.

{\small
  \[
  \begin{prooftree}
    \Type{c} = \texttt{Type}
    \using \textsf{(Cons)}
    \justifies
    \begin{array}{c}
      \EBTOJML{\ebkeyw{constants}\;c}
      = \\
      \jmlcode{/*@ }\jmlkeyw{public model } \texttt{Type } c \jmlcode{;}\\
      \quad\quad\quad\quad
      \jmlkeyw{public constraint } c\jmlcode{.equals(\bsl }\jmlkeyw{old}\jmlcode{(}c\jmlcode{)); */}
    \end{array}
  \end{prooftree}
  \]
}

{\small
  \[
  \begin{prooftree}
    \Type{v} = \texttt{Type}
    \using \textsf{(Var)}
    \justifies
    \begin{array}{c}
      \EBTOJML{\ebkeyw{variables}\ v}
      = \\
      \jmlcode{//@ } \jmlkeyw{public model } \texttt{Type } v\jmlcode{;}
    \end{array}
  \end{prooftree}
  \]
}

As axioms are often used to specify properties of constants, they are
translated as invariants.  In \eb, theorems should be
provable from axioms, matching the semantics of the
\jmlkeyw{invariant\_redundantly} clause in
\jml\footnote{Notice that this definition does not match the
  definition of axioms in \eb\ since \jml\ \jmlkeyw{invariant}s should
  be verified, contrary to \eb\ \ebkeyw{axiom}s that are taken for granted. We
  corrected this issue in Chapter \ref{chapter:eb2java}. Rules defined in Chapter
    \ref{chapter:eb2java} for \eb\ axioms and theorems are defined as
    \javakeyw{static} making clearer that they should 
    refer just to carrier sets and constants.}. \eb\
\ebkeyw{invariants} are naturally translated as \jml\
\jmlkeyw{invariants}.   Operator \PREDN\ translates an \eb\ predicate
or expression to its \jml\ counterpart.

{\small
  \[
  \begin{prooftree}
    \PRED{X(s, c)} = \texttt{X}
    \using \textsf{(Axiom)}
    \justifies
    \begin{array}{c}
      \EBTOJML{\ebkeyw{axioms}\;X(s, c)} = \\
      \jmlcode{//@ } \jmlkeyw{public invariant } \texttt{X}\jmlcode{;}
    \end{array}
  \end{prooftree}
  \]
}

{\small
  \[
  \begin{prooftree}
    \PRED{T(s, c)} = \texttt{T}
    \using \textsf{(Theorem)}
    \justifies
    \begin{array}{c}
      \EBTOJML{\ebkeyw{theorems}\;T(s, c)} = \\
      \jmlcode{//@ } \jmlkeyw{public invariant\_redundantly } \texttt{T}\jmlcode{;}
    \end{array}
  \end{prooftree}
  \]
}

{\small
  \[
  \begin{prooftree}
    \PRED{I(s, c, v)} = \texttt{I}
    \using \textsf{(Inv)}
    \justifies
    \begin{array}{c}
      \EBTOJML{\ebkeyw{invariants}\;I(s, c, v)} = \\
      \jmlcode{//@ } \jmlkeyw{public invariant } \texttt{I}\jmlcode{;}
    \end{array}
  \end{prooftree}
  \]
}

An ${initialisation}$ event executes once to initialise the machine
variables, and so is naturally translated to a \jml\
\jmlkeyw{initially} clause. \EBTOJMLN\ is used recursively to
translate the actions of the ${initialisation}$ event.

{\small
  \[
  \begin{prooftree}
    \EBTOJML{A(s, c, v)} = \texttt{A}
    \using \textsf{(Init)}
    \justifies
    \begin{array}{c}
      \EBTOJML{\ebkeyw{events}\ initialisation\:\ebkeyw{then}\:A(s,c,v)
        \:\ebkeyw{end}}
      =\\
      \jmlcode{//@ } \jmlkeyw{public initially } \texttt{A}\jmlcode{;}
    \end{array}
  \end{prooftree}
  \]
}

Each other event is translated to two \jml\ methods: a
\javacode{guard} method that tests if the guard of the corresponding
event holds, and a \javacode{run} method that models the execution of
the corresponding event.  In Rule
\textsf{Any} below, variables bound
by an \ebkeyw{any} construct are existentially quantified in the
translation, as any values for those variables that satisfy the guards
can be chosen. The translation of the guard is included in the
post-condition of the \javacode{run} method in order to bind these same
variables, as they can be used in the body of the event. Translation
of an event defined using a \ebkeyw{when} construct (Rule
\textsf{When}) is simpler as no variables need to be
bounded. Translation of events uses an additional helper operator \MODN,
that calculates the set of variables assigned by the actions of an
event (the \jml\ \jmlkeyw{assignable} clause).  Rules \textsf{Any} and
\textsf{When} defined in Chapter \ref{chapter:eb2java} vary from the
ones presented here. Rules in Chapter
\ref{chapter:eb2java} translate an \eb\ method to a Java class that extends Java Thread so
 to simulate the execution of the system as \eb\ does. We defined the
 variables bound by an \ebkeyw{any} construct as parameter of the
 methods \javacode{guard} and \javacode{run} since is more natural to
 treat them as parameters. Thus the \jml\ spec does not define an
 quantifier existential.

{\small
  \[
  \begin{prooftree} 
    \begin{tabular}{ll}
      \Type{x} = \texttt{Type} &
      \PRED{G(s,c,v,x)} = \texttt{G} \\
      \MOD{A(s,c,v,x)} = \texttt{D} &
      \EBTOJML{A(s,c,v,x)} = \texttt{A} \\
    \end{tabular} 
    \using \textsf{(Any)}
    \justifies
    \begin{array}{c}
      \begin{array}{c}
        \EBTOJMLN(\ebkeyw{event}\ evt\:\ebkeyw{any}\;x\;\ebkeyw{where}
        \;G(s,c,v,x)\\
        \quad\quad\quad\quad\quad\quad\quad\ \ \ \ \ 
        \ebkeyw{then}\:A(s,c,v,x) \:\ebkeyw{end})
        =\\
      \end{array}
      \\
      \begin{array}{l}
        \jmlcode{/*@ } \jmlkeyw{requires true} \jmlcode{;}\\
        \quad\quad \, \jmlkeyw{assignable }\bsl\jmlkeyw{nothing}\jmlcode{;}\\
        \quad\quad \, \jmlkeyw{ensures }\bsl \jmlkeyw{result
        }\jmlcode{<==> (}\bsl \jmlkeyw{exists } \texttt{Type }x
        \jmlcode{; } \texttt{G}\jmlcode{); */}\\
        \javakeyw{public abstract boolean } \javacode{guard\_evt()}\javacode{;}\\
        \\
        \jmlcode{/*@ } \jmlkeyw{requires } \javacode{guard\_evt()}\jmlcode{;}\\
        \quad\quad \,\jmlkeyw{assignable } \texttt{D}\jmlcode{;}\\
        \quad\quad \,\jmlkeyw{ensures }\jmlcode{(}\bsl \jmlkeyw{exists }\texttt{Type }x\jmlcode{;}\;
        \bsl \jmlkeyw{old}\jmlcode{(}\texttt{G}\jmlcode{) \&\& }
        \texttt{A} \jmlcode{);}\\
        \quad \, \jmlkeyw{also}\\
        \quad\quad \, \jmlkeyw{requires } !\javacode{guard\_evt()}\jmlcode{;}\\
        \quad\quad \,\jmlkeyw{assignable }\bsl \jmlkeyw{nothing}\jmlcode{;}\\
        \quad\quad \,\jmlkeyw{ensures true}\jmlcode{; */} \\
        \javakeyw{public abstract void }\javacode{run\_evt()}\javacode{;}
      \end{array}
    \end{array}
  \end{prooftree}
  \]
}

{\small
  \[
  \begin{prooftree}
    \begin{array}{l}
      \PRED{G(s,c,v)} = \texttt{G}
      \quad
      \MOD{A(s,c,v)} = \texttt{D} \quad
      \EBTOJML{A(s,c,v)} = \texttt{A}
    \end{array}
    \using \textsf{(When)}
    \justifies
    \begin{array}{c}
      \begin{array}{c}
        \EBTOJMLN(\ebkeyw{event}\ evt\:\ebkeyw{when}\;G(s,c,v)\\
        \quad\quad\quad\quad\quad\quad\quad\quad\quad\quad\quad\ \ 
        \ebkeyw{then}\:A(s,c,v) \:\ebkeyw{end})
        =\\
      \end{array}
      \\
      \begin{array}{l}
        \jmlcode{/*@ } \jmlkeyw{requires true} \jmlcode{;}\\
        \quad\quad \, \jmlkeyw{assignable }\bsl\jmlkeyw{nothing}\jmlcode{;}\\
        \quad\quad \,\jmlkeyw{ensures }\bsl \jmlkeyw{result } \jmlcode{<==> } \texttt{G}\jmlcode{; */}\\
        \javakeyw{public abstract boolean } \javacode{guard\_evt()}\jmlcode{;}\\
        \\
        \jmlcode{/*@ } \jmlkeyw{requires } \javacode{guard\_evt()}\jmlcode{;}\\
        \quad\quad \,\jmlkeyw{assignable } \texttt{D}\jmlcode{;}\\
        \quad\quad \,\jmlkeyw{ensures } \texttt{A}\jmlcode{ \&\& \bsl}\jmlkeyw{old}\jmlcode{(}\texttt{G}\jmlcode{);}\\
        \quad \, \jmlkeyw{also}\\
        \quad\quad \, \jmlkeyw{requires } !\javacode{guard\_evt()}\jmlcode{;}\\
        \quad\quad \,\jmlkeyw{assignable }\bsl \jmlkeyw{nothing}\jmlcode{;}\\
        \quad\quad \,\jmlkeyw{ensures true}\jmlcode{; */} \\
        \javakeyw{public abstract void } \javacode{run\_evt()}\jmlcode{;}
      \end{array}
    \end{array}
  \end{prooftree}
  \]
}

The \jml\ specification of each \javacode{run} method uses two
specification cases.  In the first case, the translation of the guard
is satisfied and the post-state of the method must satisfy the
translation of the actions.  In the second case, the translation of
the guard is not satisfied, and the method is not allowed to modify
any fields, ensuring that the post-state is the same as the pre-state.
This matches the semantics of \eb\ -- if the guard of an event is not
satisfied, the event cannot execute and hence cannot modify the
system state. 

The translation of ordinary and non-deterministic assignments
via operator \EBTOJMLN\ is presented below.
The symbol $:\!|$ represents non-deterministic
assignment. 
The translation does not generate the $\bsl$\jmlkeyw{old} operators
shown below when translating an ${initialisation}$ event to an
\jmlkeyw{initially} clause.
If variable $v$ is of a primitive type, the translation will use
\texttt{==} rather than the \texttt{equals} method. 

{\small
  \[
  \begin{prooftree}
    \PRED{E(s, c, v)} = \texttt{E}
    \using \textsf{(Asg)}
    \justifies
    \begin{array}{c}
      \EBTOJML{v\::=E(s, c, v)\:} = 
      v\jmlcode{.equals(\bsl} \jmlkeyw{old}\jmlcode{(}\texttt{E}\jmlcode{))}\\
    \end{array}
  \end{prooftree}
  \]
}

{\small
  \[
  \begin{prooftree}
    \PRED{P(s, c, v, v')} = \texttt{P}
    \quad \Type{v} = \texttt{Type}
    \using \textsf{(NAsg})
    \justifies
    \begin{array}{c}
      \EBTOJML{v:\!|\:P(s, c, v, v')} = \\
      \jmlcode{(\bsl}\jmlkeyw{exists } \texttt{Type } v'\jmlcode{; \bsl} \jmlkeyw{old}\jmlcode{(}\texttt{P}\jmlcode{) \&\&}\;
      v\jmlcode{.equals(}v'\jmlcode{))}
    \end{array}
  \end{prooftree}
  \]
}

Multiple actions in the body of an event are translated individually and
the results are conjoined.  For example, a pair of actions:

{\small
  \[
  \begin{array}{c}
    \ebtag{act1 }x := y\\
    \ebtag{act2 }y := x\\
  \end{array}
  \]
}

is translated to \jmlcode{x == \bsl}\jmlkeyw{old}\jmlcode{(y) \&\& y
  == \bsl}\jmlkeyw{old}\jmlcode{(x)} for integer
variables $x$ and $y$, that correctly models simultaneous actions
as required by the semantics of \eb.

\subsection{The helper Operators}
\label{eb2jml:translation:oper}
The semantics of operators \MODN, \PREDN, and \TypeN\ is the same as
described in Chapter \ref{chapter:b2jml} except that their parameters
are changed from \bmethod\ to \eb\
constructs. \MODN\ operator that \emph{collects} the variables
assigned by \eb\ actions defines an additional \eb\ construct
(presented below). In this rule, the $v$ is a relation and
$i$ is an element in the range of $v$.

{\small
  \[
  \begin{array}{l}
    \MOD{v(i)\::=E\:} = \{v\}
  \end{array}
  \]
}

An additional rule defining \PREDN\ (below) that translate applications of
\eb\ operators to calls of the corresponding methods of classes BSet and
BRelation is presented below. In these rules, the $x$ and $y$ are
sets, and the $r$ is a
relation, and the $rd$ and $rr$ are sets representing
the domain and range of the relation $r$.

{\small
  \[
  \begin{prooftree}
    \begin{array}{ll}
    \PRED{x}~=~\textsf{x} &\PRED{y}~=~\textsf{y}
    \end{array}
    \using \textsf{(Pow)}
    \justifies
    \begin{array}{l}
    \PRED{x \in \pow{(y)}} ~=~
    \textsf{y}\jmlcode{.pow().has(}\textsf{x}\jmlcode{)}
    \end{array}
  \end{prooftree}
  \]
}

{\small
  \[
  \begin{prooftree}
    \begin{array}{ll}
    \begin{array}{l}
    \Type{rd}~=~\textsf{T1} \\ \Type{rr}~=~\textsf{T2}
    \end{array}
&
    \begin{array}{l}
    \PRED{rd}~=~\textsf{rd} \\\PRED{rr}~=~\textsf{rr}
    \end{array}
    \end{array}
    \using \textsf{(RelHas)}
    \justifies
    \begin{array}{l}
    \PRED{r \in rd \rel rr} ~=\\
    \jmlcode{new
      BRelation<}\textsf{T1}
    \jmlcode{,}
    \textsf{T2}\jmlcode{>(}\textsf{rd}\jmlcode{,}\textsf{rr}\jmlcode{).has(}r
    \jmlcode{);}
    \end{array}
  \end{prooftree}
  \]
}

Notice that this definition allows users to define nested
relations. For instance, the translation of \PRED{f \in fd \rel (fr
  \rel fs)}, where $f$ is a relation, and $fd$, $fr$ and $fs$ are
sets, would be:

\[
\jmlcode{new BRelation<} \Type{fd} \jmlcode{,} \Type{
fr
  \rel fs
}\jmlcode{>}\\
\quad \jmlcode{(}fd
\jmlcode{, new
  BRelation<}\Type{fr}\jmlcode{,}\Type{fs}\jmlcode{>}\\
\quad \quad \jmlcode{(}fr\jmlcode{,}fs\jmlcode{)).has(}f\jmlcode{);}
\]



The \PREDN\ operator translates \eb\ predicates, boolean, relational
and arithmetic expressions in the natural way.  While some operations
on \eb\ sets, functions and relations have direct counterparts in the
model classes \texttt{JML\-Equals\-Set} and
\texttt{JML\-Equals\-To\-Equals\-Relation} that are built-in to \jml\
 (as shown in Chapter \ref{chapter:b2jml}),
many other operations do not.  To supply these operations, an
implemented (and specified in \jml) model classes is presented (see
Section \ref{eb2java:imp:prelude}) \texttt{BSet} (as a subclass
of \texttt{JML\-Equals\-Set}) and \texttt{B\-Rel\-a\-tion} (as a
subclass of \texttt{BSet}, a \texttt{BSet} of pairs).  Note that an \eb\ relation can be used
anywhere that a set can appear (a relation is a set of pairs), but
unfortunately \texttt{JML\-Equals\-To\-Equals\-Relation} is not a
subclass of \texttt{JML\-Equals\-Set}.  Particular types of \eb\ relations (total relations, functions, etc.)
are translated as \jmlcode{BRelation}s with
appropriate restrictions in the invariant as explained in Chapter
\ref{chapter:b2jml}.




The \TypeN\ operator translates the type of \eb\ variables and
constants given by Rodin to the corresponding \jml\ type, instead of
inferring the type as explained in Chapter
\ref{chapter:b2jml}.  All
integral types are
translated as type \jmlcode{Integer}, all relations and functions are
translated as type \jmlcode{BRelation}, and all other sets are
translated as type \jmlcode{BSet} (Section
\ref{eb2java:imp:prelude}
explains the implementation of \jmlcode{BSet} and \jmlcode{BRelation}).

\subsection{A Java Framework for \eb}
\label{eb2jml:translation:framework}
An \eb\ machine continues to operate until no event can be executed --
in particular, a machine can run indefinitely if the guard of at least
one event always holds\footnote{\ebkeyw{variant}s are not considering
  in this discussion.}.  Any event with a satisfied guard can be
executed, and all event executions are atomic.
Class \texttt{Framework} of
 Figure~\ref{fig:eb2jml:translation:framework} presents a
typical scheduler implementation of  this behaviour, assuming that
class \texttt{M\_impl} extends the abstract class resulting from the
translation of an \eb\ machine $M$, that overrides
all abstract methods of class \texttt{M} appropriately, and that the
events of machine $M$ are ${evt\_1}, {evt\_2}, \ldots, {evt\_n}$.
Note that the result of the translation is a \jml-annotated Java
abstract class that must extended by a non-abstract class
(\texttt{m\_impl} in this case) before the methods can be executed.

\begin{figure}
\begin{lstlisting}[frame=none]
public class Framework {
    public static void main(String[] args) {
        M M_impl = new M_Impl();
        int n = /* the number of events in M */;
        java.util.Random r = new java.util.Random();
        while (m_impl.guard_evt_1() || m_impl.guard_evt_2() 
               || ...  || m_impl.guard_evt_n()) {
            switch (r.nextInt(n)) { 
                case 0 : if (m_impl.guard_evt_1()) 
                             m_impl.run_evt_1(); break;
                ...
                case n - 1 : if (m_impl.guard_evt_n()) 
                                 m_impl.run_evt_n(); break;
            }
        }
    }
}
\end{lstlisting}
\caption{A framework for executing \jml-annotated Java classes
  translated from \eb\  machines.}
\label{fig:eb2jml:translation:framework}
\end{figure}

\section{The \ebtojml\ Tool}
\label{ebtojml:implementation}
The \ebtojml\ tool is implemented as a Rodin plug-in.  It uses the
recommended interfaces \cite{rodinPlugin}
to traverse the statically checked internal database of Rodin.
\ebtojml\ was developed in Java and has been tested on version 2.8 of
Rodin.

\ebtojml\ uses the Rodin API to collect all components of a machine
such as carrier sets, constants, axioms, variables,
invariants and events. Furthermore, it also collects necessary
information such as
the gluing invariant from the refined machines. 
The collected information is stored in the Rodin database and can be
accessed using the \javacode{org.eventb.core} library. \eb\ expressions
and statements are parsed and stored as abstract syntax trees, that
can be accessed and traversed using the AST library in the
\javacode{org.eventb.core.ast} package \cite{ASTRodin}. The AST
library provides services such as parsing a mathematical formula from
a string of characters (typically entered by the end-user), and
traversing the AST by 
implementing the \verb+Visitor+ design pattern. 

The \ebtojml\ implementation uses the visitor design pattern to
traverse the abstract syntax tree and generate \jml\
specifications as described in Section~\ref{ebtojml:translation}.
\eb\ includes set-theoretical notations that are not defined in 
\jml. Therefore, we developed additional utility classes that implement \eb\
operations on sets and relations. This  allows users to generate \jml\ specifications from
any stage of the \eb\ system (i.e. from the most abstract machine to
any refinement).
The generated specifications include the imports of the library
classes just described (in the \javacode{poporo.models.JML} package),
and is written to a file with the same name as the machine being
translated (with \verb+.java+ extension). Full source code for
\ebtojml\ is available at \cite{EventB2JML_webpage}.

\paragraph{Installing \ebtojml\ tool: } To work with the plug-in, one
must download Rodin from
\url{http://sourceforge.net/projects/rodin-b-sharp/} (\ebtojml\ has
been tested in Rodin version 2.8). Then one needs to add \ebtojml\
update site to the list of the update-sites in Rodin
\footnote{\ebtojml\ update site:
  \url{http://poporo.uma.pt/Projects/EventB2JmlUpdate}}.
More detailed instruction on how to install and use the tool can be
found at \cite{EventB2JML_webpage}. This tool is not longer maintained
since we updated the translation to include Java code (see Chapter
\ref{chapter:eb2java}).

\section{Using the \ebtojml\ Tool}
\label{eb2jml:apply}
We validated \ebtojml\ tool by applying it to a moderate complex
\eb\ model, the MIO model \cite{TeachFM-09}. Subsection
\ref{eb2jml:apply:EBexample} explains the input \eb\ model and Subsection
\ref{eb2jml:apply:ex_output} shows the output of \ebtojml\ on the \eb\ model. This subsection also shows an
excerpt of the implementation of the \jml-annotated Java abstract
class generated by the tool.

\subsection{An Example in \eb}
\label{eb2jml:apply:EBexample}
The MIO  is an \eb\ model of a transportation system. It
includes articulated buses that follow the main corridor routes of a
city \cite{TeachFM-09}. The transportation system is complemented with feeding buses
that connect the city with its outskirts. A partial \eb\ model of the
MIO is depicted in figures \ref{fig:eb2jml:apply:mioAbstract} and
\ref{fig:eb2jml:apply:mioRef1} (abstract and first refinement
machines). The abstract machine (see
left part of Figure~\ref{fig:eb2jml:apply:mioAbstract}) models the number of
parked buses
through the variable $parked$, and defines an \ebkeyw{invariant}
$parked \in 0
\upto min({n,m})$ that must hold before and after all the machine
events. The constants $n$ and $m$ represent the abstractions for the
number of buses and stations of the system. These
constants are defined in the machine context $ctx$ (see right part of 
Figure~\ref{fig:eb2jml:apply:mioAbstract}).

\begin{figure}[t]
  {\small
    \[ 
    \begin{array}{ll}
      \begin{array}{l}
        \ebkeyw{machine}~abstract~\ebkeyw{sees}~ctx1\\
        ~\ebkeyw{variables} ~~parked\\    
        ~\ebkeyw{invariant}\\
        ~~\ebtag{inv1 } parked \in 0 \upto min(\{n,m\})\\
        ~\ebkeyw{events}\\
        ~~initialisation \\
        ~~~\ebkeyw{then} \\ 
        ~~~~\ebtag{act1 }parked := 0 \\
        ~~~\ebkeyw{end} \\
        ~~leave \\
        ~~~\ebkeyw{when}\\
        ~~~~\ebtag{grd1 }parked > 0 \\
        ~~~\ebkeyw{then} \\
        ~~~~\ebtag{act1 }parked := parked - 1\\
        ~~\ebkeyw{end}\\
        \ebkeyw{end} 
      \end{array}
      &
      \begin{array}{l}
        \ebkeyw{context}~ctx1~ \\ \\
        \ebkeyw{contants}~n~m \\ \\
        \ebkeyw{axioms} \\
        ~~\ebtag{ax1} ~ n \in \nat1 \\
        ~~\ebtag{ax2} ~ m \in \nat1 \\
        \ebkeyw{end}
      \end{array}
    \end{array}
    \]
  }
  \caption{An excerpt of the abstract \eb\ machine for MIO (left: machine
    $abstract$. Right: context $ctx1$ that $M$ \ebkeyw{see}).}
  \label{fig:eb2jml:apply:mioAbstract}
\end{figure}

The abstract machine further defines the event $leave$ that models
when a bus leaves any station. At this stage, the model is abstract,
therefore it does not
represent the specific bus or the station. Unlike \bmethod,
in which operations are called, \eb\ defines events that might be
executed/triggered when the guard is true. For instance, in order for
a bus to leave a station ($leave$ event), the number of parked
buses must be greater than 0. The guard of the event is represented as $parked >
0$, specifying that there is at least one bus parked. When this guard
is satisfied, the event might trigger its actions, decreasing the
variable $parked$ by 1 ($parked := parked - 1$). This machine defines
other events, however for the sake of brevity, we are showing only few.

\begin{figure}[t]
  {\small
    \[ 
    \begin{array}{ll}
      \begin{array}{l}
        \ebkeyw{machine}~ref1~\ebkeyw{refines}~abstract~\ebkeyw{sees}~ctx2\\
        ~\ebkeyw{variables}~parked~~~busStat\\
        ~\ebkeyw{invariant} \\    
        ~~\ebtag{inv1r1 }busStat \in  \ebkeyw{BUSES} \pinj \ebkeyw{STATS} \\
        ~~\ebtag{inv2r1 }card(busStat) = parked \\
        ~~\ebtag{inv3r1 }finite(busStat) \\
        ~~\ebtag{inv4r1 }card(busStat) <= min(\{n,m\}) \\
        ~\ebkeyw{events} \\
        ~~initialisation  \ebkeyw{ extends } initialisation\\
        ~~~\ebkeyw{begin}~~~\ebtag{actr1 }busStat := \emptyset \\
        ~~~\ebkeyw{end} \\
        ~~leave~~\ebkeyw{extends}~~leave\\
        ~~~\ebkeyw{any}~b\ebkeyw{ where}\\
        ~~~~\ebtag{grdr1 }b \in dom(busStat)\\
        ~~~\ebkeyw{then}\\
        ~~~~\ebtag{actr1 } busStat :=  busStat \bsl \{b \mapsto busStat(b)\}\\
        ~~\ebkeyw{end} \\
        \ebkeyw{end}
      \end{array}
      &
      \begin{array}{l}
        \ebkeyw{context}~ctx2~\ebkeyw{extends}~ctx1~ \\ \\
        \ebkeyw{sets}~\ebkeyw{BUSES } \ebkeyw{STATS} \\
        \ebkeyw{end}
      \end{array}
    \end{array}
    \]
  }
  \caption{Part of an \eb\ machine for MIO (left: refinement 1 machine. Right: context).}
  \label{fig:eb2jml:apply:mioRef1}
\end{figure}

The refinement of this machine (see left part of
Figure~\ref{fig:eb2jml:apply:mioRef1}) introduces more details to the
system. It declares (in the context, right
Figure~\ref{fig:eb2jml:apply:mioRef1}) two sets, \ebkeyw{BUSES} and
\ebkeyw{STATS}, representing the set of all possible buses and the
set of all possible stations in the system. The refinement machine
defines
another variable $busStat$ that maps buses to stations,
representing which bus is parked at which station. The variable
$busStat$ is defined as a partial injective function (denoted in
\eb\ as $\pinj$), that
enforces that a bus in the domain of $busStat$ (buses parked) must
be in one station only and that each station can hold just one
bus. The refinement machine extends the abstract event $leave$ by
adding more details (it also extends other events from the abstract
machine not shown in the figure). Specifically, in order for a bus $b$ to leave a
station (the clause \ebkeyw{any} gives the machine implementer the
opportunity to choose any value that satisfies the predicate in the
guard), the bus $b$ must be a bus of the system and
needs to be parked at one station (\ebtag{grdr1}). If the guard holds,
the actions
might be executed. Hence, the number of parked buses is decremented by
one and the pair $\{b \mapsto busSta(b)\}$ is subtracted  to the function
$busStat$, indicating that bus $b$ left the station where it was
parked (\ebtag{actr1}).

\subsection{The \jml-annotated Java abstract class}
\label{eb2jml:apply:ex_output}
We used \ebtojml\ tool to generate a Java abstract class of the MIO
\eb\ model. Figure \ref{fig:eb2jml:apply:ex_output:mioJAVA} depicts an
excerpt of the output
generated by the tool. It defines \eb\ carrier \ebkeyw{set},
\ebkeyw{constants}, and
\ebkeyw{variables} with the \jml\ \jmlkeyw{model} clause so that a
programmer can bind those
variables to an actual implementation. Carrier \ebkeyw{sets} and
\ebkeyw{constants} are
defined with a \jml\ \jmlkeyw{constraint} that prevents them to mutate
their values. \eb\ \ebkeyw{sets} are defined as
\jmlcode{finite} since \eb\ axioms \ebtag{ax1} and \ebtag{ax2} specify
that.

\begin{center}
  \begin{figure}
    \begin{lstlisting}[frame=none]
public abstract class ref1{
   /*@ public model Integer m;
    public constraint m.equals(\old(m)); */
   // ... n
   /*@ public model BSet<Integer> BUSES;
     public constraint BUSES.equals(\old(BUSES)); */
   // ... STATS

   // ... 
   //@ public static invariant BUSES.finite();
   //@ public static invariant STATS.finite();
   //@ public static invariant BUSES.int_size() == n;
   //@ public static invariant STATS.int_size() == m;

   //@ public model BRelation<Integer,Integer> busStat;
   //@ public model Integer parked;
	
   /*@ public invariant
     (new Range(0,(new BSet<Integer>(n,m)).min())).has(parked) &&
     busStat.isaFunction() && busStat.inverse().isaFunction() && busStat.domain().isSubset(BUSES) && busStat.range().isSubset(STATS) &&
     busStat.finite() && busStat.int_size() == parked; */

   /*@ public initially  parked == 0 && busStat.isEmpty(); */ 

   /*@ requires true; assignable \nothing;
    ensures \result<==>(\exists Integer b; parked>0 && busStat.domain().has(b));*/
    public abstract boolean guard_leave();

   /*@ requires guard_leave(); assignable parked, busStat;
     ensures (\exists Integer b; 
        \old((parked>0 && busStat.domain().has(b))) &&  parked == \old(parked-1) 
       && busStat.equals(\old(busStat.difference((new BRelation<Integer,Integer>((new JMLEqualsEqualsPair<Integer,Integer>(b,busStat.apply(b)))))))));
    also
     requires !guard_leave(); assignable \nothing;
     ensures true; */
    public abstract void run_leave();
}
    \end{lstlisting}
\caption{A partial \jml\ specification of the MIO \eb\ model.}
  \label{fig:eb2jml:apply:ex_output:mioJAVA}
\end{figure}
\end{center}

Figure \ref{eb2jml:apply:ex_output:mioJAVAImpl} shows an implementation of the abstract
Java class of Figure \ref{fig:eb2jml:apply:ex_output:mioJAVA}. Carrier
sets and constants
are defined as \javakeyw{final} so that they cannot mutate their
value. All variables contain the \jml\ \jmlkeyw{represents} clause that binds
the abstract variable with the actual definition.

\begin{center}
  \begin{figure}
    \begin{lstlisting}[frame=none]
public class ref1_impl extends ref1{

   public final Integer mI = 3; //@ represents m = mI;
   public final Integer nI = 3; //@ represents n = nI;
   public final BSet<Integer> BUSESI = new BSet<Integer>(1,2,3); //@ represents BUSES = BUSESI;
   public final BSet<Integer> STATSI = new BSet<Integer>(1,2,3); //@ represents STATS = STATSI;
	
   public BRelation<Integer,Integer> busStatI; //@ represents busStat = busStatI;
   public Integer parkedI;	//@ represents parked = parkedI;  
	
   @Override
   public boolean guard_leave() {
      return (parkedI > 0 && busStatI.domain().has(b));
   }

   @Override
   public void run_leave() {
      if (guard_leave()){
            parkedI = parkedI - 1;
            busStatI = busStatI.difference((
              new BRelation<Integer,Integer>((
                new JMLEqualsEqualsPair<Integer,Integer>(b,busStatI.apply(b))))));
      }
}
    \end{lstlisting}
\caption{An implementation for the abstract Java class presented in
  Figure \ref{fig:eb2jml:apply:ex_output:mioJAVA}.}
  \label{eb2jml:apply:ex_output:mioJAVAImpl}
\end{figure}
\end{center}

As validation step, the generated \jml\ specifications (partially depicted
by Figure \ref{fig:eb2jml:apply:ex_output:mioJAVA}) was executed using
the \jmle\ tool
\cite{jmle,jmle-jcard:09}, validating the
syntax and type correctness of the generated file. The \jmle\ tool translates
\jml\ specifications to constraint programs, that can then be run using
the Java Constraint Kit (JCK) \cite{abdennadher02constraint}. 
Methods in the generated constraint programs can be called from
ordinary Java code, so the programs can be used directly as (large and
slow) Java implementations of the \jml\ specifications they were
generated from. 

\section{Conclusion}
\label{ebtojml:conclusion}

In this chapter we presented a set of syntactic rules to translate
\eb\ models to \jml\ specifications. We also introduced the
implementation of these rules as the \ebtojml\ tool that is a Rodin's
plug-in. We validated \ebtojml\ by applying it to a model of a
transportation system (MIO) written in \eb. Then, we manually wrote
Java code from the \jml\ specifications. Working with \ebtojml\
suggests us that software developers can
find the tool appealing to the development of software, specially to
develop critical software. One of the advantages of \ebtojml\ is
that it enables users to first model the system in \eb\ where the user
can prove the system consistent, to then transition to \jml\
specifications, where the user can manually write Java code. Our
experience also suggests that \ebtojml\ makes the use of \eb\ formal
method more popular since the user does not have to refine the \eb\
model until an implementation, that is heavy burden, rather the
user decides the level of detail in \eb\ and then translate
the model to \jml.

As a validation step, we applied \ebtojml\ to an \eb\ model, this gave
us the insight that \ebtojml\
provides a relatively quick and easy way to generate a Java
implementation from an \eb\ model. However, we need to apply our tool
in a wider variety of models. We found out that the process of
manually generating Java code from the \jml\
specifications generated by \ebtojml\ can be optimised. We decided to
upgrade the \ebtojml\ tool to the \ebtojava\ tool (explained it later
on this document) so that it can automatically
generate Java code along with the \jml\
specification\footnote{\ebtojml\ is not longer maintained.}. This will
free the user from the burden of manually writing the Java code. Chapter
\ref{chapter:eb2java} discusses this upgrade. 
\chapter{Translating \eb\ Machines to \jml-annotated Java Code}
\label{chapter:eb2java}

\paragraph{This chapter. } The previous chapters reflect the evolution
of this thesis so far. At
the beginning, we proposed a translation from \bmethod\ to \jml, but
eventually
we realised that \eb\ is a better starting point for the modelling
of the critical software. Accordingly, we implemented the \ebtojml\ tool
and generated Java code for the output \jml\ specification, manually. We
realised that the manual implementation of Java code
for the \jml\ specifications generated by the \ebtojml\ tool is not
only error-prone but also time consuming. Hence, this chapter presents
a translation of \eb\ machines to \jml-annotated Java classes and also the
implementation of the underlying translation rules as the \ebtojava\
tool as a Rodin plug-in.

Work done in this chapter is based on author's paper
\cite{conference:EB2Java:14}, co-authored by N. Cata\~no, a submitted
journal paper \cite{EventB2Java2014}, and a book chapter 
\cite{e-eb:to:jml-java}. The rest of this chapter is
organised as follows: Section~\ref{eb2java:translation} presents the
translation from \eb\ to \jml-annotated Java programs, and
Section~\ref{eb2java:imp} presents the implementation of the
\ebtojava\ tool. Section \ref{eb2java:apply} shows an example of
applying \ebtojava\ to an \eb\ model. Section \ref{eb2java:sw-dev} proposes
two software development strategies using the \ebtojava\ tool. Finally,
Section~\ref{eb2java:conclusion} concludes and mentions future work.

\paragraph{Contributions.}  The main contributions
of this work are 
\begin{inparaenum}[\itshape i\upshape)]
\item the definition of a full-fledged translation
  from \eb\ to \jml-annotated Java programs, and
\item the implementation of this translation as the \ebtojava\ tool.
\end{inparaenum}
The \ebtojava\ Java
code generator largely supports \eb's syntax. A first key feature of
this translation is that it can be applied to both abstract and
refinement machines.  Hence, \ebtojava\ tool users can generate code
for a very abstract (and incomplete) \eb\ model of a system, check
user's intention in Java - whether the system behaves as expected, and
then continue developing the \eb\ model to correct any issues and add
additional functionality as needed. \ebtojava\ can generate both
sequential and multi-threaded Java implementations of \eb\ models. 

A second key feature of this translation is the generation of
(\jml) formal specifications along with the Java code. This feature
enables users to write custom code that replaces the code generated by
\ebtojava, and then use existing \jml\ tools  to verify that the
custom code is correct.

\paragraph{Related work.}   In \cite{Mery:2011}, M\'ery and Singh
present the EB2ALL tool-set that includes the EB2C, EB2C\verb|++|,
EB2J, and EB2C$^{\sharp}$ plug-ins, each translating \eb\ machines to
the indicated language. Unlike \ebtojava, EB2ALL supports only a small
subset of \eb's syntax, and users are required to write a final \eb\
implementation refinement in the syntax supported by the tool. In
\cite{ehdl}, Ostroumov and Tsiopoulos present the EHDL prototype tool
that generates VHDL code from \eb\ models. The tool supports a reduced
subset of \eb's syntax and users are required to extend the \eb\ model
before it can be translated. In \cite{Wright09}, Wright defines a B2C
extension to the Rodin platform that translates \eb\ models to C code.
The Code Generation tool \cite{CodeGen10} generates concurrent Java
and Ada programs for a \emph{tasking} extension \cite{Tasking11} of
\eb. As part of the process of generating code with the Code
Generation tool, users have to decompose the \eb\ model by employing
the Machine Decomposition plug-in. The decomposed models are refined
and non-deterministic assignments are eliminated. Finally, users are
asked to model the flow of the execution of events in the tasking
extension.  \ebtojava\ differs from all of these tools in that
\ebtojava\ does not require user intervention before code generation,
and can translate a much larger subset of \eb\ syntax. In \cite{eventb2sql},
Wang and Wahls present the EventB2SQL tool that automatically
translates \eb\ models to Java classes that store all model data in a
relational database. Our work can be extended with this work to make
the values of machine variables persistent across executions of a
generated application.

In \cite{Damchoom}, Damchoom presents a set of rules that translate
\eb\ to Java. However, the rules account for only a small part of \eb's
syntax and have not been implemented. Toom 
et. al \cite{Toom08} have a similar motivation; they
present Gene-Auto, an automatic code generator toolset for translating
from high level modelling languages like Simulink/Stateflow
and Scicos to executable code for real-time embedded systems. Their
approach is to work at a higher level of abstraction when verifying a solution
(in the same way that we use \eb), and then to add implementation
details. 

Although the modelling of timing properties is not directly supported by
\eb, a discrete clock can certainly be designed and implemented in
\eb. In \cite{butler:time:eb:11,butler:disctime:eb:12}, Mohammad Reza
Sarshogh and Michael Butler introduce three \eb\ trigger-response
patterns, namely, deadlines, delays and expires, to encode discrete
timing properties in \eb. A ``deadline'' means that a set of events
must respond to a particular event within a bounded time. For a
``delay'', the set of response events must wait for a specified period
after the triggering of an event. An ``expiry''
pattern prevents response events from triggering after the occurrence
of an event. The authors translate timing properties as invariants, guards
and \eb\ actions. We are interested in investigating on how our code
generation framework can be extended to support timing properties in
\eb, and in encoding this extension in \ebtojava\ once the Rodin
platform fully supports the use of discrete timing events.

The Open Group has recently undertaken an effort to produce a Real-Time
Java programming language called Safety-Critical Java (SCJ)
\cite{SCJS:10} that augments Java with event handlers, memory areas
and a Real-Time Specification for Java \cite{Book:Wellings:2004}. The design
of SCJ is organised into levels so that it facilitates the certification of
Safety-Critical Systems. Providing support for the encoding of real
time properties in \ebtojava\ might require us to use SCJ rather than
Java as the implementation language for \eb.
In \cite{SCJ:Circus:13}, a refinement technique for developing SCJ
programs based on the Circus language is proposed. Circus is based on
Z, CSP, and Timed CSP so it can be used for the modelling of
safety-critical systems. However, code generation is not supported by
Circus.

\section{The translation from \eb\ machines to \jml-annotated Java
  Code}
\label{eb2java:translation}

\begin{figure}
  {\small 
    \[
    \begin{prooftree}
      \begin{array}{ll}
        \EBTOPROG{\ebkeyw{sets } S} = \texttt{S} \\
        \EBTOPROG{\ebkeyw{constants } c} = \texttt{C}\\
        \EBTOJML{\ebkeyw{axioms } X(s, c)} = \texttt{X}\\
        \EBTOJML{\ebkeyw{theorems } T(s, c)} = \texttt{T}\\
        \EBTOPROG{\ebkeyw{variables } v} = \texttt{V}\\
        \EBTOJML{\ebkeyw{invariants } I(s, c, v)} = \texttt{I}\\
        \EBTOPROG{\ebkeyw{variant } E(s, c, v)} = \texttt{Va}\\
        \EBTOPROG{\ebkeyw{events } e} = \texttt{E} \\
        \EBTOJML{\ebkeyw{event } initialisation \ebkeyw{ then } A(s,c,v) \ebkeyw{ end}} = \texttt{I1} \\
        \EBTOJAVA{\ebkeyw{event } initialisation \ebkeyw{ then } A(s,c,v) \ebkeyw{ end}} = \texttt{I2} \\
      \end{array}
      \using \textsf{(M)}
      \justifies
      \begin{array}{ll}
        \EBTOPROGN(\\
      \begin{array}{l}
        \quad\ebkeyw{machine } M \ebkeyw{ sees }ctx \\
        \quad~~\ebkeyw{variables } v \\
        \quad~~\ebkeyw{invariants } I(s,c,v) \\
        \quad~~\ebkeyw{variant } E(s,c,v) \\
        \quad~~\ebkeyw{event } initialisation \\
        \quad \quad\ebkeyw{ then } A(s,c,v) \ebkeyw{ end} \\
        \quad~~\ebkeyw{events } e \\
        \quad\ebkeyw{end}
      \end{array}
&
      \begin{array}{l}
\quad\ebkeyw{context } ctx \\
        \quad~~\ebkeyw{sets } S \\
        \quad~~\ebkeyw{constants } c\\
        \quad~~\ebkeyw{axioms } X(s, c)\\ 
        \quad~~\ebkeyw{theorems } T(s, c)\\
        \quad\ebkeyw{end}
      \end{array}
\\) =\\
        \texttt{E} \\
        \javakeyw{public class } \javacode{M \{}\\
        \quad \texttt{X} ~~ \texttt{T} ~~ \texttt{I} \\
        \quad \texttt{S} ~~\texttt{C} ~~ \texttt{V} \\
        \quad \texttt{Va}\\ 
        \quad \javakeyw{public }\javacode{Lock lock =}\\
        \quad~ \javakeyw{new}~ \javacode{ReentrantLock(} \javakeyw{true});\\\\
        \quad\jmlcode{/*@}~\jmlkeyw{public normal\_behavior}\\
        \quad\quad\quad\jmlkeyw{requires true}\jmlcode{;} \\
        \quad\quad\quad\jmlkeyw{assignable}~\jmlkeyw{\bsl{}everything}; \\
        \quad\quad\quad\jmlkeyw{ensures}~\jmlcode{I1; */}\\
        \quad \javakeyw{public}~\javacode{M()\{}\\
        \quad\quad \jmlcode{I2}\\
        \quad\quad \javacomment{// creation of Java Threads}\\
        \quad \}\\
        \texttt{\}}
      \end{array}
    \end{prooftree}
    \] 
  } 
  \caption{The translation of machine $M$, and the context $C$ that $M$
    \ebkeyw{sees}.}
  \label{fig:eb2java:translation:machines}
\end{figure}

We present our translation from \eb\ to Java and \jml\ using three 
operators (\EBTOPROGN, \EBTOJAVAN\ and \EBTOJMLN), that we define via
rewriting rules.  The primary operator is \EBTOPROGN, that 
translates \eb\ to \jml-annotated Java programs.
It uses \EBTOJAVAN\ to obtain the Java part of the
translation and \EBTOJMLN\ to obtain the \jml\ part (\EBTOJMLN\
operator was defined in Chapter \ref{chapter:eb2jml}. Here we will
discuss the main changes). For example, \eb\ \ebkeyw{invariants} are translated only as \jml\ specifications,
and so the definition of \EBTOJMLN\ has a rule for invariants, while
\EBTOJAVAN\ does not. On the other hand, the
translation of constants includes a Java part and a \jml\ part, so the
\EBTOPROGN\ rule for constants refers to both
\EBTOJAVAN\ and \EBTOJMLN\ rules for constants. The translation further employs
operators \MODN, \PREDN, and \TypeN\ defined in Chapter
\ref{chapter:eb2jml}. Operator \PREDN\ is used to translate \eb\
predicates and expressions to Java and \jml. This operator does not
handle some \eb\ constructs when it is used in translating to Java,
for instance, existential quantifiers. Section
\ref{eb2java:translation:helpers} lists all \eb\ constructs that are
not handle by operator \PREDN\ when it translates them to
Java. \FreeN\ operator that returns
the set of variables that occur free in an expression, and \STATONEN\ and
\STATTWON\ that are used in translating \eb\ machine \ebkeyw{variants}.

A machine is translated as a Java class. In translating an \eb\
machine, \EBTOPROGN\ not only considers the information provided by
the machine, but also the contexts the machine \ebkeyw{sees}. Figure
\ref{fig:eb2java:translation:machines} presents Rule \textsf{M} that translates a machine
$M$ that \ebkeyw{sees} context $ctx$. The machine is translated as a
Java class that includes \jml\ class and method specifications. The
translation of the machine includes the translation of the context the
machine \ebkeyw{sees}. Hence, the Java translation of the machine
includes the translation of carrier sets, constants, axioms and
theorems declared within the machine context. It also includes the
translation of variables and invariants declared within the
machine. 

Refinement machines are translated in the same way as abstract machines since
Rodin properly adds abstract machine components to the internal
representation of the refining machine. Refining and extending events
(defined using \ebkeyw{refines} and \ebkeyw{extends}, respectively)
are translated in the same manner as abstract events for the same reasons.

We translate carrier sets and constants as class attributes, and
restrict those attributes for verification purposes. As we
have no type information about carrier sets, they are simply
translated as sets of integers. \EBTOJMLN\ does not use the \jml\
keyword \jmlkeyw{model} for carrier sets, constants, or variables, as
defined in Chapter \ref{chapter:eb2jml},
since \EBTOPROGN\ does not generate a Java abstract class.

{\small
  \[
  \begin{prooftree}
    \EBTOJML{\ebkeyw{sets } S} = \texttt{SM} ~~
    \EBTOJAVA{\ebkeyw{sets } S} = \texttt{SA}
    \using \textsf{(Set)}
    \justifies
    \begin{array}{l}
      \EBTOPROG{\ebkeyw{sets } S} = \texttt{SM} ~ \texttt{SA}
    \end{array}
  \end{prooftree}
  \]
} 


{\small
  \[
  \begin{prooftree}
    \using \textsf{(Set)}
    \justifies
    \begin{array}{l}
      \EBTOJAVA{\ebkeyw{sets } S}
      = \\
      \jmlkeyw{public static final }~\javacode{BSet<Integer> S =} \\ 
      \quad\javakeyw{new}~\javacode{Enumerated(}\\
      \quad\quad \javacode{Integer.MIN\_VALUE,Integer.MAX\_VALUE);}\\
      
    \end{array}
  \end{prooftree}
  \]
}


Translation of constants follows a similar pattern to the translation
of carrier sets, except that in \eb, the values of constants are 
constrained by axioms. The helper operator \TypeN\ translates the type 
of an \eb\ variable or constant to the Java representation of that
type. Function \javacode{AxiomTheoremValue<Type>} returns a value of type
\javacode{Type} that satisfies the axioms defined in the contexts the
machine \ebkeyw{sees}\footnote{Function \javacode{AxiomTheoremValue<Type>} has
not yet been implemented in \ebtojava. The implementation of this
function is listed as future work of this thesis. There are many cases
in which this function cannot be implemented, in particular when
axioms or theorems contain infinite sets}.

{\small
  \[
  \begin{prooftree}
    \EBTOJML{\ebkeyw{constants}\;c} = \texttt{CM} ~~
    \EBTOJAVA{\ebkeyw{constants}\;c} = \texttt{CA}
    \using \textsf{(Cons)}
    \justifies
    \begin{array}{l}
      \EBTOPROG{\ebkeyw{constants}\;c} = \texttt{CM} ~ \texttt{CA}
    \end{array}
  \end{prooftree}
  \]
}


{\small
  \[
  \begin{prooftree}
    \begin{array}{l}
      \Type{c} = \texttt{Type} \\ 
      \javacode{val = AxiomTheoremValue<Type>()}
    \end{array}
    \using \textsf{(Cons)}\label{cons:label}
    \justifies
    \begin{array}{l}
      \EBTOJAVA{\ebkeyw{constants}\;c} = \\
      \javakeyw{public static final}~\javacode{Type c = val;}    
    \end{array}
  \end{prooftree}
  \]
}

\EBTOJMLN\ translates axioms as \jmlkeyw{static invariant}s, not as
\jml\ \jmlkeyw{invariant} as proposed in Chapter \ref{chapter:eb2jml},
see the rule below. A
JML \jmlkeyw{static invariant} can only refer to static fields, and so
this approach is consistent with our translation of
constants and carrier sets as \jmlkeyw{static} fields. Translating axioms to
\jmlkeyw{static invariant}s makes it clearer that they should not refer to
machine variables, for example. 

{\small
  \[
  \begin{prooftree}
    \PRED{X(s, c)} = \texttt{X}
    \using \textsf{(Axiom)}
    \justifies
    \begin{array}{l}
      \EBTOJML{\ebkeyw{axioms}\;X(s, c)} =\\
      \jmlcode{//@}~\jmlkeyw{public static invariant}~\jmlcode{X;}
    \end{array}
  \end{prooftree}
  \]
}



Machine variables are translated to class attributes. The JML keyword
\jmlkeyw{spec\_public} makes a \javakeyw{protected} or
\javakeyw{private} attribute or method \javakeyw{public} to any \jml\
specification.

{\small
  \[
  \begin{prooftree}
    \Type{v} = \texttt{Type}
    \using \textsf{(Var)}
    \justifies
    \begin{array}{l}
      \EBTOPROG{\ebkeyw{variables}\;v} = \\
      \jmlcode{/*@}~\jmlkeyw{spec\_public}~\jmlcode{*/}~~\javakeyw{private}~\javacode{Type v;}\\
    \end{array}
  \end{prooftree}
  \]
}

In \eb, every event must maintain the machine invariants. In \jml,
invariants state properties that must hold in every visible system
state, specifically after the execution of the class constructor and
after a method terminates. This is semantically equivalent to
conjoining the invariant to the post-condition of each method and the
constructor. Since the \ebkeyw{initialisation} event translates to the
post-condition of the class constructor (see below), and the actions
of each other event translate as the post-condition of an ``atomic''
\javacode{run\_evt} method (in Figure~\ref{fig-any}), \eb\
\ebkeyw{invariants} are naturally translated as \jml\
\jmlkeyw{invariants} as shown by rule \textsf{(Inv)} in Chapter
\ref{chapter:eb2jml}.



Machines include a specialised \ebkeyw{initialisation} event that
gives initial values to state variables.
This event is translated by \EBTOJMLN\ as the post-condition of the
(only) constructor for the Java class resulting from the translation
of the machine, and by \EBTOJAVAN\ as the body of that constructor.
Both translations give initial values to the translation of the 
machine variables.

{\small
  \[
  \begin{prooftree}
    \EBTOJML{A(s, c, v)} = \texttt{A}
    \using \textsf{(Init)}
    \justifies
    \begin{array}{l}
      \EBTOJML{\ebkeyw{event}\ 
        initialisation\:\ebkeyw{then}\:A(s,c,v) \:\ebkeyw{end}} = \jmlcode{A}
    \end{array}
  \end{prooftree}
  \]
}

{\small
  \[
  \begin{prooftree}
    \EBTOJAVA{A(s, c, v)} = \texttt{A}
    \using \textsf{(Init)}
    \justifies
    \begin{array}{l}
      \EBTOJAVA{\ebkeyw{event}\ initialisation\:\ebkeyw{then}\:A(s,c,v) \:\ebkeyw{end}}
      = \javacode{A}
    \end{array}
  \end{prooftree}
  \]
}

Other (non-initialisation) \eb\ events can be either
\ebkeyw{ordinary},  \ebkeyw{convergent} or
\ebkeyw{anticipated}. Convergent events are used for modelling
terminating systems. Anticipated events denote some abstract behaviour
that is to be made precise in a future refinement. Convergent events
must decrease the numeric machine \ebkeyw{variant} (or must remove
elements of the set machine \ebkeyw{variant}), and
\ebkeyw{anticipated} events must not increase the numeric machine
\ebkeyw{variant} (or must not add elements to the set machine
\ebkeyw{variant}).  Events that are \ebkeyw{convergent} or
\ebkeyw{anticipated} are only enabled if the value of the numeric
variant is non-negative (or the set variant is finite). An \eb\
\ebkeyw{variant} expression $``\ebkeyw{variant } E(s,c, v)$'' is
translated by \EBTOPROGN\ as a method that returns
the result of evaluating the translation of ${E}$\footnote{We translate \ebkeyw{variants} to \jml\ and Java since we
  came to realise that users might customise the Java
  code and after customisation users must be able to check that the
  customised \ebkeyw{convergent} event does monotonically decrease
  the machine \ebkeyw{variant} and the customised 
  \ebkeyw{anticipated} event does not increase it. Not as suggested in
  Chapter \ref{chapter:eb2jml}.}.

{\small
  \[
  \begin{prooftree}
    \PRED{E(s,c, v)} = \texttt{E} ~~~~ \Type{E(s,c, v)} = \texttt{T}
    \using \textsf{(Variant)}
    \justifies
    \begin{array}{c}
      \begin{array}{c} 
        \EBTOPROG{\ebkeyw{variant}\;E(s,c, v)}~=\\
      \end{array}
      \\
      \begin{array}{l}
        \texttt{T } \javacode{var = }\texttt{E}\javacode{;}\\
        \quad\jmlcode{/*@}~\jmlkeyw{public normal\_behavior}\\
        \quad\quad\quad\jmlkeyw{requires true;}\\
        \quad\quad\quad\jmlkeyw{assignable } \jmlkeyw{\bsl{}nothing;}\\
        \quad\quad\quad\jmlkeyw{ensures}~~\jmlcode{var
        }\jmlkeyw{instanceof } \jmlcode{BSet ==> }\\
        \quad \quad \quad\quad\jmlcode{(} \bsl\jmlkeyw{result
        }\jmlcode{== ((BSet)v).size()}\jmlcode{);} \\
        \quad\quad\quad\jmlkeyw{ensures}~\jmlcode{!(var
        }\jmlkeyw{instanceof } \jmlcode{BSet) ==> }\\
        \quad \quad \quad\quad\jmlcode{(} \bsl\jmlkeyw{result
        }\jmlcode{== ((Integer)v).intValue()}\jmlcode{);*/} \\
        \quad\javakeyw{public } \jmlcode{/*@ } \jmlkeyw{pure}\jmlcode{
          */ } \javakeyw{int}~\javacode{variant() \{}\\
        \quad \quad \javakeyw{if } (\javacode{var }\javakeyw{instanceof } \javacode{BSet} \javacode{)}\\
        \quad \quad \quad \javakeyw{if } (\javacode{var.finite())}\\
        \quad \quad \quad \quad \javakeyw{return } \javacode{var.size();}\\
        \quad \quad \quad \javakeyw{else}\\
        \quad \quad \quad \quad \javakeyw{return } \javacode{-1;}\\
        \quad \quad \javakeyw{else}\\
        \quad \quad \quad~\javakeyw{return}~\texttt{E}\javacode{;}\\
        \quad \javacode{\}}
      \end{array}
    \end{array}
  \end{prooftree}
  \]
}

Rules \textsf{Status1} and \textsf{Status2} below 
are used to impose the conditions associated with
\ebkeyw{variant}s on the guards and
actions of \ebkeyw{convergent} and \ebkeyw{anticipated} events.
Translating variant expressions in this manner allows the user to
verify that a customised method implementation is consistent with the
meaning of the translated event -- for example, since the translation
of a \ebkeyw{convergent} event refers to the translation of the
variant in the post-condition of its \jml\ specification, the user can
employ \jml\ machinery to verify that the customised implementation
does in fact decrease the variant. 

{\small
  \[
  \begin{prooftree}
    \using \textsf{(Status1)}
    \justifies
    \begin{array}{l}
      \STATONE{\ebkeyw{status ordinary}} = \\
      ~~~~~~~~\jmlkeyw{true}
    \end{array}
  \end{prooftree}
  \]
}

{\small
  \[
  \begin{prooftree}
    \using \textsf{(Status1)}
    \justifies
    \begin{array}{l}
      \STATONE{\ebkeyw{status convergent}} = \\
      ~~~~~~~~\jmlcode{m.variant()} >= 0 
    \end{array}
  \end{prooftree}
  \]
}

{\small
  \[
  \begin{prooftree}
    \using \textsf{(Status1)}
    \justifies
    \begin{array}{l}
      \STATONE{\ebkeyw{status anticipated}} = \\
      ~~~~~~~~\jmlcode{m.variant()} >= 0
    \end{array}
  \end{prooftree}
  \]
}

{\small
  \[
  \begin{prooftree}
    \using \textsf{(Status2)}
    \justifies
    \begin{array}{l}
      \STATTWO{\ebkeyw{status ordinary}} = \\
      ~~~~~~~~\jmlkeyw{true}
    \end{array}
  \end{prooftree}
  \]
}

{\small
  \[
  \begin{prooftree}
    \using \textsf{(Status2)}
    \justifies
    \begin{array}{l}
      \STATTWO{\ebkeyw{status convergent}} = \\
      ~~~~~~~~\jmlcode{m.variant()} < \jmlkeyw{\bsl{}old} \jmlcode{(m.variant())}
    \end{array}
  \end{prooftree}
  \]
}

{\small
  \[
  \begin{prooftree}
    \using \textsf{(Status2)}
    \justifies
    \begin{array}{l}
      \STATTWO{\ebkeyw{status anticipated}} = \\
      ~~~~~~~~\jmlcode{m.variant()} <= \jmlkeyw{\bsl{}old} \jmlcode{(m.variant())}
    \end{array}
  \end{prooftree}
  \]
}

Standard (non-initialisation) events are translated as Java threads. In
\eb, non-mutually exclusive event guards allow the interleaving of the
execution of events whereas mutually exclusive guards force events to
run sequentially. We translate the latter case (see Section
\ref{eb2java:translation:seq}) without overriding the \javacode{run()}
method, forcing the implementation to run sequentially. We translate
the former case by overring the method \javacode{run()} as explained
in the following. The translation of a standard event is defined by
Rules \textsf{Any} in Figure~\ref{fig-any}, and \textsf{When} (not
shown here): Rule \textsf{Any} refers to \eb\ events with
local variables bounded by the \eb\ clause \ebkeyw{any}; Rule
\textsf{When} goes in the same direction as rule \textsf{Any} without
referring local variables.  Each such event is
translated as a subclass of the Java \javacode{Thread} class that
includes a reference to the machine class implementation. The class
implementing the event contains three methods: a
\javacode{guard\_\-evt} method that tests if the guard of the event
$evt$ holds, a \javacode{run\_\-evt} method that models the execution
of $evt$, and a \javacode{run()} method that overrides the
corresponding Java \javacode{Thread} method. The guard and the actions
of an event are translated to separate methods so the execution of the
Java code matches the semantics of \eb: the events guards are evaluated
concurrently whereas the actions of just one event can be executed. Method
\javacode{run\_evt} is atomic -- it is executed within \javacode{lock}
and \javacode{unlock} instructions using a \javacode{Reentrant} lock
from the Java \javacode{concurrent} Library (Section
\ref{eb2java:imp:crital-section} explains our decision of using
\javacode{Reentrant}
lock rather than Java \javakeyw{synchronized} methods or an
implementation of the Bakery algorithm). 

Variables bounded by the \eb\ \ebkeyw{any} construct are
translated as parameters of the \javacode{run\_\-evt} and
\javacode{guard\_\-evt} methods (see Rule \textsf{Any}). The
expression
\javacode{GuardValue<Type>.\-next()} in method \javacode{run()}
returns a random value of type \javacode{Type} (that could be a
numeric value, a set, or a relation) that might satisfy the
event guard. 

The \jml\
specification of \javacode{run\_evt} uses two specification cases. In
the first case, the translation of the guard is satisfied (and the current
value of the variant is non-negative - or a finite set - for 
\ebkeyw{convergent} and \ebkeyw{anticipated} events), and the
post-state of the method must satisfy the translation of the event
actions and the translation of the \ebkeyw{variant} restriction. In
the second case, the translation of the guard is not
satisfied, and the method is not allowed to modify any fields,
ensuring that the post-state is the same as the pre-state.
This matches the semantics of
\eb: if the guard of an event is not satisfied, the event cannot
execute and hence cannot modify the system state. 

\begin{figure}
  {\small
    \[
    \begin{prooftree}
      \begin{array}{ll}
        \PRED{G(s,c,v,x)} = \texttt{G} & \MOD{A(s,c,v,x)} = \texttt{D} \\
        \EBTOJML{A(s,c,v,x)} = \texttt{A} & \ASGJAVA{A(s,c,v,x)}{\texttt{D}} = \texttt{B} \\
        \STATONE{status~ St} = \texttt{St1} &   \Type{x} = \texttt{Type} \\
        \STATTWO{status~ St} = \texttt{St2}
      \end{array}
      \using \textsf{(Any)\label{any:label}}
      \justifies
      \begin{array}{c}
        \begin{array}{c}
          \EBTOPROGN{(\ebkeyw{event}~ evt}
          \\ ~~~~~~~~~~\ebkeyw{status}~ St
          \\ ~~~~~~~~~~\ebkeyw{any}~x~\ebkeyw{where}~G(s,c,v,x)
          \\ ~~~~~~~~~~\ebkeyw{then}~A(s,c,v,x)~\ebkeyw{end})~=\\
        \end{array}
        \\
        \begin{array}{l}

          \javakeyw{public class}~\javacode{evt}~\javakeyw{extends}~\javacode{Thread \{}\\
          \quad\javakeyw{private}~\javacode{M m;}\\
          \\
          \quad\jmlcode{/*@}~\jmlkeyw{public normal\_behavior}\\
          \quad\quad\quad\jmlkeyw{requires true}\jmlcode{;} \\
          \quad\quad\quad\jmlkeyw{assignable}~\jmlkeyw{\bsl{}everything}; \\
          \quad\quad\quad\jmlkeyw{ensures}~\javakeyw{this}\jmlcode{.m ==
            m; */}\\
          \quad\javakeyw{public}~\javacode{evt(M m)} \{\\
          \quad\quad \javakeyw{this}\javacode{.m = m;}\\
          \quad{\}}\\
          \\
          \quad\jmlcode{/*@}~\jmlkeyw{public normal\_behavior}\\
          \quad\quad\quad\jmlkeyw{requires true}\jmlcode{;}\\
          \quad\quad\quad\jmlkeyw{assignable}~\jmlkeyw{\bsl{}nothing}; \\
          \quad\quad\quad\jmlkeyw{ensures}~\jmlkeyw{\bsl{}result}~
          \jmlcode{<==> G~\&\&~St1; */} \\
          \quad\javakeyw{public } \jmlcode{/*@ }
          \jmlkeyw{pure}\jmlcode{ */ } \javacode{boolean guard\_evt(Type }x\javacode{) \{}
          \\ \quad \quad\javakeyw{return }\javacode{G~\&\&~St1};\\
          \quad \javacode{\}}\\
          \\
          \quad\jmlcode{/*@}~\jmlkeyw{public normal\_behavior}\\
          \quad\quad\quad\jmlkeyw{requires}~\javacode{guard\_evt(}x\javacode{);}\\
          \quad\quad\quad\jmlkeyw{assignable}~\javacode{D;}~~\jmlkeyw{ensures}~\jmlcode{A~\&\&~St2};\\
          \quad\quad\jmlkeyw{also}\\
          \quad\quad\quad\jmlkeyw{requires}~\javacode{!guard\_evt(}x\javacode{);}\\
          \quad\quad\quad\jmlkeyw{assignable}~\jmlkeyw{\bsl{}nothing}\jmlcode{;}~~\jmlkeyw{ensures}~\javakeyw{true}\javacode{; */}\\
          \quad\javakeyw{public }~\javakeyw{void}~\javacode{run\_evt(Type }x\javacode{) \{}\\
          \quad\quad\javakeyw{if}~\javacode{(guard\_evt(}x\javacode{)) \{ B \}} \\
          \quad\javacode{\}}\\
          \\
          \quad \javakeyw{public void}~\javacode{run() \{}\\
          \quad \quad \javakeyw{while}\javacode{(}\javakeyw{true}\javacode{) \{}\\
          \quad \quad \quad \javacode{Type x = GuardValue<Type>.next();}\\
          \quad \quad \quad \javacode{m.lock.lock();}\\
          \quad \quad \quad \javacode{run\_evt(x);}\\
          \quad \quad \quad \javacode{m.lock.unlock();}\\
          \quad \quad \}\\
          \quad \}\\ \\

          \}
        \end{array}
      \end{array}
    \end{prooftree}
    \]
  } 
  \caption{The translation of a standard \eb\ event with local
    variables.}
  \label{fig-any}
\end{figure}

An event body consists of potentially many deterministic and
non-deterministic assignments. In \eb, the symbol  $:\!|\;$
represents non-deterministic assignment. The first
Rule \textsf{NAsg} and the first Rule \textsf{Asg} below translate
non-deterministic and deterministic assignments to \jml\
(respectively). They are used within \jml\ method post-conditions. The
\jml\ translation of a non-deterministic assignment $v:\!|\;P$ is a
\jml\ existentially quantified expression. The expression
\jmlkeyw{\bsl{}old}\javacode{(P)} ensures that \javacode{P} is
evaluated in the method pre-state. This matches the \eb\ semantics for
assignments, in which the left-hand side is assigned the value of the
right-hand side evaluated in the pre-state. The expressions
\javacode{v.equals(v')} and
\javacode{v.equals(}\jmlkeyw{\bsl{}old}\javacode{(E))} ensure that the
value $v'$ of a variable $v$ in the post-state is properly
characterised. 

The second Rule \textsf{NAsg} and the second Rule \textsf{Asg} below
translate non-deterministic and deterministic assignments to Java
(respectively). They are used by rules \textsf{Any} and \textsf{When}
to translate the body of an event. \javacode{PredicateValue<Type>(P)}
returns a value of type \javacode{Type} that satisfies predicate
\javacode{P}\footnote{Function \javacode{PredicateValue<Type>(P)} has
not yet been implemented in \ebtojava. The implementation of this
function is listed as future work of this thesis. There are many cases
in which this function cannot be implemented, in particular when
predicate \javacode{P} contains infinite sets.}. In Java, simultaneous actions are
implemented by first calculating the value of each right hand side of
the assignment into a temporary variable. Operator \ASGJAVAN\ servers
that purpose. It receives as first parameters the variable and its
assignment, and as second parameter the set of variables to be changed
on the right side of the assignment. The expression
$E[D/temp(D)]$ in second rule \textsf{Asg} changes all occurrence of
variable $var$ of the set $D$ in $E$ to $var\_temp$.

{\small
  \[
  \begin{prooftree}
    \PRED{P(s, c, v, v')} = \texttt{P} ~~\Type{v} = \texttt{Type}
    \using \textsf{(NAsg})
    \justifies
    \begin{array}{c}
      \EBTOJML{v:\!|\;P} = ~
      (\bsl \jmlkeyw{exists}~ \texttt{Type v';}~\bsl\jmlkeyw{old}\javacode{(P)}~\javacode{\&\&}~\javacode{v.equals(v'))}
    \end{array}
  \end{prooftree}
  \]
}

{\small
  \[
  \begin{prooftree}
    \begin{array}{l}
      \Type{v} = \texttt{Type} \\ 
      \PRED{P[D/temp(D)]} = \texttt{P}\\
      \javacode{val = PredicateValue<Type>(}\texttt{P}\javacode{)}
    \end{array}
    \using \textsf{(NAsg})
    \justifies
    \begin{array}{c}
      \ASGJAVA{v:\!|\;P}{D} = 
      ~\javacode{v = val;}    
    \end{array}
  \end{prooftree}
  \]
}

{\small
  \[
  \begin{prooftree}
    \PRED{E(s, c, v)} = \texttt{E}
    \using \textsf{(Asg)}
    \justifies
    \begin{array}{c}
      \EBTOJML{v\::=E\:} = 
      \javacode{v.equals(\bsl}\jmlkeyw{old}\javacode{(E));}\\
    \end{array}
  \end{prooftree}
  \]
}

{\small
  \[
  \begin{prooftree}
    \begin{array}{l}
      \Type{v} = \texttt{T}\\

      \PRED{E[D/temp(D)]} = \texttt{changed\_exp}
    \end{array}
    \using \textsf{(Asg)}
    \justifies
    \begin{array}{l}
      \ASGJAVA{v \::=E}{D} = \\
      \quad \texttt{T }\javacode{v\_temp = v;}\\
      \quad \javacode{v = } \texttt{changed\_exp}\javacode{;}\\
    \end{array}
  \end{prooftree}
  \]
}

Simultaneous assignments in the body of an event are translated
individually and the results are conjoined. Assignments translate to both
\jml\ and Java. For example, a pair of simultaneous actions $x := y$
$||$ $y := x$ is translated to the \jml\ post-condition \javacode{x
  ==} \jmlkeyw{\bsl{}old}\javacode{(y)} \javacode{\&\&}
\javacode{y ==} \jmlkeyw{\bsl{}old}\javacode{(x)} for
variables $x$ and $y$ of type integer.

{\small
  \[
  \begin{prooftree}
    \begin{array}{l}
      \EBTOJML{A1} = \texttt{A1}\\
      \EBTOJML{A2} = \texttt{A2}\\
    \end{array}
    \using \textsf{(ParAsg)}
    \justifies
    \begin{array}{l}
      \EBTOJML{A1\: ||\: A2\:} = \texttt{A1}\;\jmlcode{\&\&}\;\texttt{A2}
    \end{array}
  \end{prooftree}
  \]
}

{\small
  \[
  \begin{prooftree}
    \begin{array}{l}
      \ASGJAVA{A1}{D} = \texttt{A1}\\
      \ASGJAVA{A2}{D} = \texttt{A2}\\
    \end{array}
    \using \textsf{(ParAsg)}
    \justifies
    \begin{array}{l}
      \ASGJAVA{A1\: ||\: A2\:}{D} = \texttt{A1}\;\texttt{A2}
    \end{array}
  \end{prooftree}
  \]
}

For instance, the Java translation of $\ASGJAVA{x := y || y := x
  + y}{\{x,y\}}$ is:

{\small
  \[
  \begin{array}{l}
    \Type{x}~\javacode{x\_temp = x;}\\
    \Type{y}~\javacode{y\_temp = y;}\\
    \javacode{x = y\_temp;}\\
    \javacode{y = x\_temp  + y\_temp;}
  \end{array}
  \]
}

The \ebkeyw{with} construct  that is used in the definition
of a refinement event as a ``witness'' of a disappearing abstract
(refined) event variable is not taking into account on this
translation. A witness predicate specifies how the
disappearing variable is implemented by the refinement event. A
witness plays a similar role for an event as a ``gluing invariant''
does for a machine. A witness for an abstract event variable $x$ is a
predicate $P(x)$ involving $x$.  A deterministic witness for a
variable $x$ is an equality predicate $x = E$, where $E$ is an
expression free of $x$. As Rodin ensures that $x$ does not appear
in the refinement event ($x$ is replaced by $E$),
we do not need to translate witnesses to Java or \jml.

\subsection{The Helper Operators}
\label{eb2java:translation:helpers}
Operators \MODN, \PREDN, and \TypeN\ work in the same way as
explained in Chapter \ref{chapter:eb2jml}. Operator \PREDN\ is used to
translate \eb\ predicates and expressions to Java and \jml. This
operator does not handle some \eb\ constructs when it is used in
translating to Java, they are listed below\footnote{these constructs are listed as future
  work of this thesis.}

\[
\begin{array}{ll}
  Universal~Quantifier & \forall x \qdot P \\
  Existential ~ Quantifier & \exists x \qdot P\\
  Set ~ Comprehension &\{x \qdot P|F\}\\
  Quantified ~ Union & \Union x \qdot P \mid E\\
  Quantified ~ Intersection &   \Inter x \qdot P \mid E
\end{array}
\]

In general, it is not possible to translate
those constructs to Java since they might be bound to infinite sets so
the result would be an infinite set (quantified union and intersection
are generalisation of set comprehension).

\EBTOJMLN\ translates \eb\ set comprehension expressions to \jml\ (see
Rule \textsf{Set-Comp}) set comprehensions.
Operator \FreeN\ returns the set of variables
that occur free in an expression, the type \jmlcode{JMLObjectSet}
defines a set of objects in \jml. The rules shows the different ways of
expressing set comprehension in \eb\ and the translation for each. 
For simplicity, we assume that $E$ contains a single free variable
$x$ in the second rule, and that $E$ and $P$ do not contain a variable
named $e$ in either rule (i.e. $e \not \in \Free{E} \wedge e \not \in
\Free{P}$).  We do not translate set comprehensions to Java code since
it is not possible in general -- set comprehensions can denote
infinite sets. We may be able to translate set comprehension to Java
by restricting them to finite sets, as we do with the translation of
the \eb\ carrier sets\footnote{This is listed as a future work of this
thesis.}.

{\small
  \[
  \begin{prooftree}
    \begin{array}{ll}
      \PRED{E} = \texttt{E} &\Type{x} = \texttt{Type}\\
      \PRED{P} = \texttt{P} & \Type{E} = \texttt{Type\_e}
    \end{array}
    \using \textsf{(Set-Comp)}
    \justifies
    \begin{array}{c}
      \begin{array}{c}
        \EBTOJML{\{x~\qdot~P \mid E \}}~=\\
      \end{array}
      \\
      \begin{array}{l}
        \quad\javakeyw{new}~\jmlcode{BSet<}\texttt{Type}\jmlcode{>(}
        \quad\javakeyw{new}~\jmlcode{JMLObjectSet \{}\texttt{Type\_e~e} \jmlcode{ $\mid$}\\
        \quad \quad \quad ~(\bsl\jmlkeyw{exists }\texttt{Type} \jmlcode{ x;}~
        \texttt{ P} \jmlcode{; } \jmlcode{e.equals(}\texttt{E}\jmlcode{))}\})
      \end{array}
    \end{array}
  \end{prooftree}
  \]
}

{\small
  \[
  \begin{prooftree}
    \begin{array}{ll}
      \PRED{E} = \texttt{E} &\Type{x} = \texttt{Type}\\
      \Free{E} = \{x\} & \PRED{P} = \texttt{P}\\
      \Type{E} = \texttt{Type\_e}
    \end{array}
    \using \textsf{(Set-Comp)}
    \justifies
    \begin{array}{c}
      \begin{array}{c}
        \EBTOJML{\{E \mid P \}}~=\\
      \end{array}
      \\
      \begin{array}{l}
        \quad\javakeyw{new}~\jmlcode{BSet<}\texttt{Type}\jmlcode{>(}
        \quad\javakeyw{new}~\jmlcode{JMLObjectSet \{}\texttt{Type\_e}~\texttt{e} \jmlcode{ $\mid$}\\
        \quad \quad \quad~(\bsl\jmlkeyw{exists }\texttt{Type} \jmlcode{ x;} \texttt{ P} \jmlcode{; } \jmlcode{e.equals(}\texttt{E}\jmlcode{))}\})
      \end{array}
    \end{array}
  \end{prooftree}
  \]
}

\subsection{The Translation of \eb\ to Sequential Java Programs}
\label{eb2java:translation:seq}
An event is enabled only if the event guard holds in the current
state. This could be the case for several events and so the
interleaving semantics of \eb\ ensures that one of these events is
non-deterministically selected and executed, and thus there can be
just one executing at the time. On the other hand, mutually exclusive
event guards force machine events to run sequentially.

The translation rules for sequential Java implementation are similar
to the ones presented previously for multi-threaded Java, in which
events and machines are translated as standard Java classes rather
than threads\footnote{The \ebtojava\ tool permits users to select
  between a multi-threaded or sequential Java implementation.}.

For the execution of these sequential Java implementations we can use
the framework depicted in Figure
\ref{fig:eb2java:translation:framework}. Class
\javacode{Framework} instantiates the class \javacode{M} that
contains a reference to all Java classes that represent the \eb\
events (${evt1}, {evt2}, \ldots, {evtn}$), the sequence $x1$, \ldots, $xn$ 
represents \eb\ variables bounded by an \ebkeyw{any} constructs. The
\javacode{while} body evaluates which guard holds
($m.evtn.guard\_name()$) and executes the corresponding method
($m.evtn.run\_name()$).


\begin{figure}
\begin{lstlisting}[frame=none]
public class Framework {
    public static void main(String[] args) {
    M machine = new M();
    while (true) {
        // x1 ... xn are declared and given random values
        if (machine.evt1.guard_evt1(x1))
            machine.run_evt1(x1); break;
        else if (machine.evtn.guard_evtn(xn))
            machine.run_evtn(xn); break;
    }
}
\end{lstlisting}
\caption{A framework for executing Java classes translated from \eb\
machines in a sequential fashion.}
\label{fig:eb2java:translation:framework}
\end{figure}

\subsection{Support for \eb\ Model Decomposition}
\label{eb2java:translation:decomposition}

When modelling systems with \eb, one usually starts with the design of
a single closed machine that includes both the system
and the surrounding environment. The machine is then refined into a
more concrete model of the system. Abstract machines usually include 
few events, variables and invariants, whereas (advanced) refinements typically
contain many of them. 
The plethora of components in machines at later stages in the refinement
chain often makes 
the discharge of the corresponding proof obligations in Rodin rather intricate.
In certain cases  an \eb\ model may be regarded as being composed of two
semi-independent sub-models in the sense that variables and the
events affecting them in the integrated model could, in principle,
be neatly split between those two sub-models.
In this case, it would be very useful to provide a machine decomposition
mechanism that allows one to construct two independent machines whose
combined behaviour could nevertheless be provably shown to correspond to
the integrated model.
In \cite{Abrial:Ref:07}, J.-R. Abrial and
S. Hallerstede propose a technique for machine decomposition based on
shared variables in which each decomposed machines simulates the
behaviour of other decomposed machines through the use of
\emph{external} events. In \cite{Butler:Decomp:09} M. Butler proposes
a technique for machine decomposition by shared events in which
decomposed machines include copies of all of the variables that
events in that machine
use. The latter technique is implemented in Code Generation 
\cite{CodeGen10}. Both machine decomposition techniques produce independent
machines that include local copies of shared variables or local events
that simulate the effect of other decomposed machines acting on the
shared variables. Since the result of decomposing a machine are
valid machines, these are correctly translated by our tool.

\subsection{Support for Code Customisation}
\label{eb2java:translation:customisation}

The \jml\ specifications generated by \ebtojava\ enable users to replace
the generated Java code with bespoke implementations.
The user can then employ 
existing \jml\ tools~\cite{Burdy-etal05}
to verify 
the customised implementation against the \jml\
specification generated by the \ebtojava\ tool\footnote{Notice that
  the  current \jml\ tools cannot handle Java code with generics types
  as the code generated by \ebtojava}. 
For example, the code generated by \ebtojava\ will
represent an \eb\ function variable using an instance of class 
\javacode{BRelation} as described earlier in this section.
A developer may wish to represent this variable using
a Java \javacode{HashMap} instead, as this will make looking up the value
of a given domain element more efficient.  After generating this customised
implementation, the developer can verify it against the generated \jml\
specification, likely making use of the existing \jml\ specification
of the \javacode{HashMap} class~\cite{Cok2011}.






\section{Proof of Soundness}
\label{ebtojava::proof}
To gain confidence about our translation, it is necessary to prove
that the proposed rules generate \jml-annotated Java code that 
models what the user initially modelled in \eb, i.e. it is necessary
to prove the rules are indeed sound. N\'estor et. al. proposed an
initial proof of soundness of the translation from \eb\ to \jml
\cite{journ:proof:13}, the proof that does not consider the Java code.
The soundness proof ensures that
any state
transition step of the \jml\ semantics of the translation of some \eb\
construct into \jml\ can be {\it simulated} by a state transition step
of the \eb\ semantics of that construct. The work provides the proof for
invariants and the standard \eb\ initialising event. It does not
include full machines or \eb\ contexts.

They expressed \eb\ and \jml\ constructs as types in \eb, then
implemented the translation rules explained in the previous section
(denoted by operator \EBTOJMLN) as type transducer rules. They
defined a semantics of \eb\ and \jml\ types as state transducers. And
finally proved that the semantics of the \jml\ translation of \eb\
constructs is simulated by the \eb\ semantics of those constructs. The
soundness condition is stated as a theorem and proved interactively in
Rodin.

\section{The \ebtojava\ Tool}
\label{eb2java:imp}

\begin{figure}[h]
  \centering
  \includegraphics[scale=0.7]{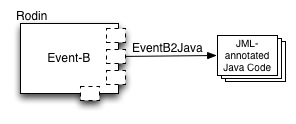}
  \caption{General structure of \ebtojava\ Rodin plug-in.}
  \label{fig:ebtojava_general}
\end{figure}

The \ebtojava\ tool is implemented as a plug-in of Rodin
\cite{rodin}. Rodin comes with
an API that provides extra functionality on top of its core platform
so as to support the implementation of applications as plug-ins. \ebtojava\
uses the Rodin API to collect the information of all the components of the machine to be
translated. Figure \ref{fig:ebtojava_general} depicts a general
structure of the \ebtojava\ tool. Rodin is composed of
several plug-ins, e.g. an editor, a proof generator, provers, and
model checkers and animators \cite{prob} (Figure
\ref{fig:ebtojava_general} depicts these plug-ins in dotted
squares). \ebtojava\ is another plug-in for Rodin. It takes an
\eb\ model and translates it to a \jml-annotated Java
program.

In the following we describe the structure of the \ebtojava\ plug-in
in full detail. We first describe the structure of the Rodin platform
and its main components, and then describe how the \ebtojava\ plug-in
interfaces with these
components to produce \jml-annotated Java code. \ebtojava\ relies on a
series of \emph{recommended} interfaces \cite{rodinPlugin} to
interface with the Rodin components.

\begin{figure*}[t]
  \centering
  \includegraphics[scale=0.5]{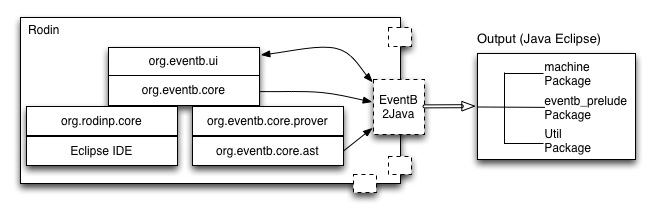}
  \caption{Specific structure of \ebtojava\ Rodin plug-in.}
  \label{fig:ebtojava_specific}
\end{figure*}

\subsection{\ebtojava\ Rodin Plug-in Structure}
Figure \ref{fig:ebtojava_specific} shows the main components of Rodin,
shown as \texttt{org.*} squares. It also shows the relation among
those components and the \ebtojava\ plug-in (the solid arrows). Rodin
is built on top of the {Eclipse IDE}. The \texttt{org.rodinp.core}
component implements the core functionality of Rodin, e.g. a database
for manipulating \eb\ models, and for storing elements such as proof
obligations and proofs. 
It further includes a static checker, a proof obligation generator and
a prover. The \texttt{org.\-eventb.\-core.\-ast} component includes a
library for manipulating mathematical formulas in the form of Abstract
Syntax Trees. It provides an abstract class (a Visitor) for parsing
the mathematical formulas. The Sequent Prover
(\texttt{org.eventb.core.seqprover}) component contains a library for
proving sequents. And the \eb\ User Interface (\texttt{org.eventb.ui})
component contains the Graphic User Interfaces that allows users to
write \eb\ models and to interact with the interactive proof engine.

\begin{figure}[h]
  \centering
  \includegraphics[scale=0.37]{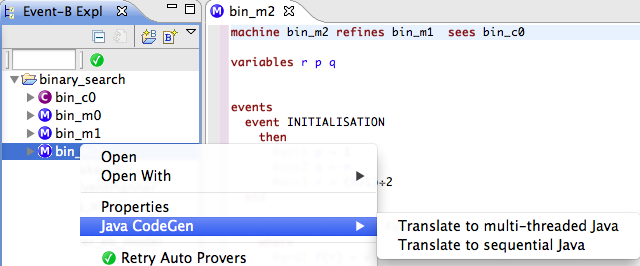}
  \caption{\ebtojava: Contextual menu in Rodin.}
  \label{fig:screenshot}
\end{figure}

\ebtojava\ uses the Rodin \texttt{org.eventb.ui} component to
manipulate context menus, e.g. to enable users to choose the type of
implementation (sequential or multi-threaded) to be generated (see
Figure \ref{fig:screenshot}). The relation between the
\texttt{org.eventb.ui} component and \ebtojava\ is depicted in Figure
\ref{fig:ebtojava_specific} with a double-headed arrow: from the
component to \ebtojava\ to capture the user's request; and from
\ebtojava\ to the component to show the code generated.

\ebtojava\ uses the Rodin
\texttt{org.eventb.core} component to collect all the information of
the machine and context to be translated, i.e. carrier sets,
constants, axioms, variables, invariants and events. Figure
\ref{fig:ebtojava_specific} represents the relation between this component
and \ebtojava\ with a single-headed arrow since our tool does not change
the \eb\ model, it just reads it.

In \eb, models are expressed using mathematical language. The
\texttt{org.\-eventb.\-core.\-ast} component encodes \eb's mathematical
language as nodes of an Abstract Syntax Tree (AST). This component
provides various services such as parsing a formula (that is,
computing its AST from a string of characters), pretty-printing a
formula, constructing new formulas directly using the API library,
type-checking formulas (that is, inferring the types of the expressions
occurring within and decorating them with their types), testing
formulas for equality, among others. \ebtojava\ uses the parsing
service provided by this component to parse mathematical formulas
to be translated to \jml-annotated Java code.

The \texttt{org.\-eventb.\-core.\-ast} component implements a
library to traverse trees (a Visitor). Figure
\ref{fig:ebtojava_specific}  uses a single-headed
arrow between the \texttt{org.eventb.core.ast} component and our plug-in since
the formulas are not changed. The input to
\texttt{org.eventb.core.ast} is part of the information collected from 
the \texttt{org.eventb.core} component. \ebtojava\ extends the Visitor
to traverse the abstract syntax trees and produce Java code and the
\jml\ specifications. Since \eb\ includes mathematical notations that
are not built-in to Java or \jml, we implemented them as Java
classes. The implementation allows \ebtojava\ to support \eb's syntax
(described in Section \ref{eb2java:imp:prelude}, and Appendix
\ref{appendix} shows for each \eb\ syntax the translation to \jml\ and
Java).

After collecting the information of the \eb\ contexts and machines and
parsing them using the Visitor implementation, \ebtojava\ 
generates an Eclipse Java project. This project contains various
packages: The \texttt{machine} package contains the translation of the
machines and contexts. This package includes a main Java class with
information about carrier sets, constants, and variables from the \eb\
model. It also contains \jml\ specifications generated from axioms and
invariants in \eb. This package also contains the translation of each
event and a test file to run the generated Java implementation.

The Eclipse project generated by \ebtojava\ further includes an 
\texttt{eventb\_\-prelude} package that contains the Java classes
necessary to support all the \eb\ syntax as explained in the next
section. Finally, the \texttt{Util} package in the Eclipse project generated by
\ebtojava\ includes utility methods. For instance, it includes an
implementation of a \javacode{SomeVal} method that returns a random
value contained within a set. It also includes the implementation of a
\javacode{SomeSet} method that returns a random subset of a set.

\ebtojava\ is available at
\url{http://poporo.uma.pt/EventB2Java}. This web
site includes detailed instructions on how to install and use the
tool. The \ebtojava\ Eclipse plug-in's {update site} is
\url{http://poporo.uma.pt/Projects/EventB2JavaUpdate}, and \ebtojava\
has been tested on Rodin version 2.8.

\paragraph{\ebtojava\ Tool Usage: } In a typical interaction with
\ebtojava, a user right-clicks an \eb\
machine in the Explorer panel of Rodin and selects ``Translate to
multi-threaded Java'' or ``Translate to sequential Java'' (as shown in Figure
\ref{fig:screenshot}). \ebtojava\
generates an Eclipse project that is available in the ``Resource''
perspective of Rodin. 

\subsection{Java Implementation of \eb\ Mathematical Notations in
  \ebtojava}
\label{eb2java:imp:prelude}
The \eb\ modelling language is composed of five mathematical languages (see
Chapter 9 of \cite{EB:Book}), namely, 
\begin{inparaenum}[\itshape a\upshape)]
\item a Propositional Language,
\item a Predicate Language,
\item an Equality Language,
\item a Set-Theoretic Language, and
\item  Boolean and Arithmetic Languages.
\end{inparaenum}
Each language defines a series of constructs to model
systems. 
To provide support for the translation from \eb, we have implemented a
series of \jml-annotated Java classes; other \eb\ constructs are
supported natively in Java.  These classes are:
\javacode{BOOL}, \javacode{INT}, \javacode{NAT}, \javacode{NAT1},
\javacode{Enumerated}, \javacode{Pair}, \javacode{BSet},
\javacode{BRelation}, and \javacode{ID} (implementing, respectively,
booleans, integers, natural numbers with and without 0, the enumerated type, 
pairs of elements, sets, relations, and the identity relation).
\javacode{BSet} is implemented as a subclass of the
standard Java class \javacode{TreeSet}, and \javacode{BRelation} as a
set of pairs.

We had previously implemented versions of these classes, for the work
described in~\cite{conference:B2Jml:12} (used also in the work
described in Chapter \ref{chapter:eb2jml}). Particular kinds of \eb\
relations (total relations, functions, etc.)
are translated as \javacode{BRelation}s with appropriate restrictions
added to the invariant  For example, \PRED{r \in D \tinj R} for sets $D$ and 
$R$
equals: \javacode{r.is\-a\-Func\-tion()} \javacode{\&\&}
\javacode{r.inverse().is\-a\-Function()} \javacode{\&\&}
\javacode{r.do\-main().equals(D)} \javacode{\&\&}
\javacode{r.range().is\-Subset(R)}, that is added to the
invariant.  We further define classes \javacode{Enumerated},
\javacode{ID}, \javacode{INT}, \javacode{NAT}, \javacode{NAT1},
\javacode{Pair} and \javacode{BOOL}.  For example, \PRED{i \in
  \mathbb{N}} = \javacode{NAT.instance.has(}$i$\javacode{)}, that
restricts $i$ to be non-negative. The \TypeN\ operator
translates the type of \eb\ variables and constants to the
corresponding Java type.  All integral types are translated as type
\javacode{Integer}, all relations and functions are translated as type
\javacode{BRelation}, and all other sets are translated as type
\javacode{BSet}.

Some of the constructs of the Propositional Language are supported
natively in Java. Negation ($\neg$) translates as \javacode{!}, 
conjunction ($\wedge$) as \javacode{\&\&}, and disjunction ($\vee$) as
\javacode{||}. Other constructs such as $\limp$ and $\leqv$ are
implemented as methods of the class \javacode{BOOL}.
The Predicate Language introduces constructs for universal and
existential quantification. Universally and existentially quantified
predicates $\forall$ $x$ $\qdot$ $(P)$ and $\exists$ $x$ $\qdot$ $(P)$ are
translated as the \jml\ universally and existentially quantified
expressions \jmlcode{(\bsl\jmlkeyw{forall } \Type{x } x; P)} and
\jmlcode{(\bsl\jmlkeyw{exists } \Type{x } x; P)} respectively, where
$P$ is the \jml\ translation of $P$\footnote{\ebtojava\ does not 
  generate Java code for quantified predicates.}.
The Predicate Language also includes a construct $e \mapsto f$ that
maps an expression $e$ of type $E$ to an expression $f$ of type
$F$. \ebtojava\ translates this construct as an instance of
\javacode{Pair<E,F>}.

The \eb\ Equality Language introduces equality predicates $E = F$ for
expressions $E$ and $F$, translated as \javacode{E.equals(F)}, if
\javacode{E} and \javacode{F} are object references, or \javacode{E ==
  F}, if they are of a primitive type. The Set-Theoretic Language
introduces sets and relations in \eb. Set operations include
membership ($\in$), cartesian product ($\cprod$), power set ($\pow$),
inclusion ($\subseteq$), union ($\bunion$), intersection ($\binter$),
and difference ($\setminus$). These operations are all implemented as
methods of the class \javacode{BSet}.
Operations on relations in \eb\ include domain restriction
($\domres$), range restriction ($\ranres$), etc. All these operations
are implemented as methods of the class
\javacode{BRelation}. Relations also include notations for surjective
relations $\srel$, total surjective relations $\strel$, functions,
etc. \ebtojava\ translates all these as instances of
\javacode{BRelation} with \jml\ \jmlkeyw{invariants} that constrain the domain
and the range of the relation, e.g. a total function is a relation in
which each element in the domain is mapped to a single element in the
range.

The Boolean and Arithmetic Languages define the set \ebkeyw{BOOL},
containing elements \ebkeyw{TRUE} and \ebkeyw{FALSE}, $\intg$,
containing the integer numbers, $\nat$, containing the natural numbers
($0$ inclusive), and $\natn$, containing the natural numbers ($0$
exclusive). \ebtojava\ includes implementations of these constructs in
Java, namely, classes \javacode{BOOL}, \javacode{INT}, \javacode{NAT},
and \javacode{NAT1}. The Arithmetic Language defines
constructs over numbers. Operators such as $\leq$, $\geq$, etc. are
directly mapped into Java operators \javacode{<=}, \javacode{>=},
etc. The construct $a \upto b$, that defines an interval between $a$
and $b$, is implemented as an appropriate instance of the class
\javacode{Enumerated}.

\subsection{Decision on using \javacode{Reentrant} lock/unlock
  methods}
\label{eb2java:imp:crital-section}
There are several ways to implement the problem of the Critical
Section in Computer Science. We compared the execution times and CPU
usage for
three methods in order to decide which method the \ebtojava\ tool
should use. We compared an implementation of the Bakery algorithm
\cite{Bakery:74}, the \javakeyw{synchronized} native Java method, and methods
\javacode{lock}/\javacode{unlock} from the \javacode{concurrent} Java
library.  All experiments are available at
\url{http://poporo.uma.pt/EventB2Java/exps.zip}

\begin{figure}
  \centering
  \begin{subfigure}[b]{0.5\textwidth}
    \includegraphics[width=\textwidth]{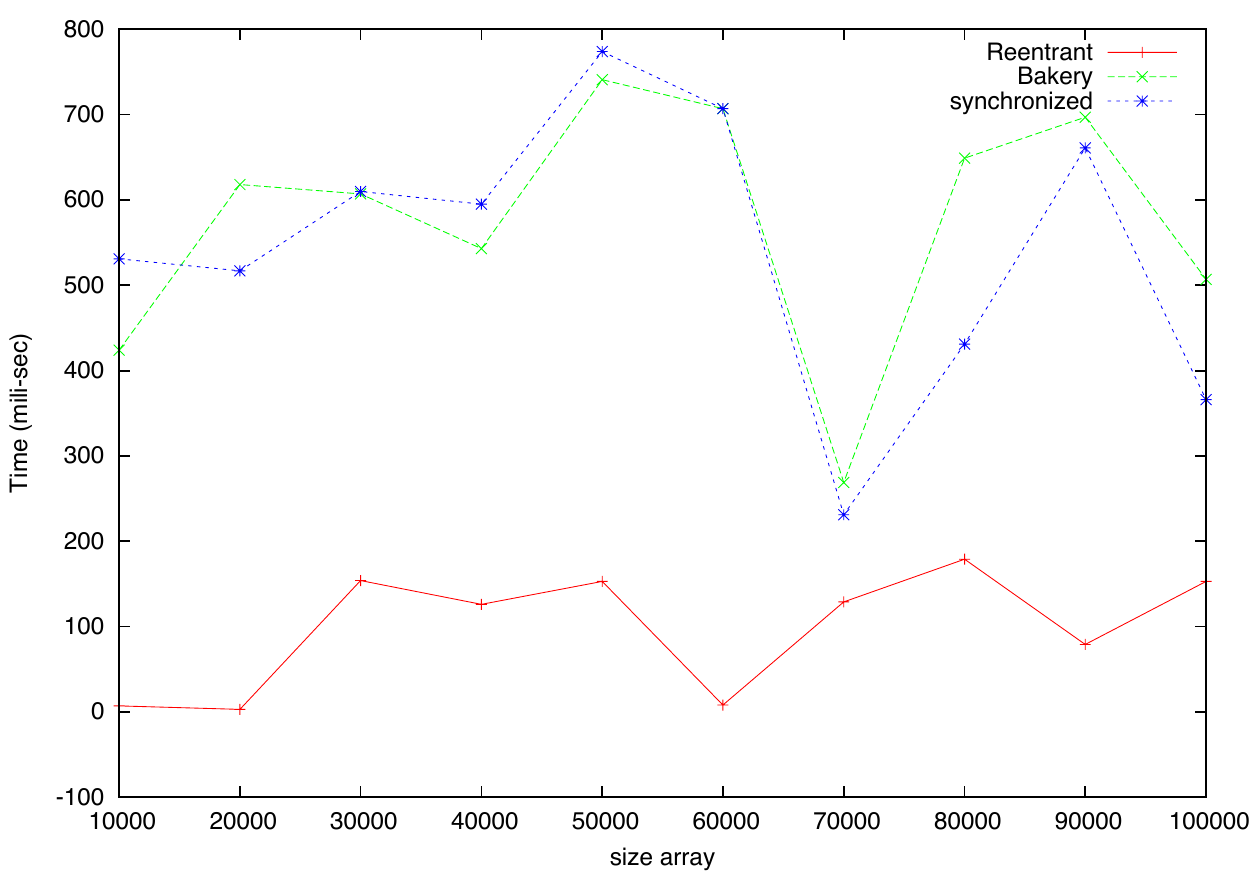}
    \caption{Binary Search}
  \end{subfigure}~
  \begin{subfigure}[b]{0.5\textwidth}
    \includegraphics[width=\textwidth]{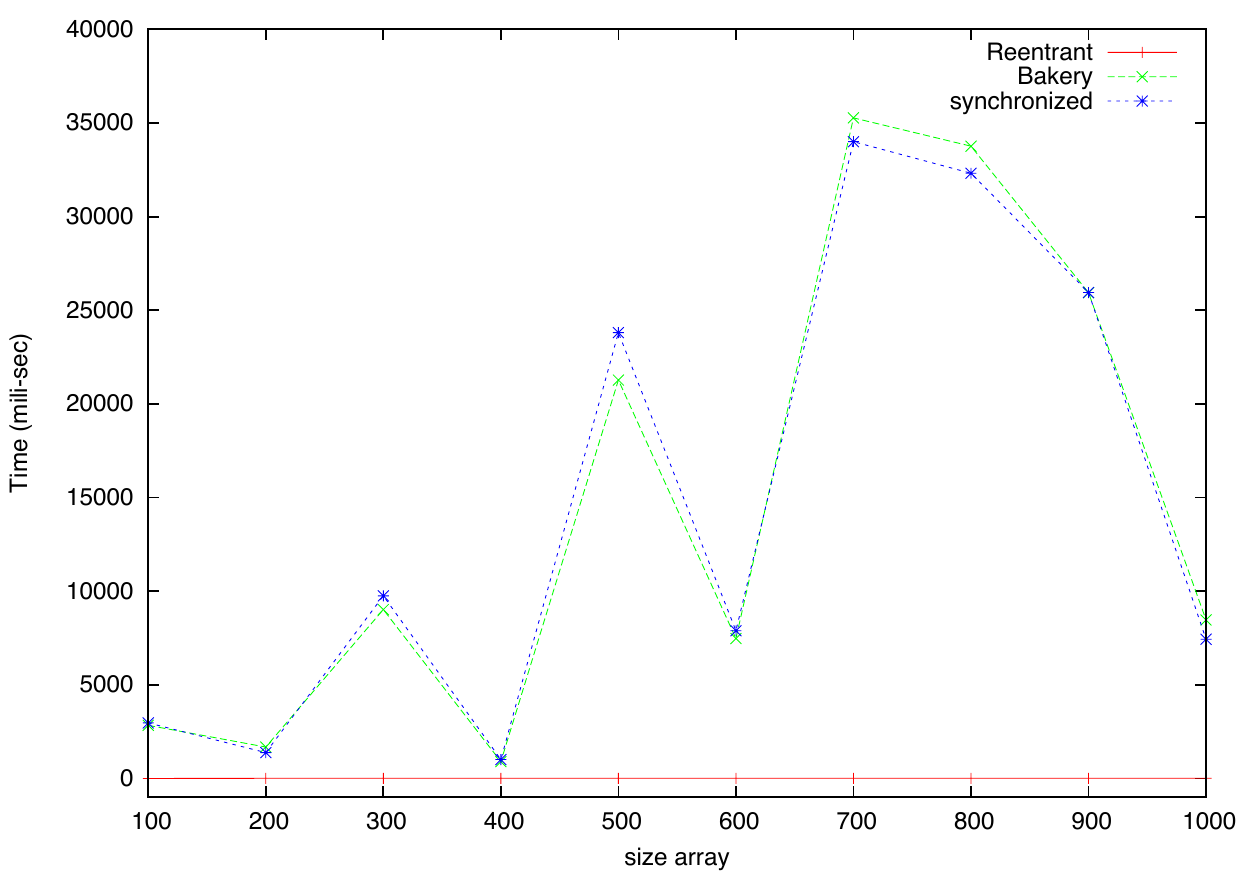}
    \caption{Linear Search}
  \end{subfigure}

  \begin{subfigure}[b]{0.5\textwidth}
    \includegraphics[width=\textwidth]{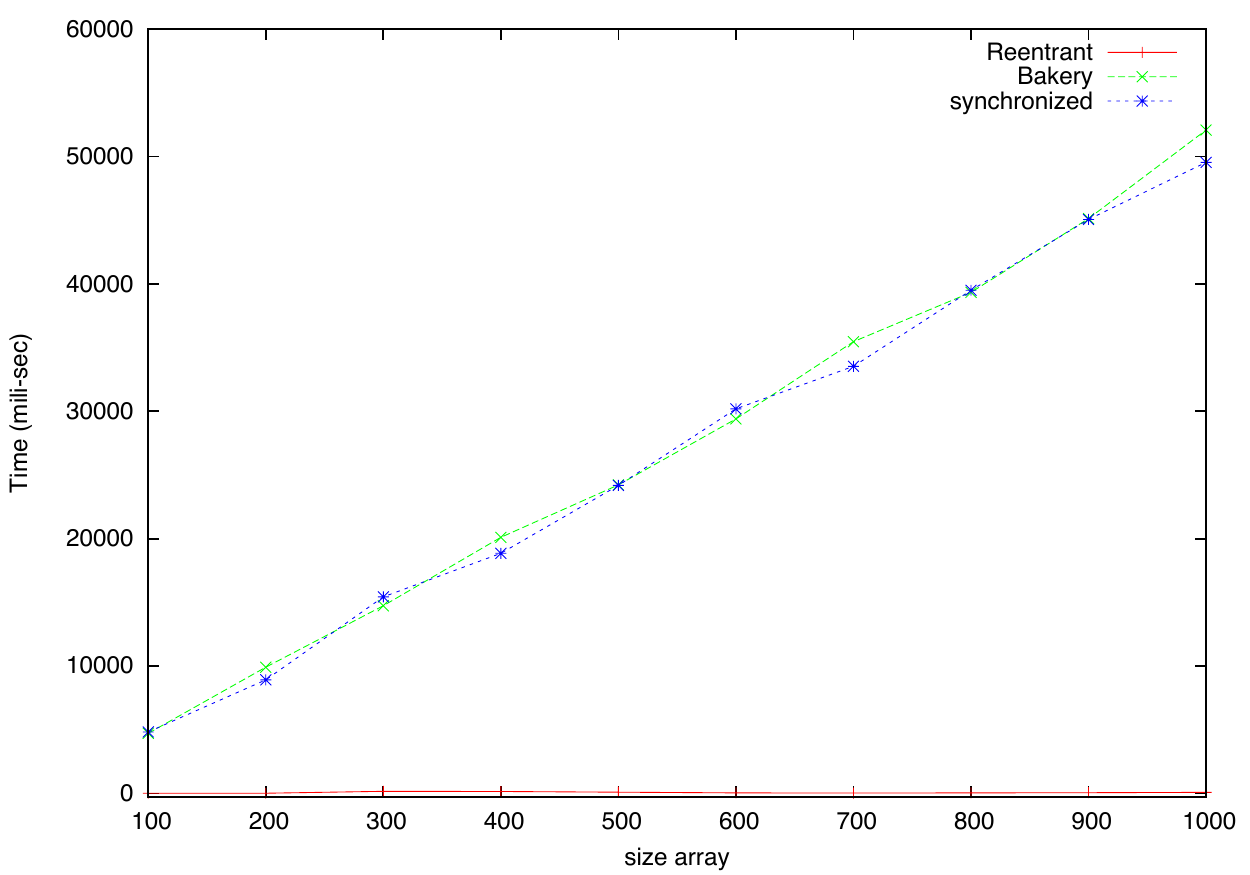}
    \caption{Minimum Element}
  \end{subfigure}~
  \begin{subfigure}[b]{0.5\textwidth}
    \includegraphics[width=\textwidth]{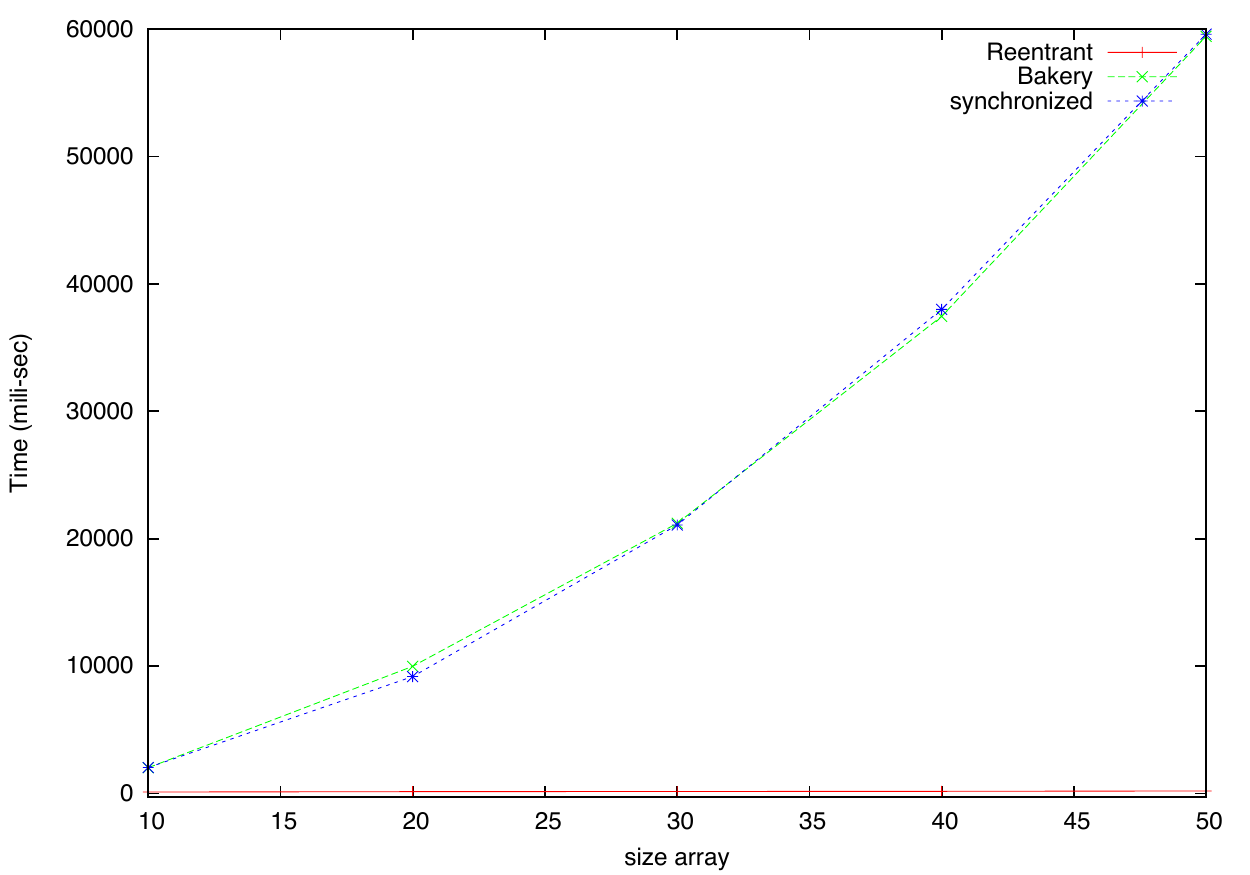}
    \caption{Sorting}
  \end{subfigure}

  \caption{Exp 1: Execution times `bakery' vs
    `\javakeyw{synchronized}' vs `\javacode{lock}/\javacode{unlock}'.}
  \label{fig:eb2java:imp:crital-section:exp1}
\end{figure}

We first compared the execution times for four multithreaded Java code
using the methods for implementing the critical section explained
above. We used a
multithreaded implementation of a Binary and Linear search in an array, the
Minimum element of an array, and the Sorting algorithm of an array. We
ran the implementation varying the size of the arrays. Figure
\ref{fig:eb2java:imp:crital-section:exp1} depicts the execution times
taken for the four algorithms. 

\begin{figure}[h]
  \centering
  \begin{subfigure}[b]{0.6\textwidth}
    \includegraphics[width=\textwidth]{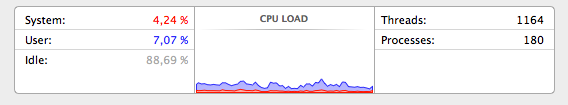}
    \caption{Bakery}
  \end{subfigure}%
  ~ 

  \begin{subfigure}[b]{0.6\textwidth}
    \includegraphics[width=\textwidth]{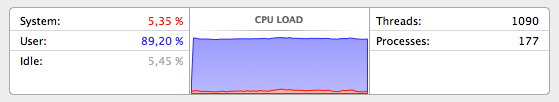}
    \caption{Synchronized}
  \end{subfigure}
  \caption{Exp 2: CPU usage `bakery' vs `\javakeyw{synchronized}'.}
  \label{fig:eb2java:imp:crital-section:exp2}
\end{figure}

We also compared the CPU usage for them. For this
experiment we used the 2 multithreaded implementations that
run forever. The code
was ran for 5 minutes, we took the CPU usage every minute
for both methods. Figure \ref{fig:eb2java:imp:crital-section:exp2}
shows the CPU usage for both
method during the 5 minutes. It can be seen that the `Bakery' method
outperforms the `\javakeyw{synchronized}' method. For the `Bakery'
algorithm, in average the CPU usage was 8\% whereas for the
`\javakeyw{synchronized}' method was 89\%.

Figure \ref{fig:eb2java:imp:crital-section:exp1} depicts that the
`\javacode{lock}/\javacode{unlock}' method outperforms both the
`Bakery' algorithm and `\javakeyw{synchronized}' method. This is due
both approaches guarantee fairness by putting each \javacode{Thread}
to sleep for a random number (we used a random number from 1 to 100),
while `\javacode{lock}/\javacode{unlock}' method guarantees fairness
by passing a boolean parameter to the constructor of the class
lock\footnote{from the documentation of the  class: ``The
  constructor for this class accepts an optional fairness
  parameter. When set true, under contention, locks favor granting
  access to the longest-waiting thread. Otherwise this lock does not
  guarantee any particular access order.''}

Our decision on using `\javacode{lock}/\javacode{unlock}' methods from
the concurrent Java library in the \ebtojava\ was based on these two
experiments: `\javacode{lock}/\javacode{unlock}' makes better use of
the CPU  and outperforms the time execution to both
``\javakeyw{synchronized}' method and `Bakery' algorithm. Another
reason is the library guarantees fairness by itself.

\section{Using the \ebtojava\ tool}
\label{eb2java:apply}

We have validated our tool by applying it to an ample set of \eb\
models. They are described in Section \ref{eb2java:comparison}. We
describe in next subsections an \eb\ model modelled by J.-R Abrial in
\cite{bs:eb:09} and show the \jml-annotated Java code generated by \ebtojava.
\subsection{An Example in \eb}

\begin{figure}[t]
  {\small
    \[
    \begin{array}{cc}
      \begin{array}{l}
        \ebkeyw{machine } bin\_m2 \ebkeyw{ sees } bin\_c0\\
        ~\ebkeyw{variables}
        ~~r~p~q \\
        ~\ebkeyw{invariant}\\
        ~~\ebtag{inv1 } p \in 1 \upto n\\
        ~~\ebtag{inv2 } q \in 1 \upto n\\
        ~~\ebtag{inv3 } r \in p \upto q\\
        ~~\ebtag{inv4 } v \in f[p \upto q] \\
        ~\ebkeyw{events}\\
        ~~initialisation \\
        ~~~\ebkeyw{then} \\ 
        ~~~~\ebtag{act1 } p := 1\\
        ~~~~\ebtag{act2 } q := n \\
        ~~~~\ebtag{act3 } r := (n+1) \div 2 \\
        ~~~\ebkeyw{end} \\
        ~~inc\\
        ~~~\ebkeyw{when}\\
        ~~~~\ebtag{grd1 } f(r) < v\\
        ~~~\ebkeyw{then}\\
        ~~~~\ebtag{act1 } r := (r+1+q) \div 2\\
        ~~~~\ebtag{act2 } p := r+1\\ 
        ~~\ebkeyw{end} \\
        ~~dec\\
        ~~~\ebkeyw{when}\\
        ~~~~~\ebtag{grd1 } f(r) > v\\
        ~~~\ebkeyw{then}\\
        ~~~~\ebtag{act1 } r := (p+r-1) \div 2\\
        ~~~~\ebtag{act2 } q := r-1\\ 
        ~~\ebkeyw{end} \\
        ~~found~\ebtag{grd1 } f(r) = v \\
        ~~\ebkeyw{end} \\
        \ebkeyw{end}\\
      \end{array}
      &
      \begin{array}{l}
        \ebkeyw{context}~bin\_c0\\
        ~\ebkeyw{constants}~n~f~v \\
        ~\ebkeyw{axioms}\\
        ~~\ebtag{axm1 } f\in 1 \upto n \tfun N\\
        ~~\ebtag{axm2 } \forall i,j.~i\in 1 \upto n \\ 
        ~~~~~~~~\wedge j \in 1 \upto n \wedge \\
        ~~~~~~~~ i \leq j \\
        ~~~~~~~~\limp \\
        ~~~~~~~~ f(i) \leq f(j) \\
        ~~\ebtag{axm3 } v \in ran(f)\\
        ~\ebkeyw{theorem}\\
        ~~\ebtag{thm1 } n \geq 1 \\
        \ebkeyw{end}\\
      \end{array}
    \end{array}
    \]
  }
  \caption{An extract of the Binary Search algorithm in \eb.}
  \label{fig:eb2java:apply:bin:EB}
\end{figure}

The Binary Search algorithm finds the index of an element within
a sorted array. It works by choosing a pivot index in the domain of
the array and comparing the value at the index with the element one is
searching for; if the value at the index is greater than the element,
then the algorithm recursively searches for the element in the
sub-array to the left of the pivot, if the value is lesser than the
element then the algorithm searches for the element in the sub-array
to the right of the pivot; otherwise it returns the index of the
element. In \cite{bs:eb:09}, J.-R. Abrial presents the full
model of the algorithm in \eb. Context $bin\_c0$ in Figure
\ref{fig:eb2java:apply:bin:EB} declares constants $f$, $n$, and $v$ to
be the array,
its size, and the value the algorithm searches for, respectively. The
correct values of these constants are axiomatised within the context
representing the preconditions of the Binary Search algorithm. Axiom
\ebtag{axm1} declares $f$ to be a total function. Axiom \ebtag{axm2}
requires $f$ to be a {sorted} array, and axiom \ebtag{axm2} requires
the value $v$ to exist within the array $f$. Theorem \ebtag{thm1} can
be deduced from axiom \ebtag{axm1}.

The current left and right indexes of the array are given by variables $p$ and
$q$, that are given initial values $1$ and $n$, respectively. The
algorithm searches for an index $p \leq r \leq q$ in the domain of $f$
such that $f(r)=v$. These conditions are modelled as machine invariants
in \eb\ (see left Figure \ref{fig:eb2java:apply:bin:EB}). The
$initialisation$ event
picks up the middle value for the pivot index $r$. The machine
$bin\_m2$ declares three standard events $inc$, $dec$, and
$found$. Event $inc$ models the case when the value is to the right of
the pivot, $dec$ the case when the value is to the left of the pivot,
and $found$ when the value is at the pivot so no further actions are
made. 

\subsection{The Generated \jml-annotated Java code}
\label{eb2java:apply:generated}

Figures \ref{fig:eb2java:apply:bin:java}, and
\ref{fig:eb2java:apply:bin:java:inc} show excerpts of the
\jml-annotated Java code
generated by \ebtojava\ for the \eb\ model depicted in Figure
\ref{fig:eb2java:apply:bin:EB}. Figure
\ref{fig:eb2java:apply:bin:java} shows the translation for the
machine, and Figure \ref{fig:eb2java:apply:bin:java:inc} depicts the
event $inc$ (events $dec$ and $found$ are not shown). \ebtojava\
translates \eb\
constants such as $f$, $n$ and $v$, directly into Java as
\javakeyw{static final} variables. However, the tool does not generate
initial values for these variables. The initial values depend on the
constraints imposed by \ebkeyw{axioms} and \ebkeyw{theorems}. For
instance, $f$ must be
a sorted array, as described by axiom \ebtag{axm2}, that contains a
value $v$, as described by axiom \ebtag{axm3}. Nonetheless, \ebtojava\
generates \jml\ specifications for these axioms that one can use to
verify whether the initial values one conjectured for these constants
are correct or not. \ebtojava\ translates variables as
\javakeyw{private} class fields with the respective
\jmlkeyw{spec\_public} \jml\ clause so variables can be used for
verification. It also defines the corresponding getter and mutator
methods (not shown in the figure). \ebtojava\ translates \eb\
\ebkeyw{invariants} as \jml\ \jmlkeyw{invariants}. Finally, defines
the initial values of variables according to the $initialisation$ \eb\
event, and creates the corresponding threads (i.e. each \eb\ event is
translated as a Java class that extends \javacode{Thread}).

\begin{center}
  \begin{figure}
    \begin{lstlisting}[frame=none]
public class bin_m2{
   public Lock lock = new ReentrantLock(true);
   /******Constant definitions******/
   //@ public static constraint f.equals(\old(f)); 
   public static final BRelation<Integer,Integer> f = Test_bin_m2.random_f;
   // ...
   /******Axiom definitions******/
   /*@ public static invariant  f.domain().equals(new Enumerated(new Integer(1),n)) && f.range().isSubset(NAT.instance) && f.isaFunction() && BRelation.cross(new Enumerated(new Integer(1),n),NAT.instance).has(f); */
   /*@ public static invariant  
      (\forall Integer i;  (\forall Integer j; ((
         new Enumerated(new Integer(1),n).has(i) && 
         new Enumerated(new Integer(1),n).has(j) && 
         i.compareTo(j) <= 0) ==> (f.apply(i).compareTo(f.apply(j)) <= 0)))); */
   // ...
   /******Variable definitions******/
   /*@ spec_public */ private Integer p;
   // ... 
   /******Invariant definition******/
   /*@ public invariant
        NAT.instance.has(r) && new Enumerated(new Integer(1),n).has(p) && ...*/
   /*@ public normal_behavior
      requires true; assignable \everything;
   ensures p.equals(1)&&q.equals(n)&&r.equals(new Integer(new Integer(n+1)/2));*/
   public bin_m2(){
      p = 1;
      q = n;
      r = new Integer(new Integer(n + 1) / 2);
      // Threads initialisation
   }
    \end{lstlisting}
\caption{\jml-annotated Java code generated by \ebtojava\ from the
  $bin\_m2$ depicted in Figure \ref{fig:eb2java:apply:bin:EB}.}
  \label{fig:eb2java:apply:bin:java}
\end{figure}
\end{center}

Figure \ref{fig:eb2java:apply:bin:java:inc} shows the \jml-annotated
Java code that
\ebtojava\ generates for event $inc$. It declares a variable
\javacode{machine} that is a reference to the main Java class that
contains the definition of carrier sets, constants, and variables. It
defines 3 methods as explained in previous sections: the
\javacode{guard\_inc} method that returns \javakeyw{true} if
the evaluation of applying the variable \javacode{r} to the relation
\javacode{f} is less than the value \javacode{v} (the value that the
algorithm is looking for). That corresponds to guard of the event
$inc$ in the \eb\ model; the
\javacode{run\_inc} method that performs the actions of the event
updating the variable \javacode{r} to \javacode{(r+1+q) / 2}, and
variable \javacode{p} to \javacode{r + 1}. Notice that it is
important to get the value of variables before the assignment
(e.g. \javacode{r\_tmp}), if that was not the case the variable
\javacode{p} would have been assigned to a wrong value; and a method
that overrides the \javacode{run} method from the Java library
\javacode{Thread}. It
implements a critical section for the execution of the method
\javacode{run\_inc} simulating the behaviour of executing events in
\eb.

\begin{center}
  \begin{figure}
    \begin{lstlisting}[frame=none]
public class inc extends Thread{
   /*@ spec_public */ private bin_m2 machine;
   // class constructor
   /*@ public normal_behavior
      requires true; assignable \nothing;
      ensures \result <==> 
                machine.f.apply(machine.get_r()).compareTo(machine.v) < 0; */
   public /*@ pure */ boolean guard_inc() {
      return machine.f.apply(machine.get_r()).compareTo(machine.v) < 0;
   }

   /*@ public normal_behavior
      requires guard_inc(); assignable machine.p, machine.r;
      ensures guard_inc() &&  
            machine.get_r(),equals(\old(new Integer(new Integer(machine.get_r() + 1 + machine.get_q()) / 2))) &&  
            machine.get_p().equals(\old(new Integer(machine.get_r() + 1))); 
    also requires !guard_inc(); assignable \nothing; ensures true; */
   public void run_inc(){
      if(guard_inc()) {
         Integer r_tmp = machine.get_r();
         Integer p_tmp = machine.get_p();

         machine.set_r(new Integer(new Integer(r_tmp + 1 + machine.get_q()) / 2)); 
         machine.set_p(new Integer(r_tmp + 1));
      }
   }

   public void run() {
      while(true) {
         machine.lock.lock(); // start of critical section
         run_inc();
         machine.lock.unlock(); // end of critical section
      }
   }
}
\end{lstlisting}
\caption{Binary Search: code generated for the $inc$ event.}
\label{fig:eb2java:apply:bin:java:inc}
\end{figure}
\end{center}

Notice that the execution of the \eb\ model described in Figure
\ref{fig:eb2java:apply:bin:EB} is sequential: the guards of the events are
mutually exclusive. In this case, it is not necessary to translate the
model to a multithreaded Java version. The \ebtojava\ tool allows user
to generate sequential Java code for those models that their execution
is sequential. Figures \ref{fig:eb2java:apply:bin:java:seq} and
\ref{fig:eb2java:apply:bin:java:inc:seq} present the sequential
\jml-annotated Java
code generated by \ebtojava. Notice that the translation of the
machine define neither threads nor the reentrant lock, and  it
declares the classes corresponding to the translation of the events as
class fields, so they can be accessed from an external class that
implements the execution of the system (e.g. the \javacode{Framework}
class presented in Figure
\ref{fig:eb2java:apply:bin:seq:framework}). Notice also that the
translated Java class of the event
$inc$ neither extends \javacode{Thread} nor overrides the
\javacode{run} method.

\begin{center}
  \begin{figure}
    \begin{lstlisting}[frame=none]
public class bin_m2{
   ...
   public inc evt_dec = new dec(this);
   public dec evt_inc = new inc(this);
   public found evt_found = new found(this);

   ...

   public bin_m2(){
      p = 1;
      q = n;
      r = new Integer(new Integer(n + 1) / 2);
   }
}
    \end{lstlisting}
\caption{Excerpt of the sequential \jml-annotated Java code generated by \ebtojava\ from the
  $bin\_m2$ depicted in Figure \ref{fig:eb2java:apply:bin:EB}.}
  \label{fig:eb2java:apply:bin:java:seq}
\end{figure}
\end{center}

\begin{center}
  \begin{figure}
    \begin{lstlisting}[frame=none]
public class inc extends Thread{
   /*@ spec_public */ private bin_m2 machine;

   ...
   public inc(bin_m2 m) {
      ...
   }

   ...
   public /*@ pure */ boolean guard_inc() {
      ...
   }

   ...
   public void run_inc(){
      ...
   }
}
\end{lstlisting}
\caption{Excerpt of the sequential Java code generated for the $inc$
  event.}
\label{fig:eb2java:apply:bin:java:inc:seq}
\end{figure}
\end{center}

\ebtojava\ generates an additional Java class defining a framework (as
described in Section \ref{eb2java:imp:prelude}) that executes the logic of the system,
Figure \ref{fig:eb2java:apply:bin:seq:framework} depicts it. The user can customise
the proposed Framework by changing the condition of the
\javakeyw{while} for \javacode{machine.evt\_found.guard\_found()}.

\begin{center}
  \begin{figure}
    \begin{lstlisting}[frame=none]
public class Framework{
   public static void main(String[] args){
      bin_m2 machine = new bin_m2();
      while (true){
         if (machine.evt_found.guard_found())
            machine.evt_found.run_found(); break;
         else if (machine.evt_inc.guard_inc())
            machine.evt_inc.run_inc(); break;
         else if (machine.evt_dec.guard_dec())
            machine.evt_dec.run_dec(); break;
      }
   }
}
\end{lstlisting}
\caption{Binary Search: Sequential Java code generated for the $inc$
  event.}
\label{fig:eb2java:apply:bin:seq:framework}
\end{figure}
\end{center}

\section{Software Development with \ebtojava}
\label{eb2java:sw-dev}
We have validated the \ebtojava\ tool by generating \jml-annotated Java
code for several \eb\ models, by comparing our tool with different
Java code generators for \eb, and by applying it in two case studies
presented in next chapter (see Chapter
\ref{chapter:case-studies}). The case studies follow two strategies
for software development describe in the following. The first one,
(described in Section \ref{eb2java:sw-dev:stra:mvc}), shows how \ebtojava\
can be used as part of a software development strategy to generate the
core functionality (the Model) of an Android application that is
organised following the MVC (Model-View-Controller) design pattern
\cite{Patterns:Gamma:95}. The second one (described in Section
\ref{eb2java:sw-dev:stra:testing}) shows how \ebtojava\ and Java Unit (JUnit)
testing can be used to refine (improve) an \eb\ model system to
conform to an existing System Test Specification (STS) document.

\subsection{Strategy on Software Development using MVC design pattern}
\label{eb2java:sw-dev:stra:mvc}
Typical software applications include an interface (the View) that
interacts with the user, a functional core (the Model) that implements
the basic functionality of the application, and a linking part (the
Controller) that \emph{disguises} all requests made by the user so that
they can be understood by the Model. The Model implements methods to
edit data and to access the internal state of the application. It
might also include a registry of dependent Views to notify when data
changes occur during the \emph{rendering} of the interface. The Controller
implements wrapping code that transforms mouse input and keyboard
shortcuts to commands in the Model.

\ebtojava\ can be used to develop a system that follows the MVC design
pattern. The View-Controller components are developed using \emph{Usability
  Engineering} techniques as advocated by J. Nielsen in
\cite{UsabEng}; the core functionality is modelled in \eb\ and the
\ebtojava\ tool is used to generate the Model in Java.

The strategy comprises the following steps:

\begin{enumerate}
\item a system is modelled in \eb\ and it is refined to the desired
  level of abstraction via a hierarchy of machine refinements.

\item all proof obligations of the above \eb\ model are discharged in
  Rodin, so one can be sure about the correctness of the system
  modelled.

\item the \eb\ model is automatically translated to Java using
  the \ebtojava\ tool.

\item the View part of the system is developed using \emph{Usability
    Engineering} techniques. \ebtojava\ generates getter and setter
  methods for machine variables in the Java code generated enable
  communication between the Model and the View.
\end{enumerate}

This strategy has been successfully applied for the development of a
Social-Event Planner explained in  Section
\ref{case-studies:sep}.

\subsection{Strategy on Software Testing}
\label{eb2java:sw-dev:stra:testing}
The use of a formal specification language to model software
requirements eliminates ambiguity and reduces the chance of
errors that can be introduced during software development. Still it
remains the issue, coming up with the formal specification that
matches customer expectations, and an implementation that matches
the formal specification. 

We propose a strategy in which \ebtojava\ is used for testing the
behaviour of a reactive system modelled in \eb\ by simulating it in
Java.

The strategy comprises the following steps:

\begin{enumerate}

\item one models the system in \eb\ following an existing
  System Requirements Specification (SRS) document.

\item all proof obligations of the above \eb\ model are discharged
  in Rodin, so one can be sure about the correctness of the model of
  the system.

\item the \eb\ model is translated to Java code using \ebtojava.

\item based on an existing System Test Specification (STS) document,
  one can write Java Unit  (JUnit)~\cite{Link2003} tests manually that
  exercise the functionality of the generated Java code. 
\end{enumerate}

If a JUnit test fails, the \eb\ model is inspected
and evolved to conform to the STS document, and \ebtojava\ is used
again to generate Java code. This process will be repeated until it
covers all of the JUnit tests.
The correctness of the \eb\ model can then be verified by discharging
proof obligations using standard \eb\ tool-kits.  After making
changes to the existing model, if required, it is translated to Java using
\ebtojava,  and the test suite runs once more to check if the
behaviour of the model as expected. 

Note that the steps that describe our strategy are
complementary as they address very different questions about the
model. The purpose of testing the Java code generated by
\ebtojava\ is to check whether the model behaves as the user's
expectation while the correctness of the \eb\ model is verified by
discharging proof obligations.  In particular, a correct model might
not have the behaviour that the developer actually intended. Regarding
this direction, one could find some similarities between ProB
\cite{prob} model
checker and our strategy. Using ProB, the user can translate the STS
document as
predicates in \eb\ and then simulate the \eb\ system to conform to the
translated STS document to the model. The main difference between our
strategy and ProB is
that ProB works directly on the \eb\ model whereas the proposed
strategy gives an opportunity to the developer (not to an
expert in \eb) to test the Java code. Also, our strategy
allows users to use an ample set of third party
software to experiment further on the generated Java code, like the
user can use a third party Java program that automatically generates
inputs for testing.

Our strategy incorporates the use of a STS document with software
testing conducted in Java. That is, the strategy uncovers
inconsistencies in the \eb\ model vis a vis the STS (detecting whether
the \eb\ model captures user's intentions). This strategy has been
successfully applied for the testing of the Tokeneer reactive system
explained in  Section \ref{case-studies:tok}.

\section{Conclusion}
\label{eb2java:conclusion}
In this chapter we presented a series of rules to generate
\jml-annotated Java code from \eb\ models. We also introduced the
implementation of these rules as the \ebtojava\ tool that is a
Rodin plug-in. \ebtojava\ enables users to generate an actual
implementation of \eb\ models in Java. The Java code contains \jml\
specifications that enable users to
customise the Java code to check further if the customised code
validates the initial model in \eb. \ebtojava\ generates both
sequential and multithreaded implementations of the models. We have
validated the tool by applying it to an ample set of \eb\ models
to generate Java implementations. 

Software developers will be benefited by EventB2Java tool in the
following way, they can start the software development by proving the
system consistent by modelling it in Event-B. Once the model is proven
correct and has the necessary details, developers can translate it to
Java using EventB2Java. Developers can customise the Java code to
check if the customised code validate the initial model in Event-B
using the JML specifications. Although, using Rodin users
can obtain a final implementation of the models by refining the model to
a very close machine implementation, but the process
of refining involves a lot of mathematical expertise as every
refinement machine needs to be proven consistent with the previous
one. \ebtojava\ makes this process easy since users can decide the
level of abstraction in the \eb\ model and use the tool to generate
Java code. 
\chapter{Translating \eb\ Machines POs to \dafny}
\label{chapter:eb2dafny}

\paragraph{This chapter. } \eb\ is used for the formal modelling of
critical software. When modelling in \eb, one needs to prove the system
consistent by discharging proof obligations (POs). Rodin, the Eclipse
IDE for working with \eb,
automatically generates the POs, and provides different ways
to discharge them: 
\begin{itemize}
\item Rodin comes with an automatic prover (\textsf{New PP}).
\item Rodin also comes with third-party automatic provers (they need to be
installed as plug-ins).
\item Users can attempt to discharge proof
obligations
interactively using the interactive proving that Rodin provides. 
\end{itemize}
The work described in this chapter seeks to help users to discharge
\eb\ POs by translating them to the input language of \dafny. Program
verification with \dafny\ works by translating the program written in
\dafny\ to the Boogie proving engine \cite{boogie-leino-06}. Boogie
generates verification conditions (VC) to then pass them to
first-order automated theorem prover (e.g. Z3
\cite{Z3:overview}, Simplify \cite{simplify}, Zap \cite{zap}) to
determine the validity  of the VC. In this sense, \ebtodafny\ equips
Rodin with other theorem provers for discharging proof obligations,
e.g. Simplify or Zap theorem provers. Having \eb\ PO translated to
Dafny and then to Boogie has some advantages: Boogie encodes the VC to
the theorem prover in such a way to make it possible to reconstruct
from a failed proof an error trace; Boogie dynamically adjust the
number of threads to depending on the verification tasks
\cite{dafnyDevEnv}. It allows the verification process to run faster;
Boogie comes with a debugger, the Boogie Verification Debugger
\cite{BVC}, that helps \dafny\ users to understand the output of the
program verifier whenever a verification could not be made. This is
useful since understanding why the theorem prover could not prove a
proof obligation gives an insight to the user on where the problem
might be in the model.

This chapter presents a set of rewriting rules to translate \eb\ proof
obligations to
the input language of \dafny\ and the implementation of the
rewriting rules as the \ebtodafny\ Rodin plug-in. We do not use the
programming language of \dafny\ but the automatic verifier associated
to it to discharge the generated PO. \ebtodafny\ supports the
full \eb\ syntax. 
The work presented in this chapter is published in
\cite{conference:TOPI:12}. I
participated in the definition of the translation rules from \eb\
Proof Obligations to \dafny\ and its implementation. The rest of this
chapter is
organised as follows. The description of the type of proof
obligations generated by the Rodin platform are shown in Section
~\ref{ebtodafny:po}. Then, the translation of Rodin
proof-obligations to \dafny\ programs is explained in
Section~\ref{ebtodafny:translation}. Section~\ref{ebtodafny:implementation}
shows the implementation of the \ebtodafny\ tool. Finally,
Section~\ref{ebtodafny:conclusion} gives conclusions.

\paragraph{Contributions.}  
\begin{inparaenum}[\itshape i\upshape)]
\item The definition of a translation from
  \eb\ proof obligations to \dafny\ through a collection of rules, one for each component
  of an \eb\ proof obligation, 
\item the implementation of these translation rules as
  the \ebtodafny\ tool. The translation allows stepwise refinement based
  approaches to have access to an ample set of formal
  techniques. 
\item The implementation of a prelude in
  \dafny\ language that implements sets (e.g. sets, relations,
  functions) as datatypes and operations over these structures as
  functions.
\end{inparaenum}

\paragraph{Related work.} In~\cite{EventBSMT:11}, David D{\'e}harbe
presents an approach to translate Rodin platform proof-obligations to
the input language of the SMT-solvers. The approach handles proof
obligations that include boolean expressions, integer arithmetic
expressions, and basic sets and relations. The \ebtodafny\ plug-in
works in a similar direction. Yet, by generating bespoke \dafny/Boogie
proof obligations, we can improve the performance of the Simplify
\cite{simplify}, Zap \cite{zap} on which Boogie works.

In~\cite{eventb2why:12}, D. Mentr\'e, C. March\'e,
J. Filli\^atre and M. Asuka present the
\textsf{bpo2why} tool that translates proof-obligations generated by
the Atelier~B suite into Why~\cite{WhySpecAndTool03}
programs. Therefore, proof-obligations can be discharged using the
Krakatoa tool~\cite{WhyKrakatoaCaduceus07} or other automatic provers
like Z3~\cite{Z3:overview}. The \ebtodafny\ plug-in works in a
similar direction as the \textsf{bpo2why} tool, but our target
language is \dafny\ rather than Why.

\section{Rodin Proof Obligations}
\label{ebtodafny:po}
The Rodin proof-obligation generator automatically generates proof
obligations (POs) based on both the underlying machines and contexts as explained in
Section \ref{background:pos}. This section presents the type of POs
generated
by Rodin in a detailed way and explains some of them through an example
presented by J.-R. Abrial in~\cite{EB:Book} (Chapter 5).
We present proof obligations (POs) as \textsf{Sequents}:
given the set of hypotheses $H$ and the goal $G$, a \textsf{Sequent}
is represented by
\[
H\\
\vdash \\
G
\]
A sequent reads as follows: under the hypotheses $H$, prove the goal $G$.

Figures \ref{fig:ebtodafny:po:Abrial_abstract}
and~\ref{fig:ebtodafny:po:Abrial_Ref1} (taken from
the example presented in \cite{EB:Book} - Chapter 5)
depict an \eb\ example (both machines see the context shown in
Figure~\ref{fig:ebtodafny:po:Abrial_abstract} right). The example
presents an \eb\ model for searching an element in a sequence of
integers. The model finds an index $i$ of an element $v$ in a sequence
$f$. Context $ctx0$ in Figure \ref{fig:ebtodafny:po:Abrial_abstract}
defines a constant $n$ to be a natural number. It represents the size
of the sequence $f$. Sequence $f$ is defined as a total function
(axiom \ebtag{ax2}) that maps natural numbers (from $1$ to $n$) to
elements of set \ebkeyw{D}. Constant $v$ is defined as a natural
number and axiom \ebtag{ax3} states that $v$ is present in the range
of function $f$ \footnote{The purpose of theorem \ebtag{thm1} is to
  introduce a theorem proof obligation later on.}. The most abstract machine
(Figure~\ref{fig:ebtodafny:po:Abrial_abstract}) models the
search process in one step. The
$search$ event defines a local variable $k$ 
that takes a value from $1$ to the number of elements in the
sequence. Additionally, the evaluation of the function $f$ in $k$ is
equal to $v$, so the variable $i$ takes the value of $k$ (notice that
$v$ is always presented in the sequence $f$, axiom
\ebtag{ax3} in the context $ctx0$ states that). 

\begin{figure}[t]
  {\small
    \[
    \begin{array}{c@{\hspace*{30pt}}c}
      \begin{array}{l}
        \ebkeyw{machine}~m0\_a~\ebkeyw{sees}~ctx0\\
        ~\ebkeyw{variables} \\
        ~~i\\    
        ~\ebkeyw{invariants}\\
        ~~\ebtag{inv1 } i \in 1 \upto n\\
        ~\ebkeyw{events}\\
        ~~initialisation \\
        ~~~\ebkeyw{then} \\ 
        ~~~~\ebtag{act1 }i := 1\\
        ~~~\ebkeyw{end} \\
        ~~search \\
        ~~~\ebkeyw{any}\\
        ~~~~k
        ~~~\ebkeyw{where}\\
        ~~~~\ebtag{grd1 }k \in 1 \upto n \\
        ~~~~\ebtag{grd2 } f(k) = v\\
        ~~~\ebkeyw{then} \\
        ~~~~\ebtag{act1 } i := k\\
        ~~\ebkeyw{end}\\
        \ebkeyw{end} 
      \end{array}
      &
      \begin{array}{l}
        \ebkeyw{context}~ctx0~ \\ \\
        \ebkeyw{sets}~\ebkeyw{D} \\
        \ebkeyw{constants}~n~f~v \\
        \ebkeyw{axioms} \\
        ~~\ebtag{ax1} ~ n \in \nat \\
        ~~\ebtag{ax2} ~ f \in 1 \upto n \tfun \ebkeyw{D} \\
        ~~\ebtag{ax3} ~ v \in ran(f) \\
        \ebkeyw{theorem} \\
        ~~\ebtag{thm1} ~ n \in \nat1 \\
        \ebkeyw{end}
      \end{array}
    \end{array}
    \]
  }
  \caption{An abstract and context machine in \eb.}
  \label{fig:ebtodafny:po:Abrial_abstract}
\end{figure}

The first refinement of the model (see
Figure~\ref{fig:ebtodafny:po:Abrial_Ref1}),
introduces the search strategy. The strategy introduces variable $j$ that 
starts at value $0$ (see event $initialisation$) and it is then incremented by
$1$ (see event $progress$). According to theorem \ebtag{thm1\_r1}, the
value $v$ is in the evaluation from $j+1$ to $n$, of the
sequence $f$. Once variable $j$ takes the value in which the evaluation
of $f(j+1)$ is equal to $v$ (guard \ebtag{grd1\_r1}), the $search$
event might be triggered and $i$ will take the value $j+1$.

\begin{figure}[h]
  {\small
    \[
    \begin{array}{l}
      \ebkeyw{machine}~m1\_a~\ebkeyw{refines}~m0\_a~\ebkeyw{sees}~ctx0\\
      ~\ebkeyw{variables} ~~i ~j\\    
      ~\ebkeyw{invariants}\\
      ~~\ebtag{inv1\_r1 } j \in 0 \upto n\\
      ~~\ebtag{inv2\_r1 } v \not \in f[i \upto j]\\
      ~~\ebtag{thm1\_r1 } v \in f[j+1 \upto n]\\
      ~\ebkeyw{variant}~~n-j \\
      ~\ebkeyw{events}\\
      ~~initialisation \ebkeyw{ extends } initialisation\\
      ~~~\ebkeyw{then} \\ 
      ~~~~\ebtag{act1\_r1 }j := 0\\
      ~~~\ebkeyw{end} \\
      ~~search ~\ebkeyw{refines } search\\
      ~~~\ebkeyw{when}\\
      ~~~~\ebtag{grd1\_r1 } f(j+1) = v \\
      ~~~\ebkeyw{with} ~~\ebtag{k: } j + 1 = k\\
      ~~~\ebkeyw{then} \\
      ~~~~\ebtag{act1\_r1 } i := j + 1\\
      ~~\ebkeyw{end}\\ 
      ~~progress\\
      ~~~\ebkeyw{status} ~~convergent \\
      ~~~\ebkeyw{when}\\
      ~~~~\ebtag{grd1\_r1 } f(j+1) \not = v \\
      ~~~\ebkeyw{then} \\
      ~~~~\ebtag{act1\_r1 } j := j + 1\\
      ~~\ebkeyw{end}\\
      \ebkeyw{end} 
    \end{array}
    \]
  }
  \caption{Refinement  machine in \eb.}
  \label{fig:ebtodafny:po:Abrial_Ref1}
\end{figure}

The set of proof obligations generated by Rodin to prove the
consistency of a system is:

\paragraph{The Invariant Preservation Proof Obligation: } it states
that events must conform to machine invariants. For instance, for the
PO generated by Rodin
``$evt$/\ebtag{inv}/INV'', one needs to prove that event
$evt$ conforms to the invariant \ebtag{inv}. Let $evt$ be as follows

{\small
  \[
  \begin{array}{l}
    evt \\
    ~~~\ebkeyw{any } x\\
    ~~~\ebkeyw{where}\\
    ~~~~~~G(s,c,v,x) \\
    ~~~\ebkeyw{then} \\
    ~~~~~~v :| ~BA(s,c,v,x,v')\\
    \ebkeyw{end}
  \end{array}
  \]
}

The sequent takes as hypotheses the axioms and theorems, the
invariant, the guard of the event, and the before-after predicate, and
one needs to prove that the modified invariant holds

{\small
  \[
  \begin{array}{ll}
    Axioms~and~theorems & A(s,c) \\
    Invariants & I(s,c,v)\\
    Guards~of ~the ~event & G(s,c,v,x)\\
    Before-after ~predicate ~of~ the~ event & BA(s,c,v,x,v')\\
    \vdash &\vdash\\
    Modified~ specific~ invariant & inv(s,c,v')
  \end{array}
  \]
}

For the example presented in Figure
\ref{fig:ebtodafny:po:Abrial_abstract} Rodin generates a
\textsf{$search$/\ebtag{inv1}/INV} proof obligation that states that
the new value of $i$ (denoted $i'$) must respect the machine invariant

{\small
  \[
  \begin{array}{ll}
    ~~n \in \nat & ~~f \in 1 \upto n\tfun \ebkeyw{D} \\
    ~~v \in ran(f)&
    ~~n \in \nat1 \\
    ~~i \in 1 \upto n &
    ~~k \in 1 \upto n \\
    ~~f(k) = v &
    ~~i' = k \\
    \vdash &\\
    ~~i' \in 1 \upto n&
  \end{array}
  \]
}
\paragraph{The Guard strengthening Proof Obligation: } it states that
the guards of the refinement machine must be
stronger than the abstract event's guards. For instance, for the PO
generated by Rodin ``$evt$/\ebtag{grd}/GRD'', one needs to prove that
the guards of the refinement event $evt$ are stronger that the guard
\ebtag{grd} of event $evt$. Let event $evt0$ and its refinement $evt$
be as follows 

{\small
  \[
  \begin{array}{ll}
    \begin{array}{l}
      evt0 \\
      ~~~\ebkeyw{any } x\\
      ~~~\ebkeyw{where}\\
      ~~~~\ebtag{grd } g(s,c,v,x) \\
      ~~~~\ldots\\
      ~~~\ebkeyw{then} \\
      ~~~~\ldots\\
      \ebkeyw{end}
    \end{array}
    &
    \begin{array}{l}
      evt \ebkeyw{ refines } evt0\\
      ~~~\ebkeyw{any } y\\
      ~~~\ebkeyw{where}\\
      ~~~~~~H(y,s,c,w) \\
      ~~~\ebkeyw{with}\\
      ~~~~~~x  : W(x,s,c,w,y) \\
      ~~~\ebkeyw{then} \\
      ~~~~\ldots\\
      \ebkeyw{end}
    \end{array}
  \end{array}
  \]
}

The sequent takes as hypotheses the axioms and theorems, the abstract
and concrete invariants, the concrete event guards, and the witness
predicates for parameters, and one needs to prove that the guard of
the event of the abstract machine holds

{\small
  \[
  \begin{array}{ll}
    Axioms~and~theorems & A(s,c) \\
    Abstract~invariants & I(s,c,v)\\
    Concrete~invariants & J(s,c,v,w)\\
    Concrete~Event~Guards & H(y,s,c,w)\\
    Witness ~predicates ~for~ parameters & W(x,s,c,w,y)\\
    \vdash &\vdash\\
    Abstract~event~ specific~ guard & g(s,c,v,x)
  \end{array}
  \]
}

For the example presented in Figures
\ref{fig:ebtodafny:po:Abrial_abstract} and
\ref{fig:ebtodafny:po:Abrial_Ref1}, Rodin generates a 
``$search$/\ebtag{grd2}/GRD'' proof obligation that states that the
guards of the concrete event $search$ are stronger than the guard
\ebtag{grd2} in the abstract event $search$:

{\small
  \[
  \begin{array}{l}
    ~~n \in \nat \\
    ~~f \in 1 \upto n\tfun \ebkeyw{D} \\
    ~~v \in ran(f)\\
    ~~n \in \nat1 \\
    ~~i \in 1 \upto n \\
    ~~j \in 0 \upto n \\
    ~~v \not \in f[i \upto j] \\
    ~~v \in f[j+1 \upto n] \\
    ~~f(j+1) = v \\
    ~~j+1 = k \\
    \vdash \\
    ~~f(k) = v
  \end{array}
  \]
}

\paragraph{The Simulation Proof Obligation: } it states that the
execution of a concrete event $evt$ is not contradictory with the
execution of the abstract event that $evt$ is refining. For instance,
for the PO generated by Rodin ``$evt$/\ebtag{act}/SIM'', one needs to
prove that the
execution of the actions of the concrete event $evt$ does not contradict
the execution of the action \ebtag{act} from the $evt0$ abstract
event. The follows illustrates an event $evt0$ and its refinement $evt$

{\small
  \[
  \begin{array}{ll}
    \begin{array}{l}
      evt0 \\
      ~~~\ebkeyw{any } x\\
      ~~~\ebkeyw{where}\\
      ~~~~\ldots\\
      ~~~\ebkeyw{then} \\
      ~~~~\ebtag{act } v :| ~BA1(s,c,v,x,v')\\
      \ebkeyw{end}
    \end{array}
    &
    \begin{array}{l}
      evt \ebkeyw{ refines } evt0\\
      ~~~\ebkeyw{any } y\\
      ~~~\ebkeyw{where}\\
      ~~~~~~H(y,s,c,w) \\
      ~~~\ebkeyw{with}\\
      ~~~~~~x  : W1(x,s,c,w,y,w') \\
      ~~~~~~v'  : W2(v',s,c,w,y,w') \\
      ~~~\ebkeyw{then} \\
      ~~~~\ebtag{act2 } w  :|~ BA2(s,c,w,y,w') \\
      \ebkeyw{end}
    \end{array}
  \end{array}
  \]
}

The sequent takes as hypotheses the axioms and theorems, the abstract
and concrete invariants, the concrete event guards, the witness
predicates for variables and parameters,
and the concrete before-after predicate, and
one needs to prove that the before-after predicate of the
abstract machine event holds

{\small
  \[
  \begin{array}{ll}
    Axioms~and~theorems & A(s,c) \\
    Abstract~invariants & I(s,c,v)\\
    Concrete~invariants & J(s,c,v,w)\\
    Concrete~Event~Guards & H(y,s,c,w)\\
    Witness ~predicates ~for~ parameters & W1(x,s,c,w,y,w')\\
    Witness ~predicates ~for~ variables & W2(v',s,c,w,y,w')\\
    Concrete ~before-after ~predicate & BA2(s,c,w,y,w')\\
    \vdash &\vdash\\
    Abstract ~before-after ~predicate & BA1(s,c,x,v')\\
  \end{array}
  \]
}

For the example presented in Figures \ref{fig:ebtodafny:po:Abrial_abstract} and
\ref{fig:ebtodafny:po:Abrial_Ref1}, Rodin generates a
``$search$/\ebtag{act1}/SIM'' proof obligation:

{\small
  \[
  \begin{array}{l}
    ~~n \in \nat \\
    ~~f \in 1 \upto n\tfun \ebkeyw{D} \\
    ~~v \in ran(f)\\
    ~~n \in \nat1 \\
    ~~i \in 1 \upto n \\
    ~~j \in 0 \upto n \\
    ~~v \not \in f[i \upto j] \\
    ~~v \in f[j+1 \upto n] \\
    ~~f(j+1) = v \\
    ~~j+1 = k \\
    ~~i=j+1\\
    \vdash \\
    ~~i = k
  \end{array}
  \]
}

\paragraph{The numeric variant Proof Obligation: } it states that
under the guards of each
\ebkeyw{convergent} or \ebkeyw{anticipated} event, the 
numeric \ebkeyw{variant} is a natural number. For instance,
for the PO generated by Rodin ``$evt$/NAT'', one needs to
prove that under the guards of
event $evt$, the \ebkeyw{variant} is a natural number. Let machine $m$
and event $evt$ be as follows

{\small
  \[
  \begin{array}{ll}
    \begin{array}{l}
      \ebkeyw{machine}~m\\
      ~\ebkeyw{variables} \\
      ~~v\\    
      ~\ebkeyw{invariants}\\
      ~~I(sc,v)\\
      ~\ebkeyw{variant}\\
      ~~n(sc,v)\\
      ~\ebkeyw{events}\\
      ~~~\ldots \\
      ~\ebkeyw{end}\\
      \ebkeyw{end} \\
    \end{array}
    &
    \begin{array}{l}
      evt \\
      ~~~\ebkeyw{status convergent} \\
      ~~~\ebkeyw{any } x\\
      ~~~\ebkeyw{where}\\
      ~~~~~~G(s,c,v,x) \\
      ~~~\ebkeyw{then} \\
      ~~~~~~\ldots\\
      \ebkeyw{end}\\
    \end{array}
  \end{array}
  \]
}

The sequent takes as hypotheses the axioms and theorems, the invariant, and the guards of the
event, and one needs to prove that the \ebkeyw{variant} is a
natural number

{\small
  \[
  \begin{array}{ll}
    Axioms~and~theorems & A(s,c) \\
    Invariants & I(s,c,v)\\
    Guards~of ~the ~event & G(s,c,v,x)\\
    \vdash &\vdash\\
    A~ numeric~variant~is~a~natural~number & n(s,c,v') \in \nat
  \end{array}
  \]
}

For the example presented in Figures
\ref{fig:ebtodafny:po:Abrial_abstract} and
\ref{fig:ebtodafny:po:Abrial_Ref1} Rodin generates a
\textsf{$progress$/NAT} proof obligation

{\small
  \[
  \begin{array}{l}
    ~~n \in \nat \\
    ~~f \in 1 \upto n\tfun \ebkeyw{D} \\
    ~~v \in ran(f)\\
    ~~n \in \nat1 \\
    ~~i \in 1 \upto n \\
    ~~j \in 0 \upto n\\
    ~~v \not \in f[i \upto j] \\
    ~~v \in f[j+1 \upto n] \\
    ~~f(j+1) \not = v \\
    \vdash \\
    ~~n - j \in \nat
  \end{array}
  \]
}

\paragraph{The Variant Proof Obligation: } it states that each
\ebkeyw{convergent} event
decreases a numeric (or remove elements from a set) \ebkeyw{variant}, and each \ebkeyw{anticipated}
event does not increase a numeric (or does not add elements to a set) \ebkeyw{variant}. For instance,
for the PO generated by Rodin ``$evt$/VAR'', one needs to
prove that a \ebkeyw{convergent}
event $evt$ decreases the (or remove elements from
the set) \ebkeyw{variant}, also 
that an \ebkeyw{anticipated} event $evt$ does not increase the
(or does not add elements to the set) \ebkeyw{variant}. Let $m$ be a
machine and $evt$ an event as follows

{\small
  \[
  \begin{array}{ll}
    \begin{array}{l}
      \ebkeyw{machine}~m\\
      ~\ebkeyw{variables} \\
      ~~v\\    
      ~\ebkeyw{invariants}\\
      ~~I(sc,v)\\
      ~\ebkeyw{variant}\\
      ~~n(sc,v)\\
      ~\ebkeyw{events}\\
      ~~~\ldots \\
      ~\ebkeyw{end}\\
      \ebkeyw{end} \\
    \end{array}
    &
    \begin{array}{l}
      evt \\
      ~~~\ebkeyw{status convergent} \\
      ~~~\ebkeyw{any } x\\
      ~~~\ebkeyw{where}\\
      ~~~~~~G(s,c,v,x) \\
      ~~~\ebkeyw{then} \\
      ~~~~~~\ldots\\
      \ebkeyw{end}\\
    \end{array}
  \end{array}
  \]
}

The sequent takes as hypotheses the axioms and theorems, the
invariant, the guards of the event,
and the before-after predicate of the event, and one needs to
prove that the value of the modified \ebkeyw{variant} is smaller (or a
subset) than
the value of \ebkeyw{variant} without modification

{\small
  \[
  \begin{array}{ll}
    Axioms~and~theorems & A(s,c) \\
    Invariants & I(s,c,v)\\
    Guards~of ~the ~event & G(s,c,v,x)\\
    Before-after ~predicate ~of~ the~ event & BA(s,c,v,x,v')\\
    \vdash &\vdash\\
    Modified~ variant~smaller~than~variant & n(s,c,v') < n(s,c,v) \\
    & or (n(s,c,v') \subset n(s,c,v))
  \end{array}
  \]
}

For \ebkeyw{anticipated} events, one needs to prove that the modified
variant is not greater than the \ebkeyw{variant} ($\ldots \vdash n(s,c,v') \leq
n(s,c,v)$) or it does not add elements to the set \ebkeyw{variant} 
($\ldots \vdash n(s,c,v') \subseteq n(s,c,v)$). For instance, Rodin
generates this \textsf{$progress$/VAR}
proof obligation for the \ebkeyw{convergent} event presented in Figure
\ref{fig:ebtodafny:po:Abrial_Ref1}

{\small
  \[
  \begin{array}{l}
    ~~n \in \nat \\
    ~~f \in 1 \upto n\tfun \ebkeyw{D} \\
    ~~v \in ran(f)\\
    ~~n \in \nat1 \\
    ~~i \in 1 \upto n \\
    ~~j \in 0 \upto n\\
    ~~v \not \in f[i \upto j] \\
    ~~v \in f[j+1 \upto n] \\
    ~~f(j+1) \not = v \\
    ~~ j' = j+1\\
    \vdash \\
    ~~n - (j + 1) < n - j
  \end{array}
  \]
}

\paragraph{The non-deterministic witness Proof Obligation: } it states
that each witness of a concrete event indeed exists. For instance,
for the PO generated by Rodin ``$evt$/$x$/WFIS'', one needs to
prove that a witness $x$ of
a concrete event $evt$ exists. Let $evt$ be a concrete event

{\small
  \[
  \begin{array}{l}
    evt \ebkeyw{ refines } evt0\\
    ~~~\ebkeyw{any } y\\
    ~~~\ebkeyw{where}\\
    ~~~~~~H(y,s,c,w) \\
    ~~~\ebkeyw{with}\\
    ~~~~~~x  : W(x,s,c,w,y,w') \\
    ~~~\ebkeyw{then} \\
    ~~~~~~BA2(s,c,w,y,w') \\
    \ebkeyw{end}
  \end{array}
  \]
}

The sequent takes as hypotheses the axioms and theorems, the abstract
and concrete invariants, the
concrete event guards, and the concrete before-after predicate, and
one needs to prove that the witness exists

{\small
  \[
  \begin{array}{ll}
    Axioms~and~theorems & A(s,c) \\
    Abstract~invariants & I(s,c,v)\\
    Concrete~invariants & J(s,c,v,w)\\
    Concrete~Event~Guards & H(y,s,c,w)\\
    Concrete ~before-after ~predicate & BA2(s,c,w,y,w')\\
    \vdash &\vdash\\
    Existance ~of~witness & \exists x \qdot W(x,s,c,w,y,w')\\
  \end{array}
  \]
}

For the example presented in Figures \ref{fig:ebtodafny:po:Abrial_abstract} and
\ref{fig:ebtodafny:po:Abrial_Ref1}, Rodin generates a
``$search$/$k$/WFIS'' proof obligation

{\small
  \[
  \begin{array}{l}
    ~~n \in \nat \\
    ~~f \in 1 \upto n\tfun \ebkeyw{D} \\
    ~~v \in ran(f)\\
    ~~n \in \nat1 \\
    ~~i \in 1 \upto n \\
    ~~j \in 0 \upto n\\
    ~~v \not \in f[i \upto j] \\
    ~~v \in f[j+1 \upto n] \\
    ~~f(j+1) = v \\
    ~~i' = j+1\\
    \vdash \\
    ~~\exists k \qdot j + 1 = k
  \end{array}
  \]
}

\paragraph{The theorem Proof Obligation: } it states that a context or
a machine theorem is indeed provable. For instance,
for the PO generated by Rodin ``\ebtag{thm1}/THM'', in the example
presented in Figure \ref{fig:ebtodafny:po:Abrial_abstract}, one needs
to prove that the theorem \ebtag{thm1} is indeed provable

{\small
  \[
  \begin{array}{l}
    ~~n \in \nat \\
    ~~f \in 1 \upto n\tfun \ebkeyw{D} \\
    ~~v \in ran(f)\\
    \vdash \\
    ~~n \in \nat1
  \end{array}
  \]
}

\paragraph{The Well-Definedness Proof Obligation: } it states that axioms,
theorems, invariants, guards, actions, variants, and witnesses are well
defined. For instance,
for the PO generated by Rodin ``$search$/\ebtag{grd2}/WD'', in the
example presented in Figure \ref{fig:ebtodafny:po:Abrial_abstract},
one needs to prove that guard \ebtag{grd2} of event
$search$ is well defined

{\small
  \[
  \begin{array}{l}
    ~~n \in \nat \\
    ~~f \in 1 \upto n\tfun \ebkeyw{D} \\
    ~~v \in ran(f)\\
    ~~n \in \nat1 \\
    ~~i \in 1 \upto n \\
    ~~k \in 1 \upto n\\
    \vdash \\
    ~~k \in dom(f) \wedge f \in \mathbb{Z} \pfun \ebkeyw{D}
  \end{array}
  \]
}

\paragraph{The Feasibility Proof Obligation: } it states that a
non-deterministic action is
feasible. For instance,
for the PO generated by Rodin ``$evt$/\ebtag{act}/FIS'',
one needs to prove that the non-deterministic assignment in action
\ebtag{act} of event $evt$ is feasible. Let $evt$ be an event

{\small
  \[
  \begin{array}{ll}
    evt\\
    ~~~\ebkeyw{any } x\\
    ~~~\ebkeyw{where}\\
    ~~~~~~G(s,c,v,x) \\
    ~~~\ebkeyw{then} \\
    ~~~~~~\ebtag{act } v :|~BA(s,c,v,x,v') \\
    \ebkeyw{end}
  \end{array}
  \]
}

The sequent takes as hypotheses the axioms and theorems, the
invariants, and the
event guards, and one needs to prove the existence of $v'$

{\small
  \[
  \begin{array}{ll}
    Axioms~and~theorems & A(s,c) \\
    Invariants & I(s,c,v)\\
    Event~Guards & G(s,c,v,x)\\
    \vdash &\vdash\\
    \exists v' \qdot before-after~predicate & \exists v' \qdot BA(s,c,v,x,v')\\
  \end{array}
  \]
}

\paragraph{The Guard Merging Proof Obligation: } it states that the
guards of a concrete event that
merges two abstract events are stronger that the disjunction abstract
events' guards. For instance,
for the PO generated by Rodin ``$evt$/MRG'',
one needs to prove that the $evt$'s guards are stronger than the
disjunction of the
abstract events that $evt$ is refining. Let $evt$ be an event that refines
both $evt01$ and $evt02$ abstract events

{\small
  \[
  \begin{array}{lll}
    \begin{array}{l}
      evt01\\
      ~~~\ebkeyw{any } x\\
      ~~~\ebkeyw{where}\\
      ~~~~~~G1(s,c,v,x) \\
      ~~~\ebkeyw{then} \\
      ~~~~~~S \\
      \ebkeyw{end}
    \end{array}
    &
    \begin{array}{l}
      evt02\\
      ~~~\ebkeyw{any } x\\
      ~~~\ebkeyw{where}\\
      ~~~~~~G2(s,c,v,x) \\
      ~~~\ebkeyw{then} \\
      ~~~~~~S \\
      \ebkeyw{end}
    \end{array}
    \begin{array}{l}
      evt \ebkeyw{ refines } \\
      ~~~~~~~evt01\\
      ~~~~~~~evt02\\
      ~~~\ebkeyw{any } x\\
      ~~~\ebkeyw{where}\\
      ~~~~~~H(s,c,v,x) \\
      ~~~\ebkeyw{then} \\
      ~~~~~~S \\
      \ebkeyw{end}
    \end{array}
  \end{array}
  \]
}

The sequent takes as hypotheses the axioms and theorems, the abstract
invariants, and the
concrete event guards, and one needs to prove the disjunction
of the abstract guards

{\small
  \[
  \begin{array}{ll}
    Axioms~and~theorems & A(s,c) \\
    Abstract~Invariants & I(s,c,v)\\
    Concrete~Event~Guards & H(s,c,v,x)\\
    \vdash &\vdash\\
    Disjunction~of~abstract~guards & G1(s,c,v,x) \vee G2(s,c,v,x)\\
  \end{array}
  \]
}

Next section shows how each proof obligation generated by Rodin is
translated to the input language of \dafny.

\section{Expressing \eb\ Proof Obligations in \dafny}
\label{ebtodafny:translation}
\dafny\ programming language (described in Section
\ref{background:dafny}) does not natively support all structures
for predicate and set theory used in \eb. Sets, relations, and other
\eb\ structures were defined in \dafny\ as datatypes, and
operations over these structures as functions. A relation 
\dafnycode{Relation<D,R>} between a set of type \dafnycode{D} and a
set of type \dafnycode{R} is a constructed type in \dafny, formalised
with the aid of a \dafnycode{Rel} type constructor, that has three
parameters, \dafnycode{domain} of type \dafnycode{D},
\dafnycode{range} of type \dafnycode{R}, and a \dafnycode{map} between
the domain and the range,
formalised as a set of pairs (as depicted
in Figure \ref{fig:eb2dafny:translation:rel_set}). For the \dafnycode{Rel} type constructor, in
the style of other languages like Objective Caml or the PVS
language~\cite{PVSLanguageReference}, \dafny\ implicitly declares a
\dafnycode{Rel?} predicate that returns \dafnykeyw{true} of any
constructed element formed with the type constructor \dafnycode{Rel}.

\begin{figure}
  {\small
    \[
    \begin{tabular}{l}
      \dafnykeyw{datatype } \dafnycode{Pair<S,T> = Pr(x: S, y: T);} \\ \\
      \dafnykeyw{datatype } \dafnycode{Relation<D,R> =} \\
      \quad\dafnycode{Rel(domain: set<D>, range: set<R>,}\\
      \quad\quad\quad~~\dafnycode{map: set<Pair<D,R>>);}
    \end{tabular}
    \]
  }
\caption{Formalising relation structures in \dafny.}
\label{fig:eb2dafny:translation:rel_set}
\end{figure}

Modelling relations and sets as datatypes rather than as classes has
two main advantages in \dafny. First, instances of classes require new
allocations, and second, their fields would need method frame
declarations (the \dafnykeyw{modifies} clause of \dafny), that can
degrade the performance of \dafny/Boogie/Z3. However, note that an
\eb\ relation can be used anywhere that a set can appear (a
relation is a set of pairs), but unfortunately datatypes cannot be
inherited. Therefore, the translation from \eb\ to \dafny\ makes sure
that operations are called on the right datatype.

The following sections present the translation of contexts and machines,
proof obligations and \eb\ operators.

\subsection{Translating \eb\ machines}
\label{ebtodafny:translation:eb-machine}
It is necessary to translate \eb\ machine and context information to
\dafny\ for
\dafny\ to be able to discharge the proof obligations: definition of
carriers sets, constants, and variables are necessary since sequents
use them and they need to be defined in \dafny; definition of axioms,
theorems, and invariants are also necessary since sequents are
composed of them as explained in the previous section. The translation
of \eb\ machines to \dafny\ uses operators that were defined via
syntactic rules.  \EBTODAFNYN\ operator translates \eb\ machines and
contexts to \dafny\ programming language. \EBTODAFNYN\ is
helped by operator \DTypeN\ that translates the type of \eb\ variables
and constants to the corresponding type in \dafny, using the datatypes
explained above. It also uses the operator \DPREDN\ to translate any
\eb\ predicate or expression to \dafny.

\begin{figure}
  {\small
    \[
    \begin{prooftree}
      \begin{array}{ll}
        \EBTODAFNY{\ebkeyw{constants}\;c} = \texttt{C} & \EBTODAFNY{\ebkeyw{sets}\;S} = \texttt{S}\\
        \EBTODAFNY{\ebkeyw{axioms}\;X(s, c)} = \texttt{X} &\EBTODAFNY{\ebkeyw{theorems}\;T(s, c)} = \texttt{T}\\
        \EBTODAFNY{\ebkeyw{invariants}\;I(s, c, v)} = \texttt{I}&\EBTODAFNY{\ebkeyw{variables}\;v} = \texttt{V}\\
      \end{array}
      \using \textsf{(Prel)}
      \justifies
      \begin{array}{ll}
        \EBTODAFNYN(\\
        \begin{array}{l}
          \ebkeyw{machine } M \\
          ~~\ebkeyw{sees } ctx \\
          ~~\ebkeyw{variables } v \\
          ~~\ebkeyw{invariants } I(s,c,v) \\
          ~~\ebkeyw{events } e \\
          \ebkeyw{end}
        \end{array}
        &
        \begin{array}{l}
          \quad\ebkeyw{context}~ctx \\
          \quad~~\ebkeyw{sets}~S \\
          \quad~~\ebkeyw{constants}~c\\
          \quad~~\ebkeyw{axioms}~X(s, c)\\
          \quad~~\ebkeyw{theorems}~T(s, c)\\
          \quad\ebkeyw{end} \\ \\
        \end{array}
        \\) =\\
        \texttt{S}\quad\texttt{C}\quad\texttt{X}\\ \\
        \texttt{T}\quad\texttt{V}\quad\texttt{I}
      \end{array}
    \end{prooftree}
    \]
  }
  \caption{Translation rule for an \eb\ machine and its context to \dafny.}
  \label{fig:ebtodafny:translation:eb-machine:RulePrel}
\end{figure}

Figure \ref{fig:ebtodafny:translation:eb-machine:RulePrel} depicts the
syntactic rule
\textsf{Prel} to translate \eb\ machine $M$ and the context it sees to
\dafny. This information is necessary to discharge the proof
obligations. Translation of refinement machines follows the same
structures as rule $M$ adding the gluing invariant and the new
variables.

Carrier sets are being modelled as set of integers. 

{\small
  \[
  \begin{prooftree}
    \using \textsf{(Sets)}
    \justifies
    \begin{array}{c}
      \EBTODAFNY{\ebkeyw{Sets } S} 
      = \\
      \dafnykeyw{var}~S:~\dafnycode{Set<Integer>;}
    \end{array}
  \end{prooftree}
  \]
}

As \dafny\ does
not include constants or axioms, constants are being modelled as 0-ary
integer functions and axioms as boolean functions with a
post-condition that introduces the axiom. Constants are assumed in the
translation of a proof obligation. Theorems are translated similar to
axioms, but they are checked (the clause \dafnykeyw{assert}) instead. The operator
\DTypeN\ in rules \textsf{Constants} and \textsf{Var} returns the
corresponding \dafny\
datatype as explained at the beginning of this section (Section
\ref{ebtodafny:translation})

{\small
  \[
  \begin{prooftree}
    \DType{c} = \texttt{Type}
    \using \textsf{(Constants)}
    \justifies
    \begin{array}{c}
      \EBTODAFNY{\ebkeyw{constants } c} = \\
      \dafnykeyw{function}~c():~\texttt{Type}\dafnycode{;}
    \end{array}
  \end{prooftree}
  \]
}

{\small
  \[
  \begin{prooftree}
    \DPRED{A} = \texttt{A}
    \using \textsf{(Axioms)}
    \justifies
    \begin{array}{c}
      \EBTODAFNY{\ebkeyw{axioms }A} = \\
      \dafnykeyw{function}~ axm():\dafnycode{bool}\\
      ~~\dafnykeyw{ensures}~ \texttt{A}
    \end{array}
  \end{prooftree}
  \]
}

\eb\ variables are modelled as variables in \dafny\

{\small
  \[
  \begin{prooftree}
    \DType{v} = \texttt{Type}
    \using \textsf{(Var)}
    \justifies
    \begin{array}{c}
      \EBTODAFNY{\ebkeyw{variables } v} 
      = \\
      \dafnykeyw{var}~v:~\texttt{Type}\dafnycode{;}
    \end{array}
  \end{prooftree}
  \]
}
Invariants are modelled as boolean functions with 
post-conditions that can be assumed or asserted. In Rule
\textsf{Invariants}, the operator \textsf{DafnyVar} collects all
variables from the invariant and imposes a non-nullness condition over
them that is a required condition in \dafny. In case there are not
variables in the \eb\ invariant, the \textsf{DafnyVar} returns a
\dafnykeyw{ensures} \dafnycode{true;}.

{\small
  \[
  \begin{prooftree}
    \DPRED{I} = \texttt{I} ~~~ \textsf{DafnyVar}(I) = \texttt{nonNullVars}
    \using \textsf{(Invariants)}
    \justifies
    \begin{array}{l}
      \EBTODAFNY{\ebkeyw{Invariants } I} = \\
      \dafnykeyw{function}~nonNullnessCond():~\dafnycode{bool} \dafnycode{\{} \\
      ~~~~~ \texttt{nonNullVars} \\ 
      \dafnycode{\}}\\ \\
      \dafnykeyw{function } inv() :\dafnycode{ bool};\\
      ~~~~\dafnykeyw{requires}~nonNullnessCond()\dafnycode{;} \\
      ~~~~\dafnykeyw{ensures}~\texttt{I}\dafnycode{;} 
    \end{array}
  \end{prooftree}
  \]
}

\subsection{Translating \eb\ proof obligations}
\label{ebtodafny:translation:po}
The translation of \eb\ proof obligations to \dafny\ uses operators
that were defined via syntactic rules, one for each proof obligation
generated by Rodin
(e.g. \textsf{INV}, \MRGN, \GRDN, \SIMN, \NATN, \FISN, \WFISN,
{\textsf{VAR}). The parameters for each operator depend on the type of
  the PO, for instance, the operator \FISN\ for the feasibility proof
  obligation takes one parameter, an \eb\ machine, whereas the
  operator \GRDN\ takes two, an \eb\ machine and its refinement.
  Depending on the type of proof obligation generated by
  Rodin, a \dafnykeyw{ghost method} is declared that might or might
  not assume local invariants, theorems or axioms. An
  additional operator (\CtxN) is defined. It translates \eb\
  \ebkeyw{axioms} and \ebkeyw{theorems} presented in a context to
  \dafny. The rules are presented as follows:
  
  \paragraph{The Invariant Preservation Proof Obligation: } rule
  \textsf{INV} generates a method in \dafny\ that
  assumes the translation of the invariants of the abstract and
  concrete machines, the translation non-nullness axiom and theorems,
  the translation of the predicate related to the witness,
  the translation of the before-after predicate of the concrete event,
  and the translation of the guards of
  the refined event. The method finally asserts the result of the translation of the
  modified invariant.

  {\small
    \[
    \begin{prooftree} 
      \begin{array}{ll}
        \Ctx{Ctx} = \texttt{AT} \\
        \EBTODAFNY{\ebkeyw{invariant } I(s,c,v)} = \texttt{I} & \EBTODAFNY{\ebkeyw{invariant } J(s,c,v,w)} = \texttt{J} \\ 
        \DPRED{H(y,s,c,w)}=\texttt{H} & \DPRED{W2(v',s,c,w,y,w')}=\texttt{W2}\\
        \DPRED{BA2(s,c,w,y,w')}=\texttt{BA2} & \EBTODAFNY{\ebkeyw{invariant } J(s,c,v',w')} = \texttt{J'}
      \end{array}
      \using \textsf{(Inv)}
      \justifies
      \begin{array}{l}
        \InvConcreteN ( \\
        \begin{array}{ll}
          \begin{array}{l}
            \ebkeyw{machine } M \ebkeyw{ sees } Ctx\\
            \ebkeyw{variables } v\\
            \ebkeyw{invariant } I(s,c,v)\\
            \ebkeyw{event } evt0\\ 
            \quad \ldots\\
            \quad \ebkeyw{end} \\
            \ebkeyw{end } , 
          \end{array}
          & 
          \begin{array}{l}
            \ebkeyw{machine } N \ebkeyw{ refines } M\\
            \ebkeyw{variables } w\\
            \ebkeyw{invariant } J(s,c,v,w)\\
            \ebkeyw{event } evt \ebkeyw{ refines } evt0\\ 
            \quad \ebkeyw{any } y  \ebkeyw{ where} \\
            \quad\quad H(y,s,c,w) \\
            \quad \ebkeyw{with} \\
            \quad\quad v' :|~ W2(v',s,c,w,y,w') \\
            \quad \ebkeyw{then} \\
            \quad\quad w :|~ BA2(s,c,w,y,w') \\
            \quad \ebkeyw{end}\\
            \ebkeyw{end } ) 
          \end{array}\\
        \end{array}
        \\= \\
        \begin{array}{l}
          \dafnykeyw{ghost method }evt\_inv\_INV() \\
          ~~\dafnykeyw{assume}~\texttt{AT}~\dafnykeyw{\&\&}~\texttt{I}~\dafnykeyw{\&\&}~\texttt{J}~\dafnykeyw{\&\&}~\texttt{H}~\dafnykeyw{\&\&}~\texttt{W2}~\dafnykeyw{\&\&}~\texttt{BA2}
          \dafnycode{;}\\
          ~~\dafnykeyw{assert}~\texttt{J'} \dafnycode{;}
        \end{array}
      \end{array}
    \end{prooftree}
    \]
  }

Rule \textsf{Inv} also takes into account the invariant preservation
proof obligation for just an abstract machine. Rule \textsf{Inv} showed
below, generates a \dafny\ method that assumes the translation of the
abstract invariant, the translation of the non-nullness axiom and
theorems, the translation of the before-after predicate and guards of
the abstract event. The method finally asserts the result of the
translation of the modified invariant.

 {\small
    \[
    \begin{prooftree} 
      \begin{array}{ll}
        \Ctx{Ctx} = \texttt{AT} \\
        \EBTODAFNY{\ebkeyw{invariant } I(s,c,v)} = \texttt{I} & \DPRED{G(s,c,v,x)}=\texttt{G}\\
        \DPRED{BA(s,c,v,x,v')}=\texttt{BA} & \EBTODAFNY{\ebkeyw{invariant } I(s,c,v')} = \texttt{I'}
      \end{array}
      \using \textsf{(Inv)}
      \justifies
      \begin{array}{l}
        \InvAbstractN ( \\
        \ebkeyw{machine } M \ebkeyw{ sees } Ctx\\
        \ebkeyw{variables } v\\
        \ebkeyw{invariant } I(s,c,v)\\
        \ebkeyw{event } evt\\ 
        \quad \ebkeyw{any } x  \ebkeyw{ where} \\
        \quad\quad G(s,c,v,x) \\
        \quad \ebkeyw{then} \\
        \quad\quad v :|~ BA(s,c,v,x,v') \\
        \quad \ebkeyw{end} \\
        \ebkeyw{end })
        \\= \\
        \dafnykeyw{ghost method } evt\_inv\_INV() \\ 
        ~~\dafnykeyw{assume}~\texttt{AT}~\dafnykeyw{\&\&}~\texttt{I}~\dafnykeyw{\&\&}~\texttt{G}~\dafnykeyw{\&\&}~\texttt{BA}\dafnycode{;}\\
        ~~\dafnykeyw{assert}~\texttt{I'}\dafnycode{;}
      \end{array}
    \end{prooftree}
    \]
  }

  \paragraph{The Feasibility Proof Obligation: } rule
  \textsf{Feas} generates a \dafny\ method that assumes the
  translation of the invariants, the translation of the non-nullness
  axioms and theorems, and the translation of the guard of the
  event. The
  method finally asserts the result of the translation of the existence
  of the witness value ensuring the before-after predicate of the event.

  {\small
    \[
    \begin{prooftree} 
      \begin{array}{ll}
        \Ctx{Ctx} = \texttt{AT} &
        \EBTODAFNY{\ebkeyw{invariants } I(s,c,v)} = \texttt{I} \\ 
        \DPRED{G(s,c,v,x)}=\texttt{G} &
        \DPRED{BA(s,c,v,x,v')}=\texttt{BA}
      \end{array}
      \using \textsf{(Feas)}
      \justifies
      \begin{array}{l}
        \FISN ( \\
        \ebkeyw{machine } M \ebkeyw{ sees } Ctx\\
        \ebkeyw{variables } v\\
        \ebkeyw{invariant } I(s,c,v)\\
        \ebkeyw{event } evt\\ 
        \quad \ebkeyw{any } x  \ebkeyw{ where} \\
        \quad\quad G(s,c,v,x) \\
        \quad \ebkeyw{then} \\
        \quad\quad v :|~ BA(s,c,v,x,v') \\
        \quad \ebkeyw{end} \\
        \ebkeyw{end })
        = \\
        \dafnykeyw{ghost method } evt\_act\_FIS() \\
        ~~\dafnykeyw{assume } \texttt{AT}~\dafnykeyw{\&\&}~\texttt{I}~\dafnykeyw{\&\&}~\texttt{G}\dafnycode{;}\\
        ~~\dafnykeyw{assert }~\dafnycode{(}\dafnykeyw{exists}~v' \dafnycode{::}~\texttt{BA}\dafnycode{);}
      \end{array}
    \end{prooftree}
    \]
  }

  \paragraph{The Guard Strengthening Proof Obligation: } rule
  \textsf{Grd} generates a method in \dafny\ that assumes the
  translation of the abstract and concrete invariants, the translation
  of the non-nullness axioms and theorems, the translation of the
  guard of the refined
  event, and the translation of the predicate related to the
  witness. The method finally
  asserts the result the abstract event's guard translation.

  {\small
    \[
    \begin{prooftree} 
      \begin{array}{ll}
        \EBTODAFNY{\ebkeyw{invariants } I(s,c,v)} = \texttt{I} &
        \Ctx{Ctx} = \texttt{AT}\\
        \EBTODAFNY{\ebkeyw{invariant } J(s,c,v,w)} = \texttt{J} &
        \DPRED{H(y,s,c,w)}=\texttt{H}\\ 
        \DPRED{W(x,s,c,w,y)}=\texttt{W} &  \DPRED{g(s,c,v,x)}=\texttt{G}
      \end{array}
      \using \textsf{(Grd)}
      \justifies
      \begin{array}{l}
        \GRDN ( \\
        \begin{array}{ll}
          \begin{array}{l}
            \ebkeyw{machine } M \ebkeyw{ sees } Ctx\\
            \ebkeyw{variables } v\\
            \ebkeyw{invariant } I(s,c,v)\\
            \ebkeyw{event } evt0\\ 
            \quad \ebkeyw{any } x  \ebkeyw{ where} \\
            \quad\quad \ebtag{grd } g(s,c,v,x) \\
            \quad \ebkeyw{then} \\
            \quad\quad \ldots \\
            \quad \ebkeyw{end} \\
            \ebkeyw{end } , 
          \end{array}
          & 
          \begin{array}{l}
            \ebkeyw{machine } N \ebkeyw{ refines } M\\
            \ebkeyw{variables } w\\
            \ebkeyw{invariant } J(s,c,v,w)\\
            \ebkeyw{event } evt \ebkeyw{ refines } evt0\\ 
            \quad \ebkeyw{any } y  \ebkeyw{ where} \\
            \quad\quad H(y,s,c,w) \\
            \quad \ebkeyw{with} 
            \quad\quad x :|~ W(x,s,c,w,y) \\
            \quad \ebkeyw{then} 
            \quad\quad \ldots \\
            \quad \ebkeyw{end}\\
            \ebkeyw{end } ) 
          \end{array}\\
        \end{array}
        \\= \\
        \begin{array}{l}
          \dafnykeyw{ghost method } evt\_grd\_GRD() \\
          ~~\dafnykeyw{assume }~\texttt{AT}~\dafnykeyw{\&\&}~\texttt{I}~\dafnykeyw{\&\&}~\texttt{J}~\dafnykeyw{\&\&}~\texttt{H}~\dafnykeyw{\&\&}~\texttt{W}\dafnycode{;}\\
          ~~\dafnykeyw{assert}~\texttt{G}\dafnycode{;}
        \end{array}
      \end{array}
    \end{prooftree}
    \]
  }

  \paragraph{The Guard Merging Proof Obligation: } rule \textsf{MRG}
  generates a method in \dafny\ that assumes the translation of the
  abstract invariant, the translation of the
  non-nullness axioms and theorems, and the translation of the guard
  of the concrete event. The method finally asserts the result of 
  the disjunction of the abstract events being merged guards translation.

  {\small
    \[
    \begin{prooftree} 
      \begin{array}{ll}
        \Ctx{Ctx} = \texttt{AT}\\
        \EBTODAFNY{\ebkeyw{invariants } I(s,c,v)} = \texttt{I} &
        \DPRED{H(s,c,v,x)}=\texttt{H} \\
        \DPRED{G1(s,c,v,x)}=\texttt{G1} & \DPRED{G2(s,c,v,x)}=\texttt{G2}
      \end{array}
      \using \textsf{(MRG)}
      \justifies
      \begin{array}{l}
        \MRGN ( \\
        \begin{array}{ll}
          \begin{array}{l}
            \ebkeyw{machine } M \ebkeyw{ sees } Ctx\\
            \ebkeyw{variables } v\\
            \ebkeyw{invariant } I(s,c,v)\\
            \ebkeyw{event } evt01\\ 
            \quad \ebkeyw{any } x  \ebkeyw{ where} \\
            \quad\quad G1(s,c,v,x) \\
            \quad \ebkeyw{then} \\
            \quad\quad S \\
            \quad \ebkeyw{end} \\ \\
            \ebkeyw{event } evt02\\ 
            \quad \ebkeyw{any } x  \ebkeyw{ where} \\
            \quad\quad G2(s,c,v,x) \\
            \quad \ebkeyw{then} \\
            \quad\quad S \\
            \quad \ebkeyw{end} \\  
            \ebkeyw{end } , 
          \end{array}
          & 
          \begin{array}{l}
            \ebkeyw{machine } N \ebkeyw{ refines } M\\
            \ebkeyw{variables } w\\
            \ebkeyw{invariant } J(s,c,v,w)\\
            \ebkeyw{event } evt \ebkeyw{ refines}\\
            \quad\quad evt01\\
            \quad\quad evt02\\ 
            \quad \ebkeyw{any } x  \ebkeyw{ where} \\
            \quad\quad H(s,c,v,x) \\
            \quad \ebkeyw{then} \\
            \quad\quad S \\
            \quad \ebkeyw{end}\\
            \ebkeyw{end } ) 
          \end{array}\\
        \end{array}
        \\= \\
        \begin{array}{l}
          \dafnykeyw{ghost method } evt\_MRG() \\
          ~~\dafnykeyw{assume }~\texttt{AT}~\dafnykeyw{\&\&}~\texttt{I}~\dafnykeyw{\&\&}~\texttt{H}\dafnycode{;}\\
          ~~\dafnykeyw{assert}~\texttt{G1 } \vee \texttt{ G2}\dafnycode{;}
        \end{array}
      \end{array}
    \end{prooftree}
    \]
  }

  \paragraph{The Simulation Proof Obligation: } rule \textsf{Sim}
  generates a \dafny\ method that assumes the translation of the
  abstract and concrete invariants, the translation of the
  non-nullness axioms and theorems,  the translation of the guard of
  the concrete event, the translation of the predicate related to the
  witness, and the translation of the
  before-after predicate of the concrete event. The method finally
  asserts the result of the abstract event's before-after predicate translation.

  {\small
    \[
    \begin{prooftree} 
      \begin{array}{l}
        \EBTODAFNY{\ebkeyw{invariants } I(s,c,v)} = \texttt{I} \\ \Ctx{Ctx} = \texttt{AT} \\
        \EBTODAFNY{\ebkeyw{invariants }\:J(s,c,v,w)} = \texttt{J} \\ \DPRED{H(y,s,c,w)}=\texttt{H} \\
        \DPRED{W1(x,s,c,w,y,w')}=\texttt{W1} \\ \DPRED{BA2(s,c,w,y,w)}=\texttt{BA2} \\
        \DPRED{W2(v',s,c,w,y,w')}=\texttt{W2} \\ \DPRED{BA1(s,c,v,x,v')}=\texttt{BA1}
      \end{array}
      \using \textsf{(Sim)}
      \justifies
      \begin{array}{l}
        \SIMN ( \\
        \begin{array}{ll}
          \begin{array}{l}
            \ebkeyw{machine } M \\ \ebkeyw{ sees } Ctx\\
            \ebkeyw{variables } v\\
            \ebkeyw{invariant } I(s,c,v)\\
            \ebkeyw{event } evt0\\ 
            \quad \ebkeyw{any } x \\ \ebkeyw{ where} \\
            \quad\quad \ldots\\
            \quad \ebkeyw{then} \\
            \quad\quad v :|~ BA1(s,c,v,x,v') \\
            \quad \ebkeyw{end} \\
            \ebkeyw{end } , 
          \end{array}
          &
          \begin{array}{l}
            \ebkeyw{machine } N \\ \ebkeyw{ refines } M\\
            \ebkeyw{variables } w\\
            \ebkeyw{invariant } J(s,c,v,w)\\
            \ebkeyw{event } evt \\ \ebkeyw{ refines } evt0\\ 
            \quad \ebkeyw{any } y \\ \ebkeyw{ where} \\
            \quad\quad H(y,s,c,w) \\
            \quad \ebkeyw{with} \\
            \quad\quad x :|~ W1(x,s,c,w,y,w') \\
            \quad\quad v' :|~ W2(v',s,c,w,y,w') \\
            \quad \ebkeyw{then} \\
            \quad\quad w :|~ BA2(s,c,w,y,w') \\
            \quad \ebkeyw{end}\\
            \ebkeyw{end } ) 
          \end{array}\\
        \end{array}
        \\= \\
        \begin{array}{l}
          \dafnykeyw{ghost method } evt\_act\_SIM() \\
          ~~\dafnykeyw{assume } \texttt{AT}~\dafnykeyw{\&\&}~\texttt{I}~\dafnykeyw{\&\&}~\texttt{J}~\dafnykeyw{\&\&}~\texttt{H}~\dafnykeyw{\&\&}~\texttt{W1}~\dafnykeyw{\&\&}~\texttt{W2}~\dafnykeyw{\&\&}~\texttt{BA2}\dafnycode{;}\\
          ~~\dafnykeyw{assert } \texttt{BA1}\dafnycode{;}
        \end{array}
      \end{array}
    \end{prooftree}
    \]
  }

  \paragraph{The Numeric Variant Proof Obligation: } rule \textsf{Nat}
  generates a \dafny\ method that assumes the translation of the machine
  invariant, the translation of the non-nullness axioms and theorems,
  and the translation of the
  guard of the event. The method finally asserts the result of the
  translation when the evaluation of the variant is a natural
  number. In rule \textsf{Nat}, $Nat$ is defined as set in \dafny\
  that contains natural numbers. 

  {\small
    \[
    \begin{prooftree} 
      \begin{array}{ll}
        \EBTODAFNY{\ebkeyw{invariants } I(s,c,v)} = \texttt{I} &
        \Ctx{Ctx} = \texttt{AT}\\
        \DPRED{G(s,c,v,x)}=\texttt{G} & \DPRED{n(s,c,v)}=\texttt{n} 
      \end{array}
      \using \textsf{(Nat)}
      \justifies
      \begin{array}{l}
        \NATN ( \\
        \ebkeyw{machine } M \ebkeyw{ sees } Ctx\\
        \ebkeyw{refines } \ldots\\
        \ebkeyw{variables } v\\
        \ebkeyw{invariant } I(s,c,v)\\
        \ebkeyw{variant } n(s,c,v)\\
        \ebkeyw{event } evt\\ 
        \quad \ebkeyw{status convergent } //or \ebkeyw{ anticipated} \\ 
        \quad \ebkeyw{any } x  \ebkeyw{ where} \\
        \quad\quad G(s,c,v,x) \\
        \quad \ebkeyw{then} \\
        \quad\quad \ldots \\
        \quad \ebkeyw{end} \\
        \ebkeyw{end })
        = \\
        \dafnykeyw{ghost method } evt\_NAT() \\
        ~~\dafnykeyw{assume } \texttt{AT}~\&\&~\texttt{I}~\&\&~\texttt{G}\dafnycode{;}\\
        ~~\dafnykeyw{assert } Nat.\dafnycode{has}(\texttt{n}) \dafnycode{;}
      \end{array}
    \end{prooftree}
    \]
  }

  \paragraph{The Variant (VAR): }
  There exist two different proof obligations related to the
  variant. It regards on the status of the event (i.e. \ebkeyw{convergent} or
  \ebkeyw{anticipated}). The following is the rule translation for variant proof
  for a \ebkeyw{convergent} event, the method asserts the result of the
  translation when the evaluation of the variant with the new values of
  variables is lower than the previous evaluation

  {\small
    \[
    \begin{prooftree} 
      \begin{array}{ll}
        \EBTODAFNY{\ebkeyw{invariants } I(s,c,v)} = \texttt{I} &\DPRED{G(s,c,v,x)}=\texttt{G}\\
        \Ctx{Ctx} = \texttt{C} & \DPRED{BA(s,c,v,x,v')}=\texttt{BA} \\
        \DPRED{n(s,c,v)} = \texttt{n} & \DPRED{n(s,c,v')} = \texttt{n'} 
      \end{array}
      \using \textsf{(Conv)}
      \justifies
      \begin{array}{l}
        \VARConvN ( \\
        \ebkeyw{machine } M \ebkeyw{ sees } Ctx\\
        \ebkeyw{variables } v\\
        \ebkeyw{invariant } I(s,c,v)\\
        \ebkeyw{variant } n(s,c,v)\\
        \ebkeyw{event } evt\\ 
        \ebkeyw{status convergent}\\ 
        \quad \ebkeyw{any } x  \ebkeyw{ where} \\
        \quad\quad G(s,c,v,x) \\
        \quad \ebkeyw{then} \\
        \quad\quad v :|~ BA(s,c,v,x,v') \\
        \quad \ebkeyw{end} \\
        \ebkeyw{end })
        = \\
        \dafnykeyw{ghost method } evt\_VAR() \\
        ~~\dafnykeyw{assume }\texttt{AT}~\&\&~\texttt{I}~\&\&~\texttt{G}~\&\&~\texttt{BA}\dafnycode{;}\\
        ~~\dafnykeyw{assert }\texttt{n'}~ < ~ \texttt{n}\dafnycode{;}
      \end{array}
    \end{prooftree}
    \]
  }

If the \eb\ variant
  is defined as a set, the method asserts the result of the
  translation when the evaluation of the variant with the new values
  of variables is a proper subset of the previous evaluation ($\texttt{n'} \subset \texttt{n}$).

  For an \textsf{anticipated} status in an event, the method asserts the result of the
  translation when the evaluation of the variant with the new values of
  variables is lower than equal the previous evaluation

  {\small
    \[
    \begin{prooftree} 
      \begin{array}{ll}
        \EBTODAFNY{\ebkeyw{invariants } I(s,c,v)} = \texttt{I} &\Ctx{Ctx} = \texttt{C}\\
        \DPRED{G(s,c,v,x)}=\texttt{G} & \DPRED{BA(s,c,v,x,v')}=\texttt{BA} \\
        \DPRED{n(s,c,v)} = \texttt{n} & \DPRED{n(s,c,v')} = \texttt{n'}
      \end{array}
      \using \textsf{(Ant)}
      \justifies
      \begin{array}{l}
        \VARAntN ( \\
        \ebkeyw{machine } M \ebkeyw{ sees } Ctx\\
        \ebkeyw{variables } v\\
        \ebkeyw{invariant } I(s,c,v)\\
        \ebkeyw{variant } n(s,c,v)\\
        \ebkeyw{event } evt\\ 
        \ebkeyw{status anticipated}\\ 
        \quad \ebkeyw{any } x  \ebkeyw{ where} \\
        \quad\quad G(s,c,v,x) \\
        \quad \ebkeyw{then} \\
        \quad\quad v :|~ BA(s,c,v,x,v') \\
        \quad \ebkeyw{end} \\
        \ebkeyw{end })
        = \\
        \dafnykeyw{ghost method } evt\_VAR() \\
        ~~\dafnykeyw{assume }\texttt{AT}~\&\&~\texttt{I}~\&\&~\texttt{G}~\&\&~\texttt{BA}\dafnycode{;}\\
        ~~\dafnykeyw{assert }\texttt{n'}~ \leq ~ \texttt{n}\dafnycode{;}
      \end{array}
    \end{prooftree}
    \]
  }

If the \eb\ variant
  is defined as a set, the method asserts the result of the
  translation when the evaluation of the variant with the new values
  of variables is a subset of the previous evaluation ($\texttt{n'} \subseteq \texttt{n}$).

  \paragraph{The non-deterministic witness (WFIS): } The translated
  method assumes the invariants of the abstract and concrete machines,
  the non-nullness axioms and theorems, the guard of the refined event, and the
  before-after predicate of the refined event. The method finally
  asserts the result of the translation of the existence of a value $x$
  that ensures the witness' predicate

  {\small
    \[
    \begin{prooftree} 
      \begin{array}{ll}
        \EBTODAFNY{\ebkeyw{invariants } I(s,c,v)} = \texttt{I} &\Ctx{Ctx} = \texttt{AT} \\
        \EBTODAFNY{\ebkeyw{invariants } J(s,c,v,w)} = \texttt{J} &
        \DPRED{H(y,s,c,w)}=\texttt{H} \\
        \DPRED{BA2(s,c,w,y,w')}=\texttt{BA2} &
        \DPRED{W(x,s,c,w,y,w')}=\texttt{W}
      \end{array}
      \using \textsf{(With)}
      \justifies
      \begin{array}{l}
        \WFISN ( \\
        \begin{array}{ll}
          \begin{array}{l}
            \ebkeyw{machine } M \ebkeyw{ sees } Ctx\\
            \ebkeyw{variables } v\\
            \ebkeyw{invariant } I(s,c,v)\\
            \ebkeyw{event } evt0\\ 
            \quad \quad \ldots\\
            \quad \ebkeyw{end} \\
            \ebkeyw{end } , 
          \end{array}
          & 
          \begin{array}{l}
            \ebkeyw{machine } N \ebkeyw{ refines } M\\
            \ebkeyw{variables } w\\
            \ebkeyw{invariant } J(s,c,v,w)\\
            \ebkeyw{event } evt \ebkeyw{ refines } evt0\\ 
            \quad \ebkeyw{any } y \ebkeyw{ where} \\
            \quad\quad H(y,s,c,w) \\
            \quad \ebkeyw{with} \\
            \quad\quad x :|~ W(x,s,c,w,y,w') \\
            \quad \ebkeyw{then} \\
            \quad\quad w :|~ BA2(s,c,w,y,w') \\
            \quad \ebkeyw{end}\\
            \ebkeyw{end } ) 
          \end{array}\\
        \end{array}
        \\ = \\
        \begin{array}{l}
          \dafnykeyw{ghost method } evt\_witness\_WFIS() \\
          ~~\dafnykeyw{assume } \texttt{AT}~\&\&~\texttt{I}~\&\&~\texttt{J}~\&\&~\texttt{H}~\&\&~\texttt{BA2}\dafnycode{;}\\
          ~~\dafnykeyw{assert } \dafnycode{(}\dafnykeyw{exists}~x~ \dafnykeyw{::}~ \texttt{W}\dafnycode{);}
        \end{array}
      \end{array}
    \end{prooftree}
    \]
  }

  \section{The \ebtodafny\ Tool}
  \label{ebtodafny:implementation}
  The \ebtodafny\ tool is integrated to Rodin as an Eclipse
  plug-in. Full source code for \ebtodafny\ is available
  in~\cite{EventB2Dafny_webpage}. The \ebtodafny\ tool parses Rodin
  proof obligations into a \dafny\ program. The proof obligation can
  include information about a machine context, e.g. sets and axioms, or
  machine variables and invariants, that might be conjoined with
  machine variables and invariants from a refinement
  machine. 

  Figure~\ref{fig:ebtodafny:implementation:dafnyEX} shows a partial
  output of applying 
  \ebtodafny\ to the \eb\ model depicted in
  Figure~\ref{fig:ebtodafny:implementatio:poEX}. The input is the invariant
  preservation proof obligation generated by Rodin for the \eb\ model
  showed in Figure~\ref{fig:ebtodafny:po:Abrial_abstract}
  (page~\pageref{fig:ebtodafny:po:Abrial_abstract}) regarding $search$ event with
  respect to invariant $\ebtag{inv1}:~i \in 1 \upto n$.

  \begin{figure}[h]
    {\small
      \[
      \begin{array}{l}
        ~~n \in \nat1 \\
        ~~f \in 1 \upto n \tfun \ebkeyw{D} \\
        ~~v \in ran(f) \\
        ~~i \in 1 \upto n \\
        ~~k \in 1 \upto n \\
        ~~f(k) = v \\
        \vdash \\
        ~~k \in 1 \upto n
      \end{array}
      \]
    }
    \caption{Proof Obligation generated by Rodin.}
    \label{fig:ebtodafny:implementatio:poEX}
  \end{figure}

  The user has to prove  that  the invariant \ebtag{inv1}
  holds after the action \ebtag{act1} ($i := k$ in event $search$)
  given the axioms, theorems, invariants, and event $search$'s guards.

\begin{figure}
  \begin{lstlisting}[frame=none]
var D : Set<Integer>;
function f() : Relation<Integer, Integer>; 
function n() : Integer;
function v() : Integer;
var i : Integer;

function ax1() : bool 
     ensures Nat.has(n());

function ax2() : bool
     ensures f().isTotalFunction() &&
       f().domain == Int.Init().upto(Integer.Init(1), n()).instance &&
       f().range == D.instance; 

function ax3() : bool
     ensures f().range.has(v()); 

function thm1() : bool
     ensures Nat1.has(n());

function nonNullnessCond() : bool
    ensures i!= null && D!= null && f!= null && n!= null && v!= null; 

function inv1() : bool
     requires nonNullnessCond();
     ensures Int.Init().upto(Integer.Init(1), n()).has(i);

ghost method search_inv1_INV(){ 
     assume nonNullnessCond(); 
     assume inv1();
     assume ax1();
     assume ax2();
     assume ax3();
     assume thm1();
     assume f().funcImage(k).equals(v());
     assert Int.Init().upto(Integer.Init(1), n()).has(k);
}
\end{lstlisting}
\caption{Partial \ebtodafny\ output.}
\label{fig:ebtodafny:implementation:dafnyEX}
\end{figure}

\ebtodafny\ defines the carrier set \ebkeyw{D} as a set of
integers. Constants are defined as 0-ary integer functions. Axioms are
translated as 0-ary boolean functions where the axiom is taken as a
post-condition. For instance, axiom \ebtag{ax2} (defined in \eb\ as $f
\in 1 \upto n \tfun \ebkeyw{D}$) is translated as function
\dafnycode{ax2} that ensures that the variable $f$ is a total
function where its domain is equal to the set of number from $1$ to
$n$ (denoted by \dafnykeyw{upto}). And its range is equal to
\dafnycode{D} (type \dafnycode{Int} is the representation of
integers). The translation  also defines the non-nullness condition
that stands that all variables and constants cannot be null values. The invariant \ebtag{inv1} is
translated as a boolean function that requires the non-nullness
condition and ensures the translation of the invariant. Finally, a
\dafnykeyw{ghost} method is defined (\dafnycode{search\_inv1\_INV}) that
assumes the non-nullness condition, the invariants, and the guard of
the event. It asserts on the evaluation of the invariant after the
execution of the event (where variables have new values) (i.e. $k \in
1 \upto n$).

\section{Conclusion}
\label{ebtodafny:conclusion}
In this chapter we presented a series of translation rules to generate
\dafny\ code from \eb\ proof obligations. We also introduced the
implementation of the rules as the \ebtodafny\ tool that is a Rodin's
plug-in. We have validated \ebtodafny\ by applying it to an \eb\ model. 

Users of \eb\ could benefit from \ebtodafny\ since the tool might help
them in the process of discharging proof obligations. Proof
obligations for complex systems are complex and difficult to
discharge. \ebtodafny\ gives additional alternatives in the process of
discharging a proof obligation by porting it to \dafny. Once in
\dafny, a software developer expert in this language can use the
automatic provers (e.g. Z3, Simplify or Zap) that come with it to prove
the proof obligations. The software developer might give insights
about any problem in the \eb\ model if the proof fails. This is
possible since Boogie comes with a debugger, the Boogie Verification
Debugger, that helps \dafny\ users to understand the output of the
program verifier. \ebtodafny\ also extends the set provers that comes
with Rodin. For instance, \dafny\ works with Simplify or Zap that are
not available for Rodin.

Discharging proof obligations in \dafny\ is semi-automatic and relies on
\dafny’s provers performance. We plan to integrate \ebtodafny\ to \dafny\
so that discharging proof obligations would include the following
steps

\begin{inparaenum}[\itshape i\upshape)]
\item to choose a proof obligation in \eb,
\item to use \ebtodafny\ to translate it to \dafny, 
\item to receive feed-back directly from \dafny\ in Rodin (\ebtodafny\ will automatically run Microsoft
Visual Studio to discharge the proof, and feed Rodin with the
feed-back that \dafny\ provides).
\end{inparaenum}
We also plan to characterise which kind of proof obligations \dafny\
is good for and which kind of proof obligations \dafny's provers
outperforms Rodin's. Finally, we plan to prove the soundness
of the rules presented in this chapter, so user can be confident that
the proof obligation generated by Rodin is indeed the same proof
obligation represented in \dafny.

\chapter{Case Studies}
\label{chapter:case-studies}


\paragraph{This chapter. } Throughout this thesis we have defined a
series of tools to work with different formal methods. The work has
ended in the translation of \eb\ models to \jml-annotated Java
code. We have also proposed two different techniques on software
development where \ebtojava\ can be used. This chapter shows two case
studies on the use of \ebtojava\ as part of two different software
developments. The chapter also includes a benchmark that compares
\ebtojava\ with two
existing tools for generating Java code from \eb\ models. 

The first case study (see Section
\ref{case-studies:sep}) describes the development of a Social-Event
Planner using \ebtojava\ and the Model-View-Controller design
pattern.  The Social-Event Planner is an Android \cite{Android}
application that can be used for planning a social event. Basically, a user can create
a social event and send invitations to his own list of people. The second case
study (see Section \ref{case-studies:tok}) presents the use of
\ebtojava\ in testing the generated Java code by \ebtojava\ for a
security-critical access control system modelled in \eb.

\paragraph{Contributions.}  The main contributions of this chapter are
\begin{inparaenum}[\itshape i\upshape)]
\item to
show how the \ebtojava\ tool can be incorporated in two
different software developments. It shows how people from
different domains can work together in the development of software.
\item  To
compare our tool with existing tools for generating Java code from
\eb\ models.
\end{inparaenum}

\paragraph{Related work.}   A preliminary version of an \eb\ model of the Tokeneer ID Station
(TIS) is presented in \cite{Padidar:Tok_EB}. This is a reduced
model of the TIS. It consists of a single abstract machine, and no machine
refinement was defined and a few proof obligations remained
undischarged. For this reason, we decided to write our own \eb\ model of Tokeneer
afresh.

In \cite{Bouquet07}, an automatic approach to provide correct testing
inputs for parameters associated to axioms is described. We can use
this work to extend \ebtojava\ to automatically assign values for
constants in \eb\ contexts. We can also use the work presented in
\cite{Berno:testing:FS:91} that constructs test data sets from formal
specifications to construct test data sets from \eb.

In \cite{Catano:jfly:10}, a strategy called JFly is proposed to evolve
informal (written in natural language) software requirements into
formal requirements written in \jml. This work can be reused to
structure the writing of JUnit tests from a STS document.

\section{The Social-Event Planner}
\label{case-studies:sep}

This section describes the development of a Social-Event Planner
following a Model-View-Controller (MVC) design pattern. Sections
\ref{case-studies:sep:reqdoc}, \ref{subsec:planner:eb}, and
\ref{case-studies:sep:generated} describe the implementation of the Model
of the system by modelling it in \eb\ to then transition to Java code
using the \ebtojava\ tool. Section \ref{case-studies:sep:generated:VC}
describes the implementation of the View and Controller of the system.

\subsection{Requirement Document for the Social-Event Planner}
\label{case-studies:sep:reqdoc}
The Social-Event Planner was modelled as a piece of software that
runs over the existing Social Network in \cite{TeachFM-09} (the Social
Networking model is briefly explained Section
\ref{eb2jml:apply:EBexample}). For
modelling the Social-Event
Planner, We followed the ``parachute'' software development strategy
of \eb\ proposed by J.-R Abrial in \cite{TheBBook}.  We classify the
requirements within two categories:

\begin{itemize}
\item Those concerned with the functionalities of the application
  labeled {\it FUN}.
\item Those concerned with the decision making labeled {\it DEC}
  (e.g. when a user has to make a decision on either going to a
  social-event or declining).
\end{itemize}

The main functionality of the Social-Event Planner application is to
allow a user to create a social-event and invite other users to it. A
social-event shall consist of the content visible to
any invited user. The user creating the social-event might enforce a specific privacy
policy over the social-event. Such a privacy policy shall consist of a
set of restricted users from which the creator of the social-event
wants to keep the social event hidden. The creator of the social-event
can allow other invited users to further invite additional users to
the social-event. These additional users must not belong to the
aforementioned set of restricted users. To illustrate this, a
user ({\it UserA}) within the social network creates an event `Picnic in the
Park'. He invites the lists of users `Close Friends' and `Trekking
Friends'. He decides he doesn't want any user belonging to the list
`Professors' to be invited. An invite message will be sent from {\it UserA} to every
member of `Close Friends' and `Trekking
Friends' that do not belong to list `Professors'. He grants the users in list `Close
Friends' the `Invite' privilege. A user ({\it UserB}) from the list `Close
Friends' decides to invite the users in the list `Institute'. The
social-event will be shared with (an invite will be sent to) every user in the
list `Institute' that doesn't belong to the list `Professors'. In the
following, we present the requirements of the social-event planner.

\begin{center}
  {
    \begin{tabular}{|p{7cm}|c|}
      \hline
      The users of the Social Network can create or delete
      social-events. & FUN-1 \\
      \hline
      The creator of a social-event can (un-)associate content with
      it. & FUN-2 \\
      \hline
    \end{tabular}
  }
\end{center}

The invited user to a social-event
can either `Join' the event if he is certain to go the social-event,
`Decline' the invitation if he is certain that he cannot go to the social-event,
or reply with a `Maybe' when he is not certain if he can make it. 

\begin{center}
  {
    \begin{tabular}{|p{7cm}|c|}
      \hline
      A user (with privileges - explain later on)  can invite a list
      of users to a social-event. & FUN-3 \\ \hline
      The `invited' users can reply to the
      social-event. & FUN-4 \\
      \hline
      A reply to an invitation shall be either 
      \begin{inparaenum}[\itshape 1\upshape)]
      \item  Join 
      \item Decline
      \item  Maybe
      \end{inparaenum}
      or the user can choose not to reply.
      & DEC-1\\ \hline 
      A user invited to a social-event can swap
      their reply between Join, Decline or Maybe. & FUN-5\\ \hline 
      A user who has been invited to a social-event can view all the content
      associated with the social-event. & FUN-6\\ \hline
    \end{tabular}
  }
\end{center} 

The users of the Social Network invited to a social-event can be
granted permissions to View or Edit content associated with the
social-event. Additionally specific users can be granted  permission
to invite additional users.

\begin{center}
  {
    \begin{tabular}{|p{7cm}|c|}
      \hline
      The users shall be able to view or edit a social-event or invite other
      users to the social-event based on permissions. & FUN-7 \\
      \hline 
      The following permissions can be awarded over a social-event to a
      user on the Social Network: 1) View 2) Edit 3) Invite. & DEC-2\\ \hline 
    \end{tabular}
  }
\end{center} 

The creator of the event, called the `owner', might allow
the invited users to further invite users by explicitly granting them
the permission to do so.

\begin{center}
  {
    \begin{tabular}{|p{7cm}|c|}
      \hline
      The user that creates a social-event shall be designated the owner of
      the social-event. & FUN-8 \\
      \hline
      The owner of any social-event shall be granted all privileges over
      it. & DEC-3\\ \hline 
      The owner of a social-event can grant `Invite' permissions to any user
      that has been invited. & FUN-9 \\ \hline
      A user with an `Invite' permission to a social-event shall be
      allowed to invite any
      users to the social-event. & FUN-10 \\ \hline
      The users that has been invited to a social-event can add content to
      the social-event in the form of comments. & FUN-11\\ \hline 
    \end{tabular}
  }
\end{center}

\paragraph{Refinement Strategy. } Below is listed the order in which
the various proposed  requirements were taken into account. 

\begin{description}
\item [ref\_socialevents.] Once we have the social networking
  core we incorporate the possibility for a user to create
  social-events and associate content with it. In this direction we
  take care of requirements FUN-1, FUN-2.
  
\item [ref\_socialinvite.]  This refinement includes the
  functionality for the owner of a social-event to invite other
  users to the social-event. The owner can grant permissions to
  specific invited users allowing them to invite other users. An
  invited user can reply to an invitation with Join/Decline or
  Maybe or change and existing reply. This refinement satisfies
  requirements FUN-3,FUN-4,FUN-7, FUN-8,FUN-9,FUN-10, FUN-11,
  DEC-1 and DEC-2.
\item [ref\_socialpermissions.] Then we add privileges in order
  for people to view and edit content associated with social-events.
  It takes into account requirements FUN-5 and FUN-6.
\end{description}

\subsection{The \eb\ Model of the Social-Event Planner}
\label{subsec:planner:eb}
The Social-Event Planner works on top of the \eb\ model for
the Social Network presented in \cite{matelas:2010}. The
Social Network
specifies a social network as composed of people
and content (e.g. photos, videos, comments). People within the social
network can share their own content. For that, the model defines
permissions over the content. The Social-Event Planner can be regarded as a plug-in of
the Social Network. The Social-Event Planner is composed of three machines
($ref6\_socialevents$, $ref7\_socialinvite$ and
$ref8\_socialpermissions$) that constitute refinements of the Social Networking
model. The follow explains part of the machines.  The full source
(the Social Network and the Social-event Planner) is available at \cite{Planner}.

In the following, we explain the refinements of the Social-Event
Planner :

\paragraph{ref6\_socialevents: } This machine represents the core of the
Social-Event Planner. A user in the social network can create
social-events and upload information to
it. Figure \ref{fig:chapter:case-studies:sep:socialevent} depicts part
of the \eb\ machine.

Machine $ref6\_socialevents$ \ebtag{see}s context $ctx\_event$
depicted in the right of Figure
\ref{fig:chapter:case-studies:sep:socialevent}. The context defines a carrier set
\ebkeyw{EVENTS} containing all possible social-events. The machine
defines variable $events$ representing the actual social-events
created within the Social Network. Variable $scontents$ is a set of contents
present in the social-event (e.g. a picture or a comment within a specific
social-event). Variable $eventcontents$ is a relation that maps contents to
social-events. This relation allows the system to know which content
belongs to which social-event. The relation is defined as a total relation so that a
social-event can contain several contents and a content must be in
at least one social-events. The variable $eventowner$ defined is defined a total
function that maps social-events to person. It models each social-event
has an unique owner. Variables $contents$, and $person$ were defined
in previous refined machine. They describe the contents in the social
network, and the set of actual people in the social network, respectively.

\begin{figure}[h]
  {\small
    \[
    \begin{array}{ll}
      \begin{array}{l}
        \ebkeyw{machine }ref6\_socialevents\\
        ~~~~~~\ebkeyw{refines } ref5\_lists \ebkeyw{ sees } ctx\_events\\
        ~\ebkeyw{variables } ~sevents~ scontents\\
        ~~~~~~ eventcontents~eventowner\\    
        ~\ebkeyw{invariant}\\
        ~~\ebtag{invr6\_1 } sevents \subseteq \ebkeyw{EVENTS} \\
        ~~\ebtag{invr6\_2 } scontents \subseteq contents \\ 
        ~~\ebtag{invr6\_3 } eventcontents \in scontents \trel sevents\\
        ~~\ebtag{invr6\_4 } eventowner \in sevents \tfun persons\\
        ~\ebkeyw{events}\\
        ~~create\_social\_event \\
        ~~~\ebkeyw{any } pe ~ se \\
        ~~~\ebkeyw{where}\\
        ~~~~\ebtag{grdr6\_1 } pe \in persons \\
        ~~~~\ebtag{grdr6\_2 } se \in \ebkeyw{EVENTS} \bsl sevents \\
        ~~~\ebkeyw{then} \\
        ~~~~\ebtag{actr6\_1 } sevents := sevents \bunion \{se\}\\
        ~~~~\ebtag{actr6\_2 } eventowner(se) := pe \\
        ~~\ebkeyw{end}\\
        \ebkeyw{end} \\
        ~~upload\_principal\_content\_planner\\
        ~~~~~~\ebkeyw{extends}~upload\_principal \\
        ~~~\ebkeyw{any}~~se ~~\ebkeyw{where}\\
        ~~~~\ebtag{grdr6\_1 } se \in sevents \\
        ~~~\ebkeyw{then} \\
        ~~~~\ebtag{actr6\_1 }scontents := scontents \bunion \{c2\}\\
        ~~~~\ebtag{actr6\_2 }eventcontents :=  \\
        ~~~~~~ eventcontents \bunion
        \{c2\mapsto se\} \\
        ~~\ebkeyw{end}\\
        \ebkeyw{end} 
      \end{array}
      &
      \begin{array}{l}
        \ebkeyw{context } ctx\_events \\
        ~~~\ebkeyw{extends } ctx\_lists \\
        \ebkeyw{sets } \ebkeyw{EVENTS} \\
        \ebkeyw{end}
      \end{array}
    \end{array}
    \]
  }
  \caption{Excerpt of the $ref6\_socialevents$ (and the context it
    \ebkeyw{sees}) \eb\ machine for the Social-Event Planer model.}
  \label{fig:chapter:case-studies:sep:socialevent}
\end{figure}

The $create\_social\_event$ event models the creation of a new
social-event $se$ (not presented already in the social-event Planner)
by a user $pe$ that belong to the Social Network.  The execution of
this event adds the new social-event $se$ to the set of existing
social-events and defines $pe$ as the owner. Event
$upload\_principal\_content\_planner$ allows users of the social-event
to upload a principal content in the social-event (local variable $c2$
correspond to the content and it is defined in the abstract
event). Symbols $\mapsto$, $\bunion$, $\bsl$, and $\subseteq$ model a
pair of elements, set union, set difference, and set subset in \eb,
respectively. The
machine defines more events not shown in the figure.

\paragraph{ref7\_socialinvite: } This machine specifies the users
invited to a social-event. Figure
\ref{fig:chapter:case-studies:sep:socialinvite} depicts
part of the \eb\ refinement. The machine \ebtag{refines} machine
$ref6\_socialevents$ and \ebtag{sees} the same context, it models the
set of invited people
to a social-event by using the
variable $invited$, that is a relation that maps social-events to
invited people. The variable is defined as a relation, so that, a user
can be invited to several social-events and a social-event can contain
several invited people. An invited user can reply to an
invitation. The user can either `join' the event, or reply as a `maybe'
or `decline'. This is represented by variables $join$, $maybe$ and
$decline$ that are modelled as relations that map social-events to
person. A user can be in just one of these states. This is ensured by invariants
\ebtag{invr7\_5}, \ebtag{invr7\_6}, and \ebtag{invr7\_7}. Invariant
\ebtag{invr7\_8}  states that a user can reply to one of these states if the
user is invited to the social-event. The owner of the social-event and
a set of invited people to that
social-event have the privilege to invite more people. This is
modelled using the variable $populate$. 

\begin{figure}[h]
  {\small
    \[
    \begin{array}{l}
      \ebkeyw{machine } ref7\_socialinvite~\ebkeyw{ refines }
      ref6\_socialevents \ebkeyw{ sees } ctx\_events\\
      ~\ebkeyw{variables} 
      ~~invited~ populate~join~maybe~decline\\    
      ~\ebkeyw{invariant}\\
      \begin{array}{ll}
        ~~\ebtag{invr7\_1 } invited \in sevents \rel persons & \ebtag{invr7\_2 } join \in sevents \rel persons\\
        ~~\ebtag{invr7\_3 } maybe \in sevents \rel persons  & \ebtag{invr7\_4 } decline \in sevents \rel persons \\
        ~~\ebtag{invr7\_5 } join \inter maybe = \emptyset&\ebtag{invr7\_6 } join \inter decline = \emptyset\\
        ~~\ebtag{invr7\_7 } maybe \inter decline = \emptyset&\ebtag{invr7\_8 } join \bunion maybe \bunion decline \subseteq invited\\
        ~~\ebtag{invr7\_9 } populate \in sevents \rel persons & \ebtag{invr7\_10 } eventowner \subseteq populate\\
        ~~\ebtag{invr7\_11 } populate \subseteq invited & \ebtag{invr6\_3 } eventcontents \in scontents \trel sevents \\
      \end{array}\\
      ~\ebkeyw{events}\\
        ~~create\_social\_event \ebkeyw{ extends } create\_social\_event~~\ebkeyw{any } pe ~ se ~~\ebkeyw{where}\\
        ~~~~\ebtag{grdr6\_1 } pe \in persons ~~\ebtag{grdr6\_2 } se \in \ebkeyw{EVENTS} \bsl sevents \\
        ~~~\ebkeyw{then} \\
        ~~~~\ebtag{actr6\_1 } sevents := sevents \bunion \{se\}~~\ebtag{actr6\_2 } eventowner(se) := pe \\
        ~~\ebkeyw{end}\\
      ~~sent\_invitation \\
      ~~~\ebkeyw{any}
      ~~~~pe ~ se ~ l1 ~~\ebkeyw{where}\\
      ~~~~\ebtag{grdr7\_1 } l1 \in dom(listpe) ~~~\ebtag{grdr7\_2 } se \in sevents \\
      ~~~~\ebtag{grdr7\_3 } se\mapsto pe \in populate \\
      ~~~\ebkeyw{then} \\
      ~~~~\ebtag{actr7\_1 } invited := invited \bunion (\{se\} \cprod listpe[\{l1\}])\\
      ~~\ebkeyw{end}\\
      ~~grant\_populate ~~\ebkeyw{any}~~ow~pe ~ se ~~\ebkeyw{where}\\
      \begin{array}{ll}
        ~~~~\ebtag{grdr7\_1 } ow \in persons &
        ~~~~\ebtag{grdr7\_2 } se \in sevents \\
        ~~~~\ebtag{grdr7\_3 } ow = eventowner(se) &
        ~~~~\ebtag{grdr7\_4 } pe \in persons \\
        ~~~~\ebtag{grdr7\_5 } se\mapsto pe \in invited&
        ~~~~\ebtag{grdr7\_6 } ow \not = pe
      \end{array}\\
      ~~~\ebkeyw{then} \\
      ~~~~\ebtag{actr7\_1 } populate := populate \bunion \{se\mapsto pe\}\\
      ~~\ebkeyw{end}\\
      \ebkeyw{end} 
    \end{array}
    \]
  }
  \caption{Excerpt of the $ref7\_socialinvite$ \eb\ machine for the
    Social-Event Planer model.}
  \label{fig:chapter:case-studies:sep:socialinvite}
\end{figure}

Invariant \ebtag{invr7\_10} states that the owner of a
social-event has the right to invite any one to it. Finally, invariant
\ebtag{invr7\_11} states that people with the right to invite other
people are invited to the social-event. 

The machine also defines the event $sent\_invite$ allows the
invited people to the social-event $se$ who are in the
$populate$ variable ($\ebtag{grdr7\_3 } se \mapsto pe \in populate$)
to invite more people from a list $l1$. Variable $listpe$ is defined
in previous refined machines. It is defined as a relation
that maps a list identifier to a set of people. Event $grant\_populate$
grants permission to an invited user to invite more people. Just the
owner of the social-event can grant such permission. Events
$reply\_with\_join$, $reply\_with\_maybe$, and
$reply\_with\_decline$ model the possibility to reply a social-event
with `join', `maybe' or `decline', respectively (these events are not
shown in Figure
\ref{fig:chapter:case-studies:sep:socialinvite}). Symbols $\cprod$,
and $\emptyset$ model a cartesian product, and empty set in \eb,
respectively.

\paragraph{ref8\_socialpermission: } Finally,  machine $ ref\_8socialpermission$
specifies the permission over the content involved in a specific
social-event. An invited user can have permission to view or edit a
specific content. The machine models these permissions by using the variables
$socialview$, that is defined as a relation that maps social content
within the social-event to
person that has privilege to view that content. And the variable
$socialedit$ that is defined as a relation that maps social content to
person that has privilege to edit that
content. Figure \ref{fig:chapter:case-studies:sep:socialpermissions}
depicts part of the \eb\ machine.

\begin{figure}[h]
  {\small
    \[
    \begin{array}{l}
      \ebkeyw{machine}~ref8\_socialpermissions~\ebkeyw{refines}~ref7\_socialinvite~\ebkeyw{sees}~ctx\_events\\
      ~\ebkeyw{variables} \\
      ~~socialviewp~ socialeditp\\    
      ~\ebkeyw{invariant}\\
      ~~~~\ebtag{invr8\_1 }socialviewp \in scontents \rel persons\\
      ~~~~\ebtag{invr8\_2 } socialeditp \in scontents \rel persons\\
      ~~~~\ebtag{invr8\_3 } eventcontents;invited \subseteq socialviewp\\
      ~~~~\ebtag{invr8\_4 } eventcontents;eventowner \subseteq socialeditp\\
      ~~~~\ebtag{invr8\_5 }socialeditp \subseteq socialviewp\\
      ~\ebkeyw{events}\\
      ~~create\_social\_event~\ebkeyw{extends}~create\_social\_event\\
      ~~~\ebkeyw{then} \\
      ~~~~\ebtag{actr8\_1 } socialviewp := socialviewp \bunion
      (eventcontents^{-1}[\{se\}] \cprod \{pe\})\\
      ~~~~\ebtag{actr8\_2 } socialeditp := socialeditp \bunion
      (eventcontents^{-1}[\{se\}] \cprod \{pe\})\\
      ~~\ebkeyw{end}\\
      \ebkeyw{end} 
    \end{array}
    \]
  }
  \caption{Excerpt of the $ref8\_socialpermission$ \eb\ machine for the
    Social-Event Planer model.}
  \label{fig:chapter:case-studies:sep:socialpermissions}
\end{figure}

The invariant \ebtag{invr8\_3} specifies that any invited person to a
social-event has privilege to view the content on it. Invariant
\ebtag{invr8\_4} states that the owner of the social-event has privilege to
edit the content on it. Finally, invariant \ebtag{invr8\_5} models
those that have privileges to edit a content, also have privileges to view
it.

This machine does not define any new event, but it extends the
previous ones. For instance, when a user $pe$ creates a social-event,
that user has permission to view and edit the content involved in that
social-event. Symbols $\ranres$, and $;$ model a range restriction, and
forward composition in \eb, respectively.

We model the Social-Event Planner in Rodin. All proof obligations
generated by Rodin were discharged.

\subsection{Generating \jml-annotated Java code for the Social-Event
  Planner \eb\ model}
\label{case-studies:sep:generated}
We used \ebtojava\ to generate \jml-annotated Java code for the last refinement of the
Social-Event Planner. The \ebtojava\ tool
generates one Java class (see an excerpt of the Java class in Figure
\ref{fig:case-studies:sep:generated:Javamachine}) containing the
translation of the carrier sets, constants
and variables (with their respective initialisations), and the \eb\
invariant. 

\begin{figure}
\begin{lstlisting}[frame=none]
public class ref8_socialpermissions{
  public Lock lock = new ReentrantLock(true);
  //@ public static constraint PERSON.equals(\old(PERSON)); 
  public static final BSet<Integer> PERSON = new Enumerated(min_integer,max_integer);

  //@ public static constraint EVENTS.equals(\old(EVENTS)); 
  public static final BSet<Integer> EVENTS = new Enumerated(min_integer,max_integer);
  ...

  /*@ spec_public */ private BSet<Integer> persons;
  /*@ spec_public */ private BSet<Integer> sevents;
  /*@ spec_public */ private BRelation<Integer,Integer> eventowner;
  ...

  /******Invariant definition******/
  /*@ public invariant
    persons.isSubset(PERSON) &&
    sevents.isSubset(EVENTS) &&
    eventowner.domain().equals(sevents) &&
    eventowner.range().isSubset(persons) && eventowner.isaFunction() && BRelation.cross(sevents,persons).has(eventowner) &&
    ... */

  // ... getter and mutator method definition

  /*@ public normal_behavior
    requires true;
    assignable \everything;
    ensures
       persons.isEmpty() &&
       sevents.isEmpty() &&
       eventowner.isEmpty() &&
       ... */
  public ref8_socialpermissions(){
    persons = new BSet<Integer>();
    sevents = new BSet<Integer>();
    eventowner = new BRelation<Integer,Integer>();
    ...

    // Thread initialisation
  }
}
\end{lstlisting}
\caption{Excerpt of the translation of machine $ref8\_socialpermissions$
  to Java.}
\label{fig:case-studies:sep:generated:Javamachine}
\end{figure}

The tool also generates a Java \javacode{Thread}
implementation for each machine event. Figure
\ref{fig:case-studies:sep:generated:Javaevent} shows
the translation of one event: $create\_social\_event$, where
\javacode{m} is a reference to the machine
class implementation (used to access machine variables via  getter and
setter methods).
Methods \javacode{guard\_create\_social\_event} and
\javacode{run\_create\_social\_event}
implement the behaviour of the $create\_social\_event$ event in Java. The first
method checks the event guard, and the second may execute when that
guard holds.
Whether  \javacode{run\_create\_social\_event} executes
when \javacode{guard\_create\_social\_event} holds is determined by
the \javacode{run()} method of \javacode{create\_social\_event} in
coordination with the respective \javacode{run()} methods of all
existing events. 

\begin{figure}
\begin{lstlisting}[frame=none]
public class create_social_event extends Thread{
 /*@ spec_public */ private ref8_socialpermissions machine;

 /*@ public normal_behavior
     requires true; assignable \everything;
     ensures this.machine == m; */
 public create_social_event(ref8_socialpermissions m) {
     this.machine = m;
 }

 /*@ public normal_behavior
     requires true; assignable \nothing;
     ensures \result <==> (machine.get_persons().has(pe) 
      && !machine.get_sevents().has(se)); */
 public /*@ pure */ boolean guard_create_social_event(Integer pe, Integer se) {
  return (machine.get_persons().has(pe) 
    && !machine.get_sevents().has(se));
 }

 /*@ public normal_behavior
     requires guard_create_social_event(pe,se);
     assignable m.sevents, m.eventowner, ...;
     ensures m.get_sevents().equals(\old((m.get_sevents()
         .union(new BSet<Integer>(se))))) 
       && m.get_eventowner().equals(\old((m.get_eventowner()
         .override(new BRelation<Integer,Integer>(
          new Pair<Integer,Integer>(se,pe)))))) && ...; 
    also
     requires !guard_create_social_event(pe,se);
     assignable \nothing;
     ensures true; */
 public void run_create_social_event(Integer pe, Integer se) {
  if(guard_create_social_event(pe,se)) {
   BSet<Integer> sevents_tmp = m.get_sevents();
   BRelation<Integer,Integer> eventowner_tmp = m.get_eventowner();
   ...
   m.set_sevents((sevents_tmp.union(new BSet<Integer>(se))));
   m.set_eventowner((eventowner_tmp.override(
          new BRelation<Integer,Integer>(new Pair<Integer,Integer>(se,pe)))));
  ...
  }
 }

 public void run() { ... }
}
\end{lstlisting}
\caption{Excerpt of the translation of event $create\_social\_event$
  to Java.}
\label{fig:case-studies:sep:generated:Javaevent}
\end{figure}
 
Variables \javacode{contents\_tmp}, \javacode{pages\_tmp}, \ldots hold
temporary values of variables \javacode{contents}, \javacode{pages},
\ldots, respectively. \ebtojava\ uses these temporary values to
implement simultaneous assignment in Java. 

The \jml-annotated Java code generated by \ebtojava\ from the last
refinement of the Social-Event Planner \eb\ model represents the Model
(M) of a MVC design pattern development. We extended this core
functionality to implement a usable version of the Social-Event
Planner as an Android application.

\subsection{The View and Controller Parts of the Social-Event Planner}
\label{case-studies:sep:generated:VC}
The View and Controller part of the system were developed in Java
using the Android API. The View part allows users to interact with the
Social-Event Planner. The {\it Controller} part makes a bridge between
the View part and the Model. Figure
\ref{fig:case-studies:sep:generated:VC:screens} depicts two of the main screen
shots of the user interface for the Social-Event Planner.

\begin{figure}[h]
  \begin{subfigure}[b]{0.5\textwidth}
    \centering
    \includegraphics[width=2in]{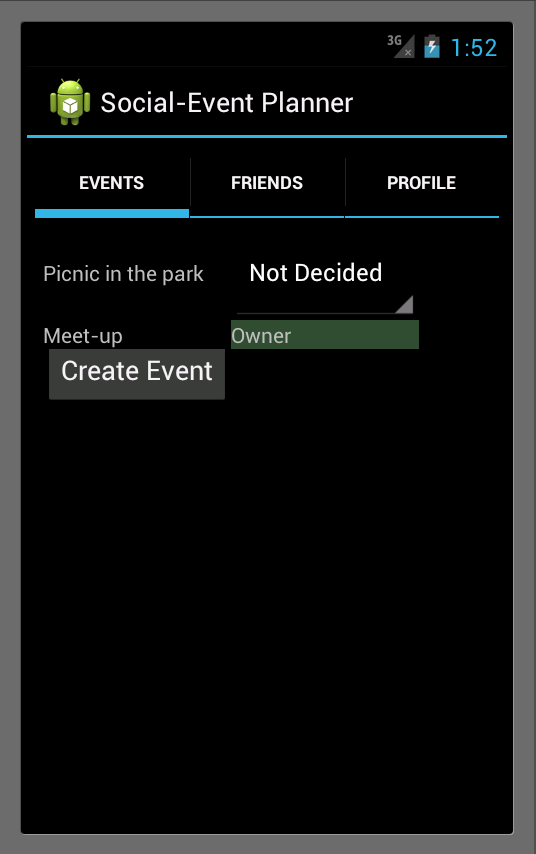} 
    \caption{Screen 1: Part of the user interface for the Social-Event Planner.}
    \label{fig:case-studies:sep:generated:VC:screenshot1}
  \end{subfigure}%
  ~ 
  \begin{subfigure}[b]{0.5\textwidth}
    \centering
    \includegraphics[width=2in]{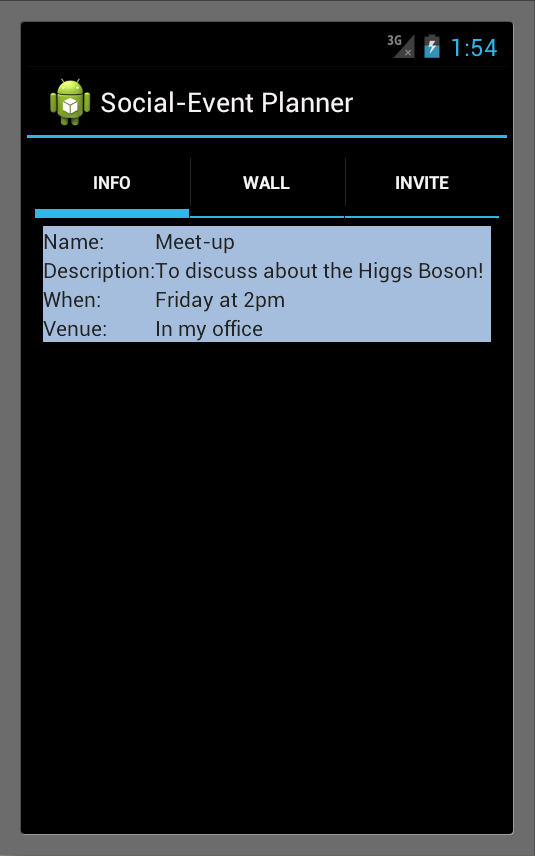} 
    \caption{Screen 2: Part of the user interface for the Social-Event Planner.}
    \label{fig:case-studies:sep:generated:VC:screenshot2}
  \end{subfigure}        
  \caption{Screenshots Social-Event Planner}
  \label{fig:case-studies:sep:generated:VC:screens}
\end{figure}

The user interface is composed of three main screens:

\begin{inparaenum}[\itshape 1\upshape)]
\item where the events (created by the user or invited for someone
  else) are displayed. The user has the option to reply to a
  social-event, to create another social-event, or to see the
  information of a specific social-event (see Figure
  \ref{fig:case-studies:sep:generated:VC:screenshot1}). The
  information of a specific social-event (see Figure
  \ref{fig:case-studies:sep:generated:VC:screenshot2}) is the name of
  the event,
  the description (e.g. date, venue), the list of the invited people
  (if the user is the owner of the event, there is an option to invite
  more people) and finally there is a `wall' where the invited people
  can comment or share content. The user can also reply to an invited
  event. 
\item The second screen allows users to see their friends, as well as
  add/delete more friends. 
\item Finally, the user has the possibility to see/change its personal
  information.
\end{inparaenum}

All the sources and the code generated and implemented for the
Social-Event Planner are available at
\url{http://poporo.uma.pt/EventB2Java/EventB2Java_studies.html}. Additionally,
Table \ref{table:eb2java:comparison:eb:st} in Section
\ref{eb2java:comparison} (page
\pageref{table:eb2java:comparison:eb:st}) presents relevant statistics
for the Social-Event Planner.

\section{Tokeneer}
\label{case-studies:tok}

The Tokeneer system was developed by Praxis High Integrity. Praxis
modelled Tokeneer in Z \cite{Z-Abrial:80,WoodcockBookOnZ} and
implemented it in Spark Ada \cite{ada}.
The Tokeneer system consists of a secure enclave and a set of system
components as shown in Figure \ref{fig:case-studies:tok:tok-system}.
The Tokeneer {ID
  Station} (TIS) is responsible for reading a fingerprint and, based
on a number of protocols and checks, ensuring that any person trying
to access the enclave is indeed permitted to enter the enclave, and
giving the corresponding grants as a user or administrator.  The
TIS communicates with a number of external components to perform its
analysis. The physical devices that are interfaced to the TIS are a
fingerprint reader, a smart-card reader, a floppy drive, and a door and a visual
display. Individuals enter the secured enclave via the {door} by
providing the credentials either to the {fingerprint reader} or the
{card reader}. The {visual display} shows messages that help to
track the progress of the user entry process into the secured
enclave. An {Audit Log} logs all events and actions performed or
monitored by the TIS. The {Token} is the card that is inserted by the user
to enter the enclave.
There are different types of certificates that are used
for verification of each {Token}, and certificates are a crucial part of
the Tokeneer system.

\begin{figure}
  \centering
  \includegraphics[scale=0.6]{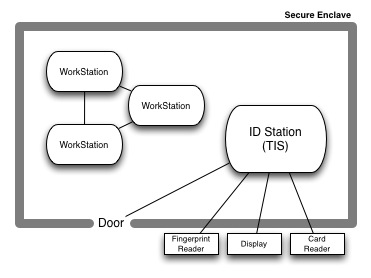}
  \caption{The Tokeneer System.}
  \label{fig:case-studies:tok:tok-system}
\end{figure}

The TIS is about 10K lines of code. Praxis wrote the software
specifications of the TIS in Z following a System Requirements
Specification (SRS) document written by
them, and manually translated the Z specification to Spark Ada. The
documents described below were written and used by Praxis for
developing the TIS and can be found at \cite{tok}.

\begin{itemize}
\item The System Requirements Specification (SRS) includes
  the TIS software requirements.

\item The Formal Specification of the TIS includes the TIS
  software requirements written in Z.

\item The specifications in the above document were later refined and
  extended in a document called the Formal Design in which
  operations in Z are extended and more system invariants are
  considered.

\item The System Test Specification (STS) presents the test
  cases for the TIS.

\end{itemize}

The following sections describe in more detail the components, and the
operations of the TIS. Very detailed information can be found in
\cite{tok}.

\subsection{TIS Components}
TIS is mainly composed of four main physical components used to
communicate the ID Station with the exterior as depicted in Figure
\ref{fig:case-studies:tok:tok-system}. The TIS contains:

\paragraph{The Door: } allows user to enter to the enclave. The
door has two possible states, it can be open or closed. It has a latch
that can be locked or unlocked, and an alarm. 

\paragraph{The Fingerprint Reader: } collects information about the
fingerprint of the users. It is used to compare if the fingerprint of
the user trying to the enclave matches the fingerprint already stored
in the system.

\paragraph{The Display: } shows short messages to the user on a
small display during the attempt to enter to the enclave. For instance
a message could be {\bf AUTHENTICATING USER}, {\bf ENTER TOKEN}.

\paragraph{Card Reader: } reads the card (token, explained later in
this section) that belongs to the user attempting to enter the
enclave. The card provides useful information about the user to the
system, the system processes this information, and allows (or not) the
entering of the user.

\subsection{TIS Operations}
TIS contains a series of operations (dis-) allowing the user of the
TIS to perform certain activities. The following presents some
concepts necessary to understand the operations:

\paragraph{Certificates: } Certificates are used for a user
validation during enrolment to the TIS (as explained later in this
section). It always contains a unique identifier, and a validity
period during which time the certificate is valid. Certificates also
have
an asymmetric key for verification, that could be optional. 

There are different types of certificates in the system. Their
hierarchy is shown in Figure
\ref{fig:case-studies:tok:certificates:types}. A
Certificate can be an ID Certificate or an Attribute Certificate.

\begin{figure}
  \centering
  \includegraphics[scale=0.5]{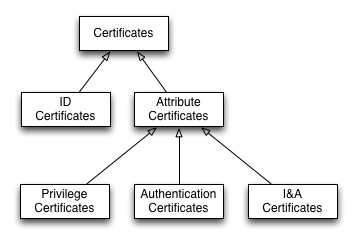}
  \caption{Hierarchy of certificate types.}
  \label{fig:case-studies:tok:certificates:types}
\end{figure}

\paragraph{ID Certificate (IDCert) } contains a reference
to a certificate of the system, the name of the user being identified,
and the asymmetric key of the user. 

\paragraph{Attribute Certificate (AttCert) } contains a reference to a
certificate of the system, and a reference of the ID Token related to
the certificate. It also contains a reference to an IDCert. An AttCert can be
private Certificate (privCert), an Identification and Authentication
Certificate (IandACert), or Authorisation Certificate (authCert).

\paragraph{Private Certificate (privCert) } contains additional
attributes: a role, it can be a user only, a guard a securityOfficer,
or an auditManager (role determines privilege over the TIS); and a
clearance that determines the ordered classifications on documents,
areas, and people. It can be: unmarked, unclassified, restricted,
confidential, secret, or top secret.

\paragraph{Identification and Authentication Certificate (iandACert) }
contains a fingerprint template that contains information reading
from a fingerprint, this information is used to compare if the
fingerprint of the user being identified matches the information in
the system.

\paragraph{Authorisation Certificate (authCert) } contains the same
structure as privCert. It is used for different check-ins to enter the
TIS. 

\paragraph{Tokens (tokens) } are smart cards belong to each user of
the system. The smart card contains a unique ID, a series of
certificates: idCert, privCert, IandACert, and an optional authCert. 

Operations over tokens are
\begin{itemize}

\item a token is {\it valid} if each certificate on it correctly cross-references to
  the IDCert, and each certificate correctly cross-references to the
  token ID.

\item if the Authorisation Certificate is present,
  it is {\it valid} if it correctly cross-references to the token ID, and
  the IDCert.

\item a token is {\it current} if all certificates on it are current,
  it means, if
  the current Time is within the validity period of each certificate. 
\end{itemize}

\subsubsection{User Entry Operations}
These operations describe the process a user needs to do to be
authenticated to further enter the enclave. It is presented as a state
transition diagram. Operations are:

\paragraph{User Token Tears: } If the user tears the Token out before
the operation is complete then the operation is terminated
unsuccessfully.

\paragraph{Reading the User Token: } This operation performs actions
that read an inserted token.

\paragraph{Validating the User Token: } Once TIS has read the user
token, the token content needs to be validated. The token passes the
validation state if

\begin{itemize}
\item the token is {\it valid} and it contains an authCert
  Certificate that cross-checks correctly with the token ID and the ID
  certificate. The token must be {\it current} and both the authCert and
  IDCert certificate can be validated. In this case Biometric checking
  is not performed, or

\item the token is {\it consistent}, {\it current}, and the IDCert,
  priviCert, and iandACert can be validated. In this case, Biometric
  checks will be required, or

\item in the case where there is a valid authCert certificate the
  biometric checks are passed.

\end{itemize}
The biometric checks are only required if the authCert Certificate is
not present or not valid. In this case the remaining certificates on
the card must be checked.

\paragraph{Reading and Validating a Fingerprint: } During the entering
process, users might be asked to provide the fingerprint as a
validation process (biometric check). This operation reads users'
fingerprint and
compares it against the fingerprint information already stored in the
system. 

\paragraph{Writing the User Token: } This operation attempts to write
an authorisation certificate in the user's token, it may
(un-)successfully written. 

\paragraph{Validating Entry and Unlocking Door: } The system will
validate the entrance of the user, if the user's token passes all
checks, the door will be unlocked.

\subsubsection{Operations within the Enclave}
Users of the TIS may have permission to operate the Enclave. Those
users are called administrator and they can perform an additional
operation with the Enclave. The process to operate the Enclave is also
presented as state transition diagram. The following describes these
operations

\paragraph{Enrolment of an ID Station: } In order for a user to
perform administrator operations he has to enrol the TIS. The user
needs to request the enrolment to the system providing 
information in a floppy, the system validates the information, and
if the data is valid, it will enrol the user as an administrator. After this
process the user (administrator) needs to provide his token to the
system.

\paragraph{Administrator Token Tear: } If the administrator tears his
token will result in his logging out from the system.

\paragraph{Administrator Log-in: } In order for an administrator to
log-in the system, the administrator needs to insert a {\it valid}
token into the token reader. If the information provided is valid, the
administrator can enter the enclave and will have the privileges
indicated in the token.

\paragraph{Administrator Log-out: } The logging-out of an
administration can happen for either the administrator removes his
token from the TIS or the authorisation certificate expires.

\paragraph{Administrator Operations: } The administrator has a set of
operations to perform. The administrator can archive the log, update
the configuration data, overrideLock, or shutdown the system. The
privileges administrators have are written in his token.

\subsection{An example of User Entry Operation}
Figure \ref{fig:case-studies:tok:state_dia:enter} is an excerpt of the
transition state
diagram for users' authentication and entry process. The figure shows
just  the transition process for a user to enter the enclave with a
token that does not contain any authCert certificate, and {\it valid}
and {\it current} privCert and iandACert Certificates. The system
requires the user to pass the biometric checks, and
finally writes its token.

\begin{figure}
  \centering
  \includegraphics[scale=0.5]{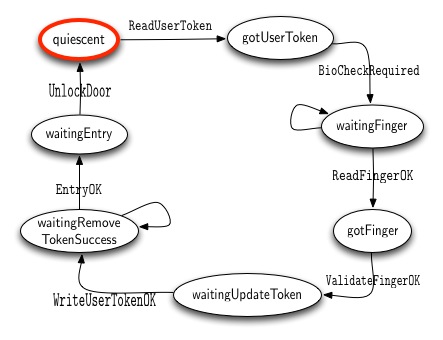}
  \caption{Excerpt State Diagram User Entry to TIS.}
  \label{fig:case-studies:tok:state_dia:enter}
\end{figure}

In Figure \ref{fig:case-studies:tok:state_dia:enter}, the ovals
represents the states
and the lines represent the operation performed (the transition). The
red oval represents the starting and ending point of the diagram. The
system starts in a \textsf{quiescent} state. Once the user puts his
token in the card reader, the operation \texttt{ReadUserToken} is
performed. This operation requires the status of the system to be in
\textsf{quiescent}, and the token to be present. The operation changes the
status to \textsf{gotUserToken} and displays a message in the display
(``AUTHENTICATING USER, PLEASE WAIT''). The system evolves and goes to
state \textsf{gotUserToken}. Since the token inserted does not contain
authCert Certificate and the privCert and iandACert certificates are
{\it valid} and {\it current}, the system requires the user for a
biometric check. The biometric check is to read the fingerprint. The
system evolves to state \textsf{waitingFinger}. The user puts the
fingerprint in the finger reader (as stated by transition
\texttt{ReadFingerOk}). The system goes to state
\textsf{gotFinger}. The operation \texttt{ValidateFingerOK} checks
that the information of the fingerprint read by the fingerprint
reader indeed matches with the information stored in the system, if so,
the system evolves to \textsf{waitingUpdateToken} that changes the
status to \texttt{waitingEntry}. Since the status is
\texttt{waitingEntry}, the token of the user is still inserted, and
the certificates on it are {\it valid} and {\it current}, the user is
granted permission to enter the enclave. The system evolves to the
state \textsf{waitingRemoveTokenSuccess} where the system waits until
the user remove the token to finally unlock the door.

\subsection{Conversion from Z to \eb}
\label{case-studies:tok:conversion}

This section discusses the strategy we followed to convert the
existing model of TIS in Z to \eb. 
Z \cite{Z-Abrial:80,WoodcockBookOnZ} is a notation for writing
specifications based on set theory and
first order logic. It includes a notation for discrete mathematics
(set theory and predicate calculus) and for describing and combining
schemas (the schema calculus) that allows one to define possible
states as well as operations that can change the state. 

In Z, one can define a variable (in capital letters) to express basic
types. For instance, the Z schema defined above defines two sets
\zkeyw{USER} and \zkeyw{TOKENID} as the sets of all possible users and
token ids of the system:

{\small
  \[
  [\zkeyw{USER}]\\
  [\zkeyw{TOKENID}]
  \]
}

That definition goes in the same direction as carrier sets in \eb. We
modelled Z variables in capital letter as carrier sets in \eb. For
instance, the definition of the Z schemas [\zkeyw{USER}] and
[\zkeyw{TOKENID}] is in \eb:

{\small
  \[
  \begin{array}{l}
    \ebkeyw{context }ctx\\
    ~~\ebkeyw{sets USERS TOKENID}\\
    \ebkeyw{end} 
  \end{array}
  \]
}

In Z, one can define a schema that defines variables. The schema
$Certificate$ (shown below) defines a common certificate in Tokeneer:
a certificate contains a unique identifier (denoted by variable $id$),
it also contains a validity period in which that certificate is valid
(denoted by variable $validityPeriod$), and a asymmetric key that
validates the certificate (denoted by variable $isValidatedBy$). 
{\small
  \begin{schema}{Certificate}
    $id $ : $ CertificateId$ \\
    $ValidityPeriod $ : \pow \zkeyw{TIME}\\
    $isValidatedBy $ : {\it optional} \zkeyw{ KEY}
  \end{schema}
}

In Z, one can instantiate schemas and refer to variables of that
schema using the notation dot ($.$). For instance one can define a
certificate $c$ and refer the variable $id$ as $c.id$. Since this is
not possible in \eb, we decided to model Z schemas in \eb\ as
several relations that map the type schema to each variable. For
instance, the following \eb\ machine defines the schema $Certificate$
in \eb:

{\small
  \[
  \begin{array}{l}
    \ebkeyw{machine }m \ebkeyw{ sees } ctx\\
    \ebkeyw{variables }
    certificates~certificateID~validityPeriods~publicKeys~isValidatedBy \\
    \ebkeyw{invariants}\\
    ~~\ebtag{inv1 } certificates \subseteq \ebkeyw{CERTIFICATES}\\
    ~~\ebtag{inv2 } publicKeys \subseteq \ebkeyw{KEYS}\\ 
    ~~\ebtag{inv3 } certificateID \in certificates \tinj \ebkeyw{CERTIFICATEID}\\
    ~~\ebtag{inv4 } validityPeriods \in certificates \trel \nat\\
    ~~\ebtag{inv5 } isValidatedBy \in certificates \pinj
    publicKeys\\
    \ebkeyw{end}
  \end{array}
  \]
}

Where context $ctx$ defines the carrier sets \ebkeyw{CERTIFICATES},
\ebkeyw{KEYS}, and \ebkeyw{CERTIFICATEID} (not shown here). The
model of variables in \eb\ follows the Z specification of them and the
System Requirement Specification (SRS) document. For instance, just from the
Z specification is not possible to deduce that id variable needs to be
defined as a total injection. However, reading the SRS document one
can realise id variable define a unique id for each certificate in the
system. In Z, one can define a variable as {\it optional}, we model
that as a partial injection ($\pinj$ in \eb), meaning there can be a certificate that
does not have any public key associated to it. To create a new
certificate one needs to update all variables as:
{\small
  \[
  \begin{array}{l}
    \ldots \\
    \ebkeyw{then}\\
    ~~\ebtag{act1 } certificates~ := ~certificates \bunion \{new\_c\}\\
    ~~\ebtag{act2 } publicKeys~ := ~publicKeys \bunion \{new\_pk\} \\
    ~~\ebtag{act3 } certificateID~ := ~certificateID \bunion \{new\_c
    \mapsto new\_certid\} \\
    ~~\ebtag{act4 } validityPeriods~ := ~validityPeriods \bunion \{new\_c
    \mapsto new\_valtime\} \\
    ~~\ebtag{act5 } isValidatedBy~ := ~validityPeriods \bunion \{new\_c
    \mapsto new\_pk\} \\
    \ebkeyw{end}
  \end{array}
  \]
}

So to access the identifier value of a certificate $c$ (as in Z
is $c.id$), in \eb\ one does $certificateID.apply(c)$.

In Z, one can also define a schema composed of two parts divided by a
line: above the line one can define variables or import another
schemas already defined, below the line one can define predicates. The
following Z schema $ReadUserToken$ depicts this kind of schema:

{\small
  $[$\zkeyw{STATUS} ::= $\{quiescent, gotUserToken, waitingFinger\}]$\\
  $[$\zkeyw{PRESENCE} = $\{absent, present\}]$\\
  $[$\zkeyw{DISPLAYMESSAGE} ::= $\{wait, welcome, insertFinger\}]$\\
  \begin{schema}{Context}
    $status $: \zkeyw{STATUS}\\
    $tokenPresence $: \zkeyw{PRESENCE}\\
    $display $: \zkeyw{DISPLAYMESSAGE}
  \end{schema}
  \vspace{-.8cm}
  \begin{schema}{ReadUserToken}
    Context\\
    \where
    $status $ = $ quiescent$\\
    $tokenPresence $ = $ present$\\
    $status' $ = $ gotUserToken$\\
    $display' $ = $ wait$
  \end{schema}
}

Z Schema $ReadUserToken$ is partially defining one of the states
defined in Figure \ref{fig:case-studies:tok:state_dia:enter}. The
schema is not
defining any new variable but it is importing schema $Context$ so 
predicate expressions like $status = quiescent$ can be used. We
translate a schema as an \eb\ event where the predicate is the guard
of the event. Notice that Schema $ReadUserToken$ defines a predicate
with the aid ($'$), that is how schemas evolve in time. For instance,
predicate $status' = gotUsertToken$ means the new value for variable
$status$ will be $gotUsertToken$. Z predicates with the aid of $'$ is
translated to \eb\ in the actions of the event. the following \eb\
machine translates the schemas:

{\small
  \[
  \begin{array}{l}
    \ebkeyw{machine } m \ebkeyw{ sees } ctx\\
    \ebkeyw{variables}~~status~tokenPresence~display\\
    \ebkeyw{invariants}\\
    ~~\ebtag{inv1 } status \in \ebkeyw{STATUS}\\
    ~~\ebtag{inv2 } tokenPresence \in \ebkeyw{PRESENCE}\\
    ~~\ebtag{inv3 } display \in \ebkeyw{DISPLAYMESSAGE}\\\\
    ~~ReadUserToken\\
    ~~~\ebkeyw{where}\\
    ~~~~\ebtag{grd1 } status = quiescent\\
    ~~~~\ebtag{grd2 }tokenPresence = present\\
    ~~~\ebkeyw{then}\\
    ~~~~\ebtag{act1 }status := gotUserToken\\
    ~~~~\ebtag{act2 } display := wait\\
    ~~\ebkeyw{end}\\
    \ebkeyw{end}
  \end{array}
  \]
}

Context $ctx$ defines carrier sets \ebkeyw{STATUS}, \ebkeyw{PRESENCE},
and \ebkeyw{DISPLAYMESSAGE}, and defines the corresponding axioms. For
instance the axiom related to \ebkeyw{STATUS} is
\[
\ebtag{ax1 } \ebkeyw{STATUS }=~\{quiescent,gotUserToken,waitingFinger\}\\
\ebtag{ax2 } quiescent \not = gotUserToken\\
\ebtag{ax3 } quiescent \not = waitingFinger\\
\ebtag{ax4 } gotUserToken \not = waitingFinger
\]
Notice that the \eb\  translation of the Z schema $Context$ does not
define variables as relation (as shown for Z schema
$Certificate$). Our decision on doing that is that schema $Context$
defines variables $status$, $tokenPresence$, and $display$ that are
always the same during the execution of the system,
whereas schema $Certificate$ can be instantiated any time a new
certificate is created.

We follow the notion of translation described in this section to
translate the TIS system modelled in Z to \eb.

\subsection{Modelling the Tokeneer ID Station (TIS) in \eb}
\label{case-studies:tok:tokEB}

We modelled the Tokeneer ID Station (TIS) in \eb\ following the Z
model of the TIS and the documentation provided by Praxis. 
We followed the
``parachute'' software development strategy of \eb\ proposed by J.-R
Abrial in \cite{TheBBook}. Table \ref{tab:case-studies:tok:srs} lists
a few software requirements of the TIS. The
table includes some functional (\texttt{FUN}) and environmental
(\texttt{ENV}) requirements. We wrote an abstract machine, six
machine refinements and an additional AuditLog machine as shown in
Table \ref{tab:case-studies:tok:refinement}. 

The abstract machine models certificates. The first and the second
refinements include specialised certificates. These first three
machines model certificates and its hierarchy as shown in Figure
\ref{fig:case-studies:tok:certificates:types}. The third refinement
models fingerprints
and the internal status to enter the enclave. The fourth refinement
models entry to the enclave and the display used by the TIS. The fifth
refinement models enrolments to the enclave using a certificate, and
the sixth refinement models some administrative functionality. 
Machine $AuditLog$ models a log of all events and actions performed or
monitored by the TIS.

\begin{table}
  \begin{tabular}{ l p{10cm}}
    \hline
    \textbf{Req.} & \textbf{Description}\\

    \hline
    \verb+FUN-1+ &  Certificates have a unique ID, a period 
    during which they are valid, and a public key of the user 
    (used to sign and verify the certificate). \vspace{.15cm}\\

    \verb+FUN-2+ &  The system contains two kinds of certificates: ID and
    Attribute certificates. They are used during enrolment and are 
    present on tokens. \vspace{.15cm}\\

    \verb+FUN-3+ &  Attribute certificates are categorised into three types as
    follows: Authorisation and Privilege certificates that have the
    same structure, and I\&A certificates. \vspace{.15cm}\\

    \verb+FUN-4+ &  Tokens are used to store all the required information of
    a user. Each token contains ID,  Privilege, I\&A,
    and Authorisation (optional) certificates. \vspace{.15cm}\\

    \verb+ENV-1+ &  Tokens are the data read from an inserted Smart
    Card. The system contains a card reader to read tokens. \vspace{.15cm}\\

    \verb+FUN-5+ &  Tokens should be \textit{valid} (certificates correctly
    cross reference to the ID Certificate) and \textit{current}
    (all included certificates are up to date) before processing. \vspace{.15cm}\\

    \verb+FUN-6+ &  TIS enrolment transition is defined by the transition
    state diagram in  \cite{tok} (41\_2.pdf/pp. 59/ Fig. 7.1). \vspace{.15cm}\\

    \verb+ENV-2+ &  The system contains a floppy drive \vspace{.15cm}\\

    \verb+FUN-7+ &  The TIS maintains an audit log with fixed size that logs all
    the actions taken within the enclave. \vspace{.15cm}\\

    \verb+FUN-8+ &  TIS administrators are users with higher security
    privileges. Administrators may log on to the TIS console, 
    log-off, or start an operation. \vspace{.15cm}\\

    \verb+ENV-3+ & The System contains a fingerprint reader. Fingerprints 
    are used if bio-metric security is required. \vspace{.15cm}\\

    \verb+ENV-4+ &  The System contains a small display outside the enclave. \vspace{.15cm} \\

    \verb+ENV-5+ &  The door controls user entry to the enclave, and is either open or closed. \vspace{.15cm}\\    

    \verb+FUN-9+ &  The user entry transition is defined by the transition
    state diagram in \cite{tok} (41\_2.pdf/pp.43/ Fig.6.1), that tracks
    progress through user entry. \vspace{.15cm}\\

    \hline
  \end{tabular}
  \centering
  \caption{System Requirements Specification of TIS.}
  \label{tab:case-studies:tok:srs}
\end{table}

\begin{table}
  \setlength{\tabcolsep}{.3em}
  \begin{tabular}{  l p{1 cm} p{5cm}}
    \hline
    \textbf{Machines} & \textbf{Level} & \textbf{Description}\\
    \hline

    {Abstract} &  \verb+FUN-1+ &  Basic certificates (Level-0). \vspace{.2cm}\\

    {Ref-1} & \verb+FUN-2+ & IDCert, AttCert (Level-1) \vspace{.2cm}\\ 

    {Ref-2} & \verb+FUN-3+ & PrivCert, AuthCert, I\&ACert (Level-2) \vspace{.2cm}\\

    {Ref-3} & \verb+ENV-1+, \verb+FUN-4+, \verb+FUN-5+ & Token, fingerprint, and internal status to enter the enclave \vspace{.2cm}\\

    {Ref-4} &  \verb+ENV-4+,  \verb+ENV-5+, \verb+FUN-9+ & Entry to the enclave, the display \vspace{.2cm}\\

    {Ref-5} & \verb+FUN-6+ & Enrolment \vspace{.2cm}\\

    {Ref-6} &   \verb+FUN-8+ & Admin \vspace{.2cm}\\

    {AuditLog} &   \verb+FUN-7+, \verb+ENV-2+ & Audit Log \vspace{.2cm}\\
    \hline
  \end{tabular}
  \centering
  \caption{Refinement Strategy for the Tokeneer system.}
  \label{tab:case-studies:tok:refinement}
\end{table}

Figure \ref{fig:tok:ref3} shows an excerpt of the third refinement of
the TIS \eb\ model. Machine $ref3\_entry\_L1$ models the entry of an user
to the enclave. The machine \ebkeyw{sees} context $ctx\_ref3$ (not
shown in the figure) that defines carrier sets \ebkeyw{CERTIFICATES}
(the set of all type of certificates), \ebkeyw{KEYS} (the set of
asymmetric keys used for signing and validating certificates),
\ebkeyw{TOKENID} (the set of all tokens), and \ebkeyw{ENTRY\_STATUS}
(the set of all possible status for entering the enclave). Variables
$certificates$, $publicKeys$, $isValidatedBy$, and $validityPeriods$
define the properties of each certificate as specified by the functional
requirement FUN-1. Variable $attCert$ defines the properties of one
kind of certificate (Attribute Certificates) as specified by the functional
requirement FUN-2. Variables $privCert$, $iandaCert$, and $authCert$
define the properties of Attribute Certificates as specified by the
functional requirement FUN-3. Variables $tokenPrivCert$,
$tokenIandaCert$, $tokenAuthCert$, $tokenID$, and $attCertTokID$
define the properties for tokens as specified by the functional
requirement FUN-4. Variable $entry\_status$ defines the status on the
entry of a user to the enclave. For instance, the diagram depicted in
Figure \ref{fig:case-studies:tok:certificates:types} shows an excerpt
of the process of
a user to enter the enclave. The states of the diagram are represented
by variable $entry\_status$, it starts with the value
\textsf{quiescent}, after having checked the user token, the system
goes to next state, giving the value of  \textsf{gotUserToken} to
$entry\_status$. Variable $currentToken$ models the token is being
read from the card reader.

\begin{figure}[h]
  {\small
    \[
    \begin{array}{l}
      \ebkeyw{machine } ref3\_entry\_L1 \ebkeyw{ sees } ctx\_ref3\\
      \ebkeyw{variables } ~certificates~publicKeys~isValidatedBy~validityPeriods~attCert\\
      ~~~privCert~iandaCert~authCert~tokenPrivCert~tokenIandaCert\\
      ~~~tokenAuthCert~entry\_status~currentToken~tokenID~attCertTokID\\
      \ebkeyw{invariants}\\
      ~~~\ebtag{inv1 } certificates \subseteq \ebkeyw{CERTIFICATES}~~~~\ebtag{inv2 } publicKeys \subseteq \ebkeyw{KEYS}\\
      ~~~\ebtag{inv3 } validityPeriods \in certificates \trel \nat\\
      ~~~\ebtag{inv4 } isValidatedBy \in certificates \pinj publicKeys\\
      ~~~\ebtag{inv5 } attCert \subseteq certificates       ~~~\ebtag{inv11 } entry\_status \in \ebkeyw{ENTRY\_STATUS}\\
      ~~~\ebtag{inv6 } \ebkeyw{partition}(attCert, privCert,iandaCert,authCert)\\
      ~~~\ebtag{inv7 } tokenID \subseteq \ebkeyw{TOKENID} ~~~\ebtag{inv12 } currentToken \in \ebkeyw{TOKENID}\\
      ~~~\ebtag{inv8 } tokenPrivCert \in tokenID \tinj  privCert\\
      ~~~\ebtag{inv9 } tokenIandaCert \in tokenID \tinj  iandaCert\\
      ~~~\ebtag{inv10 } tokenAuthCert \in tokenID \pinj authCert\\
      ~~~\ebtag{inv13 } attCertTokID \in attCert \tinj tokenID\\
      ~BioCheckRequired\\
      ~~~\ebkeyw{any }currentTime~~\ebkeyw{where}\\
      ~~~~~\ebtag{grd1 } entry\_status = gotUserToken~~~~\ebtag{grd2 } currentTime \in \nat\\
      ~~~~~\ebtag{grd3 } (currentToken \in dom(tokenAuthCert) \wedge \\
      ~~~~~~~~~~~~~~~~currentTime \not \in validityPeriods[\{tokenAuthCert(currentToken)\}])\\
      ~~~~~~~~~~~~~~~~\vee currentToken \not \in dom(tokenAuthCert)\\
      ~~~~~\ebtag{grd4 } currentToken \in ran(attCertTokID) \wedge \\
      ~~~~~~~~~~~~~~~~ attCertTokID∼(currentToken) \in dom(isValidatedBy)\\
      ~~~~~\ebtag{grd5 } currentToken \in dom(tokenPrivCert) \wedge \\
      ~~~~~~~~~~~~~~~~ tokenPrivCert(currentToken) \in dom(isValidatedBy)\\
      ~~~~~\ebtag{grd6 } currentToken \in dom(tokenIandaCert) \wedge \\
      ~~~~~~~~~~~~~~~~ tokenIandaCert(currentToken) \in dom(isValidatedBy)\\
      ~~~\ebkeyw{then}\\
      ~~~~~\ebtag{act1 } entry\_status :=  waitingFinger\\
      ~\ebkeyw{end}\\
      \ebkeyw{end}
    \end{array}
    \]
  }
  \caption{Excerpt third refinement machine TIS \eb\ model.}
  \label{fig:tok:ref3}
\end{figure}

Figure \ref{fig:tok:ref3} shows a single event of machine
$ref3\_entry\_L1$, event $Bio\-Check\-Required$. Once the user that wants
to enter to the enclave puts his token into the card reader, the
system reads the information within the token smart card. A biometric
check is required when 

\begin{itemize}
\item the status of the entry is $gotUserToken$ as stated by guard
  \ebtag{grd1}, 

\item the user token is valid for entry into the enclave, i.e if the
  token 
  \begin{itemize}
  \item is consistent (e.g. $currentToken \in dom(tokenPrivCert)$),  

  \item ID certificate, Privilege certificate and IandA certificate can be validated (e.g. $tokenPrivCert(currentToken) \in
    dom(isValidatedBy)$) as stated by guards \ebtag{grd4}, \ebtag{grd5},
    and \ebtag{grd6}, and

  \end{itemize}

\item the Authorisation Certificate is not present (e.g. $currentToken
  \not \in dom(tokenAuthCert)$) or not valid (e.g. $currentTime \not
  \in validityPeriods[{tokenAuthCert(currentToken)}])$) as stated by
  guard \ebtag{grd3}.

\end{itemize}

The biometric check consists in reading the fingerprint of the user so
the system can compare it against the fingerprint already stored in
the system. If a biometric check is required, the system goes to state
\textsf{waitingFinger} as stated by the action \ebtag{act1}.

\subsection{Generating Java code for the TIS \eb\ model}
\label{case-studies:tok:generated}
After modelling TIS in \eb\ and discharging all proof obligations, we
generated Java code of the model using \ebtojava. Figure
\ref{fig:case-studies:tok:machine:java} depicts
an excerpt of the translation of the machine $ref\_3\_entry\_L1$ and
Figure \ref{fig:case-studies:tok:biocheck:java} shows an excerpt of
the translation of the event $BioCheckRequired$. 

Figure \ref{fig:case-studies:tok:machine:java} defines carrier sets,
variables, and a constructor of the class.

\begin{figure}
  \begin{lstlisting}[frame=none]
public class ref3_entry_L1{
  BioCheckRequired evt_BioCheckRequired = new BioCheckRequired(this);

 public static final BSet<Integer> CERTIFICATES = new Enumerated(INT.min,INT.max);
  // ...  definition of the rest of carrier sets

  private BSet<Integer> attCert;
  private BRelation<Integer,Integer> isValidatedBy;
  // ...  definition of the rest of variables
  
  // ... definition of getter and mutator methods

  public ref3_entry_L1(){
    attCert = new BSet<Integer>();
    isValidatedBy = new BRelation<Integer,Integer>();
    // ... initialisation of class fields

    }
}
  \end{lstlisting}
  \caption{Partial translation of machine $ref3\_entry\_L1$.}
  \label{fig:case-studies:tok:machine:java}
\end{figure}

Figure \ref{fig:case-studies:tok:biocheck:java} shows a partial
translation of
event $BioCheckRequired$ where \javacode{machine} is a reference to
the machine class
implementation. The Java code includes methods
\javacode{guard\_BioCheckRequired} (the translation of the
event guard) and \javacode{run\_BioCheckRequired} (the
translation of the event body). 
The \jml\ specifications generated by \ebtojava\ are omitted since the
specifications are not used in generating tests or customising the code in
this example.

\begin{figure}
  \begin{lstlisting}[frame=none]
public class BioCheckRequired{
  private ref3_entry_L1 machine;

  public BioCheckRequired(ref3_entry_L1 m) {
    this.machine = m;
  }

  public boolean guard_BioCheckRequired(Integer currentTime) {
    return (
        machine.get_entry_status().equals(machine.gotUserToken) && 
        machine.get_tokenAuthCert().domain().has(machine.get_currentToken()) && 
        !machine.get_validityPeriods().image(new BSet<Integer>(machine.get_tokenAuthCert().apply(machine.get_currentToken()))).has(currentTime) || !machine.get_tokenAuthCert().domain().has(machine.get_currentToken()) && ...;
    }

  public void run_BioCheckRequired(Integer currentTime){
    if(guard_BioCheckRequired(currentTime)) {
      Integer entry_status_tmp = machine.get_entry_status();
      machine.set_entry_status(machine.waitingFinger);
    }
  }
}
  \end{lstlisting}
\caption{Partial translation of event $BioCheckRequired$.}
  \label{fig:case-studies:tok:biocheck:java}
\end{figure}

\subsection{Writing JUnit Tests}
\label{subsec:toktests}

Software Testing \cite{Black:09} can be used to validate software
requirements that are expressed in a formal language. A common way of
testing is the formulation of expected results. Hence, testing is
achieved by comparing the results from executing the system against
the expected ones.  

The System Test Specification of the TIS includes
32 test cases organised in eight categories as shown in Table
\ref{tab:case-studies:tok:sts}.  We wrote Java code for these 32 test
cases in two
steps. We first used the \ebtojava\ tool to translate the \eb\ model
of the TIS to Java.  We generated a sequential version of the model in
Java since the tests are run sequentially. We then gave initial values
for Java constants that respect the axioms on those constants defined
in the \eb\ model.

\begin{table}
  \begin{tabular}{l p{8cm}}
    \hline
    \textbf{Category Test} & \textbf{Description} \\
    \hline
    \verb+Enrolment+ & 3 tests for 
    starting an un-enrolled TIS attempting to enroll using different
    types of certificates. \vspace{.2cm}\\

    \verb+UserEntry+ & 14 tests for allowing
    administrators with different roles and users with
    different kinds of certificates to enter the enclave. \vspace{.2cm}\\

    \verb+UpdateConfig+& 5 tests for allowing different kinds of
    administrators to update the configuration of the enclave.  \vspace{.2cm}\\

    \verb+Override+& 1 test for overriding the operation of the door
    by the guard administrator. \vspace{.2cm}\\

    \verb+Admin Login+& 3 tests for allowing
    an administrator to log-in to the TIS. \vspace{.2cm}\\

    \verb+Admin Logout+& 1 test for
    logging-out of the TIS. \vspace{.2cm}\\

    \verb+Shutdown+& 2 tests for shutting down the system. \vspace{.2cm}\\

    \verb+ArchiveLog+& 3 tests for
    actions over the AuditLog component. \vspace{.2cm}\\

    \hline  
  \end{tabular}
  \centering
  \caption{System Test Specification of the TIS.}
  \label{tab:case-studies:tok:sts}
\end{table}  

We ran the 32 JUnit tests using the input data provided by
Praxis. Then we compared the obtained results against the expected
results also provided by Praxis.  As an example of a test, Praxis
defined {\it UserEntry1} as one of the test cases of the \verb+UserEntry+
category. The test allows an administrator with role ``Security
Officer'' to enter the enclave and acquire a valid Auth
Certificate. The test follows the state diagram presented in Figure
\ref{fig:case-studies:tok:state_dia:enter}, it goes through the
following steps:

\begin{itemize}
\item ReadUserToken 
\item BioCheckRequired
\item ReadFingerOK 
\item ValidateFingerOK 
\item ConstructAuthCert 
\item WriteUserTokenOK 
\item EntryOK 
\item UnlockDoorOK
\end{itemize}

Figure \ref{fig:case-studies:tok:testcases} shows the JUnit
implementation of test
{\it UserEntry1}. Variable \texttt{machine} (a reference to the machine in the
Java implementation) gives access to all the variables and events of the
model. Method \javacode{set\_\-test\_\-UserEntry1} is used to initialise
variables, variables are initialised according to initial values given
by Praxis. When executed, this test will fail if any \javacode{guard\_evt}
method returns \javakeyw{false}, or if any \javacode{run\_evt} method does not set
the proper screen and display messages, as stated by Praxis
documentation in Expected Results.
The
final result of this test matches the expected result: the messages on
the screen were correct, and Authorisation Certificate was created,
and the door is open so user can enter to the enclave.

\begin{figure}
 \begin{lstlisting}[frame=none]
    @Test
    public void test_UserEntry1(){
    set_test_UserEntry1(machine);

    // ReadUserToken
    Assert.assertTrue("Guard evt_ReadUserToken not satisfied.", machine.evt_ReadUserToken.guard_ReadUserToken(token_user_to_read));
    machine.evt_ReadUserToken.run_ReadUserToken(token_user_to_read);
    Assert.assertEquals(machine.get_displayMessage1(), ref6_admin.wait);

    // BioCheckRequired 
    Assert.assertTrue("Guard evt_BioCheckRequired not satisfied.", machine.evt_BioCheckRequired.guard_BioCheckRequired(currentTime));
    machine.evt_BioCheckRequired.run_BioCheckRequired(currentTime);
    Assert.assertEquals(machine.get_displayMessage1(), ref6_admin.waitingFinger);
	
    machine.set_FingerPresence(machine.present);
    // ReadFingerOK
    Assert.assertTrue("Guard evt_ReadFingerOK not satisfied.", machine.evt_ReadFingerOK.guard_ReadFingerOK());
    machine.evt_ReadFingerOK.run_ReadFingerOK();
    Assert.assertEquals(machine.get_displayMessage1(), ref6_admin.wait);

    Integer fingerPrint = User01fp;
    // ValidateFingerOK
    Assert.assertTrue("Guard evt_ValidateFingerOK not satisfied.", machine.evt_ValidateFingerOK.guard_ValidateFingerOK(fingerPrint));
    machine.evt_ValidateFingerOK.run_ValidateFingerOK(fingerPrint);
    Assert.assertEquals(machine.get_displayMessage1(), ref6_admin.wait);

    // ConstructAuthCert -> built-in writeUserTokenOK
    // WriteUserTokenOK
    Assert.assertTrue("Guard evt_WriteUserTokenOK not satisfied.", machine.evt_WriteUserTokenOK.guard_WriteUserTokenOK(cert_params));
    machine.evt_WriteUserTokenOK.run_WriteUserTokenOK(p_id_cert, p_priv, p_ce, p_tid, p_serial, p_issuer, p_period, p_pubkey, p_class);
    Assert.assertEquals(machine.get_displayMessage1(), ref6_admin.wait);

    // EntryOK
    Assert.assertTrue("Guard evt_EntryOK not satisfied.", machine.evt_EntryOK.guard_EntryOK(currentTime));
    machine.evt_EntryOK.run_EntryOK(currentTime);
    Assert.assertEquals(machine.get_displayMessage1(), ref6_admin.openDoor);

    machine.set_userTokenPresence(machine.absent);
    // UnlockDoorOK
    Assert.assertTrue("Guard evt_UnlockDoorOK not satisfied.", machine.evt_unlockDoorOK.guard_unlockDoorOK(currentTime));
    machine.evt_unlockDoorOK.run_unlockDoorOK(currentTime);
    Assert.assertEquals(machine.get_displayMessage1(), ref6_admin.doorUnlocked);
}
 \end{lstlisting}
    \caption{The {\it UserEntry1} Test Case in JUnit.}
    \label{fig:case-studies:tok:testcases}
  \end{figure}

  During the first round of testing, the Java code did not pass all 32
  JUnit tests.  We inspected the \eb\ model and discovered that the
  model was creating a specialised Authorisation Certificate for a user
  in the wrong event. As this error did not invalidate the model, it
  could not be detected via model verification in \eb. We corrected the
  \eb\ model, discharged all the proof obligations again, and used the
  \ebtojava\ tool to regenerate the Java code.  We repeated this process
  until the code passed all 32 JUnit tests.  Notice that this
  misbehaviour might be found using ProB on the \eb\ model, our
  strategy enables software developers that are not familiar with \eb\
  to simulate the \eb\ model by translating it to Java code and
  performing JUnit tests on the Java code generated. So the software
  developer does not have to deal with translating the STS document as
  predicates in \eb\ to conform it to the \eb\ model. Our \eb\ model
  of the TIS,  the Java code generated by the \ebtojava\ tool, and the
  32 JUnit tests  that we wrote can be found at
  \url{http://poporo.uma.pt/Tokeneer.html}.

  There are several benefits that can be obtained by applying our
  strategy for simulating an \eb\ model by translating it to Java and
  performing JUnit tests. The user gains confidence in the correctness and
  appropriateness of the modelled system by discharging all the \eb\
  proof obligations in Rodin. The JUnit tests provide an additional
  layer of confidence by checking that the behaviour of the \eb\ model in
  Java is what the user actually intended.
  The Java code generated by the \ebtojava\ tool is an actual
  initial implementation of the \eb\ model that can be used as is, or further
  refined and customised as needed.

\section{Comparing \ebtojava\ to other \eb\ code generators}
\label{eb2java:comparison}
We are interested in comparing our \ebtojava\ tool
against other tools that generate Java implementations from \eb\ models.
In particular, we have compared
\ebtojava\ with Code Generation \cite{CodeGen10,Tasking11} by A. Edmunds
and M. Butler, and EB2J \cite{Mery:2011} by D. M{\'e}ry and
N. Singh. Although Code Generation can generate Ada code in addition
to Java, we were interested in examining and analysing its ability to
generate Java code only. Likewise, EB2J is able to generate C,
C{\small\verb|++|} and \csharp{} code, but we did not consider this in
our comparison.

The comparison defines a set of six performance criteria as
follows. 
\begin{inparaenum}[\itshape i\upshape)]
\item ``Generation Process'' -- does
  the user need to adapt the \eb\ model before using the
  tool to generate Java code. It might be 
  \begin{inparaenum}[\itshape a\upshape)] 
  \item ``Automatic'', if the
    user does not need to edit or extend the \eb\ model, or 
  \item ``Assisted'', if the user does need to do so, or 
  \item ``Automatic/Assisted'', if the user needs to do so in some
    cases and does not in other. 
  \end{inparaenum}
\item  ``Executable'' -- does the generated code 
  compile and run as is. 
\item ``Support for Code Customisation''  --
  does the tool furnish a mechanism
  for the user to be able to customise the generated code and to
  verify whether the customised code is correct. 
\item  ``Support for \eb's Syntax''
  -- does the tool 
  \begin{inparaenum}[\itshape a\upshape)]
  \item  ``Fully'', 
  \item ``Largely'', or 
  \item  ``Scarcely'' support the current syntax of \eb. 
  \end{inparaenum}
\item ``Execution Time'' -- how long does it take for the generated
  code to execute and to give a result (if the execution terminates).
  Finally, 
\item ``Effective Lines of Code'' -- the actual number of lines
  of Java code generated by the tool.
\end{inparaenum}

\begin{table}
  \begin{centering}
    \begin{tabular}{|l|c|c|c|}
      \hline
      {\bf \eb\ Model} & {\bf LOC} & {\bf \# Mch } & {\bf \# Evt}\\ \hline
      {Social-Event Planner} \cite{Planner} & 1326 & 9 & 35\\ \hline
      {MIO} \cite{TeachFM-09} & 586 & 7 & 21\\ \hline
      {Heating Controller} \cite{Heater} & 458 & 15 & 32  \\ \hline
      {State Machine} \cite{StateMachine} & 86  & 2 & 5  \\ \hline
      {Binary Search} \cite{bs:eb:09} & 101 & 3  & 3\\ \hline
      {Linear Search} \cite{bs:eb:09} & 54 & 2 & 2 \\ \hline
      {Minimum Element} \cite{bs:eb:09} & 64 & 2 & 3 \\ \hline
      {Reversing Array} \cite{bs:eb:09} & 64 & 2 & 2\\ \hline
      {Sorting Array} \cite{bs:eb:09} & 137 & 3  & 4\\ \hline
      {Square Root Number} \cite{bs:eb:09} & 84 & 3  & 2\\ \hline
    \end{tabular} 
    \vspace{.1cm}
    \caption{Statistics of the \eb\ Models.}
    \label{table:eb2java:comparison:eb:st}
  \end{centering}
\end{table}

In addition to defining a set of performance criteria, we need to
provide a fair context for comparing tools. We selected the nine \eb\
models shown in Table~\ref{table:eb2java:comparison:eb:st}.  We
developed two of the
systems -- the Social-Event Planner \cite{Planner} and the MIO model
\cite{TeachFM-09}. 
The Social-Event Planner is presented as a case study in
Section~\ref{case-studies:sep}.
MIO is an \eb\ model of a massive transportation system
that includes articulated buses following the main corridor routes of
a city (briefly described in Chapter \ref{chapter:eb2jml}).  The
Heating Controller \cite{Heater} and the State Machine
\cite{StateMachine} models were developed by one of our tool
competitors. The Heating Controller is an \eb\ model
of a heating controller that provides an interface to adjust and
display a target temperature, and to sense and display the current
temperature, among other functionality. State Machine is an \eb\ model
of state machines. The rest of the examples in Table
\ref{table:eb2java:comparison:eb:st}
are sequential program developments written by J.-R. Abrial in
\cite{bs:eb:09}. Linear and Binary Search are the \eb\ models of the
respective searching algorithms. Minimum Element is an \eb\ model for
finding the minimum element of an array of integers. Reversing and
Sorting Array are \eb\ models for reversing and sorting an array
respectively. Square Root Number is an \eb\ model for calculating the
square root of a number.

Table \ref{table:eb2java:comparison:eb:st} presents some statistics
about the \eb\ models
used in the comparison. ``LOC'' stands for Lines of Code in \eb, and
``\# Machines'' and ``\# Events'' for the number of machines and
events respectively of the \eb\ model.
\ebtojava\ successfully generated \jml-annotated Java code
for all the models in Table \ref{table:eb2java:comparison:eb:st} -- we
were able to run
the Java code as generated in each
case. All of the examples in Table
\ref{table:eb2java:comparison:eb:st} are available from
\url{http://poporo.uma.pt/EventB2Java/EventB2Java_studies.html}. The site
includes the \eb\ models and the Eclipse projects with the generated
\jml-annotated Java implementations. The Eclipse projects also include
test files that can be used to run the Java code. These test files are
generated automatically by \ebtojava, except in cases where
the \eb\ models make use of axioms. In those cases, we wrote and
added the test files manually. For example, Binary Search defines a
constant $v$ to be the searched for value, and a function $f$ to be
the array containing the values, so that $v \in ran(f)$. For the
\ebtojava\ generated code to work, one needs to manually assign a
value to $v$ that is in the array $f$. In writing a file to test the
Java code of the of Binary Search algorithm, one must consider those
conditions on $v$ and $f$.

\begin{table}
  \begin{centering}
    \begin{tabular}{|l|l|l|l|l|}
      \hline
      {\bf Tool} & {\bf Gen.} & {\bf Exec.} &
      {\bf EB} & {\bf Code}\\
      & {\bf Proc.} & {\bf Code} &
      {\bf support} & {\bf Custom.}\\
      \hline
      \ebtojava\ & Aut./ Ast. & Yes & Largely & Yes \\ \hline
      Code Gen. \cite{CodeGen10} & Ast. & Yes & Fairly & No  \\ \hline
      EB2J \cite{Mery:2011} & Ast. & Yes & Scarcely & No \\ \hline
    \end{tabular} 
    \caption{Tool Comparison.}
    \label{table:eb2java:comparison:comparison}
  \end{centering}
\end{table}

Table \ref{table:eb2java:comparison:comparison} shows how the tools
considered in our 
comparison compare on the criteria of Generation Process ({\bf Gen. Proc.}),
Executable Code ({\bf Exec. Code}), Support for \eb's
Syntax ({\bf EB Support})
and Support for Code Customisation ({\bf Code Custom.}). Regarding
``Generation Process'', EB2J and Code Generation are (always)
``Assisted'' (Ast.) since tool users (always) need to modify (extend) the
\eb\ model for the tools to be able to generate code. \ebtojava\ is
``Automatic/Assisted'' (Aut/Ast.). More precisely, it is ``Automatic'' in all
cases except when the \eb\ model makes use of axioms. 
As \ebtojava\ does not yet generate Java code for axioms (that constrain
the values of constants), the user must choose values for those constants.
\ebtojava\ does generate \jml\ specifications for axioms, so the user
can employ \jml\ machinery~\cite{Burdy-etal05} to confirm that the values
chosen are valid with respect to the original \eb\ model.

The Code Generation tool (Code Gen.) is ``Assisted'' as it always requires the
user to employ the Event Model Decomposition Rodin plug-in
\cite{Abrial:Ref:07} to decompose \eb\ models into sub-models.
For example, if
the \eb\ machine models the system and the environment components of a
reactive system, then the plug-in can generate each part separately.
In addition to decomposing the model, users of Code Generation have
to explicitly specify the execution order for events in the Java
implementation.
If the \eb\
model includes axioms and constants, tool users need to conjecture
values for the constants in \eb\ and use the Rodin platform to
discharge related proof obligations.

Regarding the comparison criterion ``Support for \eb's Syntax'' ({\bf
  EB. Support}),
\ebtojava\ largely supports \eb's syntax, in part by generating and
using libraries supporting \eb\ syntax in Java as described in
Section~\ref{eb2java:imp:prelude}. 
None of the three tools in the comparison can translate non-deterministic
assignments to Java (although \ebtojava\ does generate \jml\ specifications
for them).
EB2J and Code Generation require
the user to write a final \eb\ refinement that does not include
non-deterministic assignments. The EB2J tool ``Scarcely'' provides
support for \eb's syntax and so users are required to furnish an
additional \eb\ refinement that only uses the syntax supported by the
tool. For instance, EB2J is unable to translate the invariant
$\ebtag{inv } pages \in contents \strel persons$ that states that
$pages$ is a total surjective relation that maps $contents$ to
$persons$. For EB2J to support the syntax of that invariant, the user
has to write an \eb\ model refinement that includes the
definition of a total surjective relation, e.g. through the three
invariants shown below.

{\small
  \[
  \begin{array}{l}
    \ebtag{invA } owner \in contents \rel persons \\
    \ebtag{invB } dom(owner) = contents \\
    \ebtag{invC } ran(owner) = persons
  \end{array}
  \]
}

\begin{table}
  \begin{centering}
    \begin{tabular}{|l|c|c|c|}
      \hline
      & Code Gen. & EB2J & \ebtojava\ \\ \hline
      Social Event Planner & N/A & N/A & 1531 (+391) \\ \hline
      MIO & N/A & N/A & 825 (+272) \\ \hline
      Heating Controller & 285 & N/A & 1612 (+418) \\ \hline
      State Machine & 48 & N/A & 198 (+62) \\ \hline
      Binary Search & N/A & N/A & 71 (+33) \\ \hline
      Linear Search & N/A & N/A & 48 (+31) \\ \hline
      Minimum Element & N/A & 68 & 68 (+46) \\ \hline
      Reversing Array & N/A & 66 & 55 (+39) \\ \hline
      Sorting Array & N/A & 79 & 92 (+64) \\ \hline
      Square Root Number & 60 & 51 & 53 (+31) \\ \hline
    \end{tabular} 
    \caption{eLOC for the generated Code.}
    \label{table:eb2java:comparison:loc}
  \end{centering}
\end{table}

Table~\ref{table:eb2java:comparison:comparison} indicates whether the
code generated by each
tool is executable as generated. However, there
were cases in which EB2J was incorrect.  For
example, for the Minimum Element model, the tool was unable to infer
the type of the constant $n$, that is defined as natural number greater
than $0$ and represents the number of elements in the array to be
searched. EB2J issued the message ``/* No
translatable type found for [n] */''. EB2J was also unable to infer
the types of constants $n$, $f$, and variable $g$ in the Reversing
Array example. $f$ is the array to be reversed, defined as a 
function mapping from $1 \upto n$ to the set of integers, and $g$ is the
reversed array. Finally, EB2J did not translate parallel assignments
properly for the Reversing, Sorting Array, and the Square Root Number
models. For example, EB2J translated $g := g \ovl \{i \mapsto g(j)\}
\ovl \{j \mapsto g(i)\}$ as \javacode{g[i] = g[j]; g[j] =
  g[i]}. However, this translation is incorrect since assignments in
\eb\ are to be executed simultaneously.

\ebtojava\ is the only tool that provides support for 
``Code Customisation''. The \jml\ specifications generated by
\ebtojava\ enable users to replace (parts of) the code generated by \ebtojava\
with bespoke implementations.
Thus, the user may customise the
generated implementation and then use  \jml\ machinery~\cite{Burdy-etal05}
to verify the customised
implementation against the \jml\ specification generated by
the \ebtojava\ tool.  

Table \ref{table:eb2java:comparison:loc} shows the eLOC (Effective
Lines of Code)
generated by each tool. eLOC is a
measure of all logical lines in the Java code, and does not include
blank spaces, comments, specifications, or single curly
brackets. We used the ELocEngine software \cite{elocengine} to
calculate eLOCs. As shown in the table, the Code Generation tool
was able to generate Java code for only three of the ten \eb\ models.
We were unable to decompose the remaining seven models
(marked as ``N/A'') since they included many variables, that
made it too challenging. The EB2J tool was able to generate code for
four out of the ten \eb\ models. However, the generated code contained
minor errors in Java that we were able to fix. The errors concerned
inferring the types of some variables and translating parallel
assignments as explained above. For the remaining six models, EB2J 
issued only one error message.  The Binary Search model uses universal
quantification, that is not supported by the tool.
\ebtojava\ was able to generate \jml-annotated Java code for all
models, and this code
compiled and ran in each case. In particular, the universally quantified
assertion mentioned above appeared in an axiom, which \ebtojava\ translates
to \jml\ but not Java.  In Table~\ref{table:eb2java:comparison:loc},
the number in
parentheses for \ebtojava\ gives the number of lines of \jml\ specifications
generated for each model.

\begin{table}
  \begin{centering}
    \begin{tabular}{
        |l
        |>{\centering\arraybackslash}m{1.5cm}
        |>{\centering\arraybackslash}m{1cm}
        |>{\centering\arraybackslash}m{1.5cm}
        |>{\centering\arraybackslash}m{0.8cm}
        |>{\centering\arraybackslash}m{1.5cm}
        |>{\centering\arraybackslash}m{0.6cm}|}
      \hline
      \multirow{2}{*}{Array Size} 
      & \multicolumn{2}{|c|}{{\bf Sorting Array }}
      &  \multicolumn{2}{|c|}{{\bf Reverse Array}} 
      & \multicolumn{2}{|c|}{{\bf Minimum Array}}
      \\ \cline{2-7}

      & \ebtojava\ & EB2J & \ebtojava\ & EB2J & \ebtojava\ & EB2J \\ \hline
      100,000 & 23 & 13093 &  264 & 1 & 29 & 0 \\ \hline
      200,000 & 28 & 51910 & 258 & 55 & 28 & 1\\ \hline
      300,000 & 37 & 182311 & 198 & 305 & 30 & 1\\ \hline
      400,000 & 152 & 329614 & 416 & 406  & 32 & 1\\ \hline
      500,000 & 172 & 497133 & 457 & 548 & 28 & 1\\ \hline
    \end{tabular}
    \caption{Execution times in milliseconds for the Java code generated by 
      \ebtojava\ and
      EB2J for the Sorting, Reverse and Minimum Array \eb\ models.}
    \label{table:eb2java:comparison:sort:min:rev}
  \end{centering}
\end{table} 

Finally, the \eb\ models for Binary and Linear Search, Minimum
Element, and Reversing and Sorting Arrays include events whose guards are
mutually exclusive. Hence, we used \ebtojava\ and EB2J to
generate (sequential) Java implementations for each of these models, and
because the generated implementations always complete execution,
compared the times the generated implementations took to
complete for various inputs.   In each case, we ran the implementations
10 times and
took the average time. Table \ref{table:eb2java:comparison:sort:min:rev}
shows how the
times compare for the Sorting, Reverse Array and Minimum Array models. For the
Sorting Array model, the code generated by \ebtojava\ outperformed that
generated by EB2J. For
the Minimum Array model, EB2J outperformed \ebtojava, though times
are close. For the Reverse Array model, EB2J outperformed \ebtojava\
as well, although \ebtojava\ approaches EB2J as the input size gets
larger. The experiment shows that both tools
generate runnable implementations for the considered \eb\ models. For
\ebtojava, the Java classes that implement the \eb\ mathematical
constructs exhibit good performance, especially when dealing with large
inputs. This is due to the implementation using the \javacode{TreeSet}
Java class. EB2J did outperform \ebtojava\ in some cases. We believe
that this is largely due to the implementation of method \javacode{apply}
(applying a relation to a set of elements) of class
\javacode{BRelation}. In \ebtojava, the method \javacode{apply}
iterates over each element of the relation, so searching for an
element is $O(n)$ and searching for $k$ elements is
$O(k*n)$. EB2J uses arrays to store relations, so applying a
relation to a set is linear in $k$.

All times reported in
Table~\ref{table:eb2java:comparison:sort:min:rev} were collected by
running the Java code generated by \ebtojava\ and EB2J on a Mac OS X 
laptop with an Intel Core i5 2.3 GHz processor.  The \eb\ models, 
generated code and timing harness used
are available at \url{http://poporo.uma.pt/EventB2Java/tests.zip}.

\section{Conclusion}
In this chapter we presented two case studies on software development using
\ebtojava, demonstrating the effectiveness of using the tool. The
first case study was the implementation of a Social-Event
Planner Android application developed using a Model-View-Controller
(MVC) design pattern. The second case study was the testing of
Tokeneer, a security-critical access control system. We also presented
a benchmark comparing \ebtojava\ against two existing tools for
generating Java code from \eb\ models. The benchmark was composed of
9 \eb\ models and 6 comparison criteria. 

Our experience on developing the first case study suggests that
software developers can benefit from \ebtojava\ in several
ways: modelling in \eb\ enables users to define properties that the
software needs to preserve. For instance, regarding permissions over
social-events, an interesting property is that invited people to a
social-event have permissions over that event to view and edit
its contents. We
formalised this property (and many others) as \eb\ \ebkeyw{invariants}. We
proved that the model was consistent by discharging all proof obligations. We were
sure that the Java code generated by \ebtojava\ respects those
properties; software developers can also benefit from \ebtojava\ since
they do not have to refine
the \eb\ model until it is closed to a machine implementation, whereas the
software developers
decide that the model has enough details to be translated to
Java using \ebtojava; finally, software developers can benefit from \ebtojava\ since having
a Java implementation of an \eb\ model allows the model to interact with other
implementations. For instance, the Java code generated by the tool
represents the Model in a MVC development that interacts with the
implementation of the Controller in Java, and the implementation of
the View in Android.

Our experience on developing the second case study shows us that
software developers can benefit from \ebtojava, since the tool can be used
in testing the correct behaviour of an \eb\ model by translating it to
Java and performing JUnit tests in Java. The process
allows system developers to be sure that the behaviour of the model
is indeed the
behaviour that they intended from the beginning. The process of
developing the second case study, initially
shows us that, even though the \eb\ model was correct (all proof
obligations were discharged), the model was not behaving according to
our intentions. The JUnit tests uncovered an issue in the Java code
generated by the tool for one of the \eb\ events. We inspected that
event in \eb, found and corrected the error. We discharged
the proof obligations again to be sure the model was still consistent,
and used \ebtojava\ to generate Java code of the model again. We
repeated this process until the generated code passed all 32 JUnit
test cases. The final generated code is an actual implementation of a
model in \eb\ that was proven correct, and the code is behaving
according to what we expect.

The benchmark showed us that \ebtojava\
outperforms other Java code generators for \eb\ in several ways:
\ebtojava\ generates (and embeds) \jml\
specifications in the Java code. That enables users to customise the
Java code to further check if the customised code does not invalidate
the initial model. We found out that the EB2J and Code Generation tools do
not support code customisation; the generation of Java code process in
\ebtojava\ is automatically/assisted whereas for the other two tools
it is
always assisted. This makes our tool more useful since it is easy to
use.

We are planning on undertaking a more complex case study where experts
in modelling in \eb\ and experts in developing in Java (and \jml) can
work together.
\chapter{Future Work}
\label{chapter:future-work}
The work presented in this thesis can be extended in different
ways. Figure \ref{fig:future-work} depicts my future work (in red),
that is explained below. The figure shows (in black) the work done
during this thesis.

\begin{figure}
  \centering
  \includegraphics[scale=0.55]{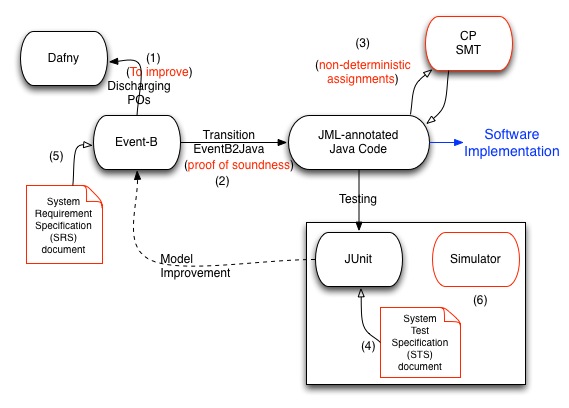}
  \caption{Future Work.}
  \label{fig:future-work}
\end{figure}

\begin{itemize}
\item We plan to investigate and to characterise
  the type of POs for which \dafny\ outperforms existing Rodin
  proof-engines. (this is depicted in Figure \ref{fig:future-work}
  as (1))

\item Currently, the \ebtodafny\ tool takes as an input an \eb\ PO and
  translates it to \dafny. It also translates the whole \eb\ machine
  necessary for discharging the PO. We plan on translating a given \eb\
  PO without translating the whole \eb\ machine (this is depicted in
  Figure \ref{fig:future-work} as (1)).

\item Currently, proof obligations generated by \ebtodafny\ are
  manually fed into \dafny. We are planning on integrating
  \ebtodafny\ to \dafny\ and Microsoft Visual Studio, so the process
  of discharging POs is automatic without leaving Rodin proof manager
  (this is depicted in Figure \ref{fig:future-work}  as (1)). 


\item In \cite{journ:proof:13}, Cata\~no et. al. proposed a proof of
  soundness of the translation from \eb\ models to \jml\
  specifications. The proof takes into account any \eb\ substitutions,
  invariants, and the standard \eb\ $initialising$ event. We are
  planning on extending the proof to fully prove the translation of
  \eb\ machines and contexts to \jml, and proving the soundness of the
  translation from \eb\ to Java code (this is depicted in Figure \ref{fig:future-work}
  as (2)). Providing a proof of soundness
  of our \ebtojava\ tool gives the
  user confidence about the generated \jml-annotated Java code.

\item One major frustration in our work is the inadequate tool support
  for verifying Java programs with respect to \jml\
  specifications. Existing verification tools such as KeY
  \cite{KeYURL} and Krakatoa \cite{Marche03thekrakatoa} can not handle
  the full syntax of Java and \jml, particularly with regard to
  generics. We would like to undertake a case study on replacing parts
  of the code generated by \ebtojava\ with bespoke implementations
  and then verifying those implementations against the generated \jml\
  specifications. However, performing such verification without
  adequate tool support is time consuming and prone to error.

\item There are several \eb\ constructs that \ebtojava\
  translates to \jml\ but not to Java. Those constructs
  are universal and existential quantifiers, set comprehension
  (quantified union and intersection that are generalisations of set
  comprehension), non-deterministic assignment (denoted in this
  document as the function \javacode{PredicateValue}), and assignments
  to \eb\ constants (denoted in this document as the function
  \javacode{AxiomTheoremValue}). In general, these constructs
  are not possible to be translated into code, particularly when they
  are referring to infinite sets. We plan to translate these \eb\
  constructs (when bound variables are restricted to a finite set) to
  the input language of a solver such as the Z3 SMT solver
  \cite{Z3Url}, and use it to find values that satisfy these \eb\
  constructs. Although, this would require writing another tool that
  translates \eb\ predicates and expressions to the solver syntax.



\item In Section \ref{case-studies:tok} we showed a case study where
  \ebtojava\ was applied for
  testing the behaviour of an \eb\ model. This was achieved by translating it to Java
  using \ebtojava\ and perform manually written JUnit test. An issue
  with this strategy is users can introduce errors while 
  writing the JUnit tests manually. We are planning on automating this process
  so to avoid human errors. In \cite{Catano:jfly:10}, a strategy called
  JFly is proposed to evolve informal (written in natural language)
  software requirements into formal requirements written in \jml. This
  work can be reused to structure the writing of JUnit tests from a
  System Test Specification document (this is depicted in Figure \ref{fig:future-work}
  as (4)). Likewise, we are planning on
  investigating a way to map \eb\ requirements with a System
  Requirements Specification document. In
  \cite{Traichaiyaporn:reqEB}, an approach is proposed to obtain a
  correct representation of requirement specifications in \eb\ from a
  specific semantics also proposed in that approach (this is depicted in Figure \ref{fig:future-work}
  as (5)).

\item As a validation step of an \eb\ model behaviour, we proposed to
  translate the model to Java and perform JUnit tests (as stated in
  the previous bullet). Another validation step is to simulate the
  \eb\ model so users can see if its behaviour matches the user's
  expectations. In \cite{Yang:simEB},
  a JavaScript simulation framework for \eb\ (JeB) is proposed. It
  translates \eb\ models to JavaScript in order to animate the \eb\
  model. The idea goes in the same direction as ProB \cite{prob}. We
  are planning on extending the \ebtojava\ tool so it can generate this
  kind of simulation. Thus users can check the behaviour of the \eb\
  model (this is depicted in Figure \ref{fig:future-work}
  as (6)). Our proposal will outperform JeB by generating
  Java code that serves as a final implementation. The idea is also to
  have a unique framework that comprises everything: users can obtain
  \jml-annotated Java code from \eb\ models, they can check if the
  behaviour of the \eb\ model is the expected by performing JUnit tests
  or simulating the model in Java. 

\item Although the modelling of timing properties is not directly
  supported by \eb, a discrete clock can certainly be designed and
  implemented in \eb. In
  \cite{butler:time:eb:11,butler:disctime:eb:12}, M. Sarshogh and
  M. Butler introduce three \eb\ trigger-response patterns, namely,
  deadlines, delays and expires, to encode discrete timing properties
  in \eb. A ``deadline'' means that a set of events must respond to a
  particular event within a bounded time. For a ``delay'', the set of
  response events must wait for a specified period after the
  triggering of an event. An ``expiry'' pattern prevents response
  events from triggering after the occurrence of an event. The authors
  translate timing properties as invariants, guards and \eb\
  actions. We are interested in investigating on how our code
  generation framework can be extended to support timing properties in
  \eb, and in encoding this extension in \ebtojava\ once the Rodin
  platform fully supports the use of discrete timing events.

\end{itemize}
\chapter{Conclusion}
\label{chapter:conclusion}
This thesis investigated the answer to the question: is it possible to
combine Stepwise refinement and Design-by-Contract formal approaches
in the development of systems, providing the user with the benefits of
both? For
this purpose, this thesis presented the 
translation from \eb\ models to
\jml-annotated Java code. The translation is defined by means of
syntactic rules. These rules were implemented as the \ebtojava\ tool
that is a Rodin plug-in (described in Chapter
\ref{chapter:eb2java}). \ebtojava\ bridges stepwise refinement method with
\eb\ to Design-by-Contract with \jml\ and Java, answering the research
question we proposed from the beginning: \ebtojava\ does allow users
to combine modelling systems by stepwise refinement and Design-by-Contract together in the
development of systems. This enables users to take advantage of the strengths
and avoiding the weaknesses of each approach. \ebtojava\ 
generates Java executable code directly from abstract (or more refined)
\eb\ models, providing the option to verify the code against the
generated \jml\ specifications whenever users decide to customise the
code. When modelling in \eb, users need to prove the system to be
consistent by discharging proof obligations. Rodin generates these
proof obligations, some of
them automatically being discharged by Rodin's provers and some others
needing the intervention of the user. The manually proving of proof
obligations can be a difficult task, so we proposed a translation from
\eb\ proof obligations to the input language of \dafny, thus users can
use Z3 (the automatic prover associated to \dafny) as a prover
(described in Chapter \ref{chapter:eb2dafny}). Our
intention is to provide tools that help users in the process of
proving an \eb\ model correct.

The findings of this investigation could be of interest to both
researchers working with \eb\ and software developers working with
\jml\ and Java. Using \eb\ in the early
stages of the software development gives developers excellent support
for modelling software systems in an abstract manner, and particularly for verifying
safety, security and correctness properties of those
models. Transitioning to \jml\ at an appropriate point (as determined by
the developers themselves, rather than being dictated by the tools
being used) allows developers to take full advantage of data
structures and APIs in the implementation language, and permits
software developers with less mathematical expertise to contribute
earlier in the development process. Furthermore, having the \jml\
specifications embedded into the Java code also gives an insight of
documentation of the code that  can be read easily.

Researchers investigating about the translation from a language A to
a language B might also find this thesis of interest. They can
see the development of this thesis as a guide for defining translation
rules and its implementation from one
language to another. 

Code generation for \eb\ is not a new concept. For instance, modelling
in \eb\ one can refine an \eb\ model until accomplishing an actual
implementation. However, this is difficult, since users
have to discharge proof obligations that become harder through
refinements. This particular issue can make the use of formal methods
less popular. Another way of code generation for \eb\ is the use of
existing tools. There are 
tools that allow users to translate \eb\ models to different
programming languages. Our tool outperforms these tools since: \ebtojava\ 
fully supports the \eb\ syntax, except for universal and
  existential quantifiers, set comprehension (quantified union and
  intersection that are generalisations of set comprehension),
  non-deterministic assignment, and assignments to \eb\
  constants for which the tool generates \jml\ specifications. Hence,
\ebtojava\ does not impose restrictions on the syntax of the model to
be translated. Users can decide the
level of abstraction in the model and then transition to Java
code. Current tools for generating Java code from
\eb\ models impose restrictions to users since the tools do not fully
support the \eb\ notation. Therefore, users need to evolve the \eb\ model
to use the syntax supported by those tools; on the other hand, as far
as I can tell, our tool is the only
one that generates \jml\ specifications embedded in the Java code. This
gives the user another layer of confidence when the user decides to
customise the code since the generated code can be verified against
the \jml\ specifications. 

In my experience on developing software without formal methods
I have seen that one can end up with a final implementation of the
software in a relatively short time. However, in many cases, the non
use of formal methods makes the implementation misbehave, so one needs
to correct it, making the maintenance of software a
difficult task and a waste of time. For instance, since we are not using formal methods,
reasoning about the model is not possible, thus finding an error is
difficult. One could use testing to uncover misbehaviours/errors, but
testing can tell about problems among the scope of the testing process but
nothing beyond. I have found the development
of software using \ebtojava\ very interesting, since the development starts in \eb,
where one can propose properties that the system needs to preserve and
one must prove that the model indeed preserves those
properties. \ebtodafny\ can help in this process. Then, using
the \ebtojava\ tool to generate Java code where the development of the
system 
continues. The generated Java code is an actual implementation of the
\eb\ model and contains two main advantages: 
\begin{inparaenum}[\itshape i\upshape)]
\item the code preserves the initial properties that the user
  defined, and
\item the code contains \jml\ specifications so users can customise
  the code being able to check if the customised code meets the \jml\
  specifications. 
\end{inparaenum}
Users can argue that the use of \eb\ can be difficult since one
needs to discharge proof obligations delaying the
development of software at early stages. However, my experience on
using \eb\ at early stages of development suggests that the
final implementation of the development needs to not be maintained,
since the code does not contain errors. So, regarding times of
development of software, the use of \ebtojava\ is more constructive
than using
just Java or any other language.

I believe that having a tool that translates stepwise refinement
models into Design-by-Contract makes formal techniques and tools more
usable. This is of paramount importance, since the popularity of formal
methods has not increased as much as researchers might want due to the
level of expertise needed to work with these methods. Having tools
that automise the process of working with these methods supply
the popularity that formal method should have. I have seen the popularity
of different methods being increased by the development of tools. For
instance, theorem proving was introduced by
Begriffsschrift in 1884, but just until a couple of years ago theorem
proving has become more popular thanks to the implementation of tools
that assist users in the process. I see the use of \eb\ and \jml\
increasing thanks to \ebtojava\ since the tool enables people to automatise the process
of combining \eb\ and Java+\jml\ in the development of
software. \ebtojava\ still
needs to be more developed,
especially in proving the soundness of the entire translation rules,
but I believe the investigation of this thesis is one step forward on
making formal methods more popular.

As a result of my study, further research might be to undertake a more
complex software development using \ebtojava, involving different
expertise researchers, experts in the notation underlying \eb\ and
Java-\jml\ developers. An interesting case
study is to implement the same software development using three
approaches: using our tool, using just Java, and using just \eb. So we
could take metrics to compare the three developments in order to have an
insight on the time to develop the software, on the effort put by
modellers and implementers, and on the time used to maintain the
software. 

Another interest further research might be to use \ebtojava\ in
Academia to help students to relate formal developments in both \eb, and
Java and \jml. Nowadays, the use of formal methods is mostly done by
theoretic researchers. Software developers are not too familiar
with mathematics and logics, so they do not use formal methods. Hence,
they are skeptical about their use. On the other hand, theoretic researchers
do not often use programming languages. Both sides can argue about
(dis-)advantages of each approach. It is quite difficult to convince
any end to use another approach. I truly believe that both ends can
work together if the teaching of these kind of approaches is part of
the future researchers education. I see \ebtojava\ can fullfil this
purpose since the tool enables students to relate formal methods and
code in software development.

\bibliographystyle{plain}
\bibliography{eventBCodeGen}
\appendix
\chapter{\eb\ syntax supported by \ebtojava}
\label{appendix}
This appendix shows the full syntax of \eb\ and the translation to
Java and \jml. The \eb\ modelling language is composed of five
mathematical languages (see Chapter 9 of \cite{EB:Book}), namely, 
\begin{inparaenum}[\itshape a\upshape)]
\item a Propositional Language,
\item a Predicate Language,
\item an Equality Language,
\item a Set-Theoretic Language, and
\item  Boolean and Arithmetic Languages.
\end{inparaenum}
The following shows the translation of each construct in the \eb\
languages to Java and \jml.

$P$ and $Q$ are predicates, $E$ and $F$ are expressions, $x$ is
a variable, $S$ and $T$ are sets, $f$ and $g$ are relations, $Pr$ is a
Pair, and $a$ and $b$ are Integers. The construct \javacode{Type(tt)}
translates the type of the \eb\ variable $tt$ to the corresponding
Java type.

\section{The Propositional Language}

\begin{tabular}{ c|c|c}
{\bf Event-B Op.} & {\bf \jml} & {\bf Java} \\ \hline
$\neg P$     &    \javacode{!P}   &     \same \\ \hline
$P \wedge Q$     &    \javacode{P \&\&  Q}   &     \same \\ \hline
$P \vee Q$     &    \javacode{P ||  Q}   &     \same \\ \hline
$P \limp Q$     &    \javacode{BOOL.implication(P,Q)}   &     \same \\ \hline
$P \leqv Q$     &    \javacode{BOOL.bi\_implication(P,Q)}   &     \same \\ \hline
\end{tabular}

\section{The Predicate Language}

\begin{tabular}{ c|c|c}
{\bf Event-B Op.} & {\bf \jml} & {\bf Java} \\ \hline
$\forall x \qdot P$     &  \jmlcode{(\bsl forall Type(x) x ; P)} &  \noS  \\ \hline
$\exists x \qdot P$     &  \jmlcode{(\bsl exists Type(x) x ; P)} & \noS \\ \hline
$E \mapsto F$    &      \javacode{new Pair(E,F)}      &      \same \\ \hline
$E = F$    &      \javacode{E.equals(F)}      &      \same \\ \hline
\end{tabular}

\section{The Set-Theoretic Language}

\begin{tabular}{ c|c|c}
{\bf Event-B Op.} & {\bf \jml} & {\bf Java} \\ \hline
$E \in F$     &    \javacode{F.has(E)}   &  \same \\ \hline
\end{tabular}

\subsection{Axioms of set theory}
\begin{tabular}{ c|c|c}
{\bf Event-B Op.} & {\bf \jml} & {\bf Java} \\ \hline
$E \cprod F$     &    \javacode{BRelation.cross(E,F)}   &  \same \\ \hline
$\pow(E)$     &    \javacode{E.pow()}   &  \same \\ \hline
$\{x \qdot P|F\}$     &   
\begin{tabular}{p{5cm}}
\jmlcode{new BSet<Type(x)>(new JMLObjectSet Type(e) e $\mid$ (\bsl
  exists Type(x) x; P; e.equals(E)))  }\\
\end{tabular}
 &\noS \\\hline
$E = F$    &      \javacode{E.equals(F)}      &      \same \\ \hline
\end{tabular}

\subsection{Elementary set operators}
\begin{tabular}{ c|c|c}
{\bf Event-B Op.} & {\bf \jml} & {\bf Java} \\ \hline
$S \subseteq T$     &    \javacode{S.isSubset(T)}   &  \same \\ \hline
$S \subset T$     &    \javacode{S.isProperSubset(T)}   &  \same \\ \hline
$S \bunion T$     &    \javacode{S.union(T)}   &  \same \\ \hline
$S \binter T$     &    \javacode{S.intersection(T)}   &  \same \\ \hline
$E \bsl T$     &    \javacode{S.difference(T)}   &  \same \\ \hline
$\emptyset$     &    \javacode{BSet.EMPTY}   &  \same \\ \hline
\end{tabular}

\subsection{Generalisation of elementary set operators}
\begin{tabular}{ c|c|c}
{\bf Event-B Op.} & {\bf \jml} & {\bf Java} \\ \hline
$union(S)$     &    \javacode{BSet.union(S)}   &  \same \\ \hline
$\Union x \qdot P \mid E$     &  
\begin{tabular}{p{5cm}}
\jmlcode{BSet.union(new BSet<Type(x)>(new JMLObjectSet Type(e) e
  $\mid$ (\bsl exists Type(x) x; P; e.equals(F))))}
\end{tabular} & \noS \\\hline
$inter(S)$     &    \javacode{BSet.intersection(S)}   &  \same \\
\hline
$\Inter x \qdot P \mid E$    &
\begin{tabular}{p{5cm}}
\jmlcode{BSet.intersection(new BSet<Type(x)>(new JMLObjectSet Type(e)
  e $\mid$ (\bsl exists Type(x) x; P; e.equals(F))))}
\end{tabular}
& \noS \\ \hline
\end{tabular}

\subsection{Binary relation operators}
\begin{tabular}{ c|c|c}
{\bf Event-B Op.} & {\bf Java} & {\bf \jml} \\ \hline
$S \rel T$     &   \javacode{BRelation<Type(S),Type(T)>}   &  * \\ \hline
$S \trel T$     &   \javacode{BRelation<Type(S),Type(T)>}   &  * \\ \hline
$S \srel T$     &   \javacode{BRelation<Type(S),Type(T)>}   &  * \\ \hline
$S \strel T$     &   \javacode{BRelation<Type(S),Type(T)>}   &  * \\ \hline
$dom(f)$     &   \javacode{f.domain()}   &  \sameJ \\ \hline
$range(f)$     &   \javacode{f.range()}   & \sameJ  \\ \hline
$f^{-1}$     &   \javacode{f.inverse()}   &  \sameJ \\ \hline
$S \domres f$     &   \javacode{f.restrictDomainTo(S)}   &  \sameJ \\ \hline
$S \ranres f$     &   \javacode{f.restrictRangeTo(S)}   &  \sameJ \\ \hline
$S \domsub f$     &   \javacode{f.domainSubtraction(S)}   &  \sameJ \\ \hline
$S \ransub f$     &   \javacode{f.rangeSubtraction(S)}   &  \sameJ \\ \hline
$f[x]$     &   \javacode{f.image(x)}   &  \sameJ \\ \hline
$f \fcomp g$     &   \javacode{f.compose(g)}   &  \sameJ \\ \hline
$f \bcomp g$     &   \javacode{f.backwardCompose(g)}   &  \sameJ \\ \hline
$f \ovl g$     &   \javacode{f.override(g)}   &  \sameJ \\ \hline
$f \dprod g$     &   \javacode{f.directProd(g)}   &  \sameJ \\ \hline
$f \pprod g$     &   \javacode{f.parallel(g)}   &  \sameJ \\ \hline
\end{tabular}

$*$ JML specifications associate to this operator is explained in \ref{sec:jml:spec}

\subsection{Functions operators}
\begin{tabular}{ c|c|c}
{\bf Event-B Op.} & {\bf Java} & {\bf \jml} \\ \hline
$id$     &   \javacode{new ID()}   &  \same \\ \hline
$S \pfun T$     &   \javacode{BRelation<Type(S),Type(T)>}   &  * \\ \hline
$S \tfun T$     &   \javacode{BRelation<Type(S),Type(T)>}   &  * \\ \hline
$S \pinj T$     &   \javacode{BRelation<Type(S),Type(T)>}   &  * \\ \hline
$S \tinj T$     &   \javacode{BRelation<Type(S),Type(T)>}   &  * \\ \hline
$S \psur T$     &   \javacode{BRelation<Type(S),Type(T)>}   &  * \\ \hline
$S \tsur T$     &   \javacode{BRelation<Type(S),Type(T)>}   &  * \\ \hline
$S \tbij T$     &   \javacode{BRelation<Type(S),Type(T)>}   &  * \\ \hline
$prj_1(Pr)$     &   \javacode{Pr.fst()}   &  \sameJ \\ \hline
$prj_2(Pr)$     &   \javacode{Pr.snd()}   &  \sameJ \\ \hline
\end{tabular}

$*$ JML specifications associate to this operator is explained in \ref{sec:jml:spec}

\section{Boolean and Arithmetic Language}

\begin{tabular}{ c|c|c}
{\bf Event-B Op.} & {\bf \jml} & {\bf Java} \\ \hline
BOOL &  \javacode{BOOL} & \same \\ \hline
TRUE &  \javacode{true} & \same \\ \hline
FALSE &  \javacode{false} & \same \\ \hline
$\intg$ &  \javacode{INT} & \same \\ \hline
$\nat$ &  \javacode{NAT} & \same \\ \hline
$\nat1$ &  \javacode{NAT1} & \same \\ \hline
$succ(a)$ &  \javacode{a+1} & \same \\ \hline
$pred(a)$ &  \javacode{a-1} & \same \\ \hline
$0$ &  \javacode{0} & \same \\ \hline
$1$ &  \javacode{1} & \same \\ \hline
\ldots & &  \\ \hline
$a + b$ &  \javacode{a + b} & \same \\ \hline
$a * b$ &  \javacode{a + b} & \same \\ \hline
$a \expn b$ &  \javacode{Math.pow(a,b)} & \same \\ \hline
\end{tabular}

\subsection{Extension of the arithmetic language}
\begin{tabular}{ c|c|c}
{\bf Event-B Op.} & {\bf \jml} & {\bf Java} \\ \hline
$a \leq b$ &  \javacode{a.compareTo(b) <= 0} & \same \\ \hline
$a < b$ &  \javacode{a.compareTo(b) < 0} & \same \\ \hline
$a \geq b$ &  \javacode{a.compareTo(b) >= 0} & \same \\ \hline
$a > b$ &  \javacode{a.compareTo(b) > 0} & \same \\ \hline
$finite(S)$ &  \javacode{S.finite()} & \same \\ \hline
$a \upto b$ &  \javacode{new Enumerated(a,b)} & \same \\ \hline
$a - b$ &  \javacode{a - b} & \same \\ \hline
$a / b$ &  \javacode{a / b} & \same \\ \hline
$a$ $mod$ $b$ &  \javacode{a \% b} & \same \\ \hline
$card(S)$ &  \javacode{S.size()} & \same \\ \hline
$max(S)$ &  \javacode{S.max()} & \same \\ \hline
$min(S)$ &  \javacode{S.min()} & \same \\ \hline

\end{tabular}

\section{Some other JML specs}
\label{sec:jml:spec}
All Event-B functions and relations are translated as
instances of \javacode{BRelation}. Restrictions made by each kind of
function/relation are translated to the JML \jmlcode{invariant}:

\begin{tabular}{ c|p{8cm}}
{\bf Event-B Op.} & {\bf JML spec} \\ \hline
$f \in S \rel T$     &  \jmlcode{f.domain().isSubset(S) \&\& f.range().isSubset(T)} \\ \hline
$f \in S \trel T$     & \jmlcode{f.domain().equals(S) \&\& f.range().isSubset(T)} \\ \hline
$f \in S \srel T$     &  \jmlcode{f.domain().isSubset(S) \&\& f.range().equals(T)} \\ \hline
$f \in S \strel T$     &  \jmlcode{f.domain().equals(S) \&\& f.range().equals(T)} \\ \hline
$f \in S \pfun T$     &  \jmlcode{f.isaFunction() \&\& f.domain().isSubset(S) \&\& f.range().isSubset(T)} \\ \hline
$f \in S \tfun T$     &  \jmlcode{f.isaFunction() \&\& f.domain().equals(S) \&\& f.range().isSubset(T)} \\ \hline
$f \in S \pinj T$     &  \jmlcode{f.isaFunction() \&\&
  f.inverse().isaFunction() \&\& f.domain().isSubset(S) \&\& f.range().isSubset(T)} \\ \hline
$f \in S \tinj T$     &  \jmlcode{f.isaFunction() \&\&
  f.inverse().isaFunction() \&\& f.domain().equals(S) \&\& f.range().isSubset(T)} \\ \hline
$f \in S \psur T$     &  \jmlcode{f.isaFunction() \&\& f.domain().isSubset(S) \&\& f.range().equals(T)} \\ \hline
$f \in S \tsur T$     &  \jmlcode{f.isaFunction() \&\& f.domain().equals(S) \&\& f.range().equals(T)} \\ \hline
$f \in S \tbij T$   &   \jmlcode{f.isaFunction() \&\&
  f.inverse().isaFunction() \&\& f.domain().equals(S) \&\& f.range().equals(T)} \\ \hline
\end{tabular}

\end{document}